\documentclass[pre,aps,twocolumn,showpacs,floatfix]{revtex4-1}

\usepackage{dcolumn}
\usepackage{amsmath}
\usepackage{amssymb}
\usepackage{bm}
\usepackage{graphicx,subfigure}
\usepackage{epsf,epsfig}
\usepackage{hyperref}
\usepackage{color}
\usepackage{enumerate}

\begin{document}

\title{Dynamic nuclear structure emerges from chromatin crosslinks and motors}
\author{Kuang Liu$^1$, Alison E. Patteson$^1$, Edward J. Banigan$^{2}$, J. M. Schwarz$^{1,3}$}
\affiliation{$^1$ Department of Physics and BioInspired Syracuse, Syracuse University, Syracuse, NY USA,
$^2$ Institute for Medical Engineering and Science and Department of Physics, MIT, Cambridge, MA $^3$ Indian
Creek Farm, Ithaca, NY, USA}
\date{\today}
\begin{abstract}

The cell nucleus houses the chromosomes, which are linked
  to a soft shell of lamin protein filaments. Experiments indicate that correlated chromosome dynamics and nuclear shape fluctuations arise from motor activity. 
  To identify the physical mechanisms, we develop a 
  model of an active,
  crosslinked Rouse chain bound to a polymeric shell. System-sized
  correlated motions occur but require both motor activity {\it and}
  crosslinks. Contractile motors, in particular, enhance chromosome
  dynamics by driving anomalous density fluctuations. Nuclear shape
  fluctuations depend on motor strength, crosslinking, and
  chromosome-lamina binding. Therefore, complex chromatin dynamics
  and nuclear shape emerge from a minimal, active
  chromosome-lamina system. 
\end{abstract}                                                                                   
\maketitle

The cell nucleus houses the genome, or the material containing instructions for building the proteins that a cell needs to function. This material is $\sim 1$ meter of DNA with proteins, forming chromatin, and it
is packaged across multiple spatial scales to fit inside a $\sim 10\ \mu\mathrm{m}$ nucleus~\cite{gibcus13}.
Chromatin is highly dynamic; for instance, correlated motion of micron-scale genomic regions over
timescales of tens of seconds has been observed in mammalian cell nuclei~\cite{zidovska13, shaban18, saintillan18, barth20, shaban20}. This correlated motion diminishes both in the absence
of ATP and by inhibition of the transcription
motor RNA polymerase II, suggesting that motor 
activity plays a key role~\cite{zidovska13,shaban18}. These dynamics
occur within the confinement of the cell nucleus, which is enclosed by
a double membrane and 10-30-nm thick filamentous layer of lamin
intermediate filaments, the lamina ~\cite{shimi15, mahamid16,
  turgay17}. Chromatin and the 
lamina interact through various proteins ~\cite{dechat08, solovei13,
  leeuw18} 
  and form structures such as lamina-associated domains
(LADs)~\cite{guelen08, vansteensel17}. 
Given the complex spatiotemporal properties of a cell nucleus, 
how do correlated chromatin dynamics emerge and what is their interplay with nuclear shape?

Numerical studies suggest several explanations for correlated chromatin
motions. 
Individual unconfined active semiflexible polymer chains with exponentially correlated noise exhibit 
enhanced displacement correlations \cite{ghosh14}. 
With confinement, a Rouse chain with long-range hydrodynamic interactions that
is driven by extensile dipolar motors can exhibit correlated motion
over long length and timescales~\cite{saintillan18}.  Correlations
arise due to the emergence of local nematic ordering within the
confined globule. However, such local nematic ordering has yet to be
observed. 
In the absence of activity, a confined heteropolymer may exhibit correlated motion, with anomalous diffusion of small
loci~\cite{liu18,dipierro18}. However, in marked contrast with
experimental results~\cite{zidovska13, shaban18}, introducing activity
in such a model does not alter the correlation length at short timescales and decreases it at longer timescales.  

Through interactions or linkages with the lamina, chromatin dynamics may influence the shape of the nuclear lamina. Experiments
have begun to investigate this notion by measuring nuclear shape fluctuations~\cite{talwar13, makhija16, chu17}.  
Depletion of ATP, the fuel for many molecular motors, diminishes the magnitude of the shape
fluctuations, as does the inhibition of RNA polymerase II transcription activity by $\alpha$-amanitin~\cite{chu17}.
Other studies have found that depleting linkages between chromatin and the nuclear lamina 
results in more
deformable nuclei~\cite{guilluy14, schreiner15}, enhanced curvature fluctuations~\cite{lionetti20}, and/or abnormal nuclear
shapes~\cite{stephens19}. 
Interestingly, depletion of lamin A in several human cell lines leads to increased diffusion of chromatin, suggesting
that chromatin dynamics is also affected by linkages to the lamina~\cite{bronshtein15}. Together, these experiments demonstrate the
critical role of chromatin and its interplay with the nuclear lamina
in determining nuclear structure. 

To understand these results mechanistically, we construct a chromatin-lamina
system with the chromatin modeled as an {\it active} Rouse chain and the lamina as an elastic, polymeric shell with
linkages between the chain and the shell.  
Unlike previous chain and shell models~\cite{banigan17, stephens17,
  lionetti20}, our model has motor activity. 
We implement a generic and simple type of motor, namely extensile and contractile monopoles, representative
of the scalar events considered in a two-fluid model of
chromatin~\cite{bruinsma14}. We also include chromatin crosslinks,
which may be a 
consequence of motors forming droplets~\cite{cisse13} and/or complexes~\cite{nagashima19}, as well as
chromatin binding by proteins, such as heterochromatin protein I (HP1)~\cite{erdel20,strom20}. Recent 
rheological measurements of the nucleus support the notion of chromatin
crosslinks~\cite{banigan17,stephens17,strom20}, as does indirect evidence from
chromosome conformation capture (Hi-C)~\cite{belaghzal19}.  In
addition, we explore how the nuclear shape and chromatin dynamics
mutually affect each other by comparing results for an elastic,
polymeric shell with those of a stiff, undeformable one.

\begin{figure}[t]
	\centering
	\includegraphics[width=0.5\textwidth]{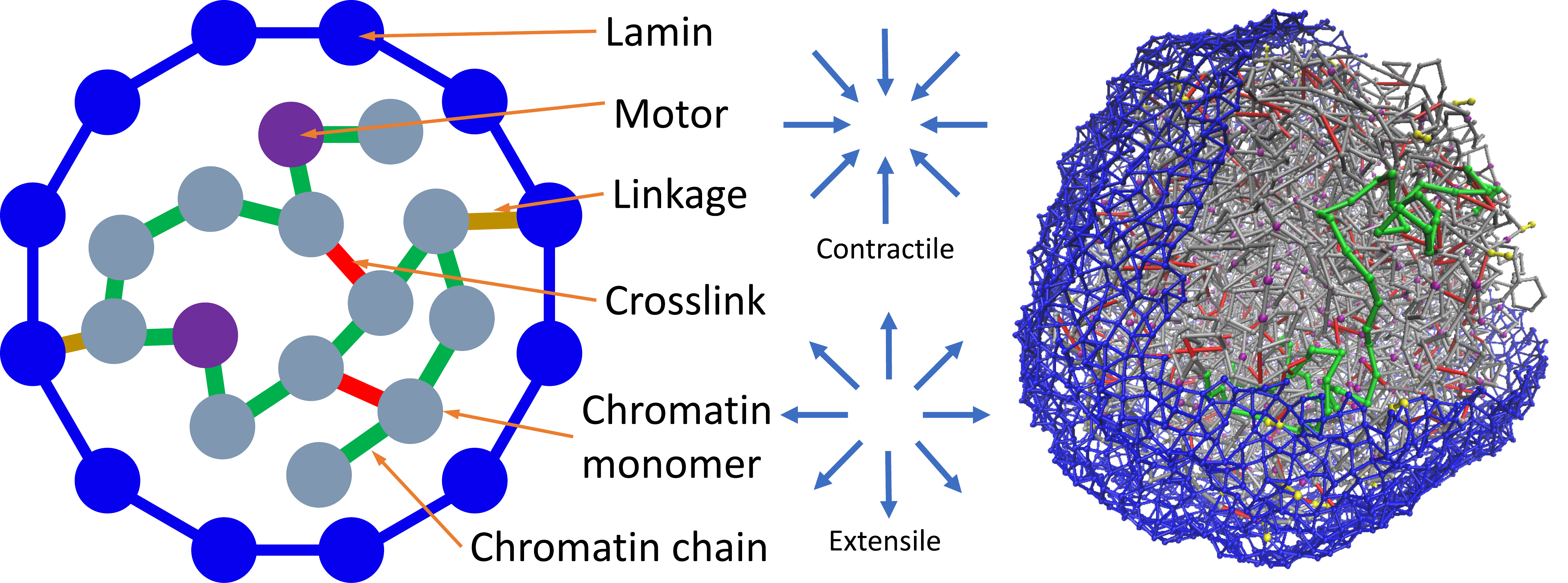}
	\caption{Left: Two-dimensional schematic of the model. Center:
          Schematic of the two types of motors. Right: Simulation snapshot. 
          The chromatin polymer is composed of linearly connected monomers, shown in gray. 
          Active chromatin subunits are shown in purple. 
          The lamina is composed of lamin subunits, shown in blue.}
	\label{fig:schematic}
\end{figure}

{\it Model:}  Interphase chromatin is modeled as a Rouse chain consisting of
5000 monomers (each representing $\lesssim 1\ \mathrm{Mb}$ of chromatin)
with radius $r_c$ connected by Hookean springs with spring constant $K$. We 
include excluded
volume interactions with a repulsive, soft-core potential between any
two monomers, $ij$, and a distance, $|\vec{r}_{ij}|$, between their centers,  
through the potential $U_{ex}=\frac{1}{2}K_{ex}(|\vec{r}_{ij}|-\sigma _{ij})^2$ for $|\vec{r}_{ij}|<\sigma_{ij}$, where $\sigma
_{ij}=r_{c_{i}}+r_{c_{j}}$, and zero otherwise. Previous mechanical experiments and modeling suggest extensive crosslinking \cite{stephens17, banigan17, strom20}, so we include $N_C \leq 2500$ crosslinks
between chromatin monomers by introducing a spring between different
parts of the chain with the same spring constant as along the chain. 

In addition to (passive) thermal fluctuations, we also allow
for explicit motor activity along the chain.  In simulations with motors, we assign $N_m=400$ chain
monomers to be active. An active monomer has motor strength $M$ and
exerts sub-pN force ${\bf F_{a}}= \pm M {\hat r_{ij}}$ on monomers
within a fixed range. Active monomers do not experience a reciprocal
force, ${\bf - F_{a}}$, so the system is out of equilibrium (see SM,
which includes Refs.~\cite{ou17,falk19,kang15,banigan20}).  
Motor forces may be attractive or ``contractile,'' drawing in chain monomers, or alternatively, repulsive or ``extensile,'' pushing them away (Fig.~\ref{fig:schematic}), similar to other explicit models of motor activity \cite{bruinsma14, saintillan18, manna19}. 
Since motors \textit{in vivo} are dynamic, unbinding or turning off after some characteristic time, we
include a turnover timescale, $\tau_m$, for the motor monomers, after which a motor moves to another position on the
chromatin. We study $\tau_m = 20$, corresponding to $\sim 10\ \mathrm{s}$, \textit{i.e.}, comparable to the timescale of experimentally observed chromatin motions \cite{zidovska13, shaban18}, but shorter than the turnover time RNA polymerase \cite{darzacq07}.

The lamina is modeled as a layer of 5000 identical
monomers connected by springs with the same radii and spring constants
as the chain monomers and an average coordination number $z\approx
4.5$, as supported
by previous modeling~\cite{banigan17, stephens17, lionetti20} and imaging experiments~\cite{shimi15, mahamid16, turgay17}. Shell monomers also have a repulsive soft core.  
We model the chromatin-lamina linkages as $N_L$ permanent springs with stiffness $K$
between shell monomers and chain monomers (Fig.~\ref{fig:schematic}). 

The system evolves via Brownian dynamics, obeying the overdamped equation of motion:
$\xi\dot{\bf{r}}_i = ( {\bf F_{br}} +  {\bf F_{sp}} + {\bf F_{ex}} + {\bf F_{a}})$, where $\bf F_{br}$ denotes the (Brownian) thermal force, $\bf F_{sp}$ denotes the harmonic forces due to chain
springs, chromatin crosslink springs, and chromatin-lamina linkage springs, and ${\bf F_{ex}}$ denotes the force due to excluded volume. 
We use Euler updating, a time step
of $d\tau=10^{-4}$, and a total simulation time of $\tau=500$. 
For the passive system, ${\bf
  F_a}=0$. In addition to the deformable shell, we also simulate a
hard shell by freezing out the motion of the shell monomers. To assess
the structural properties in steady state, we measure both the
radial globule, $R_g$, of the chain and the self-contact probability.  After these
measures do not appreciably change with time, we consider the system
to be in steady state.  See SM for these measurements, 
simulation parameters, and other simulation details.

\begin{figure*}[t]
	\centering
	\includegraphics[width=0.95\textwidth]{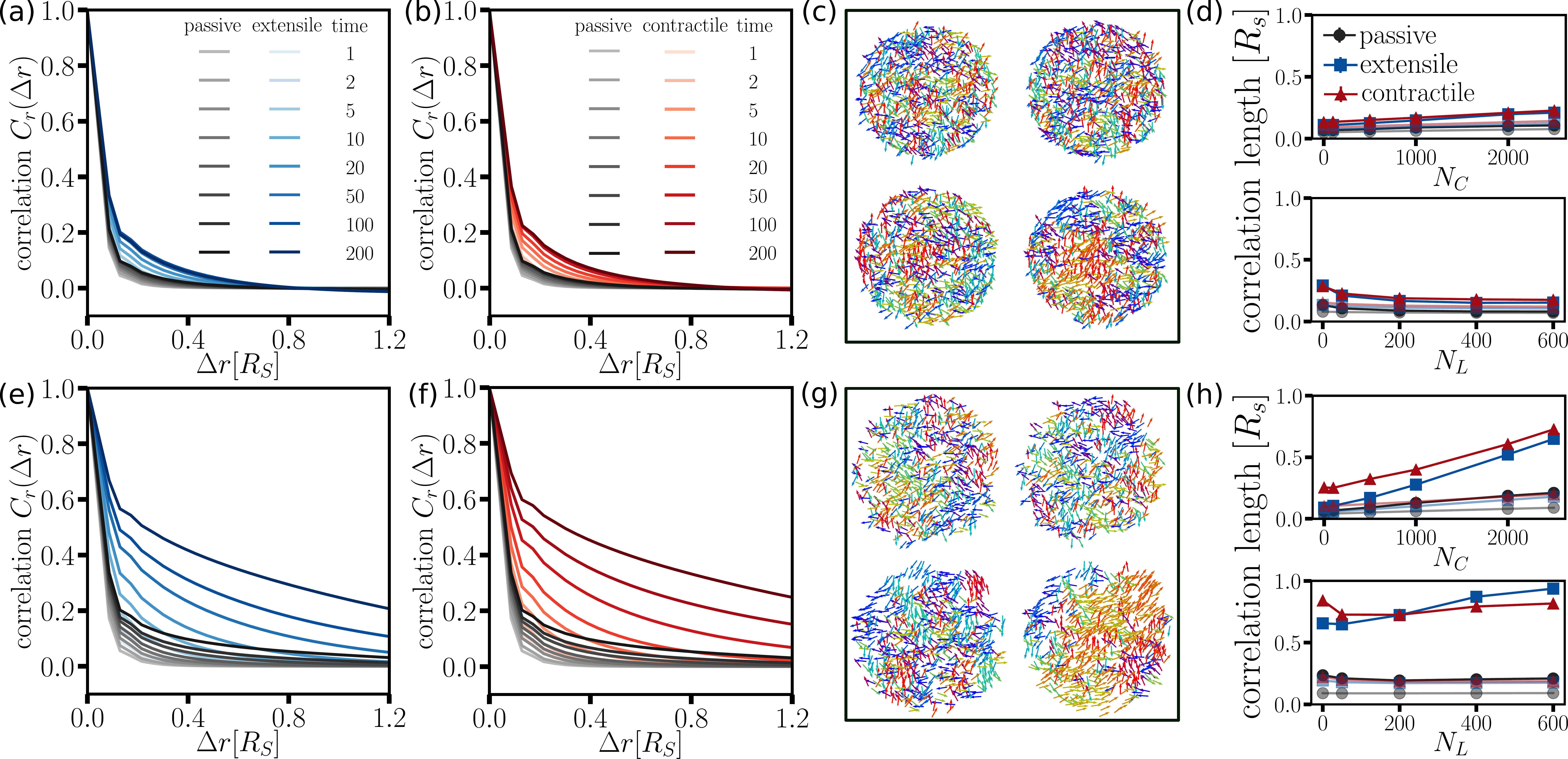}
	\caption{(a)
          The spatial autocorrelation function $C_r(\Delta r, \Delta \tau)$ for passive and extensile cases at different
          time lags, $\Delta\tau$, for the hard shell, while (b)
          shows the contractile and passive case. (c) Two-dimensional
          vector fields for $\Delta \tau=5$ (left), $50$  (right) for the passive case (top) and
          the contractile case (bottom).  (d) The correlation length as a
          function of $N_L$ and $N_C$ for the two time lags in (c). (e$\sim$h): The bottom row shows the same
          as the top row, but with a soft shell. Lengths shown in units of the hard-shell radius, $R_s = 10$. See SM for
          representative fits to obtain the correlation length.
          }
	\label{fig:features1}
\end{figure*}

{\it Results:} We
first look for correlated chromatin motion in both hard
shell and deformable shell systems. We do so by quantifying the correlations between the
displacement fields at two different points in time. Specifically, we
compute the normalized spatial autocorrelation
function defined as $C_r(\Delta r, \Delta
\tau)=\frac{1}{N(\Delta r)}\sum_{N(\Delta r)}\frac{<\bf {d}_i(\bf {r},
  \Delta \tau)\cdot \bf {d}_j(\bf{r}+\Delta r,\Delta \tau) >}{< \bf
    {d}^2(\bf{r},\Delta \tau)>},$ 
where $\Delta \tau$ is the time window, $\Delta r$ is the distance
between the centers of the two chain monomers at the beginning of the
time window, $N(\Delta r)$ is the number of $ij$ pairs of monomers within distance
$\Delta r$ of each other at the beginning of the time window, and $\bf {d}_i$ is the displacement of the $i^{\mathrm{th}}$ chain
monomer during the time window, defined with respect to
the origin of the system. Two chain monomers moving in
the same direction are positively correlated, while monomers
moving in opposite directions are negatively correlated. 

Fig.~\ref{fig:features1} shows $C_r(\Delta r, \Delta \tau)$ for passive and
active samples in both hard shell (Figs.~\ref{fig:features1} (a) and (b)) and soft shell cases for
$N_C=2500$, $N_L=50$, $M=5$, and $\tau_m=20$ (Figs.~\ref{fig:features1} (e) and (f)) (see SM for results with other parameters). Both the passive and active samples exhibit short-range correlated motion
when the time window is small, \textit{i.e.}, $\Delta \tau<5$. However, for longer time windows, both the extensile and contractile
active samples exhibit more long-range correlated motion than the
passive case.  Correlations are also stronger for longer $\tau_m$ (see SM), similar to findings for individual active polymers~\cite{ghosh14}. These correlations are visible in quasi-2d spatial maps of
instantaneous chromatin velocities, which show large regions of
coordinated motion in the active, soft shell case (Figs.~\ref{fig:features1} (c) and (g)).

To extract a correlation length to study the correlations as a function of both
$N_C$ and $N_L$, we use a Whittel-Marten (WM) model fitting function, $C_r(r)=\frac{2^{1-\nu}}{\Gamma (\nu)}\left( \frac{r}{r_{cl}}
\right) ^{\nu}K_{\nu} \left( \frac{r}{r_{cl}} \right)$, for each time
window (Fig.~\ref{fig:features1} (f))~\cite{shaban18}. The parameter
$\nu$ is approximately $0.2$ for all cases studied. For the hard shell, the
correlation length decreases with number of linkages (Fig.~\ref{fig:features1} (d)).
This trend is opposite in deformable shell case with activity and long time lags (Fig.~\ref{fig:features1} (h)). For the hard shell,
linkages effectively break up the chain into
uncorrelated regions. For the soft shell, the shell deforms
in response to active fluctuations in the chain. For both types of
shells, the correlation length increases with the number of crosslinks
(Figs.~\ref{fig:features1} (d) and (h)), with a more significant increase in the
soft shell active case.  It is also interesting to note
that the lengthscale for the contractile case is typically larger than that of the
extensile case, at least for smaller numbers of linkages.

Given the differences in correlation lengths between the hard and
soft shell systems, we looked for enhanced motion of the system in the soft
shell case.  
Enhanced 
motion has been predicted for active polymers~\cite{ghosh14, osmanovic17, chaki19} and observed in active particle systems confined by a deformable
shell~\cite{paoluzzi16}.  Similarly, we observe the active chain
system moving faster than diffusively (see SM). In the shell's
center-of-mass frame, the correlation length is decreased, but still
larger than in the hard shell simulations (see SM). Interestingly,
experiments demonstrating large-scale correlated motion measure
chromatin motion with an Eulerian specification (\textit{e.g.}, by
particle image velocimetry) and do not subtract off the global center
of mass~\cite{zidovska13,shaban18,shaban20}.  However, one experiment noted that they observed drift of the nucleus on a frame-to-frame basis, but considered it negligible over the relevant time scales~\cite{shaban18}. 
Additionally, global rotations, which we have not considered, could yield large-scale correlations.

\begin{figure}[t]
	\centering
\includegraphics[width=0.45\textwidth]{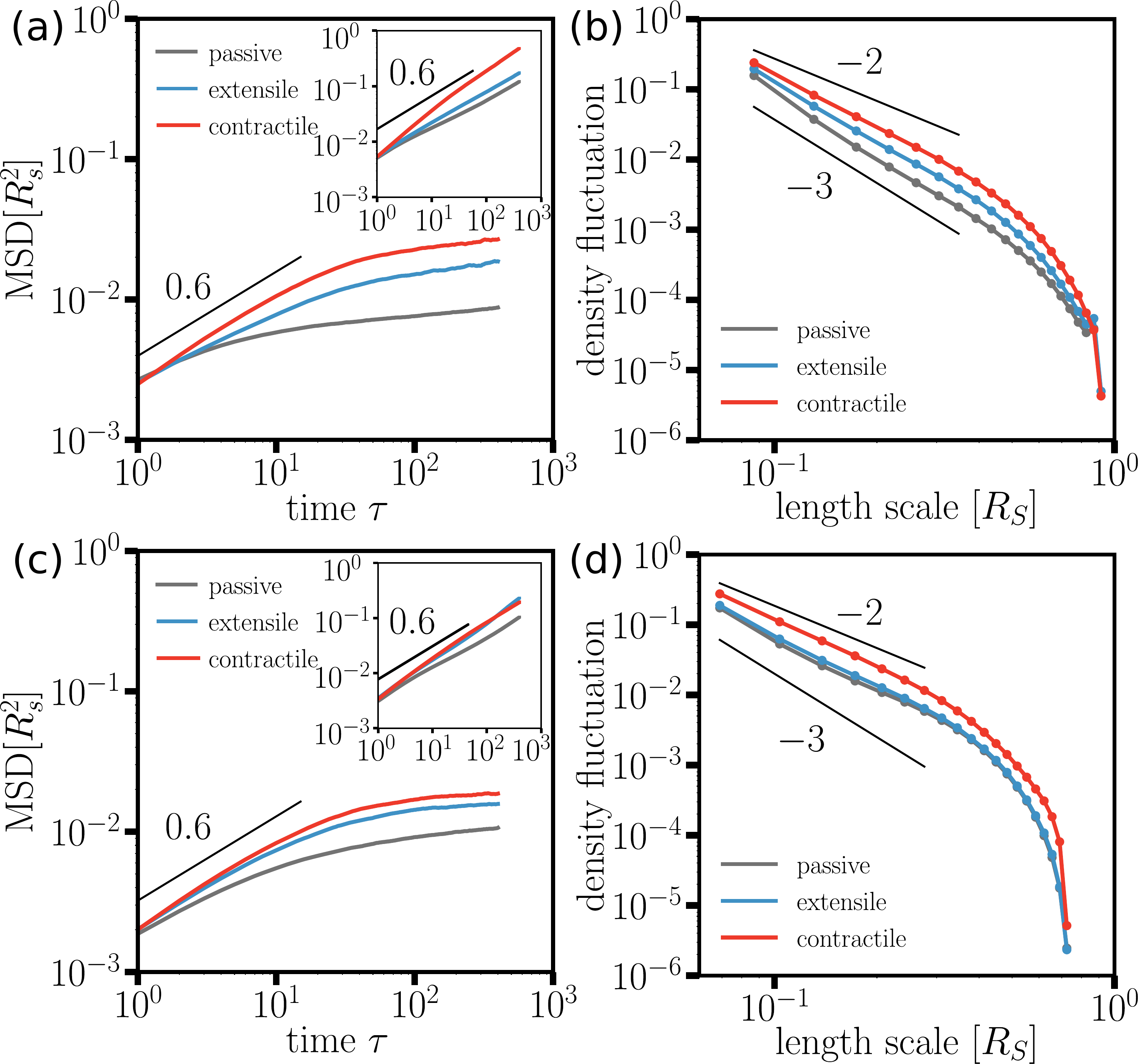}
	\caption{(a) MSD for the hard shell case with $N_C=2500$,
          $N_L=50$, and $M=5$. For the inset, $N_C=0$. (b) Density fluctuations
          for the same parameters as in (a). Figures (c) and (d) show the soft
          shell equivalent to (a) and (b). 
           }
	\label{fig:features2}
\end{figure}

We also study the mean-squared displacement of the chromatin chain to
determine if the experimental feature of anomalous diffusion is
present.  Figs.~\ref{fig:features2} (a) and (c) show the mean-squared displacement
of the chain with $N_L=50$ and $N_C=2500$ as measured with
reference to the center-of-mass of the shell for both the hard shell
and soft shell cases, respectively. For the hard shell, the
passive chain initially moves subdiffusively with an exponent of $\alpha\approx0.5$, which is consistent with an uncrosslinked Rouse chain with
excluded volume interactions~\cite{rubinstein03}. 
However, the passive system crosses over to potentially glassy
behavior after a few tens of simulation time units.  We present $N_C=0$ case in the
inset to Fig.~\ref{fig:features2} (a) for comparison to demonstrate that crosslinks potentially drive a gel-sol transition, as observed in prior experiments~\cite{khanna19}. The active hard shell 
samples exhibit larger displacements than passive
samples, with $\alpha\sim0.6$ initially before
crossing over to a smaller exponent at longer times. 

Additionally, the contractile system exhibits larger displacements than the extensile system. We found that a broader spectrum of steady-state density fluctuations for the contractile system drive this behavior (Fig.~\ref{fig:features2} (b)). This generates regions of lower density into which the chain can move, leading to increased motility.  
The active cases exhibit anomalous density fluctuations, with the variance
in the density falling off more slowly than inverse length cubed (in 3D).
Finally, the MSD in the hard shell case is suppressed by more boundary
linkages or crosslinks (see SM). For the soft shell case, we
observe similar trends as the hard shell.
\begin{figure}[t]
	\centering
	\includegraphics[width=0.46\textwidth]{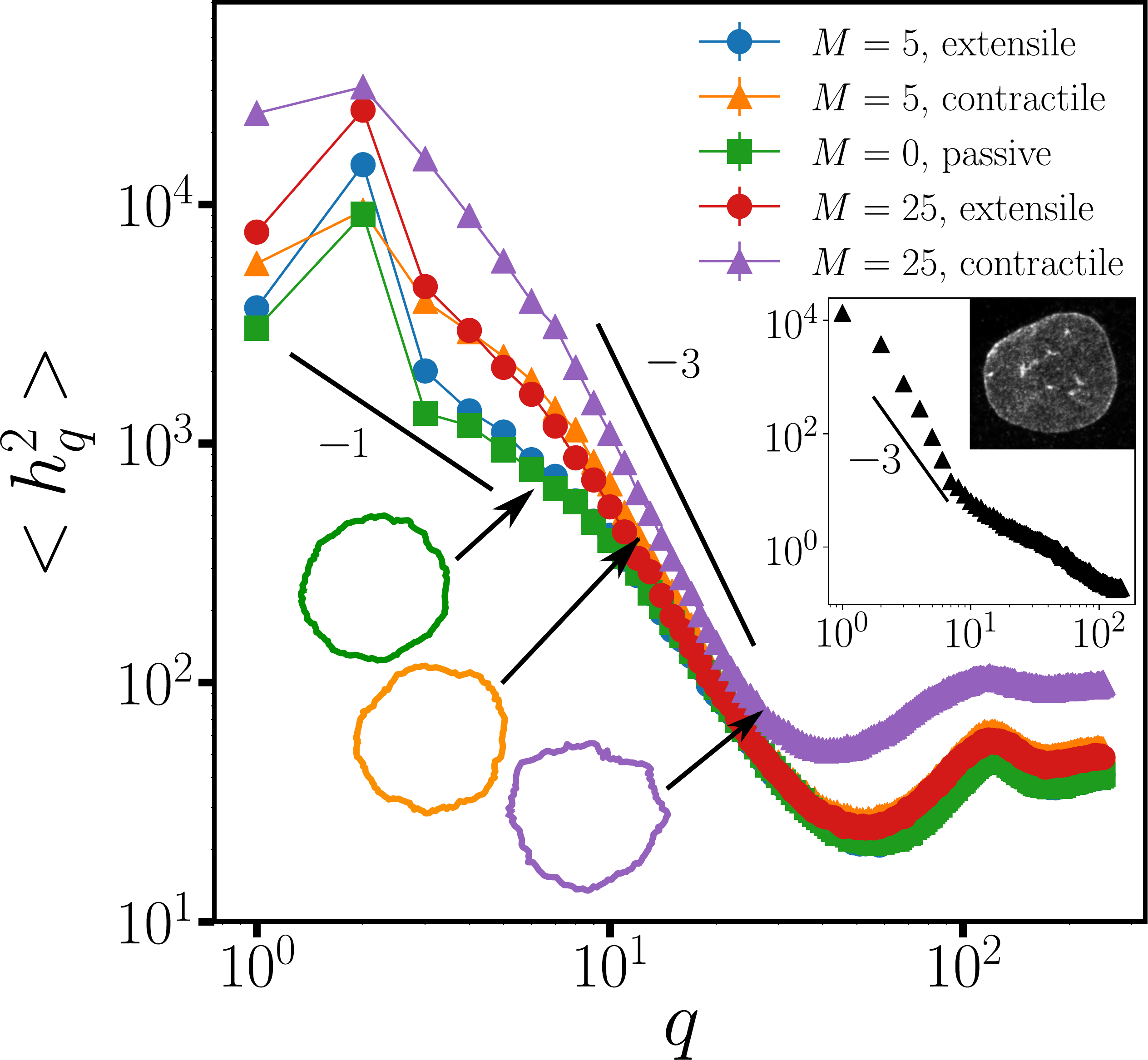}

	\caption{Power spectrum of the shape fluctuations with $N_{L}=50$
          and $N_{C}=2500$ for the passive and both active
          cases. Different motor strengths are shown.  The inset
        shows experimental data from mouse embryonic fibroblasts with
        an image of a nucleus with lamin A/C stained. 
        }
	\label{fig:features3}
\end{figure}

Next, we examine nuclear shape. In Fig.~\ref{fig:features3}, we plot the power spectrum
of the shape fluctuations of the shell for a central 
cross-section as a function of wavenumber $q$ for different
motor strengths.  
Shape fluctuations are less significant for both the passive and
extensile systems than for the contractile systems.  This difference could be due to more anomalous density fluctuations in the contractile case, demonstrating that chromatin spatiotemporal dynamics directly impacts nuclear shape.
The fluctuation spectrum is dominated by an approximate $q^{-3}$
decay, which is characteristic of bending-dominated fluctuations in a
cross-section of a fluctuating shell \cite{milner87, faucon89,
  hackl97, pecreaux04, rodriguez-garcia15}.  Bending fluctuations are
consistent with previous experimental observations~\cite{chu17} and
simulations~\cite{banigan17} of cell nuclei, theoretical
  predictions for membranes embedded with active particles
  \cite{ramaswamy00, gov04}, and our experiments measuring nuclear
shape fluctuations in mouse embryonic fibroblasts (MEFs) (inset to
Fig.~\ref{fig:features3} and see SM for materials and methods).  For
the passive case, we also observe a narrow regime of approximate $q^{-1}$ scaling at small $q$, which is characteristic of membrane tension, and saturation at large $q$ due to the discretization of the system. For the active cases, we only clearly observe the latter trend. Additionally, the amplitude of the shape fluctuations increases with motor strength, $N_C$, and $N_L$ (see SM).

{\it Discussion:} We have studied a composite chromatin-lamina system in the presence of
activity, crosslinking, and linkages between chromatin
and the lamina. Our model captures correlated
chromatin motion on the scale of the nucleus in the presence of both
activity and crosslinks (Fig.~\ref{fig:features1}). The deformability of the shell also plays a
role. We find that global translations of the composite soft shell
system contribute to the correlations.  We observe anomalous diffusion for the chromatin (Figs.~\ref{fig:features2} (a) and (c)), as
has been observed experimentally~\cite{bronshtein15}, with a crossover to a smaller anomalous
exponent driven by the crosslinking~\cite{khanna19}. Interestingly, the contractile
system exhibits a larger MSD than the extensile one, which is
potentially related to the more anomalous density fluctuations in the
contractile case (Figs.~\ref{fig:features2} (b) and (d)). Finally, nuclear 
shape fluctuations depend on motor strength and on amounts of
crosslinking and chromatin-lamina linkages (Fig.~\ref{fig:features3}). Notably, the contractile case
exhibits more dramatic changes in the shape fluctuations as a function
of wavenumber as compared to the extensile case.

Our short-range, overdamped model contrasts with an earlier confined,
active Rouse chain interacting with a solvent via long-range
hydrodynamics~\cite{saintillan18}. While both models generate correlated
chromatin dynamics, with the earlier model, such correlations are
generated only with extensile motors that drive local nematic ordering of the chromatin
chain~\cite{saintillan18}. Moreover, correlation
lengths in our model are significantly larger than those obtained in a previous confined
active, heteropolymer simulation~\cite{liu18}. Activity in this
earlier model is modeled as
extra-strong thermal noise such that the correlation length decreases
at longer time windows as compared to the passive case. This decrease
contrasts with our results (Figs.~\ref{fig:features1} (d) and (h)) and experiments~\cite{shaban18}. In addition, our model takes into account
deformability of the shell and the chromatin-lamina
linkages. Future experiments could potentially distinguish these mechanisms by looking for prominent features of our model, such as a dependence on chromatin bridging proteins and linkages to the lamina and effects of whole-nucleus motions.

Further spatiotemporal studies of nuclear shape could investigate the role of the cytoskeleton. 
Particularly interesting would be \textit{in vivo} studies with vimentin-null cells, which have minimal mechanical coupling between the cytoskeleton and the nucleus. 
Vimentin is a cytoskeletal intermediate filament that forms a protective cage on the outside of the nucleus and helps regulate the nucleus-cytoplasm
coupling and, thus, affects nuclear shape~\cite{patteson19}. 
The amplitudes of the nuclear shape fluctuations in vimentin-null
cells may increase due to a softer perinuclear shell; alternatively, they may decrease due to fewer linkages between the nucleus and the mechanically active cytoskeleton, which may impact nuclear shape fluctuations \cite{talwar13, makhija16, rupprecht18}.

There are intriguing parallels
between cell shape ~\cite{keren08, tinevez09, barnhart11} and nuclear shape with cell shape being driven by an
underlying cytoskeletal network---an active, filamentous system driven
by polymerization/depolymerization, crosslinking, and motors, both
individually and in clusters, that can
remodel, bundle and even crosslink filaments. Given the emerging
picture of chromatin motors acting collectively~\cite{cisse13,nagashima19}, just as myosin motors
do~\cite{wilson10}, the parallels are strengthened. Moreover, the more
anomalous density fluctuations for the contractile motors as compared
to the extensile motors could potentially be relevant in random actin-myosin
systems typically exhibiting contractile behavior, even though either
is allowed by a statistical symmetry~\cite{koenderink18}. On the other hand, distinct physical mechanisms may govern nuclear shape since the chromatin fiber is generally more flexible than cytoskeletal filaments and the lamina is stiffer than the cell membrane.

We now have a minimal chromatin-lamina model that can be augmented with 
additional factors, such as different types of motors---dipolar,
quadrupolar, and even chiral, such as torque dipoles. Chiral motors may readily
condense chromatin just as twirling a fork ``condenses'' spaghetti.  
Finally, there is now compelling evidence that nuclear actin exists in the cell nucleus~\cite{nuclearactin}, but its form and function are under investigation. Following reports that nuclear actin filaments may alter chromatin dynamics and nuclear shape~\cite{wang19, lamm20, takahashi20}, we propose that short, but
stiff, actin filaments acting as stir bars could potentially increase the correlation length of micron-scale chromatin dynamics, while chromatin motors such as RNA polymerase II drive the dynamics. Including such
factors will help us further quantify nuclear dynamics 
to determine, for example, mechanisms for extreme nuclear shape deformations, such as nuclear blebs~\cite{nuclearblebs, stephens19b}, and ultimately how nuclear
spatiotemporal structure affects nuclear function. 

EJB thanks Andrew Stephens for helpful discussions and critically reading the manuscript. 
JMS acknowledges financial support from NSF-DMR-1832002 and from the
DoD via an Isaac Newton Award. JMS and AEP
acknowledge financial support from a CUSE grant. EJB was supported by the NIH Center for 3D Structure and Physics of the Genome of the 4DN Consortium (U54DK107980), the NIH Physical Sciences-Oncology Center (U54CA193419), and NIH grant GM114190.


\let\oldaddcontentsline\addcontentsline
\renewcommand{\addcontentsline}[3]{}
\let\addcontentsline\oldaddcontentsline

\clearpage
\newpage

\begin{widetext}
\begin{center}
\textbf{\large Dynamic nuclear structure emerges from chromatin crosslinks and motors \\ Supplementary Material}
\end{center}

\setcounter{equation}{0}
\setcounter{figure}{0}
\setcounter{table}{0}
\setcounter{page}{1}
\makeatletter
\renewcommand{\theequation}{S\arabic{equation}}
\renewcommand{\thefigure}{S\arabic{figure}}
\renewcommand{\bibnumfmt}[1]{[S#1]}

\appendix

\tableofcontents
\section{Model}

\subsection{System and initialization}
We use a Rouse chain with soft-core repulsion between each monomer capturing excluded volume effects to represent the chromatin. Since the chromatin is contained within the lamina, modeled as a polymeric shell, we present the protocol to obtain the initial configuration for the composite system. As shown in Fig.~\ref{fig:configuration}(left), we first implement a three-dimensional self-avoiding random walk in an FCC lattice for 5000 steps to generate the chain. We then surround the chain in a large polymeric, but hard, shell. To create the shell, we generate a Fibonacci sphere with 5000 nodes and identify 5000 identical monomers with these nodes. The springs between the shell monomers form a mesh and each shell monomer is connected to $4.5$ other shell monomers on average, which models observations of the lamina in imaging experiments \cite{shimi15} and follows previous mechanical modeling \cite{stephens17, banigan17}. These monomers have same physical properties as the chain monomers in terms of size and spring strength. 

We then shrink the shell (Fig.~\ref{fig:configuration}(center)) by moving the shell monomers inwards by the same amount. During the shrinking process, chain monomers interact with the shell monomers via the soft-core repulsion and, therefore, also move inwards. In addition, every chain monomer experiences thermal fluctuations and is constrained by elastic forces and soft-core repulsion forces. Once the shell radius reaches its destination radius after some time, we then thermalize the positions of the shell monomers and adjust rest length of springs respectively to make the mesh less lattice-like. We, thus, arrive at the initial configuration of the system Fig.~\ref{fig:configuration}(right). We obtain 100 such initialized samples to obtain an ensemble average for each measurement.  The destination radius $R_s$ is $10$. We set the monomer radius to be $r_c=0.43089$ so that the packing fraction $\phi$ is approximately $0.4$ in the hard shell limit comparable to electron microscopy tomography experiments~\cite{ou17}, simulations of chromatin confined within the nucleus~\cite{falk19}, and theoretical estimates~\cite{kang15}, while $\phi$ is smaller in soft-shell cases due to expansion as the shell monomers undergo thermal fluctuations. 

\begin{figure}[h]
	\begin{center}
		\includegraphics[width=0.3\linewidth]{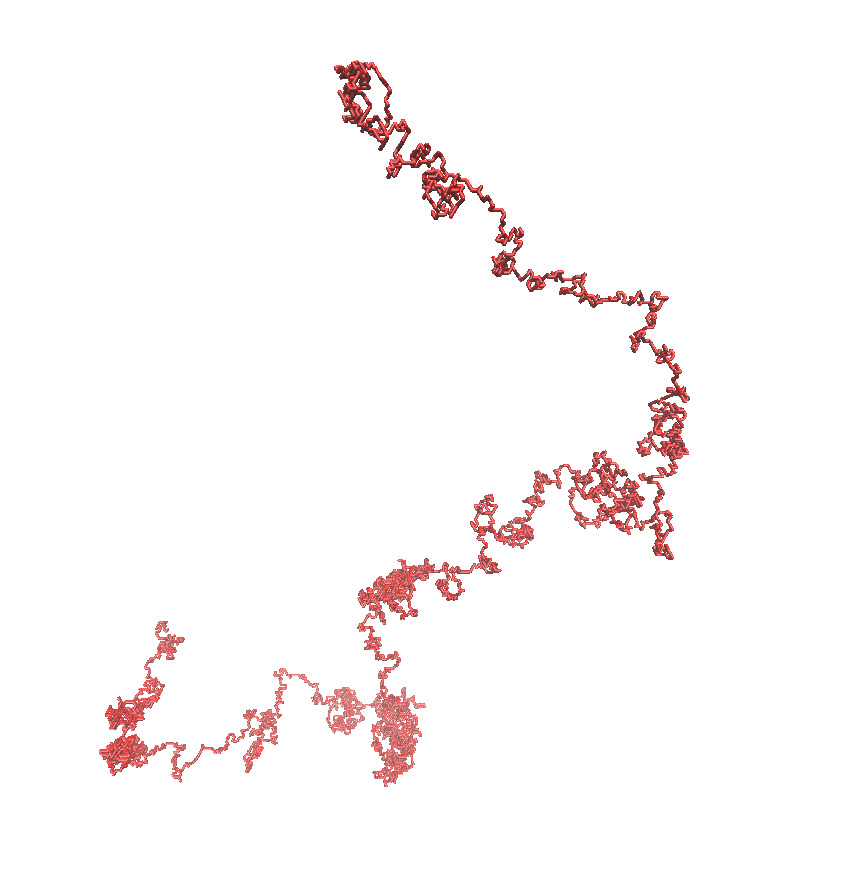}
		\includegraphics[width=0.3\linewidth]{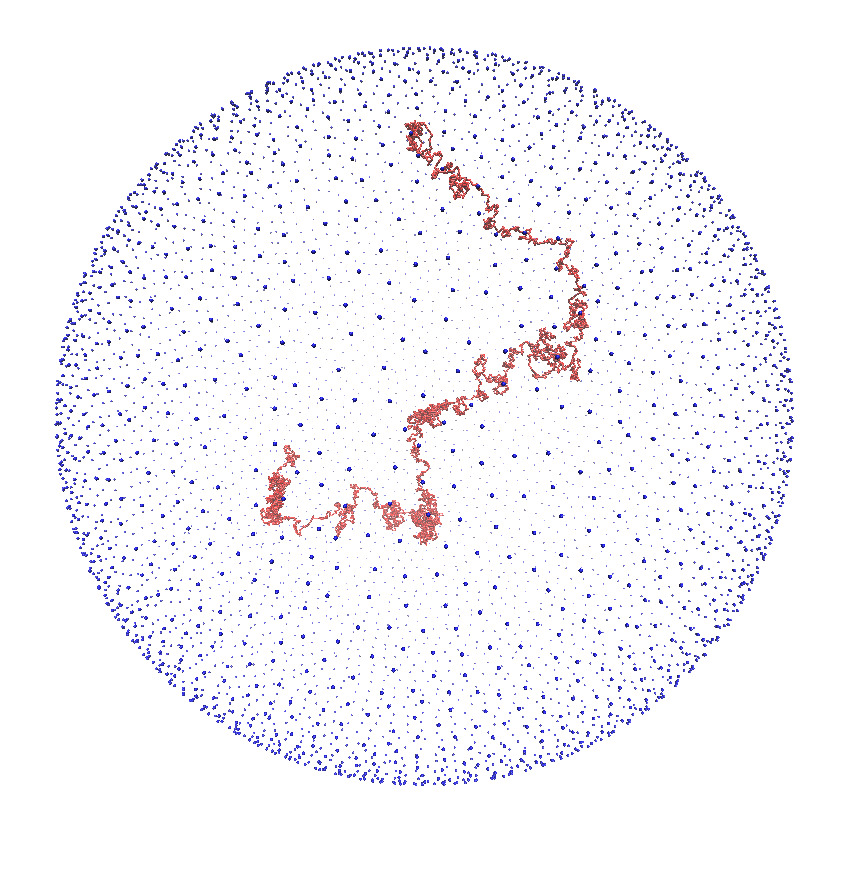}
		\includegraphics[width=0.3\linewidth]{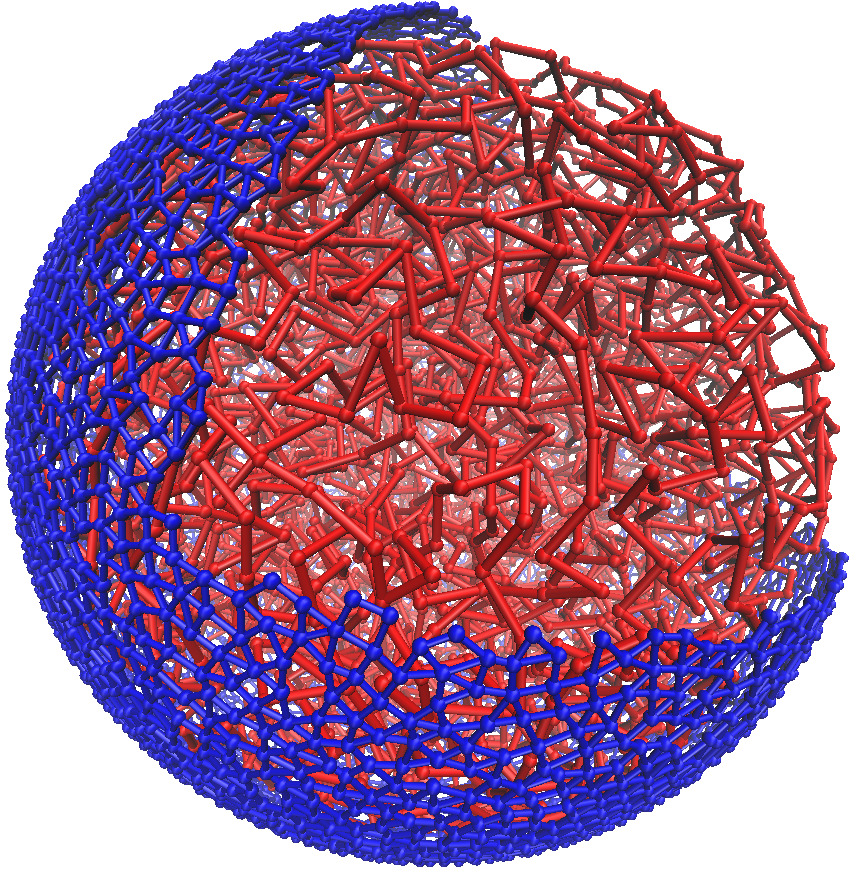}
	\end{center}
	\caption{Left: The chain is initially generated via a self-avoiding random walk on an FCC lattice. Center: The chain is then enclosed in a Fibonacci sphere. Right: Composite system at time $\tau=0$.}
	\label{fig:configuration}
\end{figure}

\subsection{Parameters}
In our simulations, we use the set of parameters shown in Table S1. We now address how the simulation parameters map to biological values. One simulation length unit corresponds to $1\,\mu \mathrm{m}$, one simulation time unit corresponds to 0.5 seconds, and one simulation energy scale corresponds to approximately $10^{-21}\, J=k_B T$, $T=300$~K.  With this mapping, the spring constant corresponds to approximately $1.4\times 10^{-4}$ $\frac{\mathrm{nN}}{\mu \mathrm{m}}$ with a Young's modulus for the chain of $0.28\, \mathrm{Pa}$. 

\begin{table}
\begin{center}
\begin{tabular}{ l c c }
 Diffusion constant & $D$ & $1$ \\ 
 Thermal energy & $k_BT$ & $1$ \\   
 Simulation timestep & $d\tau$ & $10^{-4}$ \\   
 Number of chain monomers & $N$ & $5000$ \\   
 Radii of chain monomers & $r_c$ & $0.43089$ \\   
 Number of shell monomers & $N_{s}$ & $5000$ \\ 
 Radii of shell monomers & $r_{s}$ & $0.43089$ \\  
 Radius of hard shell & $R_{s}$ & $10$ \\  
 Packing fraction & $\phi$ & $0.400$ \\
 Spring constant & $K$ & $140$ \\ 
 Soft-core repulsion strength & $K_{ex}$ & $140$ \\
 Number of motors & $N_m$ & $400$\\  
 Motor strength & $M$ & $5/25$\\
 Turnover time for motors & $\tau_m$ & $20$ \\ 
 Number of crosslinks & $N_C$ & $0/100/500/1000/2000/2500$ \\
 Number of linkages & $N_L$ & $0/50/200/400/600$ \\
Damping & $\xi$ & $ 1$\\
\end{tabular}
\caption{Table of the parameters used in the simulations.}
\end{center}
\end{table}

Motors are characterized by three parameters: the total number of motors, $N_m$; the motor strength, $M$; and the turnover time, $\tau_m$. 
We focus on systems with $N_m = 400$ motors, which is comparable to, but less than $N_m\approx 833$ motors used in a previous active chromatin model \cite{saintillan18}. To assess the dependence of the effects on the prevalence of the motors, we also considered smaller numbers, $N_m$, of motors (see Figs.~\ref{fig:correlationengthvsstiffness} and \ref{fig:shape5}).  The motor strength, $M$ spans the range $0.02<M<0.2\ \mathrm{pN}$, consistent with a previous model for nonequilibrium molecular forces acting within chromatin and DNA~\cite{saintillan18}. Importantly, this is also smaller than typical forces exerted by molecular motors within the genome~\cite{banigan20}, so molecular motor strength inside living cells is not a limiting factor for achieving the nonequilibrium effects in our model.  The turnover time $\tau_m = 20$ corresponds to an ``on'' time or residence time of $10\ \mathrm{s}$.  This is shorter than the typical lifetime of RNA polymerase which is of order $100-1000\ \mathrm{s}$~\cite{darzacq07} (the same is true of many other molecular motors acting on chromatin, such as cohesin~\cite{gerlich06}). Results for longer (and shorter) residence times are also explored (Figs.~\ref{fig:correlation_3} and \ref{fig:shape4}).

\section{Simulation results}

\subsection{Radius of gyration and radial distribution of monomers}
For a polymer, the radius of gyration is defined as $R_g = \sum (r_i-r_{cm})^2 / N$, where $N=5000$ is total chain monomer number. In the hard shell case, we fix the radius of the shell to $R_s=10$. In the soft-shell case, the shell expands due to the thermal fluctuations and due to the activity of the chain inside. Fig.~\ref{fig:radius} (top row) shows the radius of gyration of the chain (solid lines) and the average radius of shell (dashed lines) in the soft shell case as function of time. After a short-time initial expansion, both the chain's and the shell's respective radii reach a plateau by $100\,\tau$ for most parameters, indicating that the system is reaching steady state. Only for the zero crosslinks with contractile activity, does the radius of gyration continue to increase slightly over the duration of the simulation of $500\,\tau$. Fig.~\ref{fig:radius} (bottom row) shows the steady-state radial distribution of monomers within the soft shell.

\begin{figure*}[t]
	\begin{center}
		\includegraphics[width=0.19\linewidth]{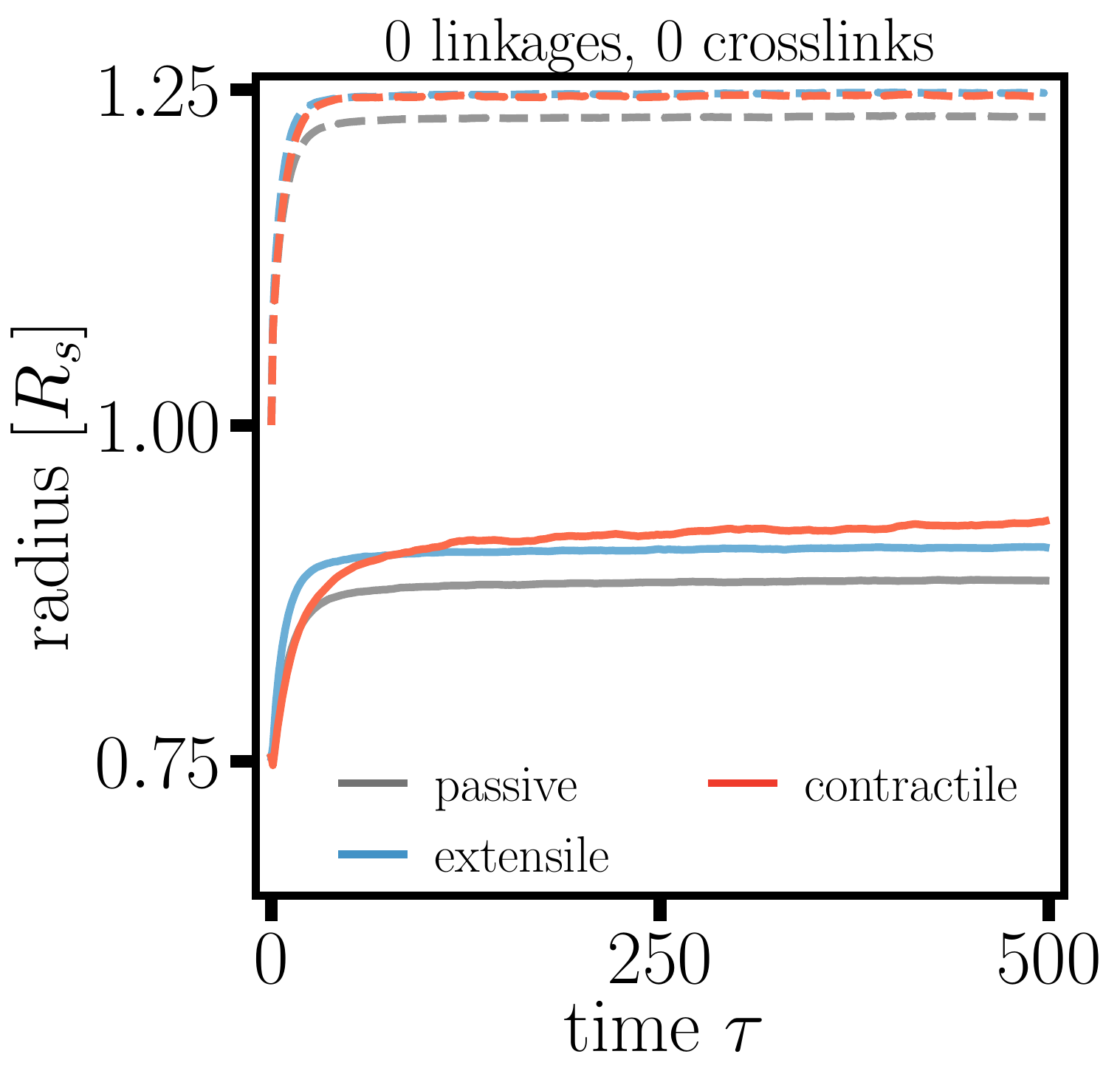}
		\includegraphics[width=0.19\linewidth]{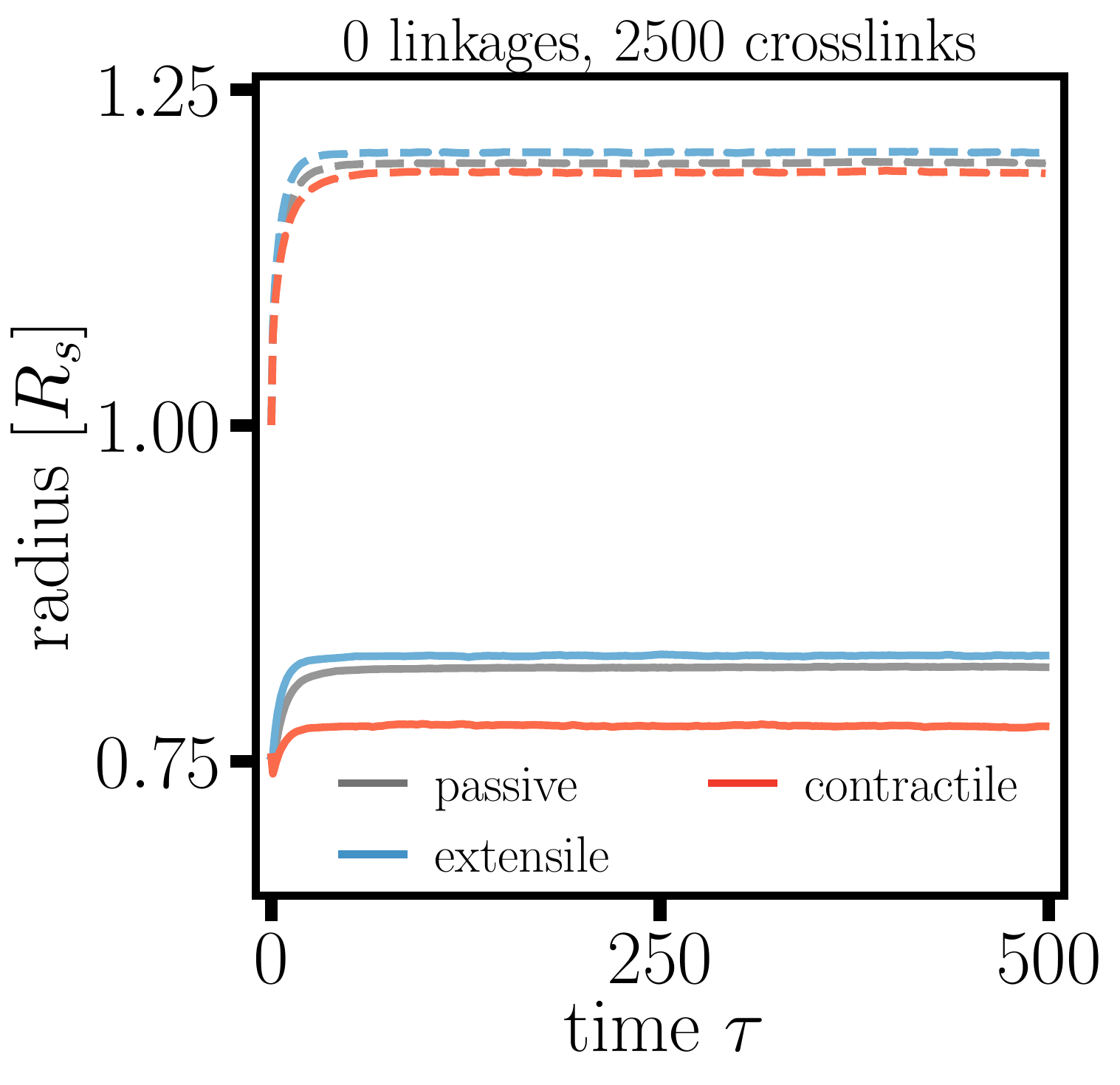}
		\includegraphics[width=0.19\linewidth]{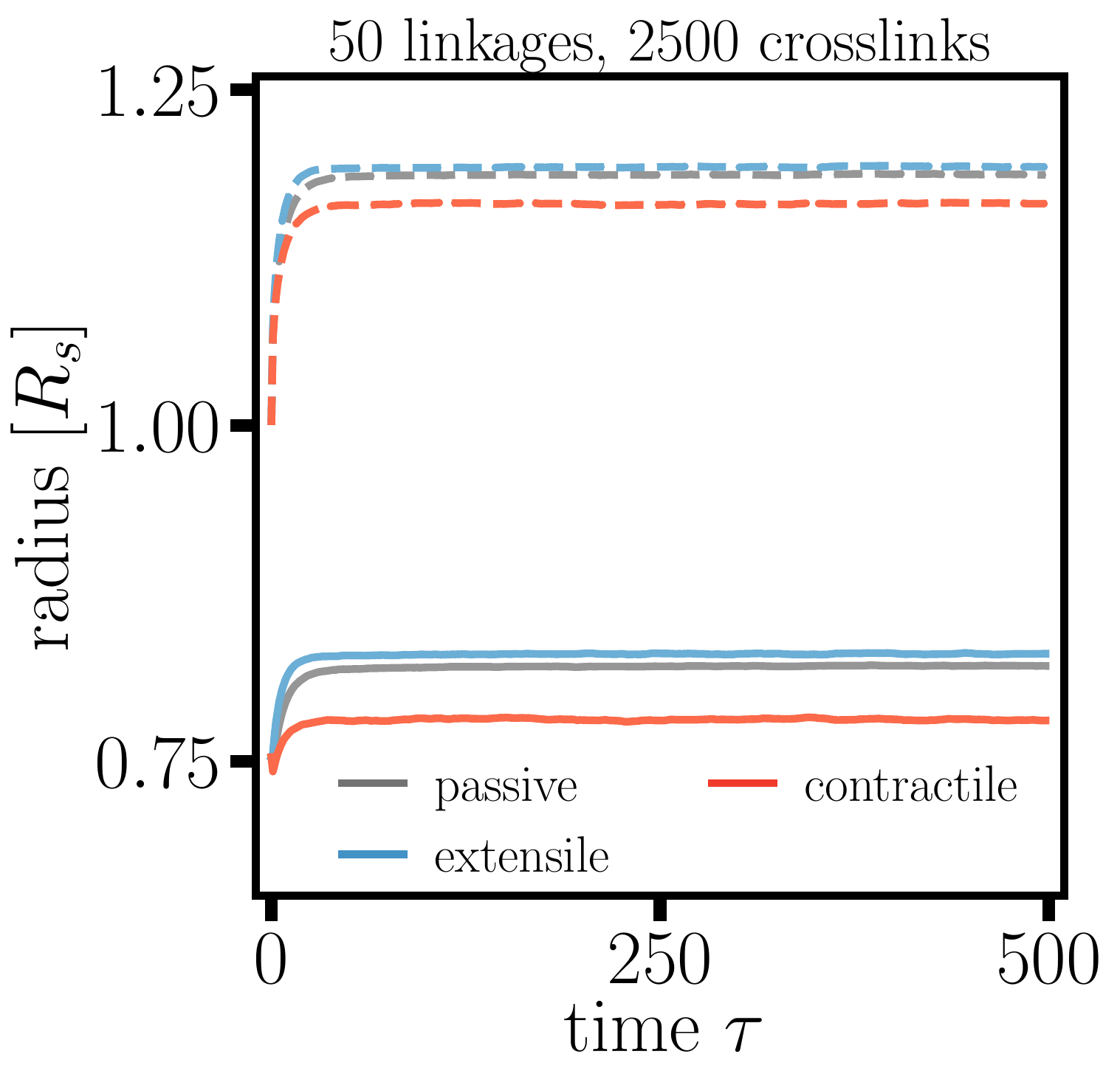}
		\includegraphics[width=0.19\linewidth]{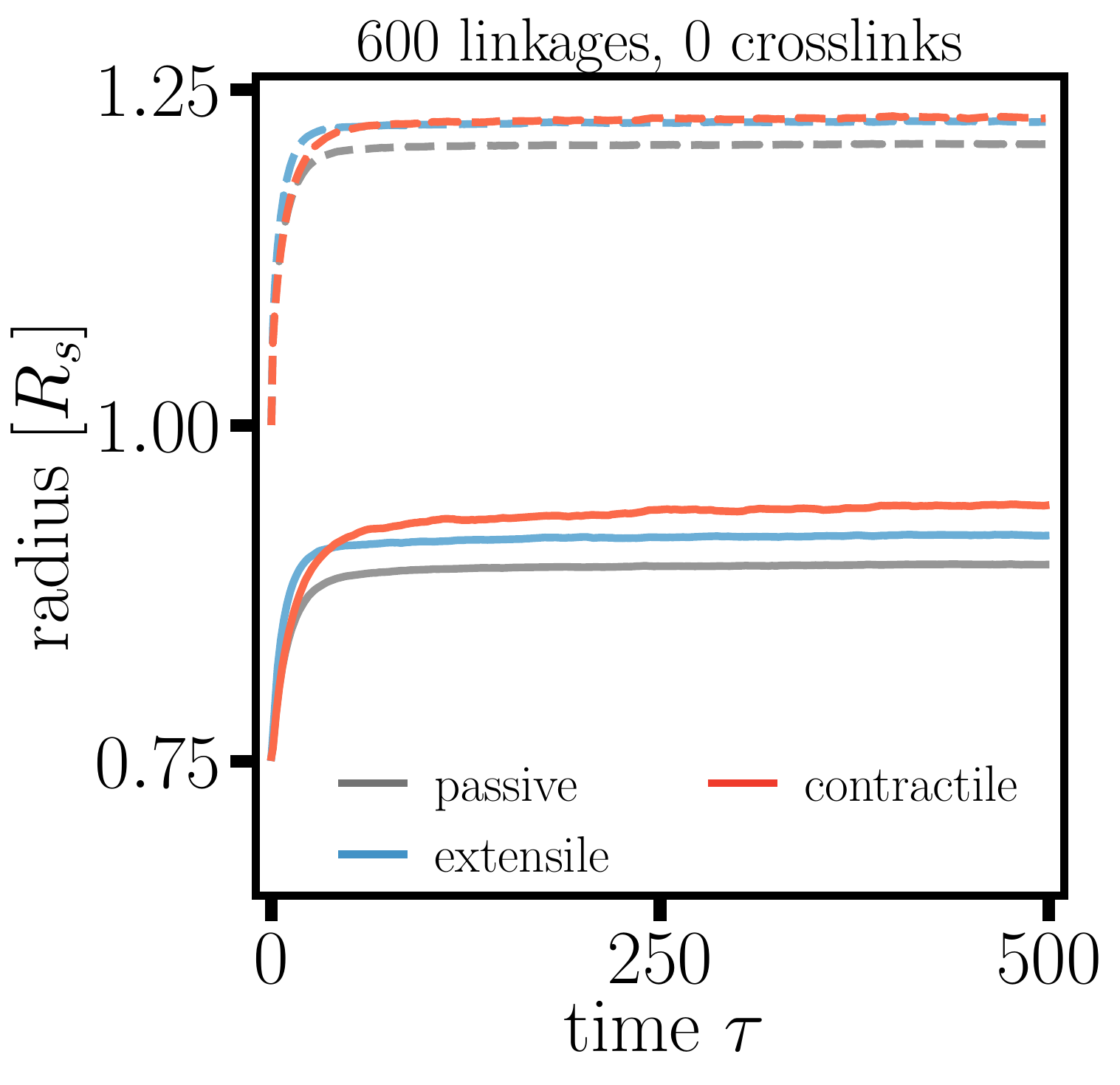}
		\includegraphics[width=0.19\linewidth]{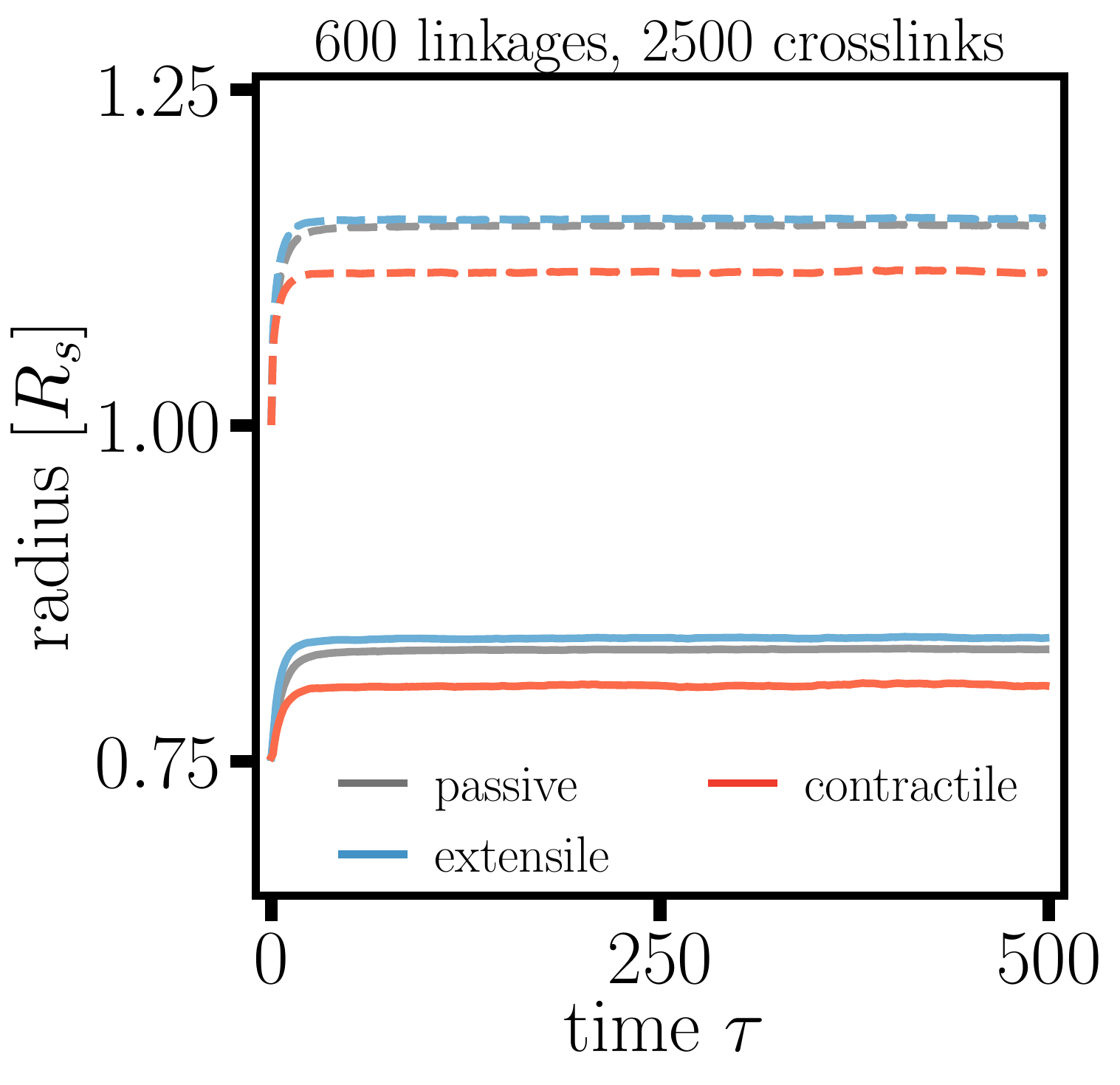}
		\includegraphics[width=0.19\linewidth]{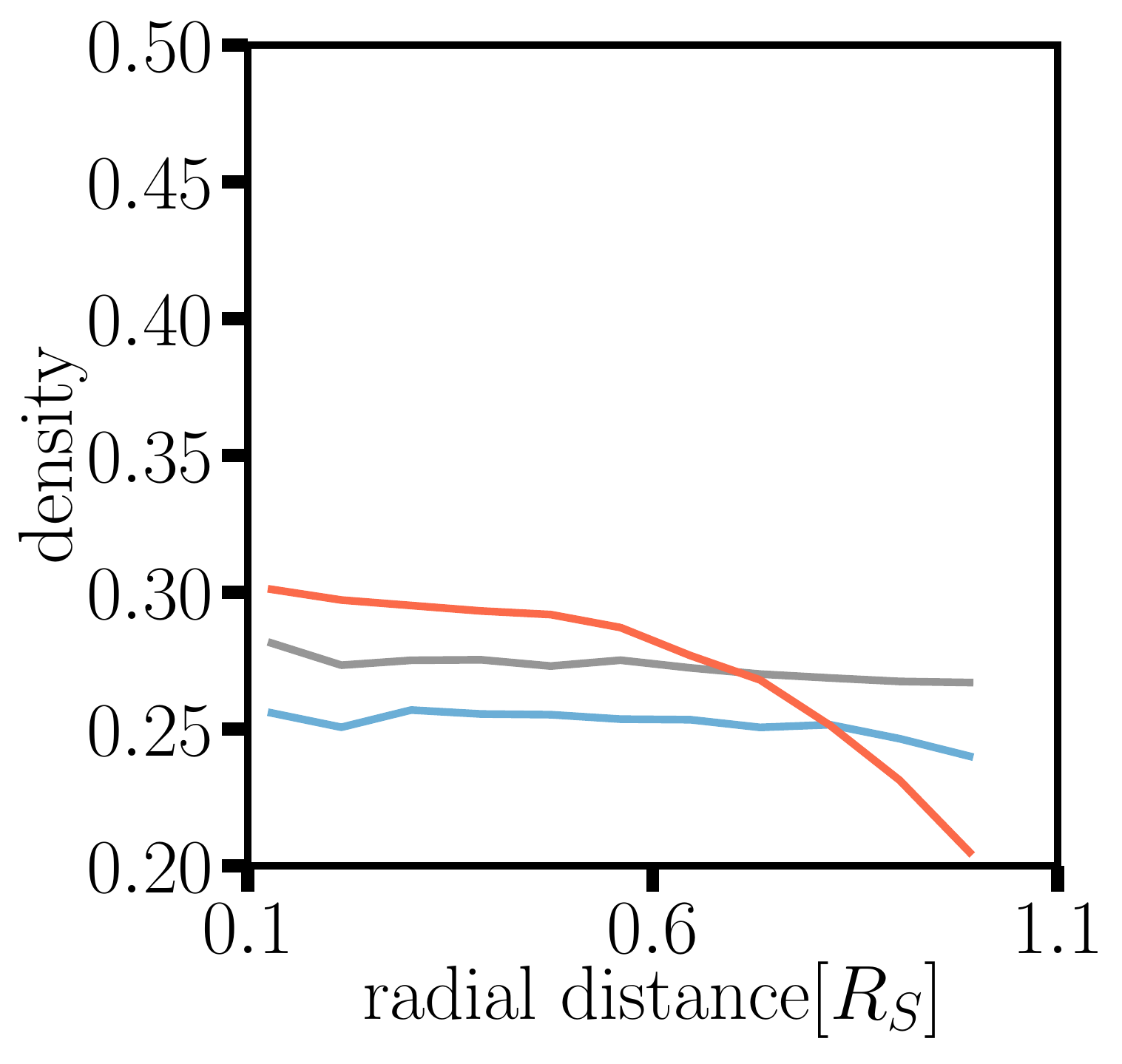}
		\includegraphics[width=0.19\linewidth]{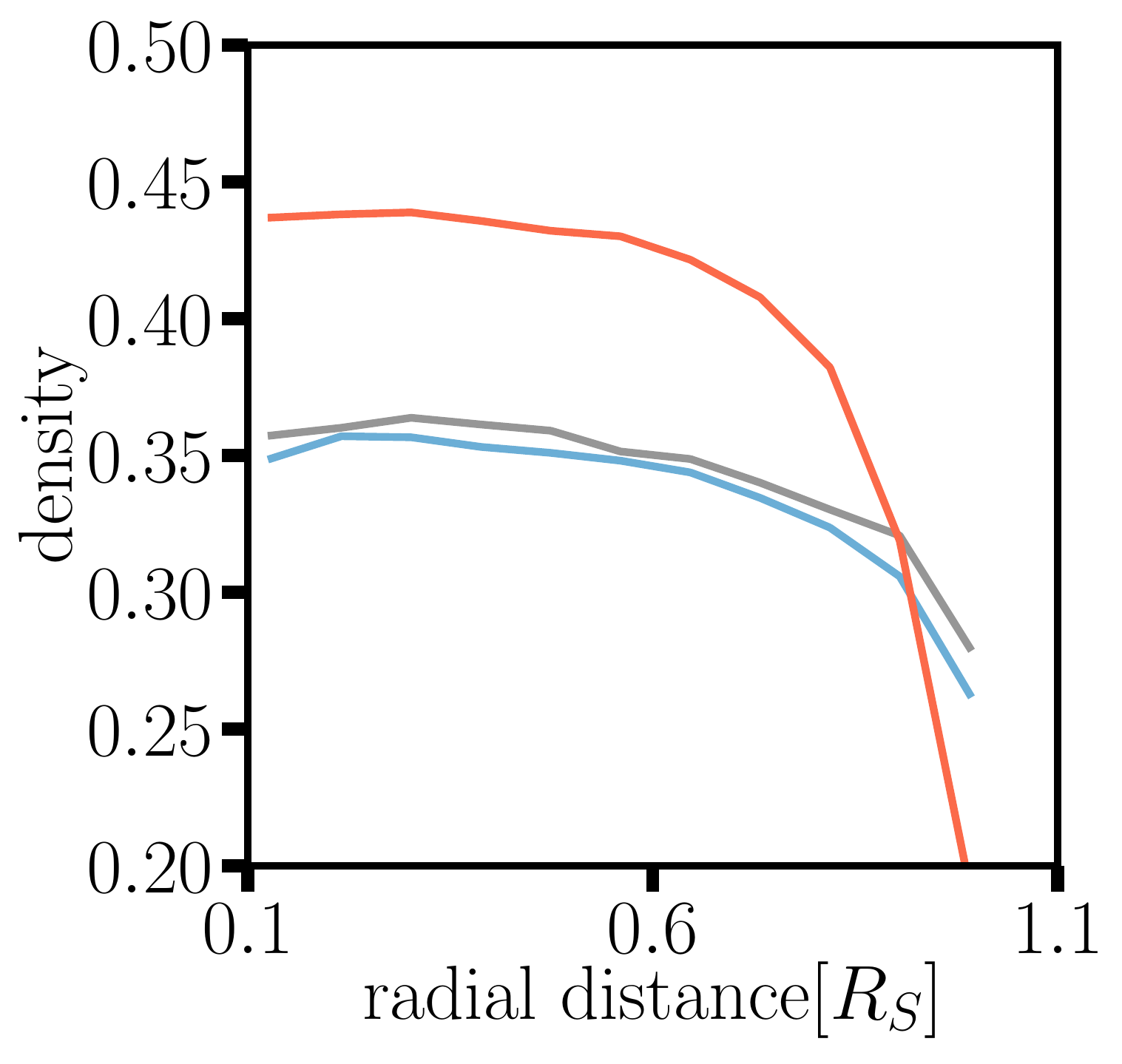}
		\includegraphics[width=0.19\linewidth]{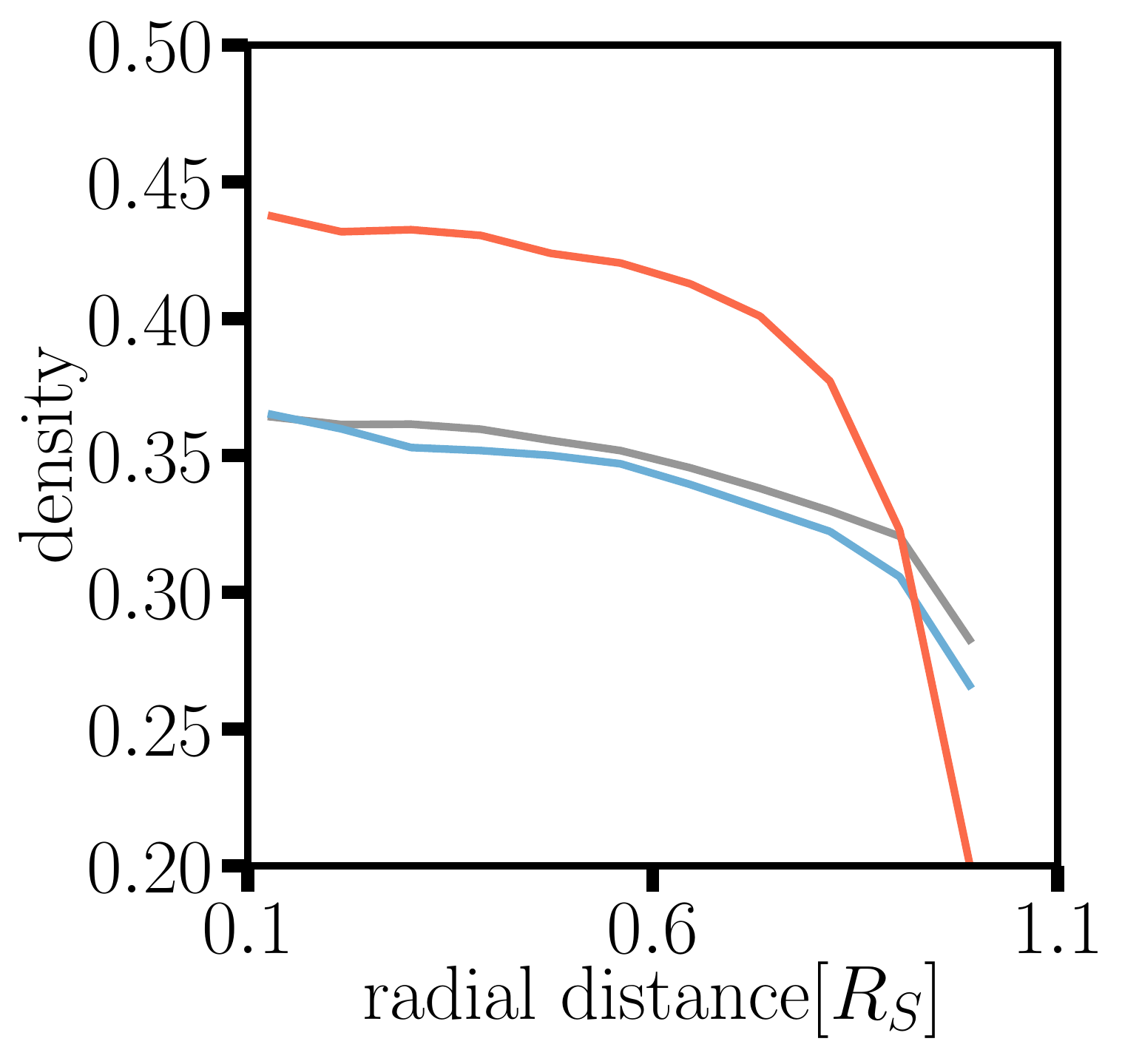}
		\includegraphics[width=0.19\linewidth]{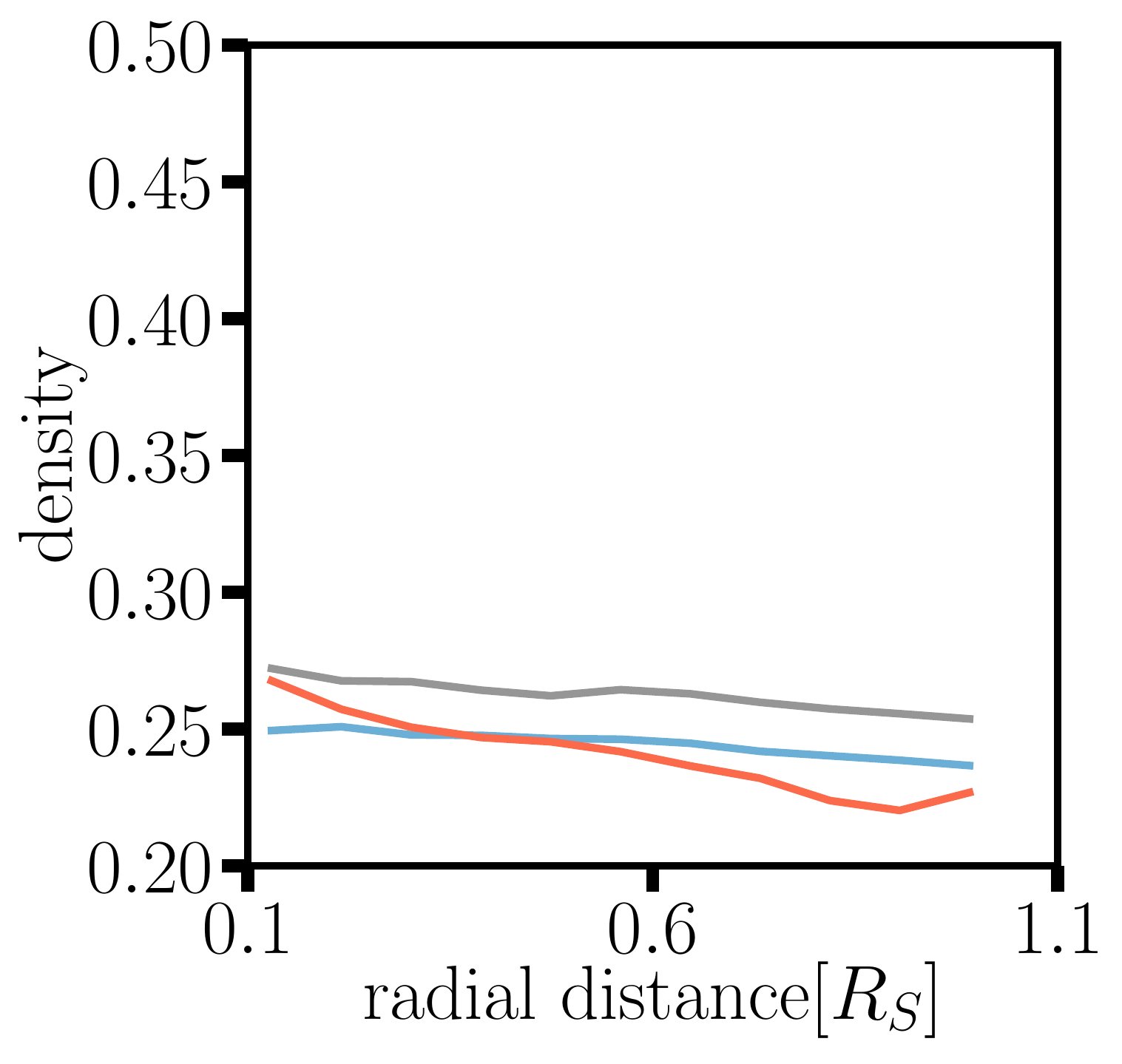}
		\includegraphics[width=0.19\linewidth]{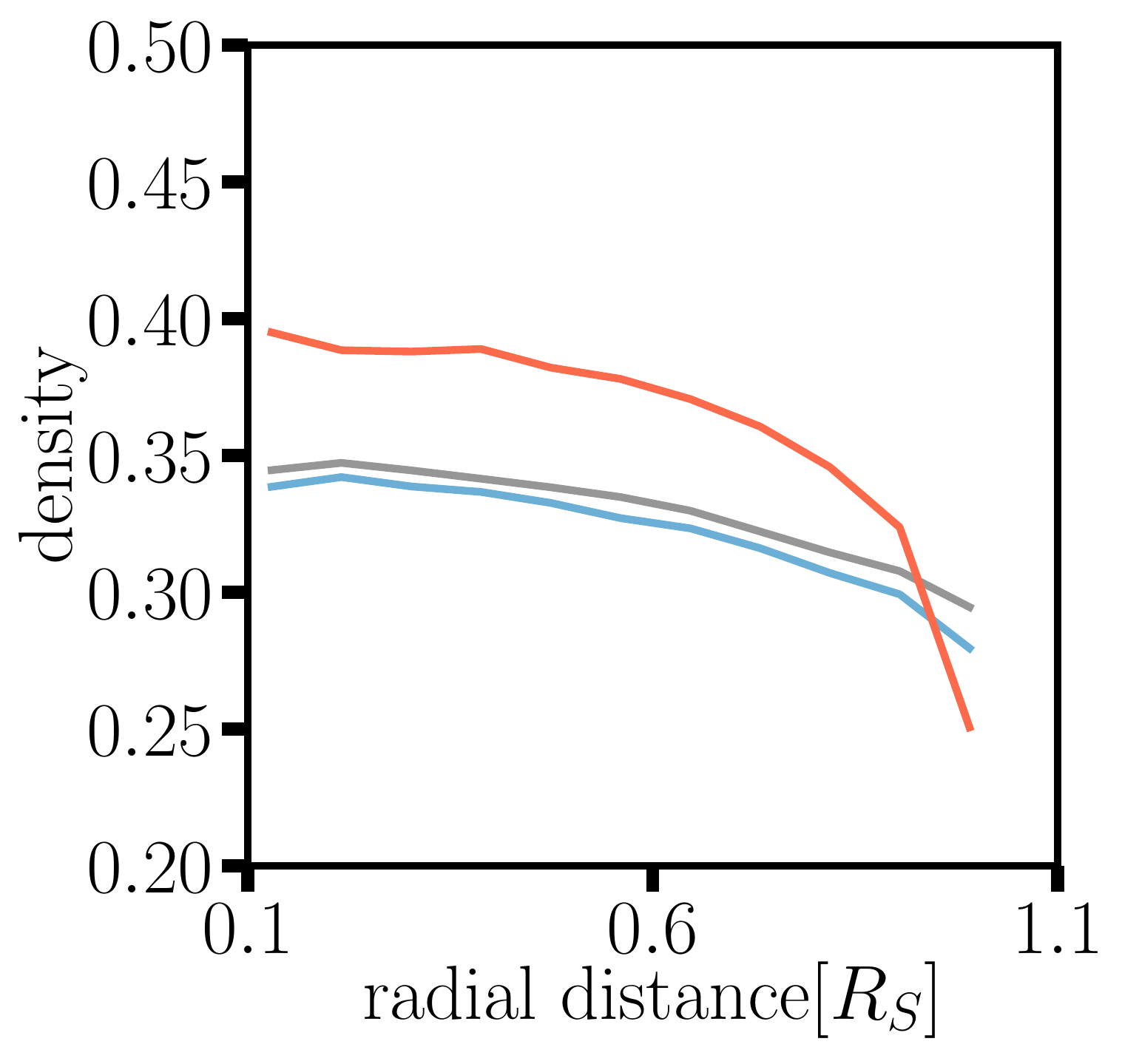}
	\end{center}
	\caption{First row: Radius of gyration of the chain (solid lines) and average radius of the shell (dashed lines) as function of simulation time for $N_C=2500$ and $N_L=50$ (middle figure).  For contrast, $R_g$ for $N_L=0,600$ and $N_c= 0,2500$ are also plotted.  Only the soft shell case are shown. Second row: corresponding radial density distributions of monomers, up to a distance $R_s$ from the center of the shell (note that overall density depends on measured soft shell radius in the top row). All lengths are shown in units of the hard-shell radius $R_s = 10$.}
	\label{fig:radius}
\end{figure*}

\subsection{Self-contact probability}
Since the globule radius is an averaged quantity, we also look for steady state signatures in the self-contact probability, which yields information about the chromatin spatial structure. More specifically, Hi-C allows one to quantify the local chromatin interaction domains at the megabase scale~\cite{lieberman-aiden09}. Such domains are stable across different eukaryotic cell types and species~\cite{dekker16}. To quantify such interactions in the simulations, one determines the number of monomers in the vicinity of the $ith$ chain monomer.  In other words, one creates an adjacency matrix. This adjacency matrix is shown Fig. \ref{fig:contact_map} for two examples. To compute the self-contact probability, one sets a threshold distance that a pair of monomers within that range is considered to be in contact. Then the fraction of contacted pairs for each polymeric distance $1,2,3,4,...$ is calculated. This fraction as a function of polymeric distance is called the self-contact probability. See Fig. \ref{fig:contact_prob} for the self-contact probability for  $N_L=N_C=0$ at the beginning and at the end of the simulation for the soft shell case. While there is some change between the two, in Fig. \ref{fig:contact_prob2}, we show the self-contact probability for different times $\tau$ to demonstrate that after $\tau=50$, the probability does not change with time, implying a steady state.  

\begin{figure}[h]
	\begin{center}
		\includegraphics[width=0.25\linewidth]{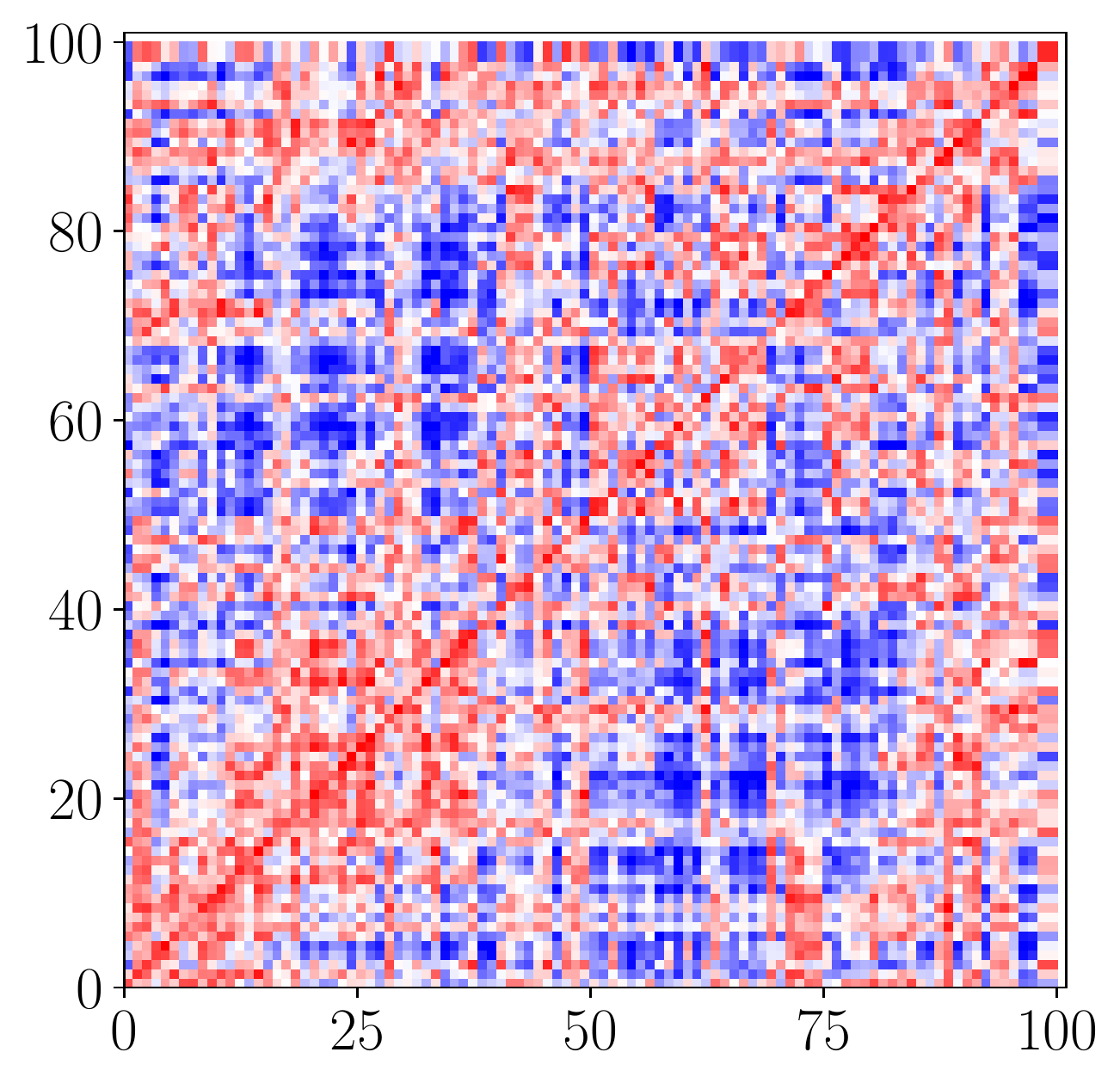}
		\includegraphics[width=0.25\linewidth]{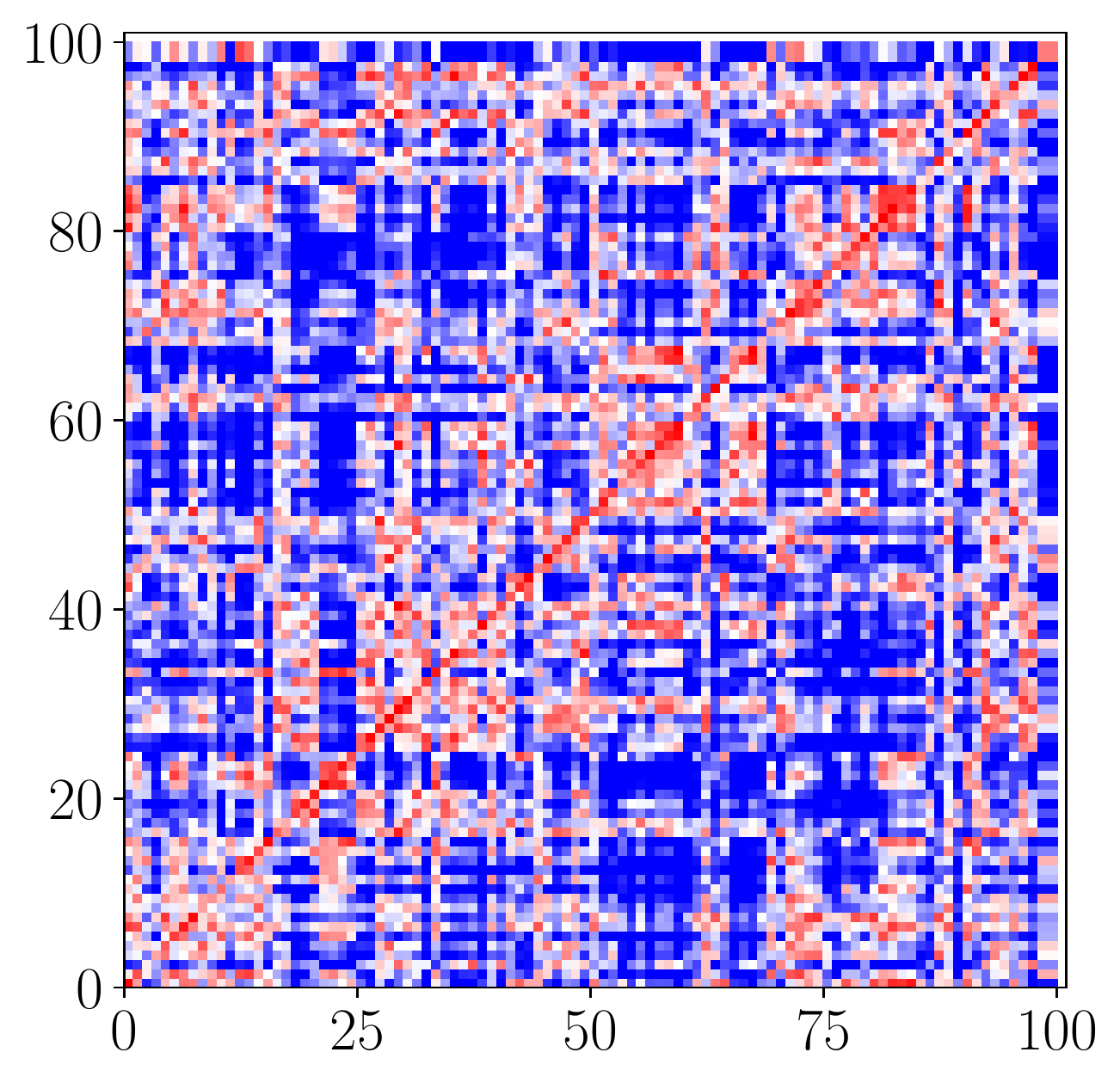}
	\end{center}
	\caption{Contact map for a soft shell, contractile system with no linkages or crosslinks at the beginning and at the end of the simulation, i.e. $\tau=0$ and $\tau=500$.}
	\label{fig:contact_map}
\end{figure}

\begin{figure}[h]
	\begin{center}
		\includegraphics[width=0.25\linewidth]{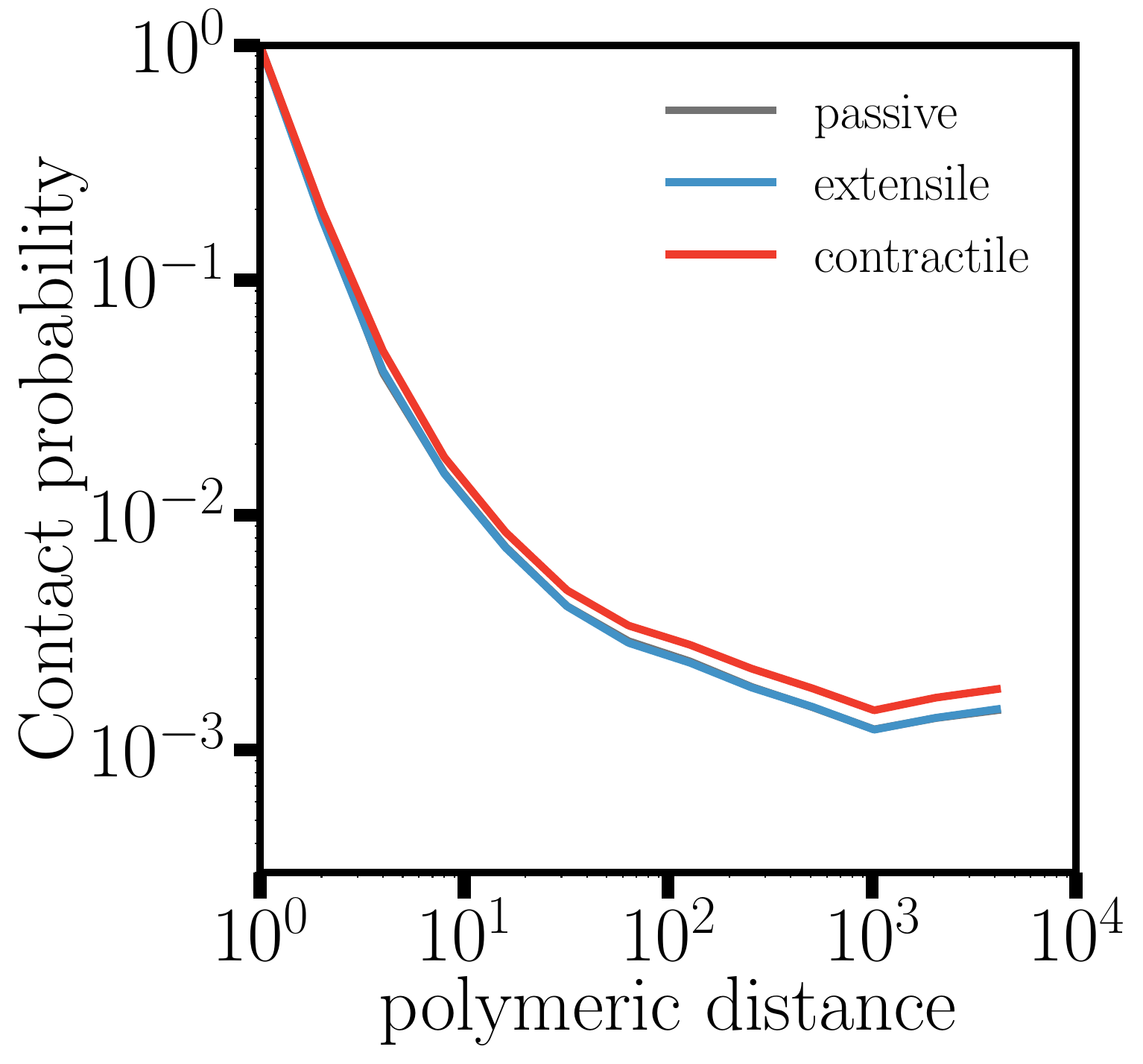}
		\includegraphics[width=0.25\linewidth]{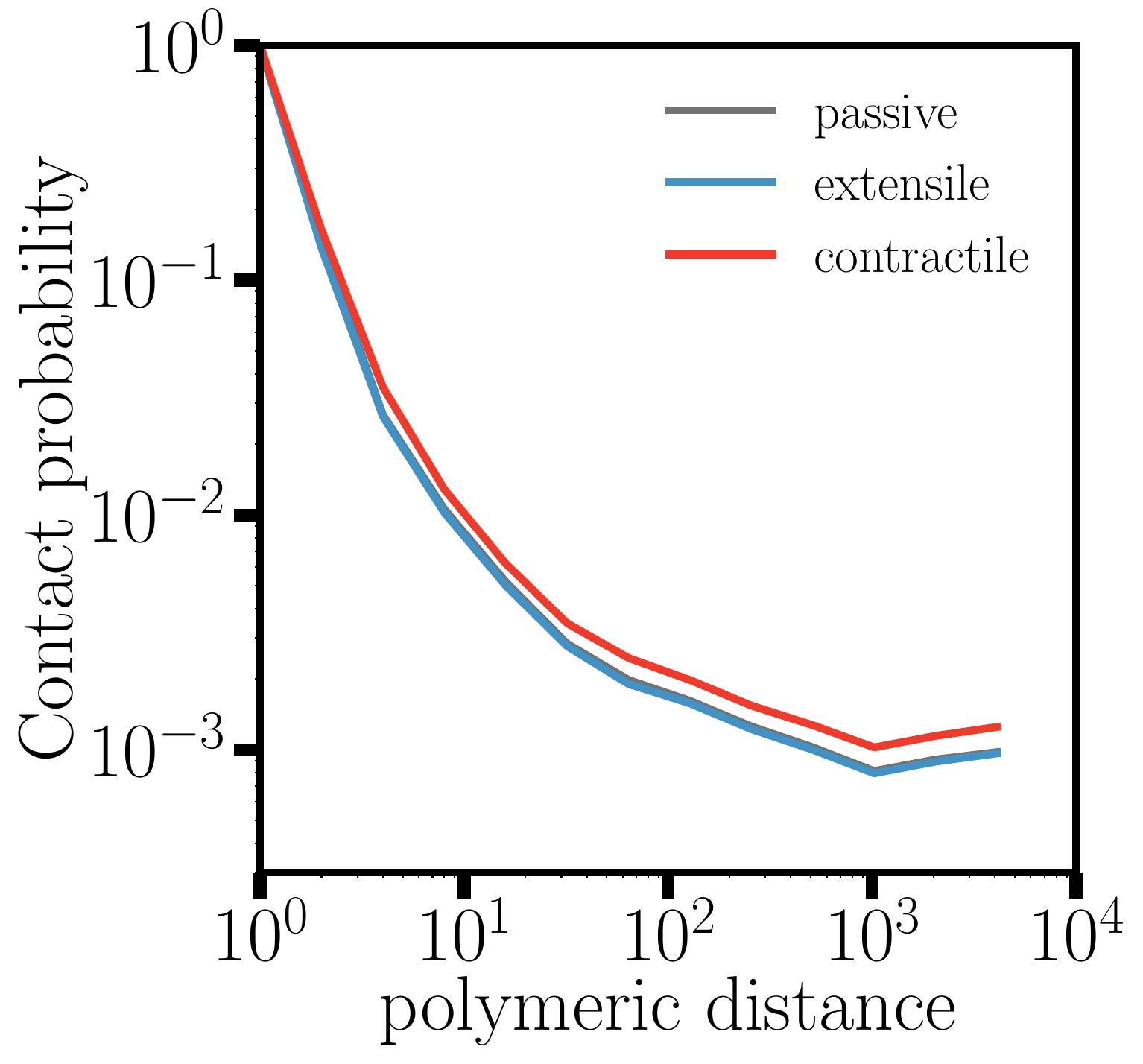}
	\end{center}
	\caption{Self-contact probability for the hard shell case (left) and the soft shell case (right) with the latter corresponding to right figure in previous Fig.~\ref{fig:contact_map}.}
	\label{fig:contact_prob}
\end{figure}

\begin{figure}[h]
	\begin{center}
		\includegraphics[width=0.25\linewidth]{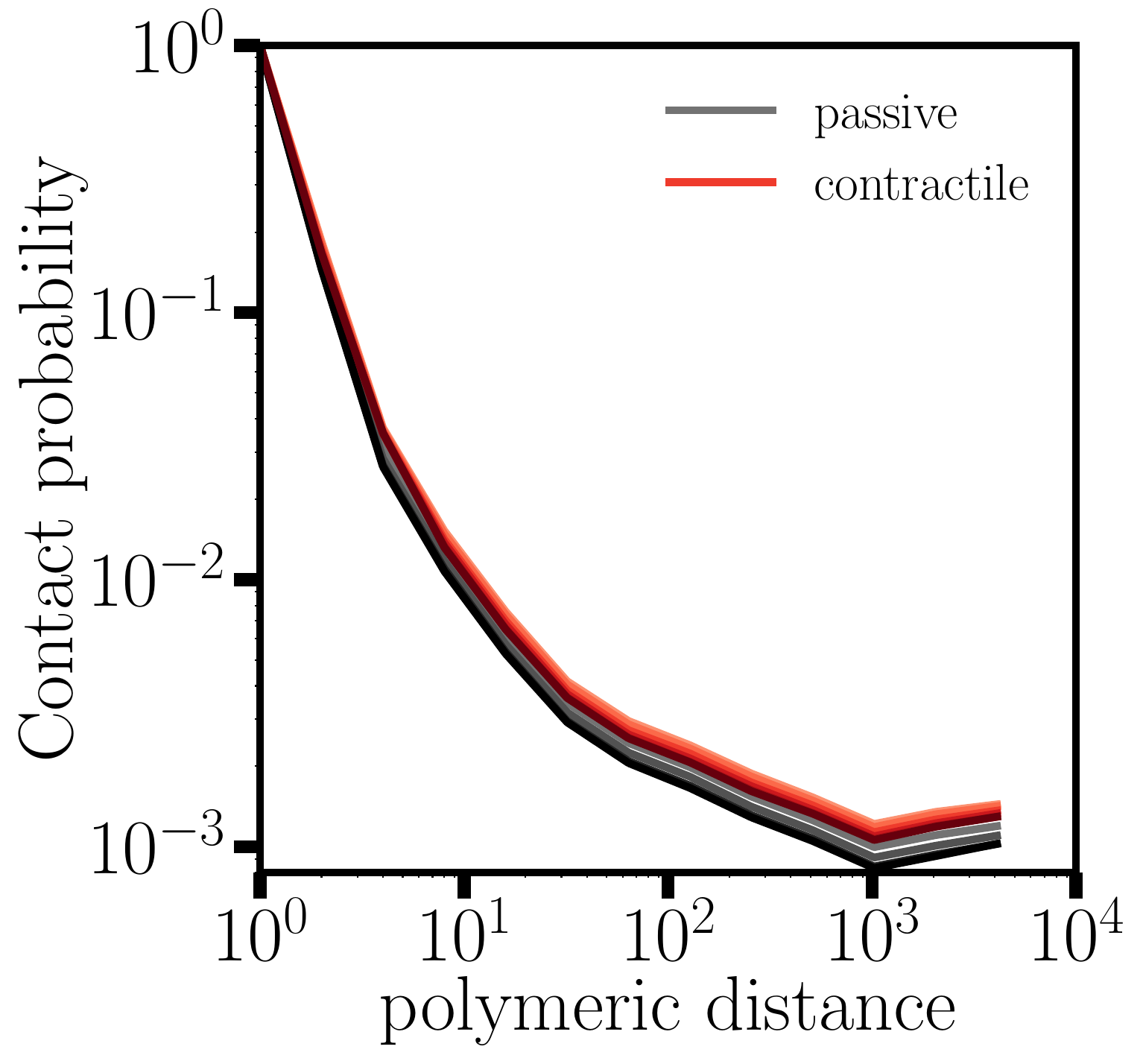}
		\includegraphics[width=0.25\linewidth]{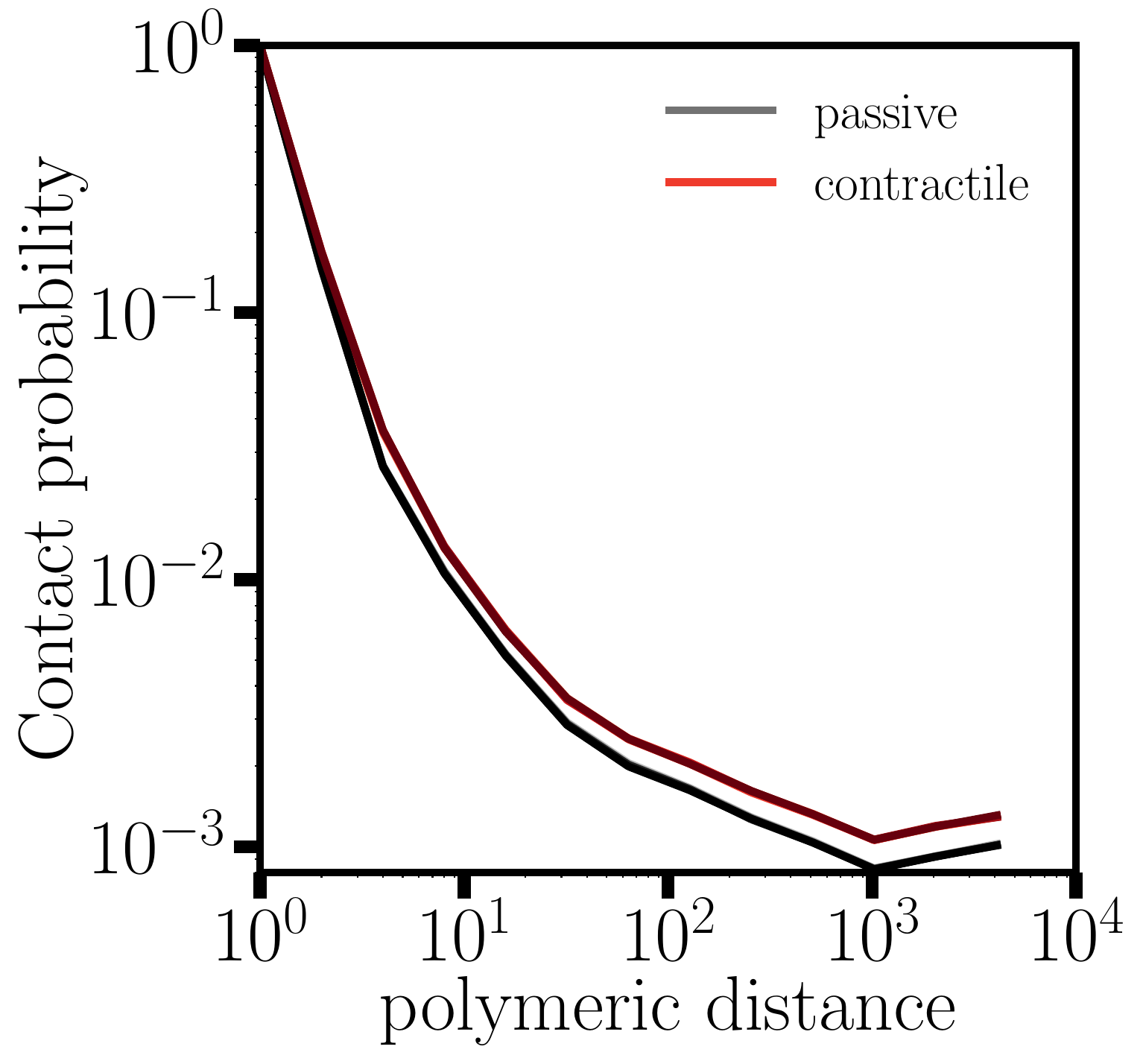}
	\end{center}
	\caption{Self-contact probability at $\tau=0,1,2,10,20,50\,\tau$ (left) and for $\tau=50$ and $\tau=300$ (right) for soft shell passive and contractile systems with $N_C=2500$ and $N_L=50$. }
	\label{fig:contact_prob2}
\end{figure}

\subsection{Mean-Squared Displacement}
To quantify the dynamics of the chain, we compute its mean-squared displacement (MSD) measured with respect to the center of mass of the shell. Fig. \ref{fig:MSD} plots the MSD of the chain during the duration of the simulation. At short time scales, the chain undergoes sub-diffusive motion and the MSD follows an exponent around $\alpha\approx0.6$ for $N_C=2500$ and $N_L=50$. At longer time scales, the MSD crosses over to a smaller exponent. The value of the exponent depends on $N_C$ and $N_L$. In all cases, the active systems diffuse faster than the passive system, and contractile motors enhance diffusion more than extensile motors. The insets in Fig.~\ref{fig:MSD} show the MSD for the center of mass of the chromatin chain for the soft shell.  For the crosslinked, active chain, this MSD is slightly faster than diffusive. 

\begin{figure*}[ht]
	\begin{center}
		\includegraphics[width=0.19\linewidth]{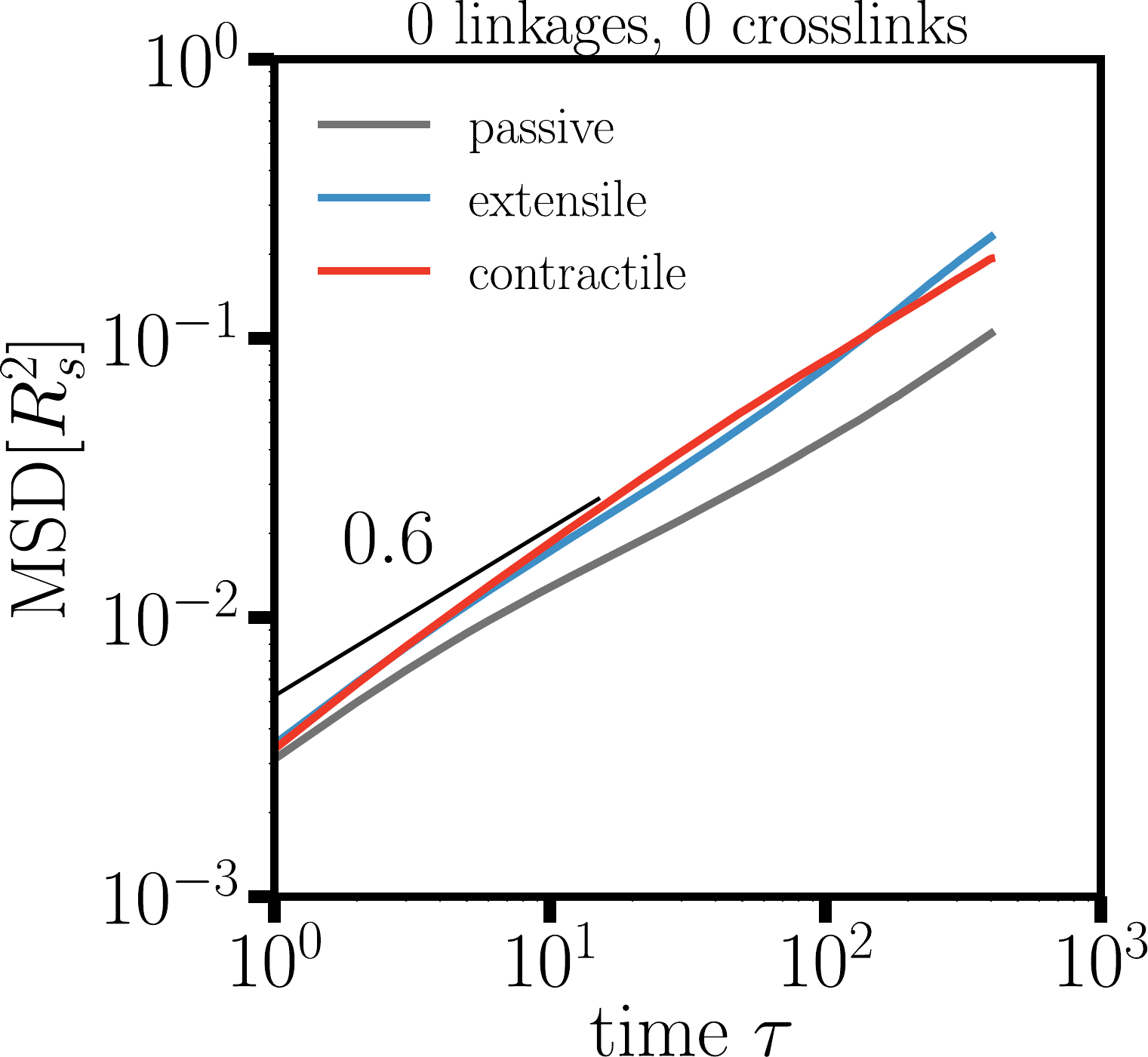}
		\includegraphics[width=0.19\linewidth]{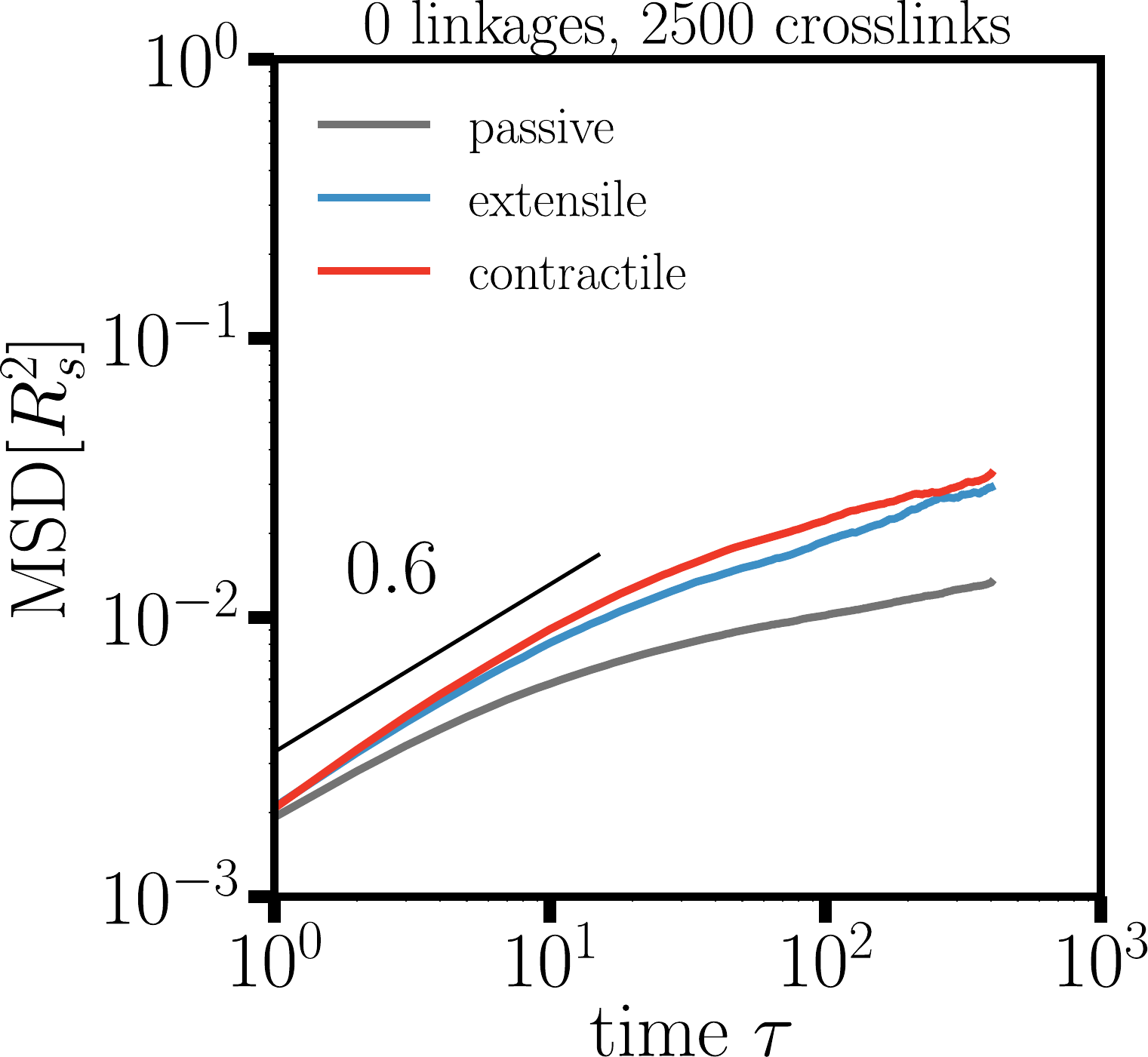}
		\includegraphics[width=0.19\linewidth]{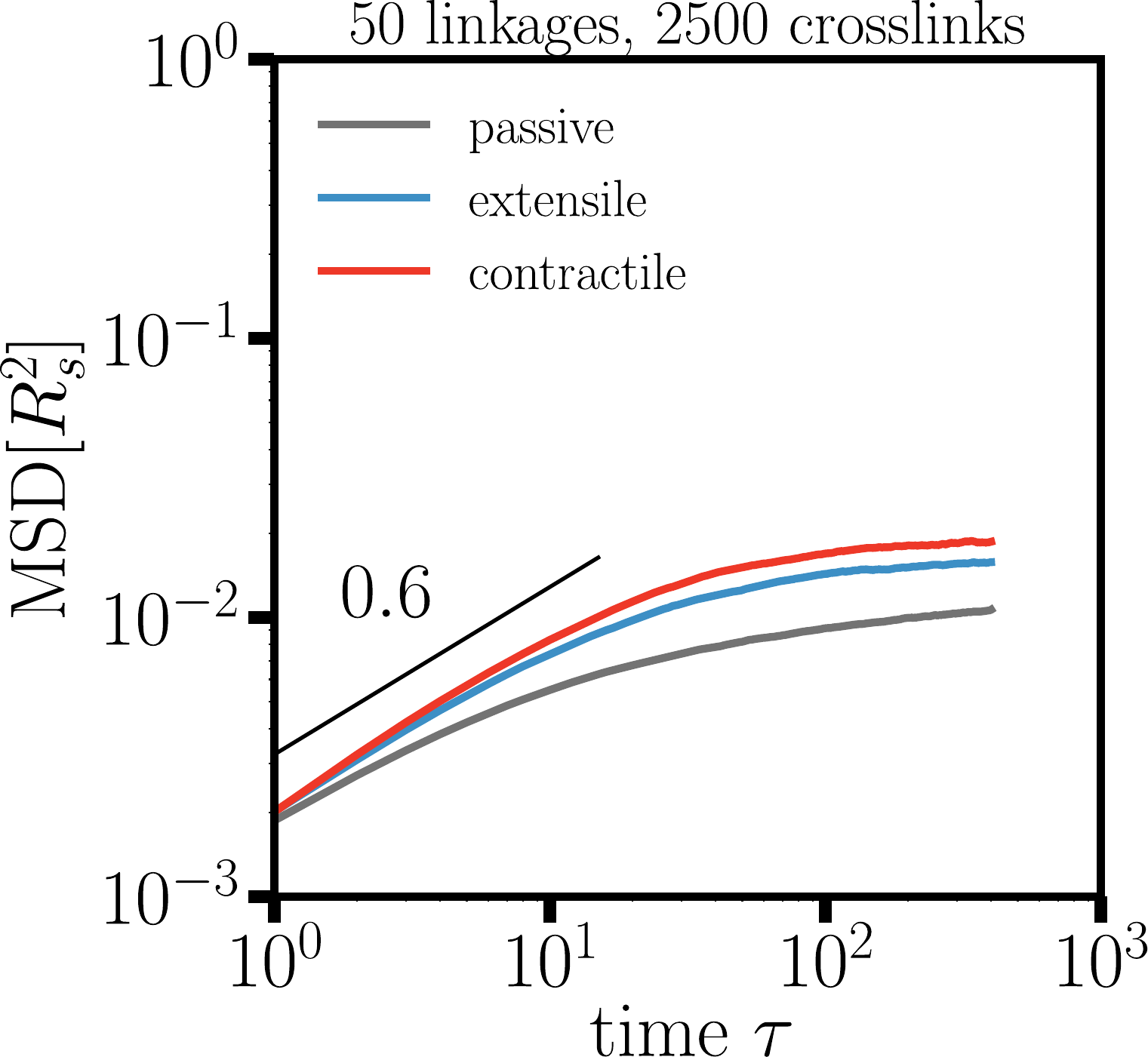}
		\includegraphics[width=0.19\linewidth]{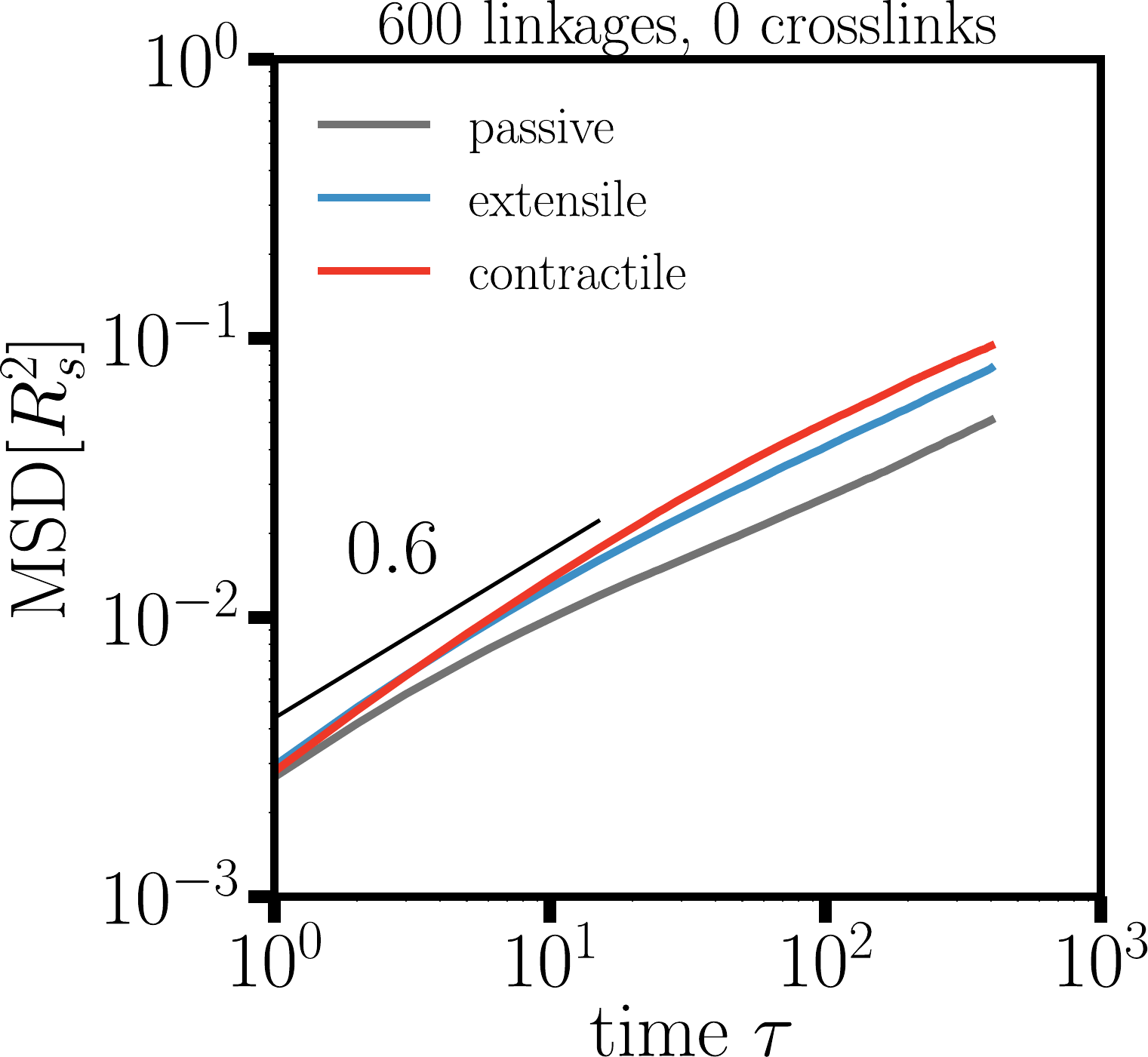}
		\includegraphics[width=0.19\linewidth]{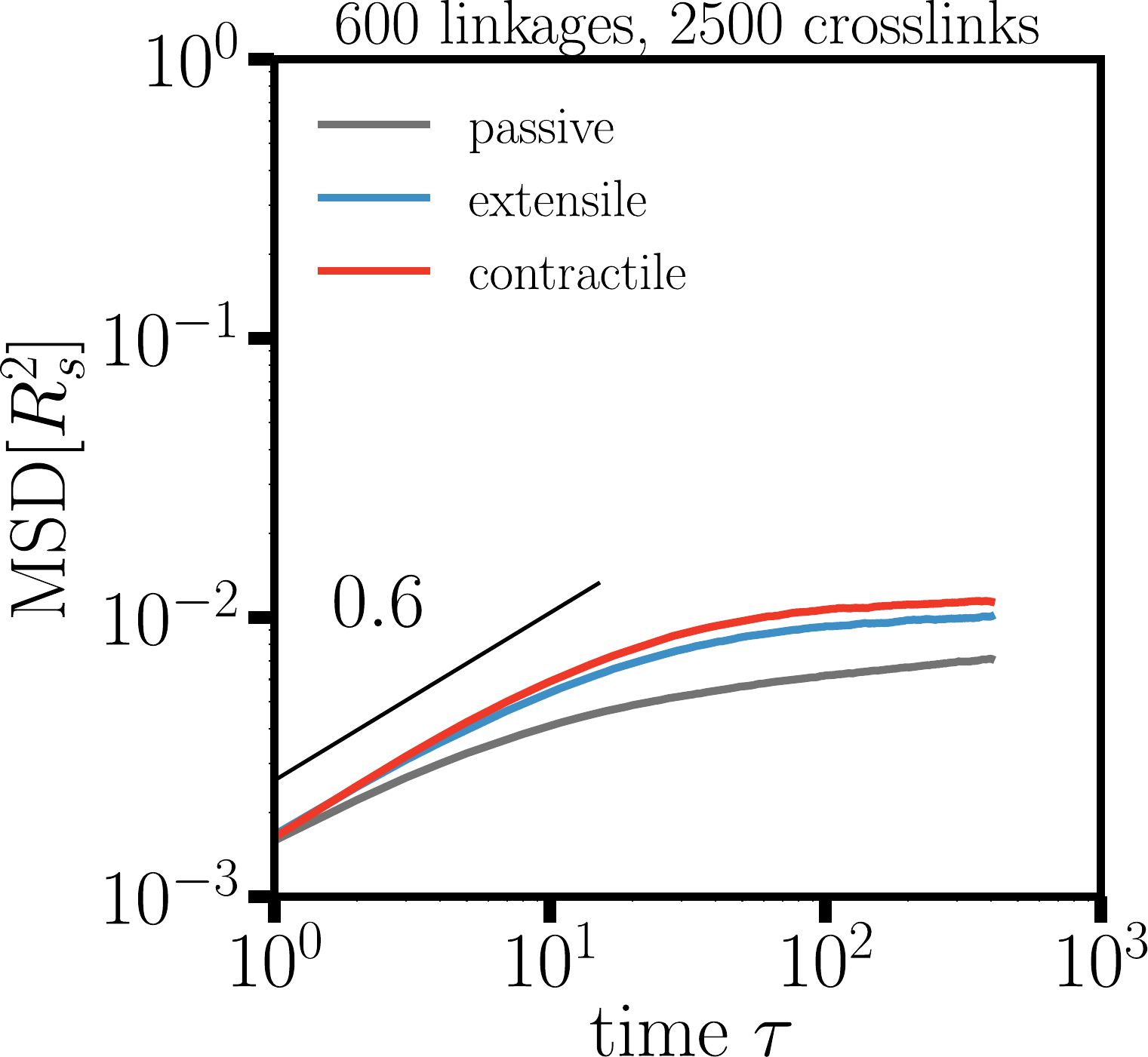}
		
		\includegraphics[width=0.19\linewidth]{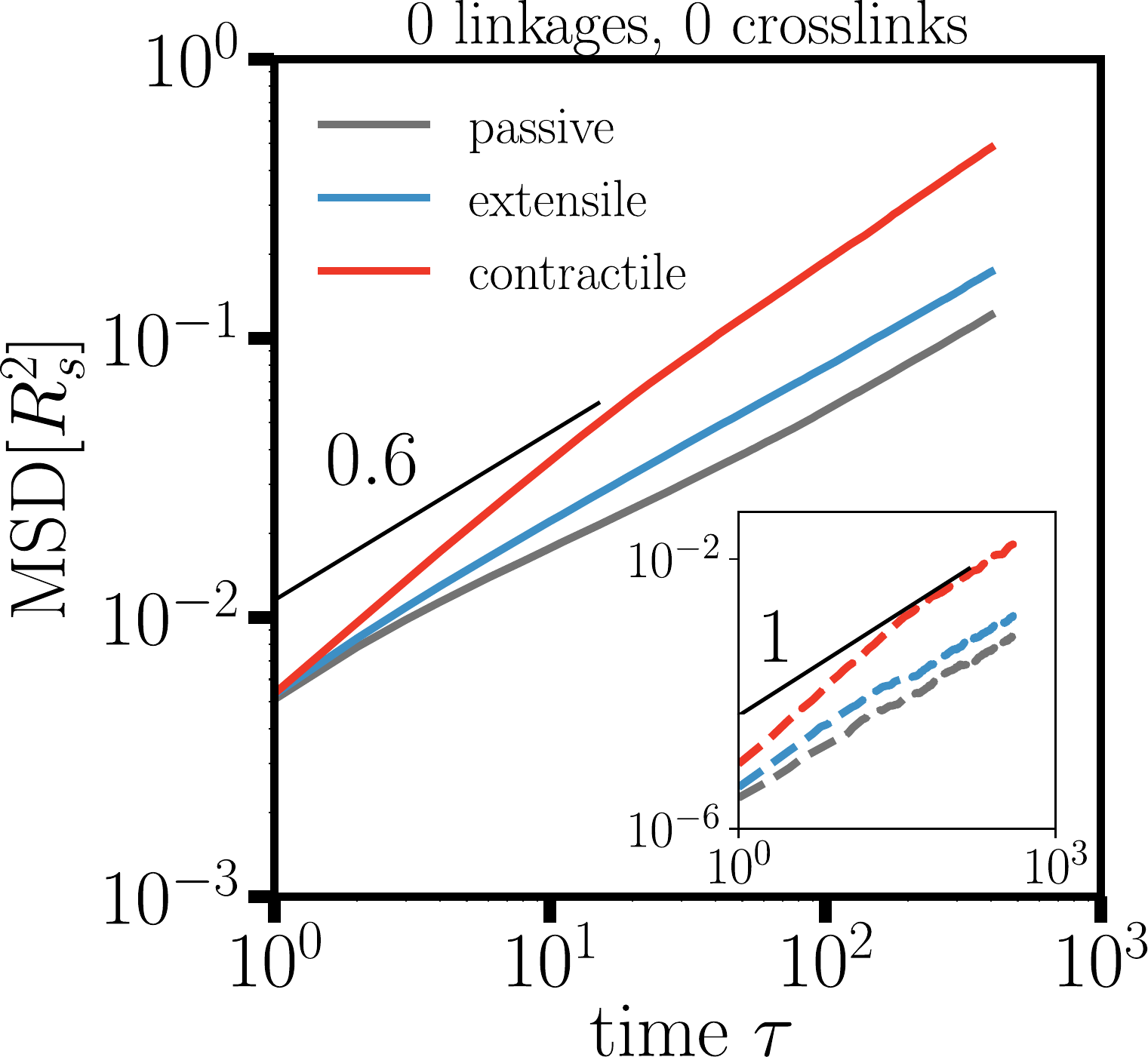}
		\includegraphics[width=0.19\linewidth]{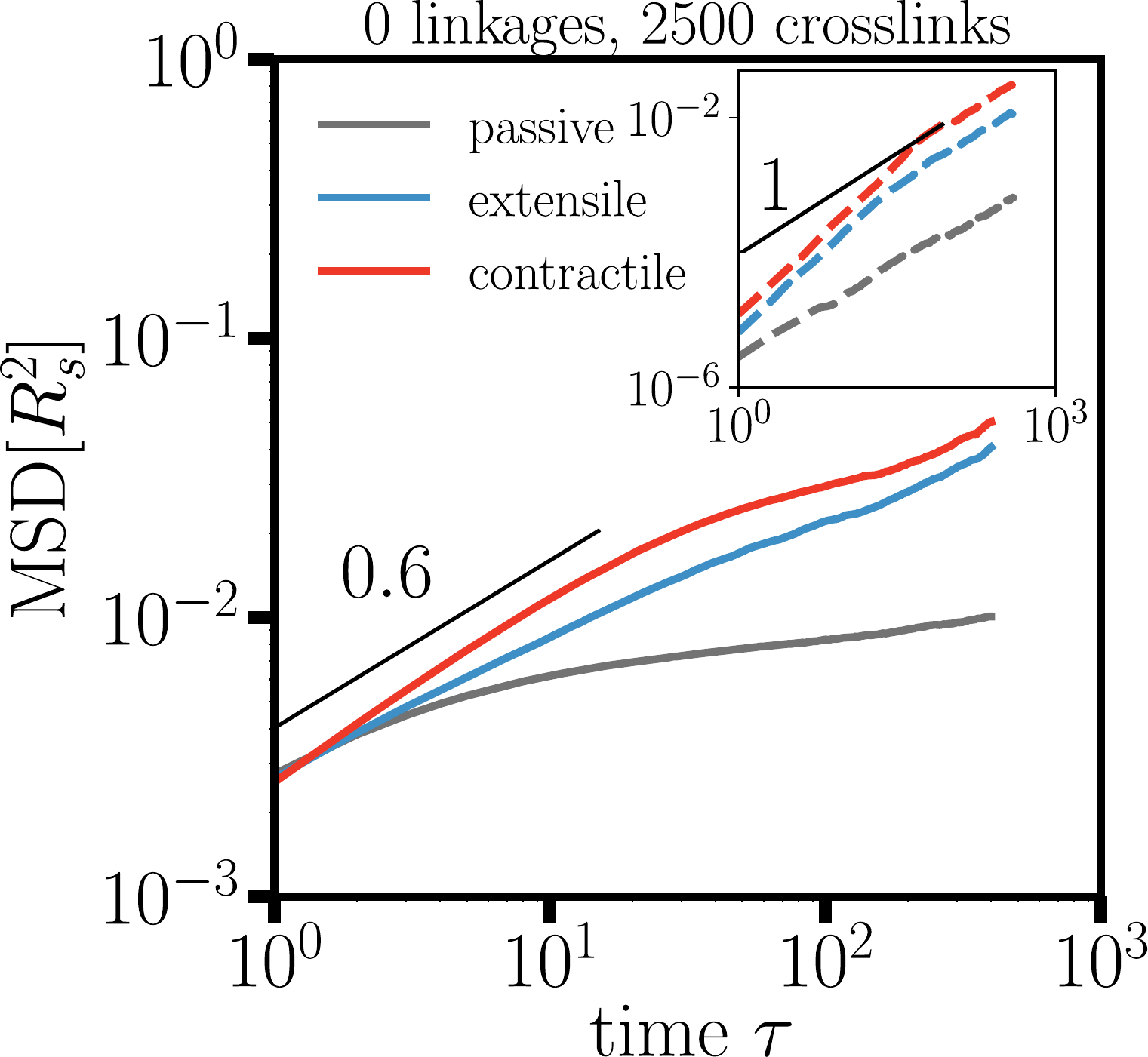}
		\includegraphics[width=0.19\linewidth]{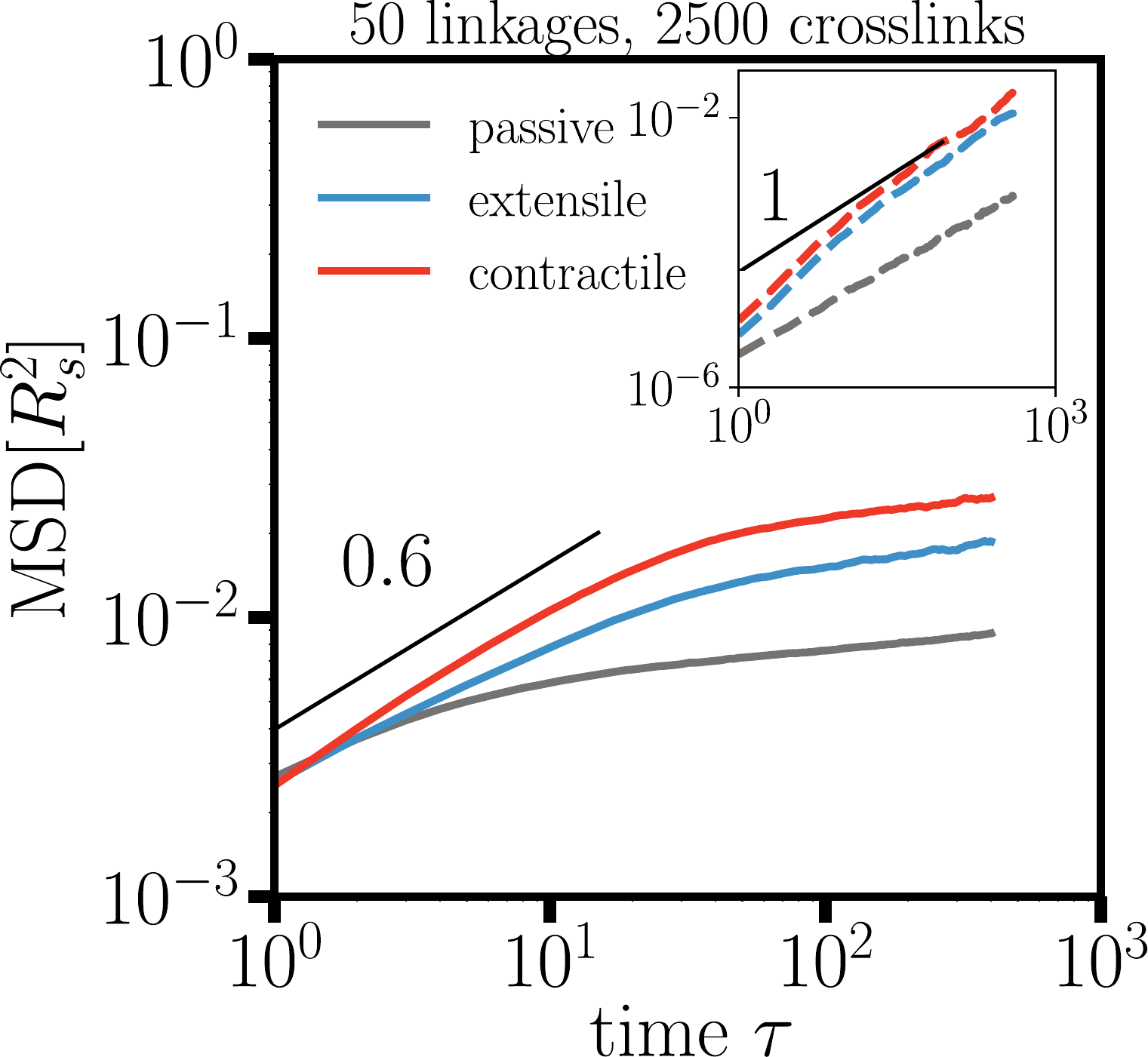}
		\includegraphics[width=0.19\linewidth]{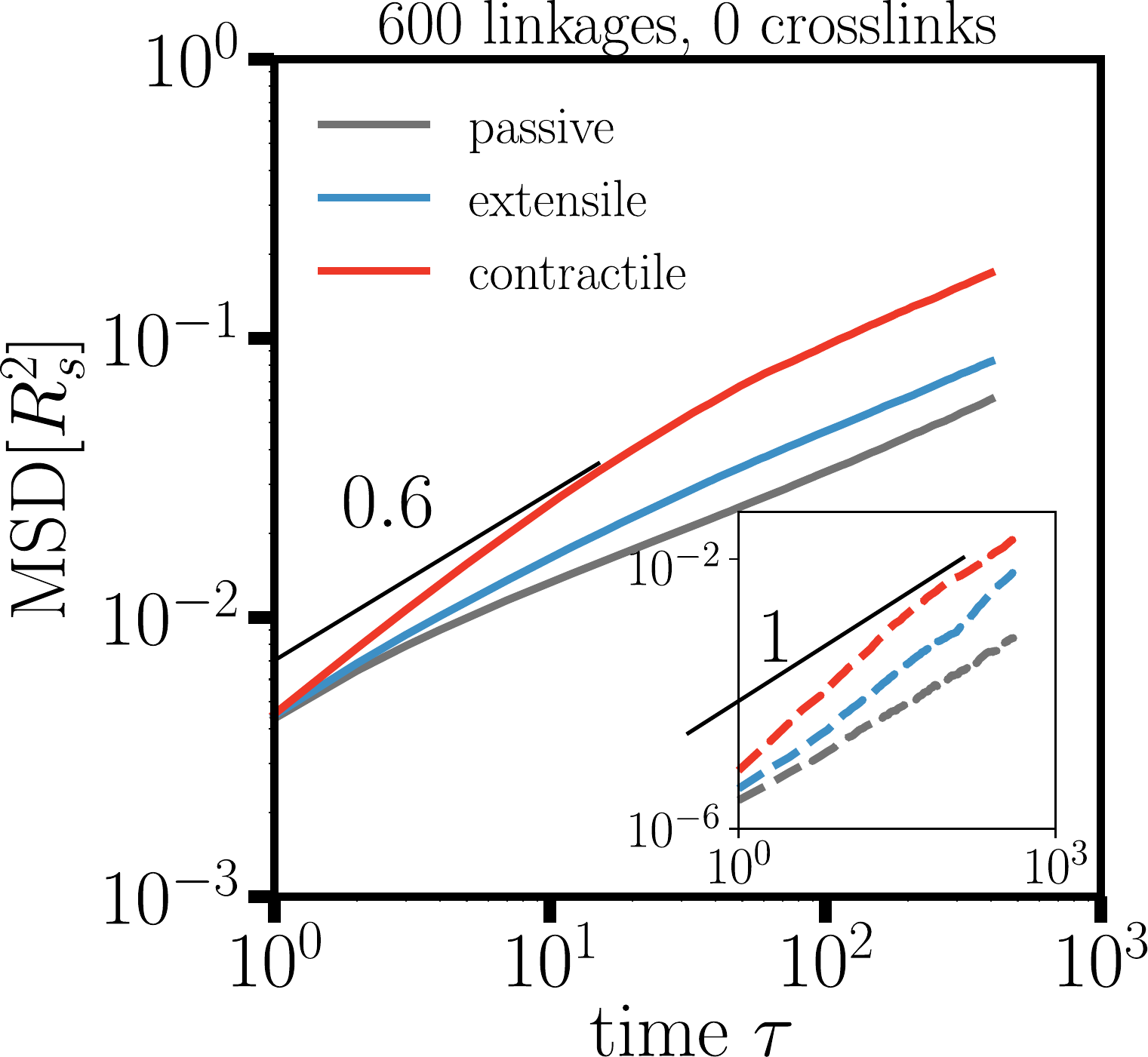}
		\includegraphics[width=0.19\linewidth]{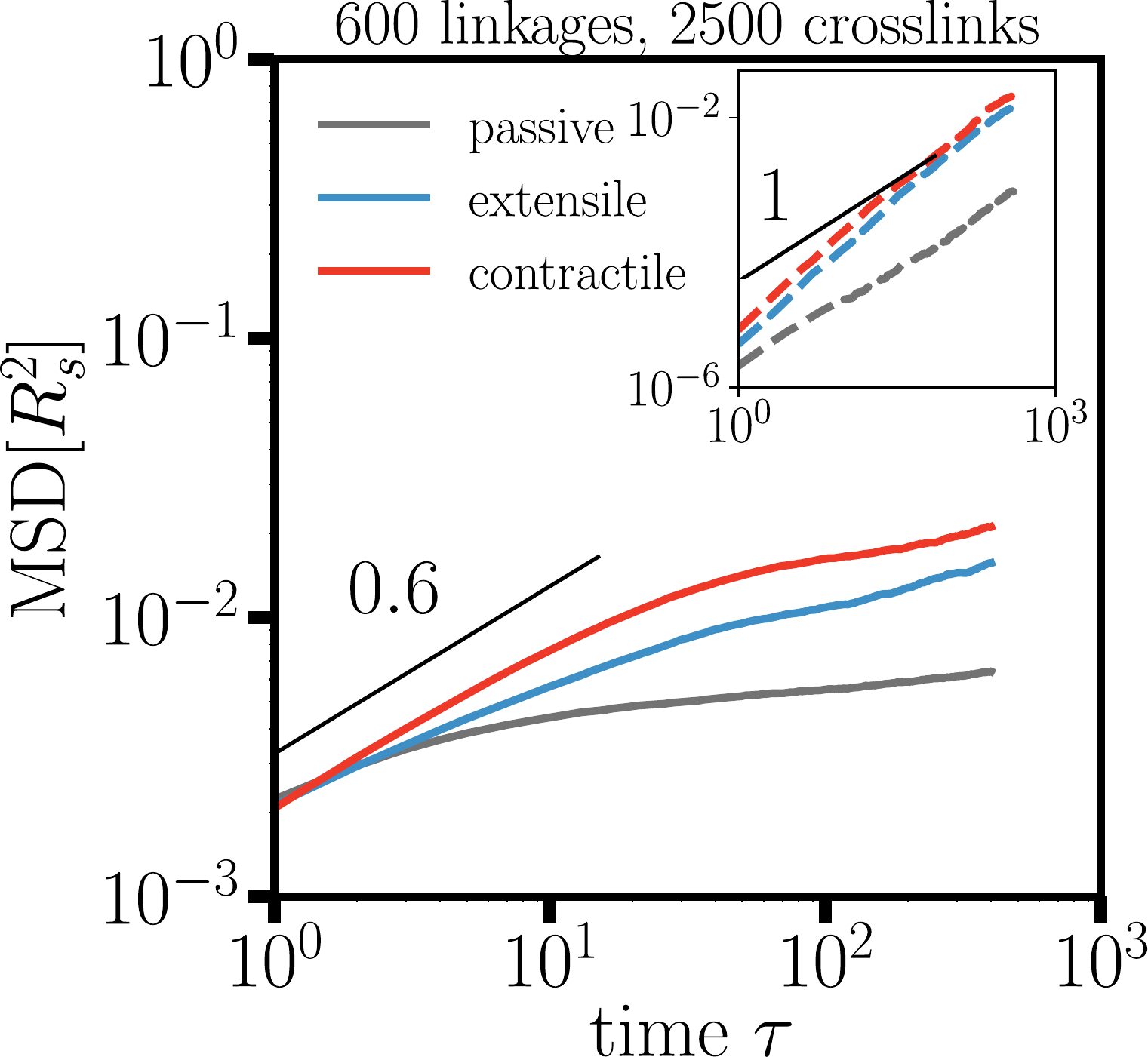}
	\end{center}
	\caption{MSD as a function of time for $N_C=2500$, $N_L=50$, and $M=5$ (middle column) and for four extreme cases (0 or 600 linkages, 0 or 2500 crosslinks) in the hard shell (top row) and the soft shell (bottom row). Insets are MSD plots of the center of mass of the chain.}
	\label{fig:MSD}
\end{figure*}

\subsection{Density fluctuations}
The density fluctuations are computed in the following way:
\begin{itemize}
 \item Select a spherical region in the system with radius $r_d$ and
   count the number of monomers in that region. 
 \item Randomly select spherical regions at other places with the same
   radius and count the monomers included.
 \item Compute the variance of counted monomer amount $\sigma ^ 2$ for this radius $r_d$.
 \item Vary $r_d$ and repeat the above three steps and obtain the variance for each $r_d$. 
\end{itemize}

We plot $\sigma ^ 2$ as a function of $r_d$. Typically, for a group of randomly distributed monomers in three dimensions, the density fluctuations scale as $\sigma ^2 \sim r_d^{-3} $. From Fig. S7 we see that the overall density fluctuations are broader in the active cases, as compared to the passive cases. Contractile motors induce more anomalous density fluctuations, particularly in the soft shell case.

\begin{figure*}[hb]
	\begin{center}
		\includegraphics[width=0.19\linewidth]{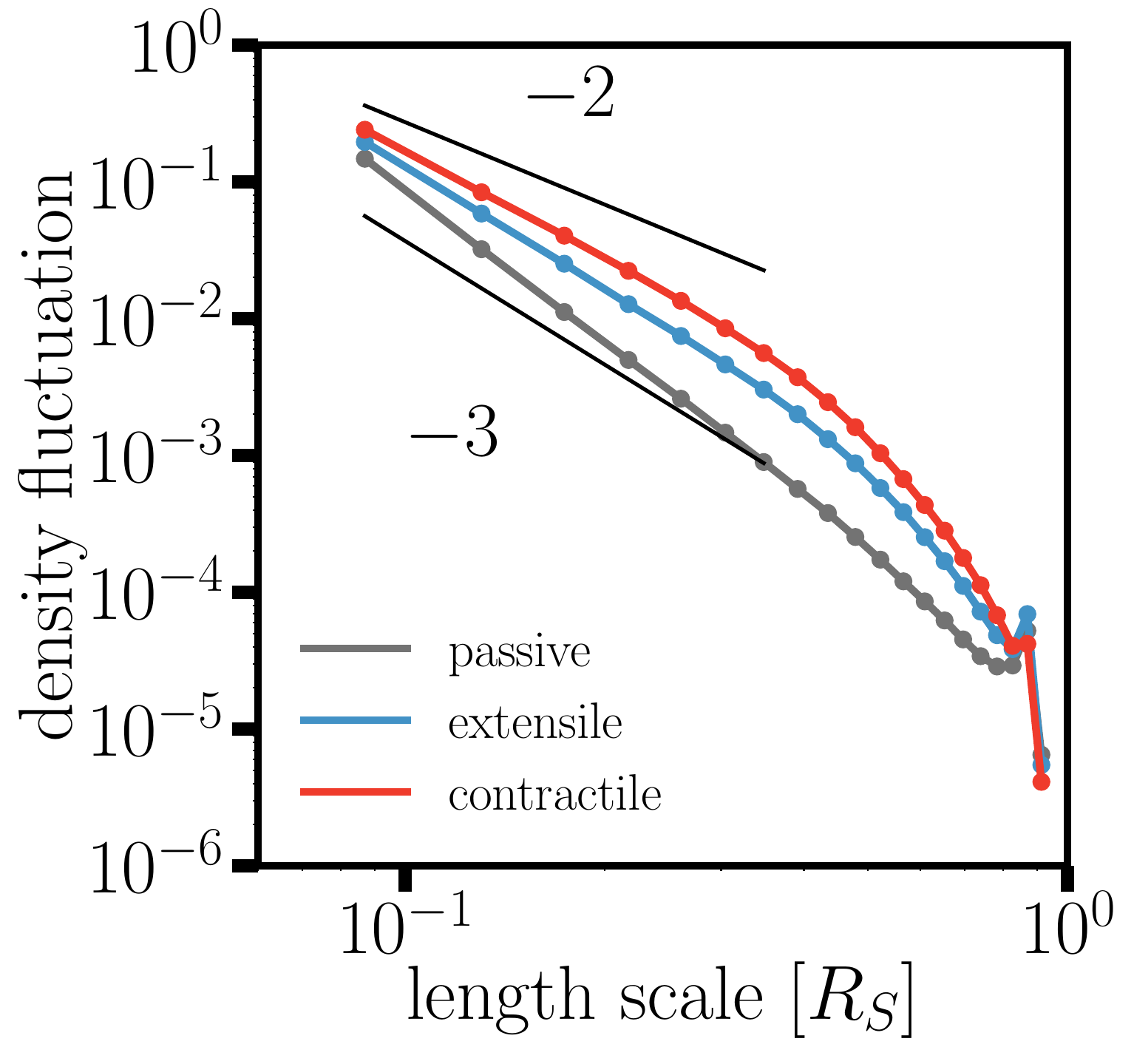}
		\includegraphics[width=0.19\linewidth]{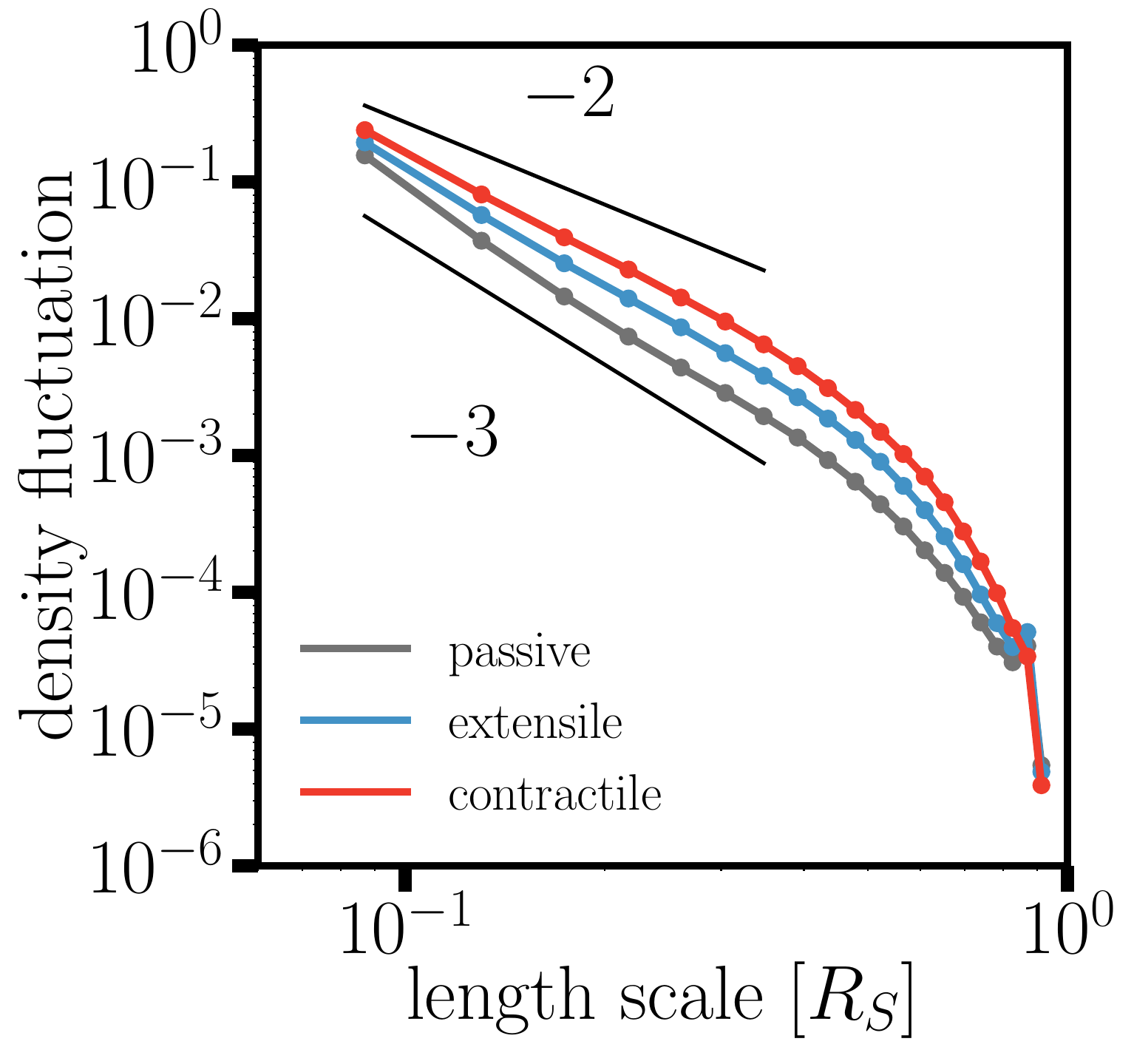}
		\includegraphics[width=0.19\linewidth]{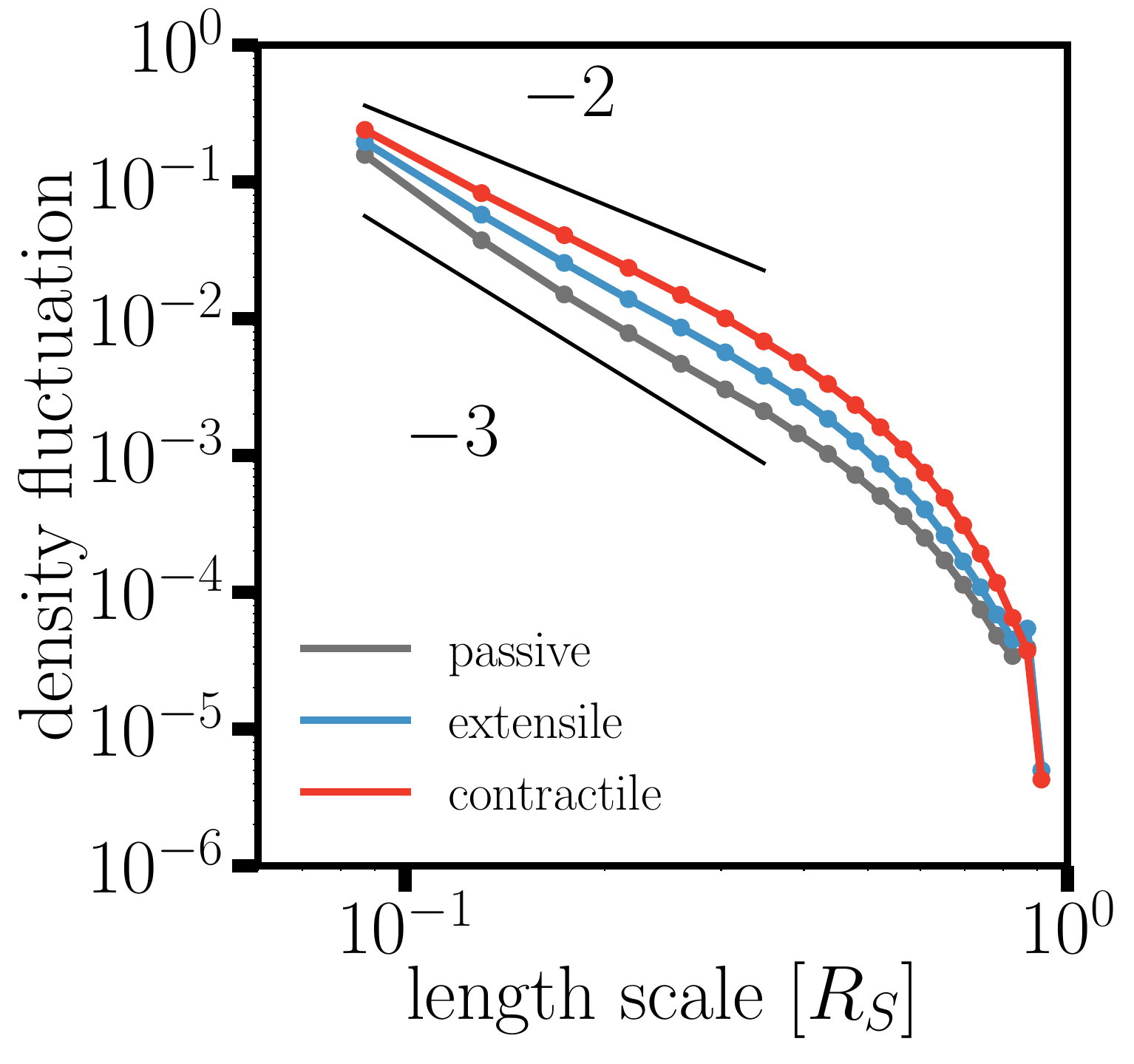}
		\includegraphics[width=0.19\linewidth]{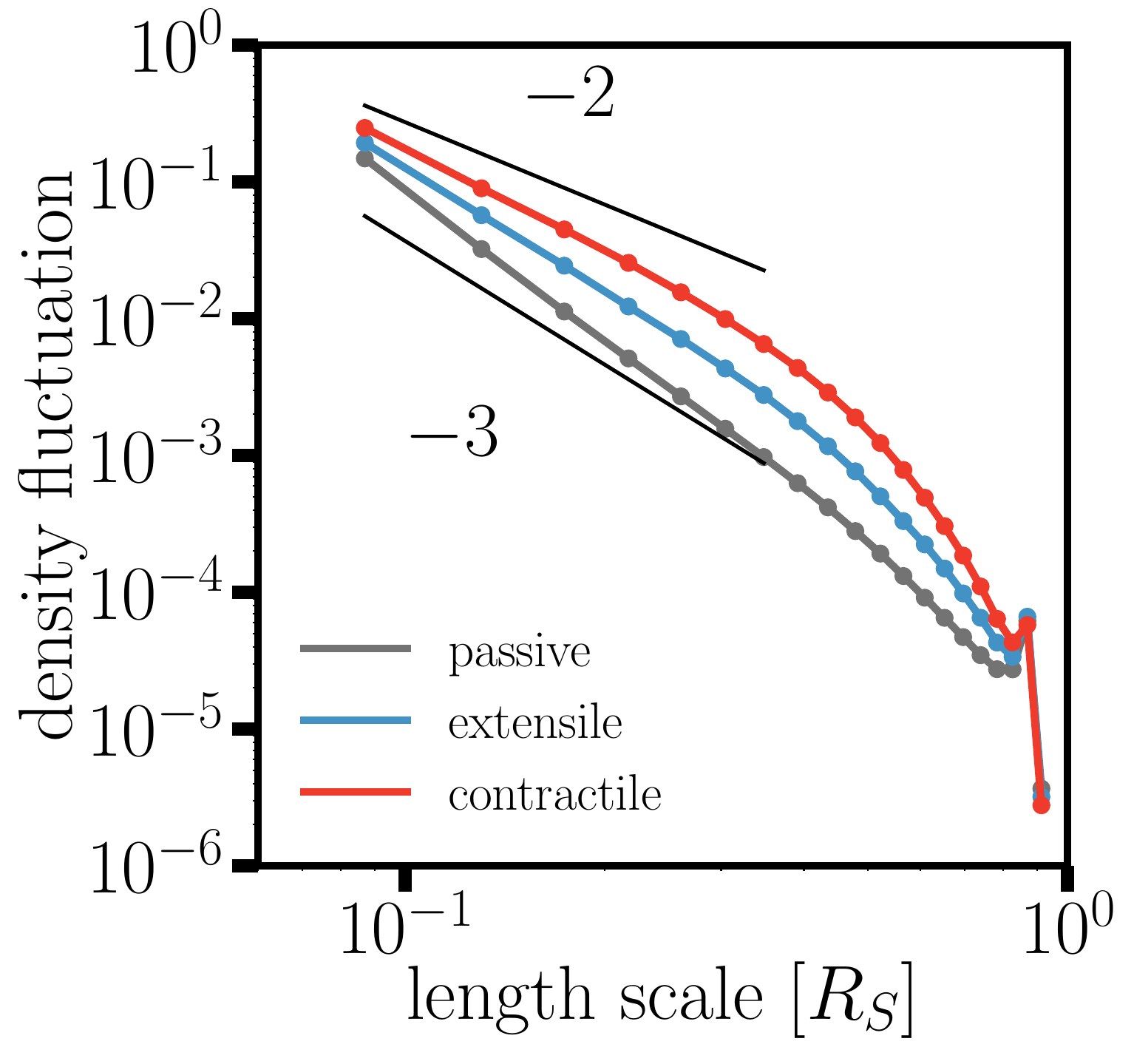}
		\includegraphics[width=0.19\linewidth]{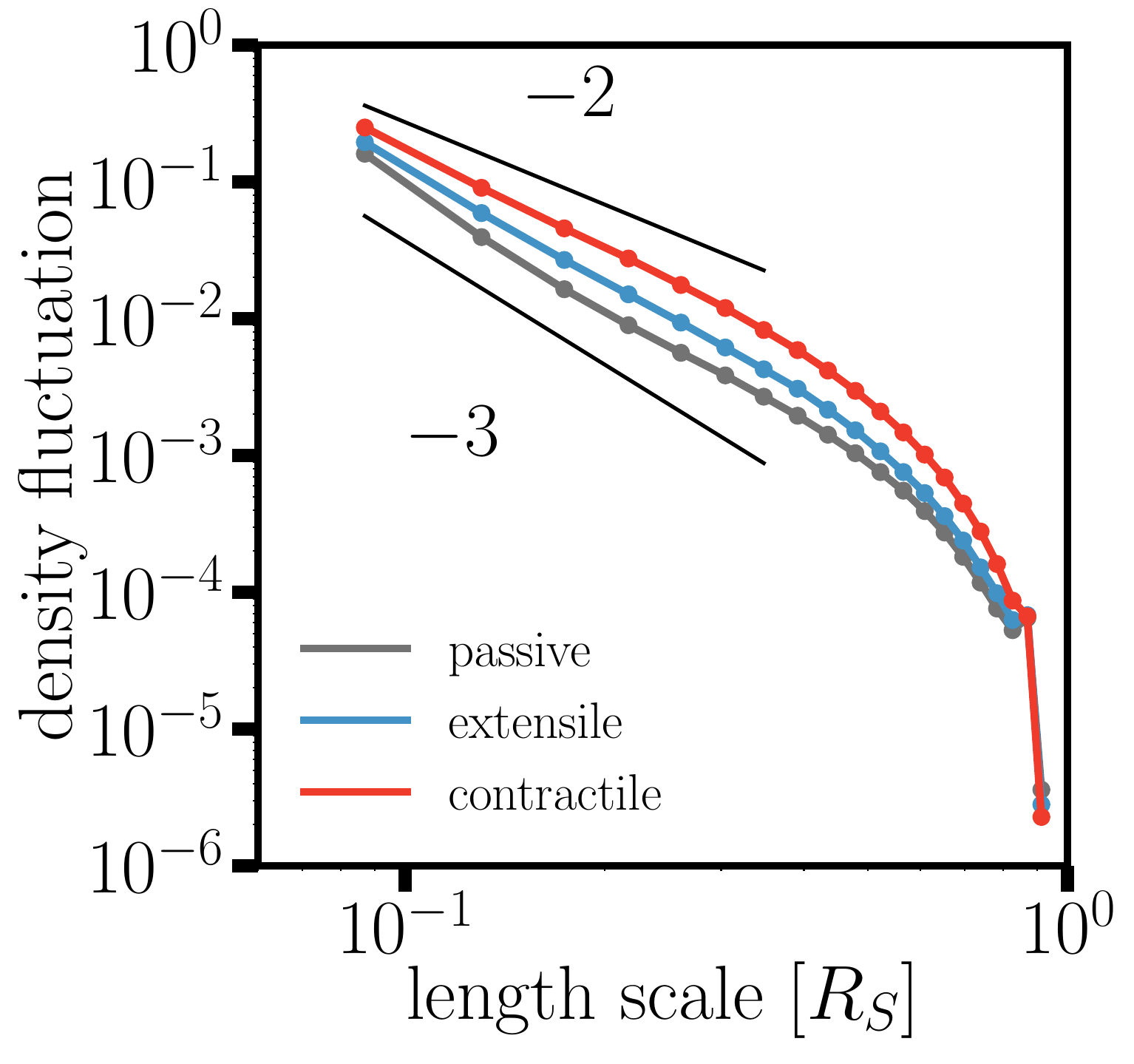}

		\includegraphics[width=0.19\linewidth]{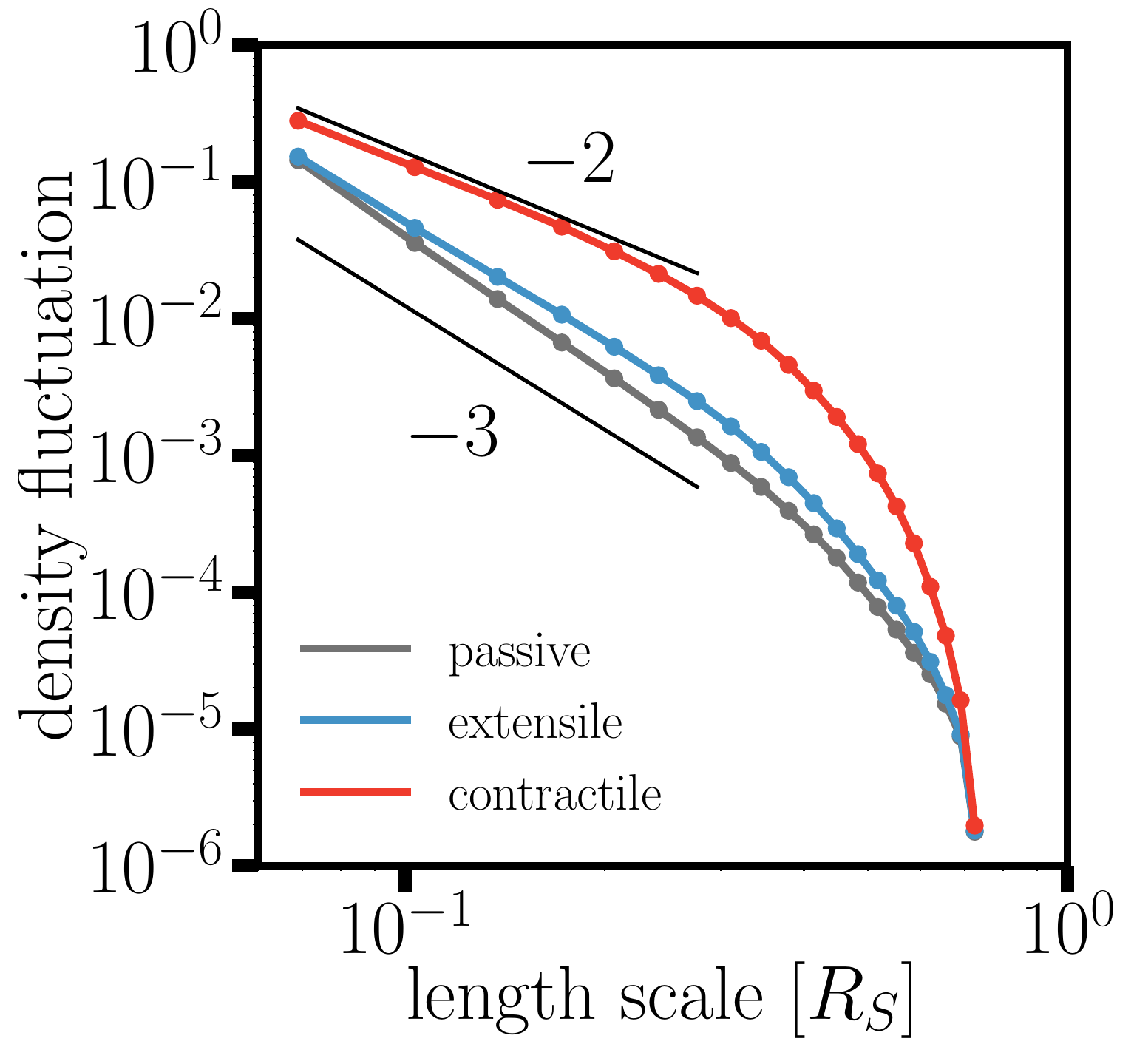}
		\includegraphics[width=0.19\linewidth]{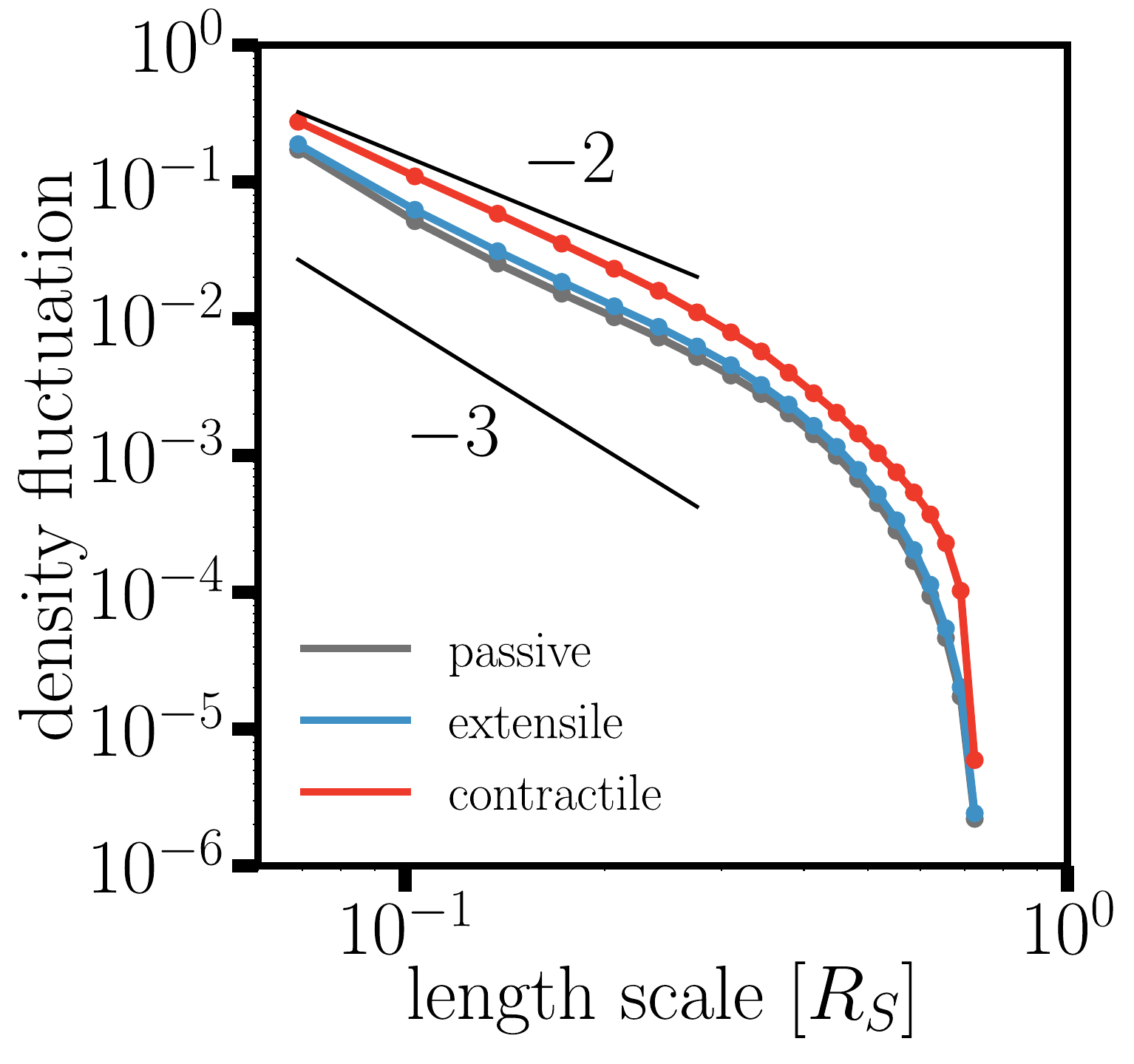}
		\includegraphics[width=0.19\linewidth]{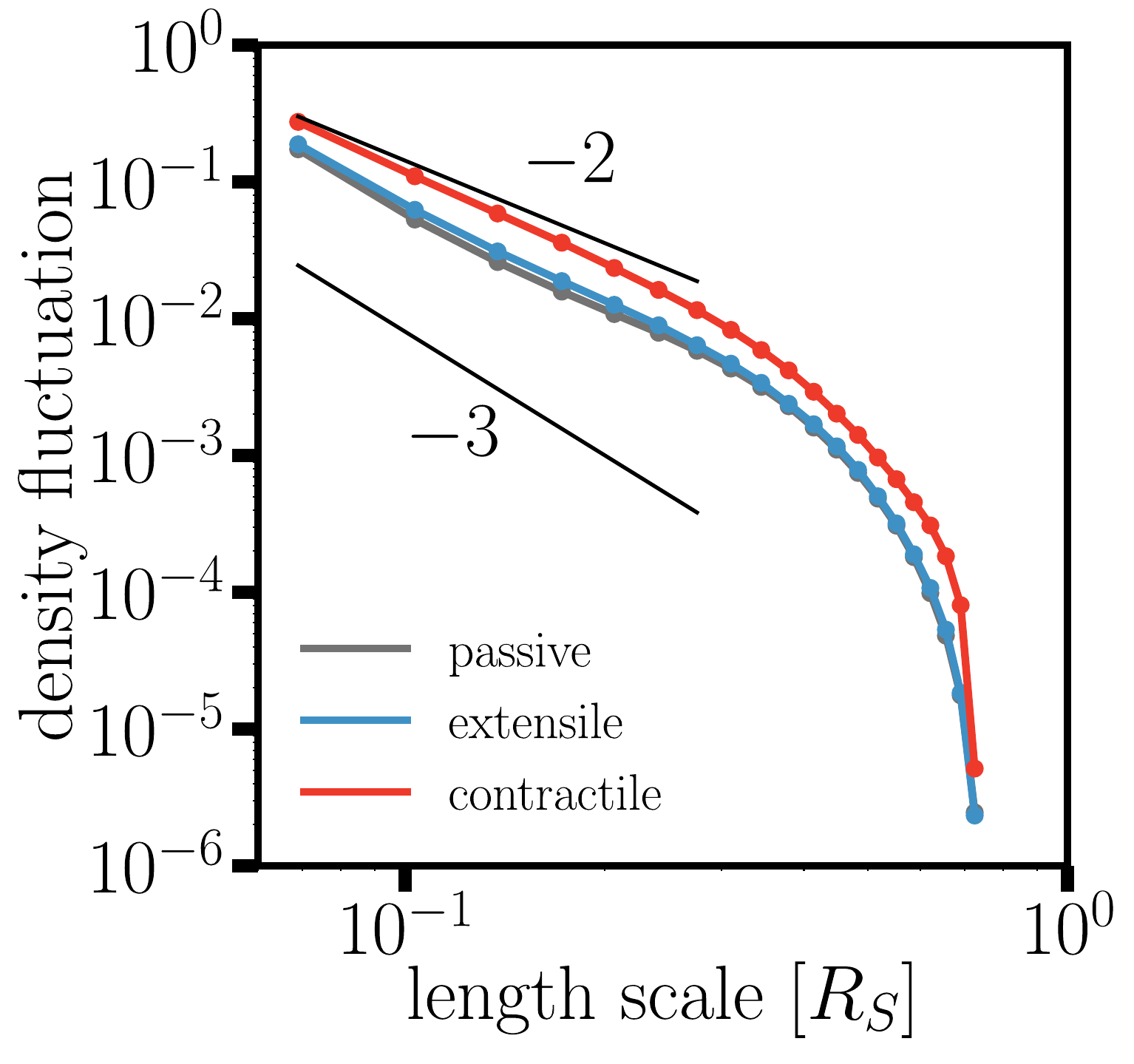}
		\includegraphics[width=0.19\linewidth]{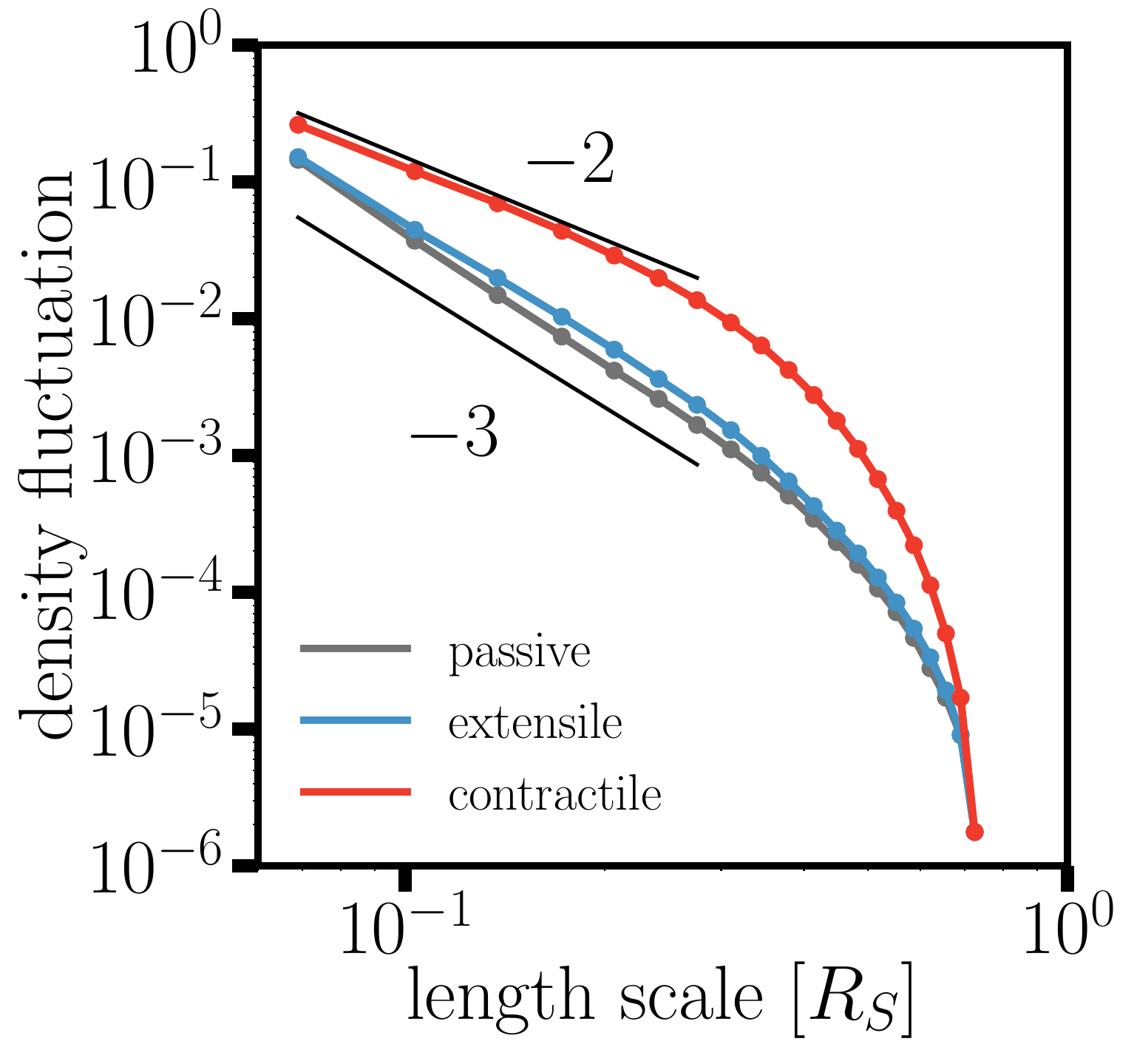}
		\includegraphics[width=0.19\linewidth]{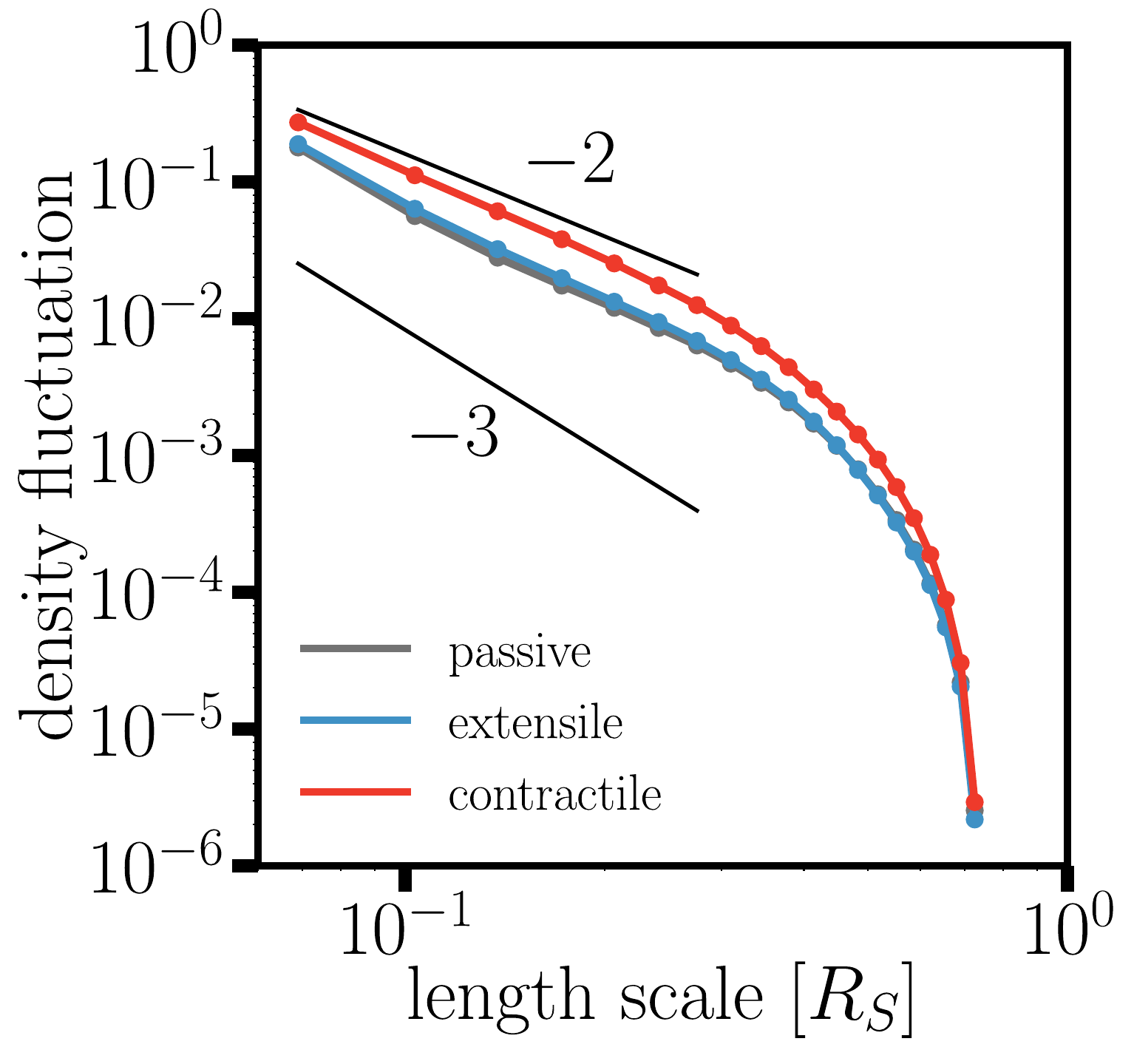}
	\caption{Density fluctuations for $N_C=2500$ and $N_L=50$ (middle column) and four extreme cases (0 or 600 linkages, 0 or 2500 crosslinks) in the hard shell (top row) and in the soft shell (bottom row). The arrangement of parameters is the same as in the previous figure.}
        \end{center}
	\label{fig:density}
\end{figure*}

\subsection{Correlation function and correlation length}
To evaluate the spatial and temporal correlation motion along the
chain, we compute the spatial autocorrelation function. Suppose $\vec d (\vec r, \Delta \tau)$ is the displacement of monomer at $\vec r$ over time, $\vec d (\vec r + \Delta r, \Delta \tau)$ is the displacement of another monomer, which is located a distance $\Delta r$ away and over the same time window. We then use the function below to compute the correlation function:
$$
C(\Delta r, \Delta \tau) = \frac{\langle \vec d (\vec r, \Delta \tau) \cdotp \vec d (\vec r + \Delta r, \Delta \tau)\rangle }{ \langle \vec d^2(\vec r, \Delta \tau) \rangle}.
$$


From Ref. ~\cite{Shaban18} we assume the correlation function follows $C_r(r)=\frac{2^{1-\nu}}{\Gamma (\nu)}\left( \frac{r}{r_{cl}}\right) ^{\nu}K_{\nu} \left( \frac{r}{r_{cl}} \right)$, where $r_{cl}$ is the extracted correlation length, $K_\nu$ is the Bessel of the second type of order $\nu$, and $\nu$ is a smoothness parameter. Larger $\nu$ denotes that the underlying spatial process is smooth, not rough, in space. In Fig.~\ref{fig:correlation_1} we show the correlated function computed from numerical simulations (dots) and the fitted correlation function from the above formula (lines) for different parameters. Lines from light to dark represent time windows from short to long ($1\,\tau$, $2\,\tau$, $5\,\tau$, $10\,\tau$, $20\,\tau$, $50\,\tau$, $100\,\tau$, $200\,\tau$ ). We see that the numerical results with shorter time windows fit the formula better.

\begin{figure*}[h]
	\begin{center}
		\includegraphics[width=0.19\linewidth]{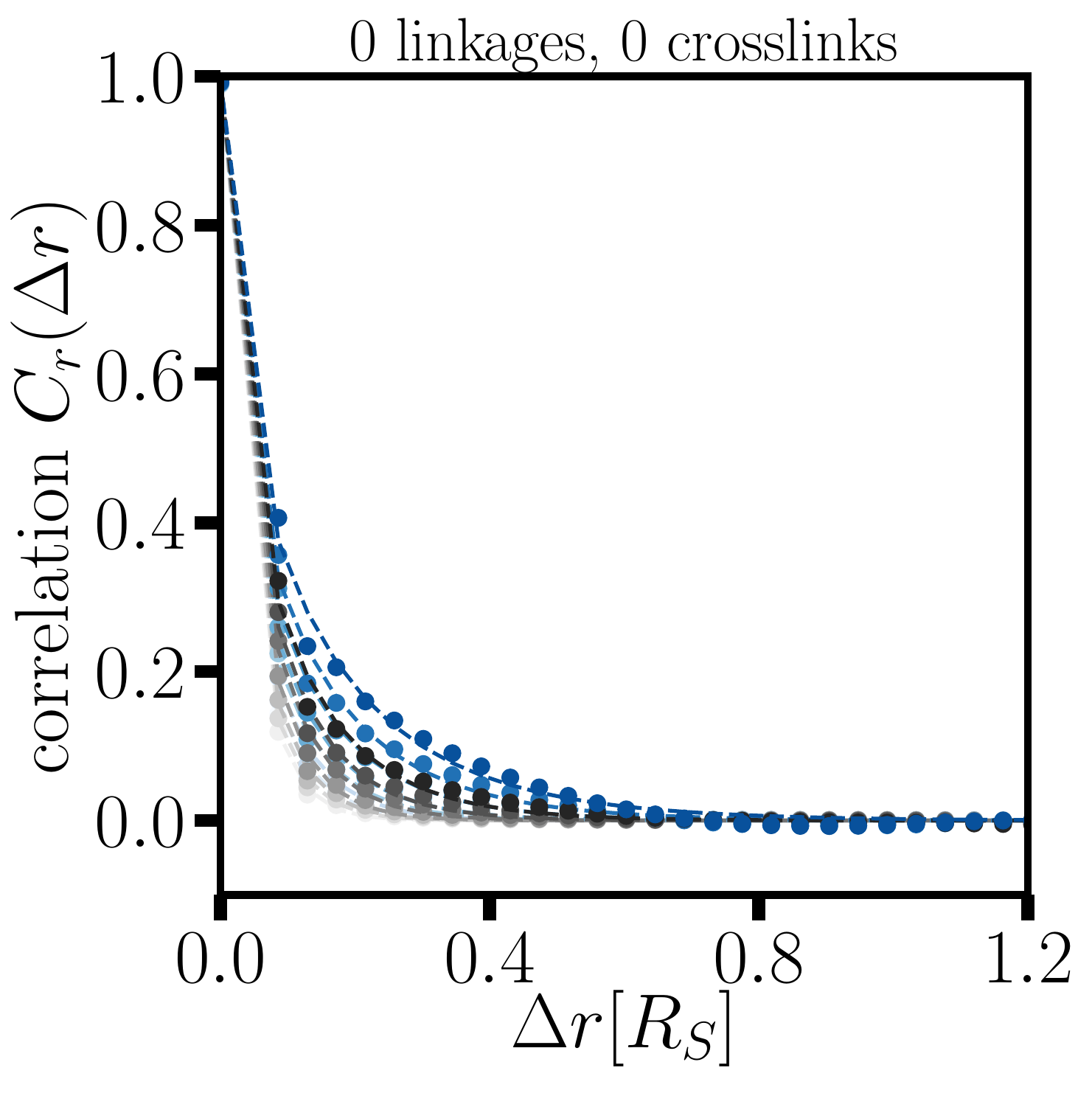}
		\includegraphics[width=0.19\linewidth]{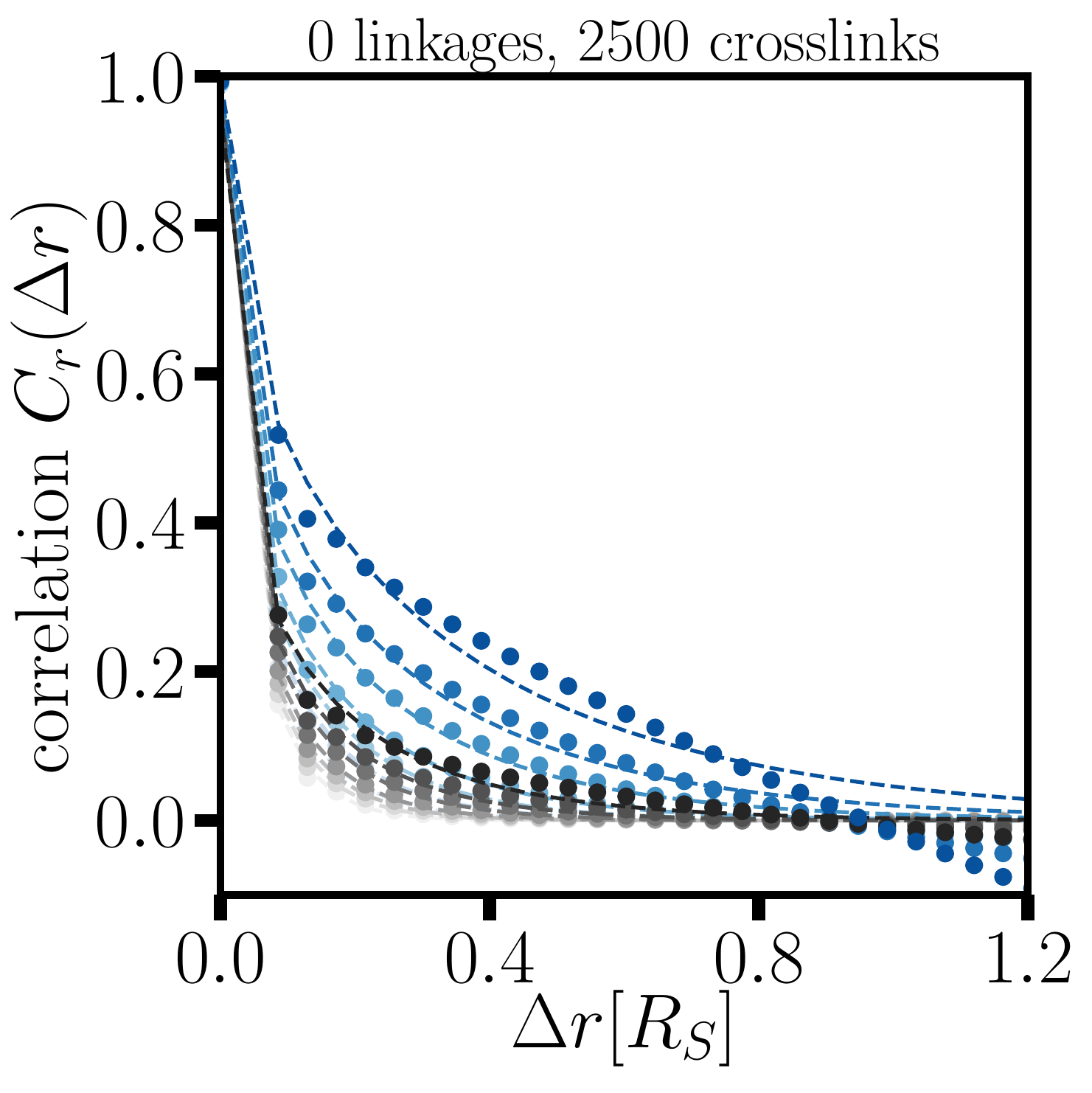}
		\includegraphics[width=0.19\linewidth]{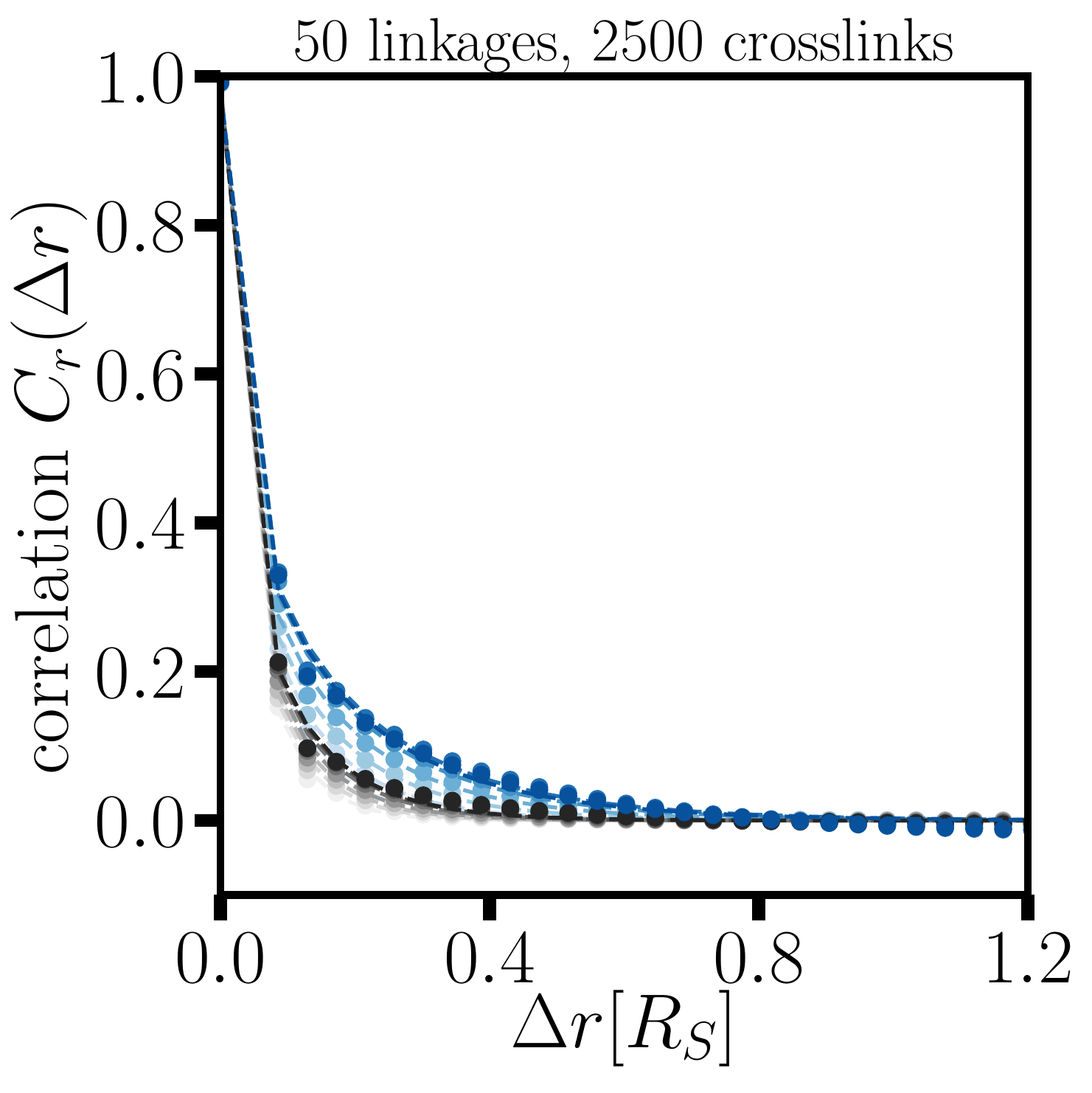}
		\includegraphics[width=0.19\linewidth]{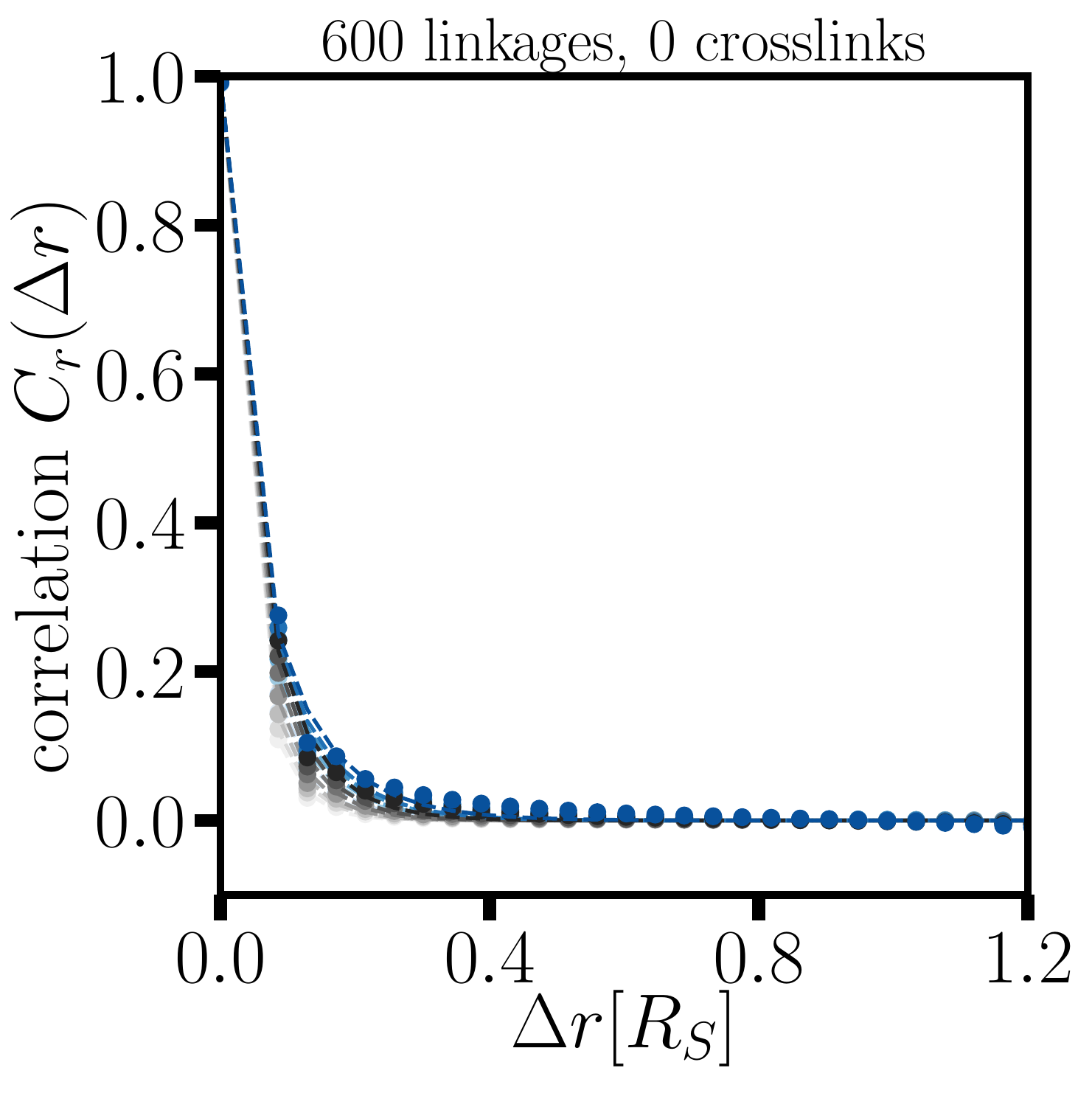}
		\includegraphics[width=0.19\linewidth]{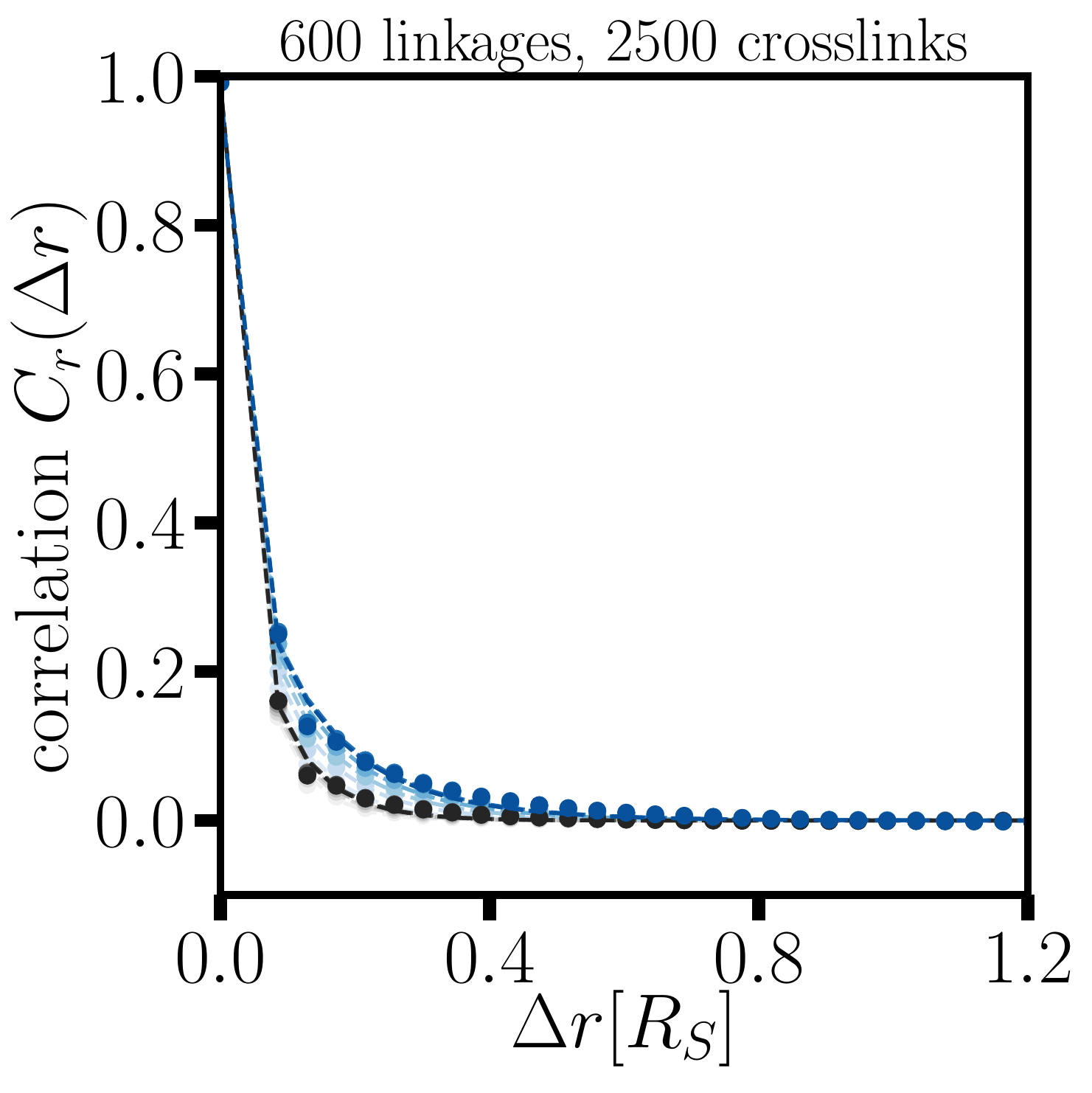}

		\includegraphics[width=0.19\linewidth]{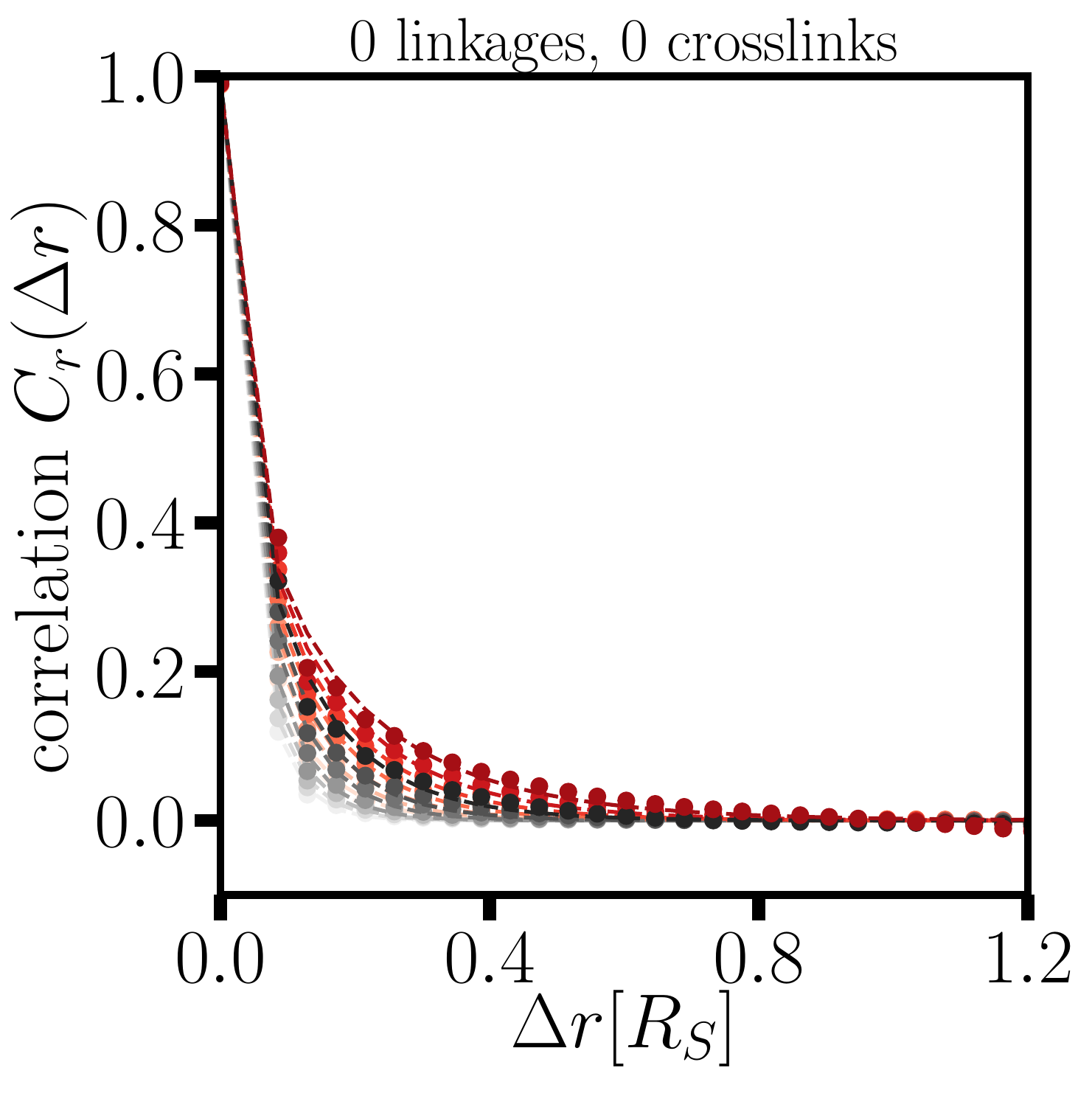}
		\includegraphics[width=0.19\linewidth]{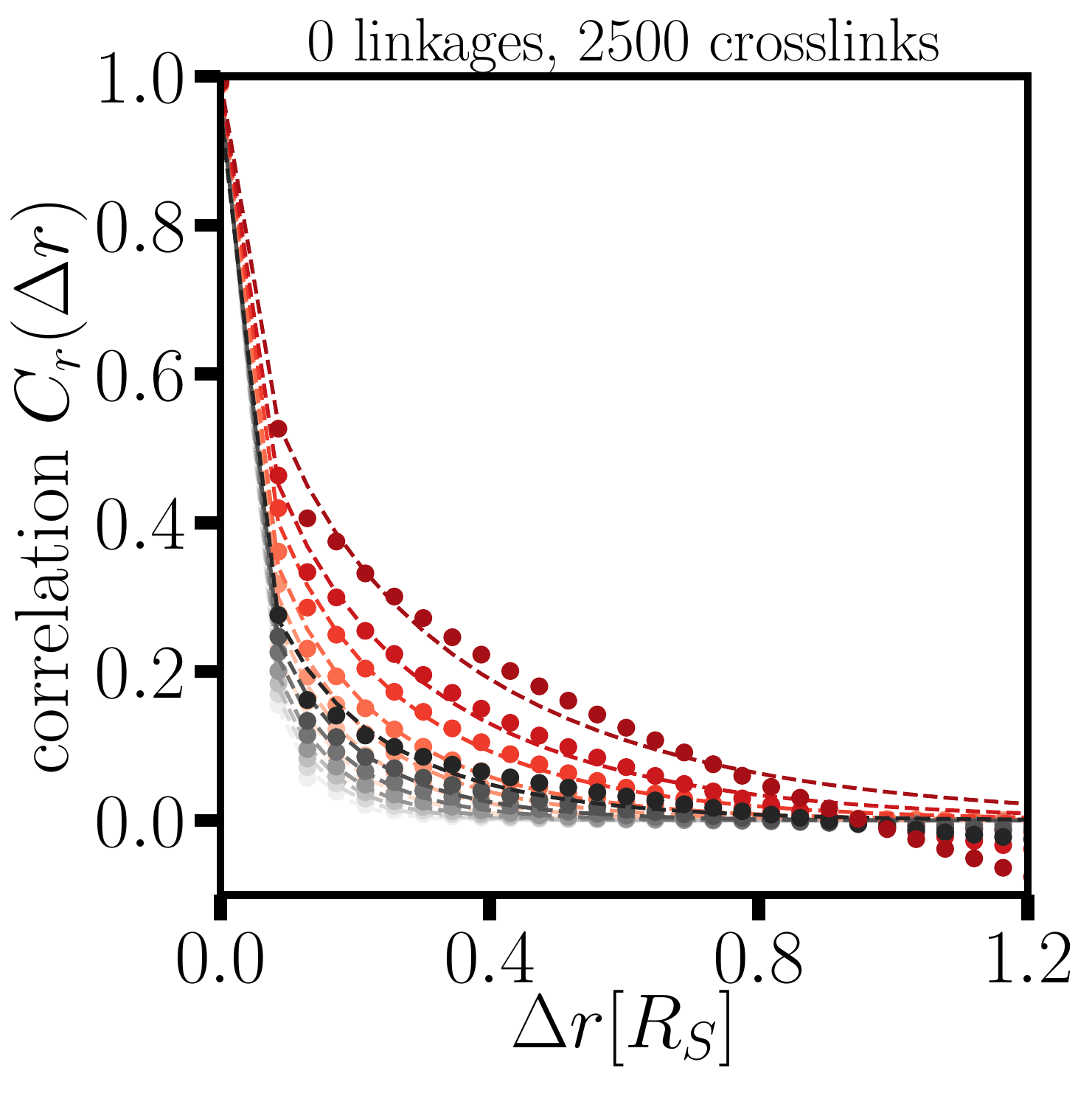}
		\includegraphics[width=0.19\linewidth]{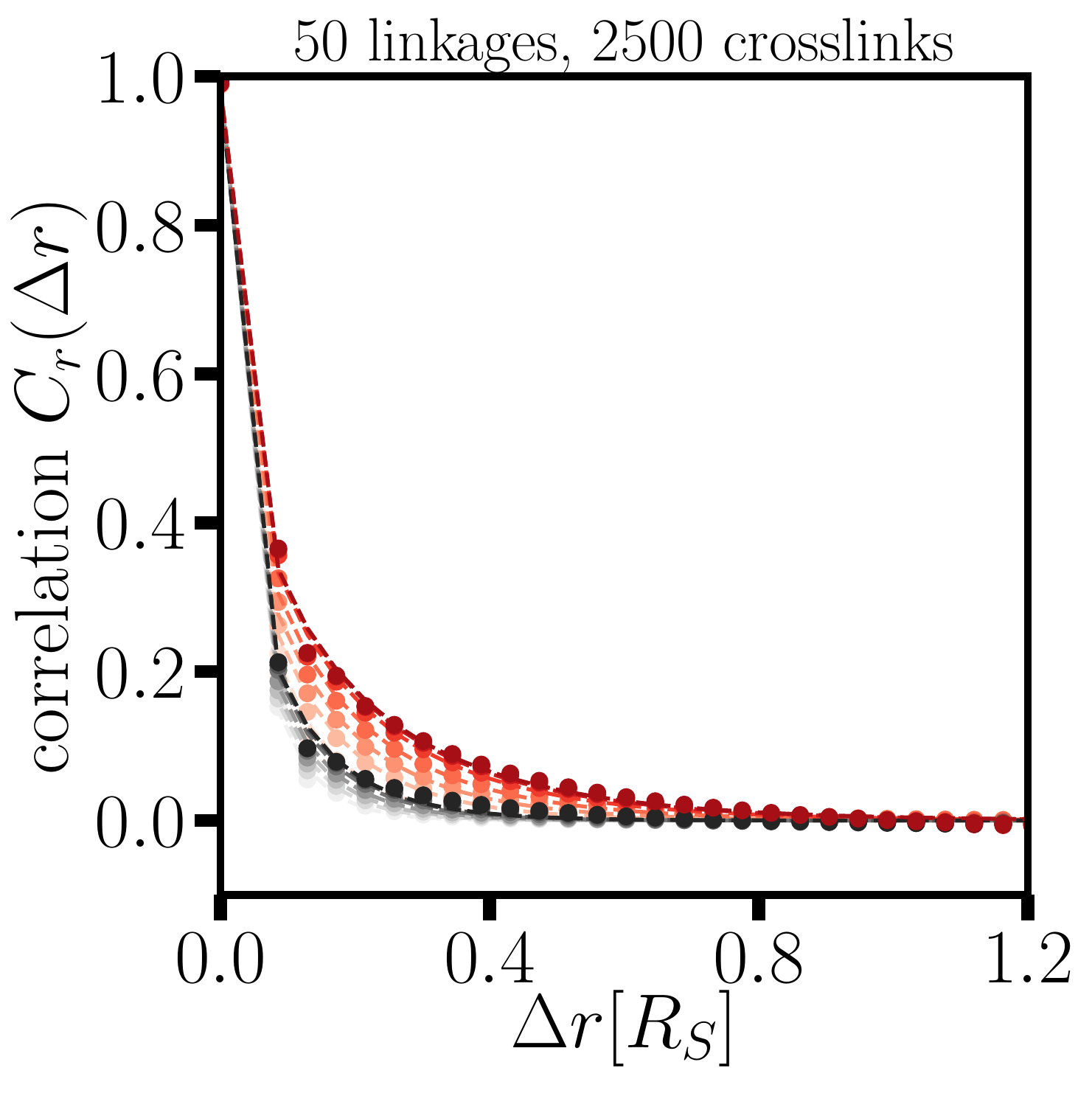}
		\includegraphics[width=0.19\linewidth]{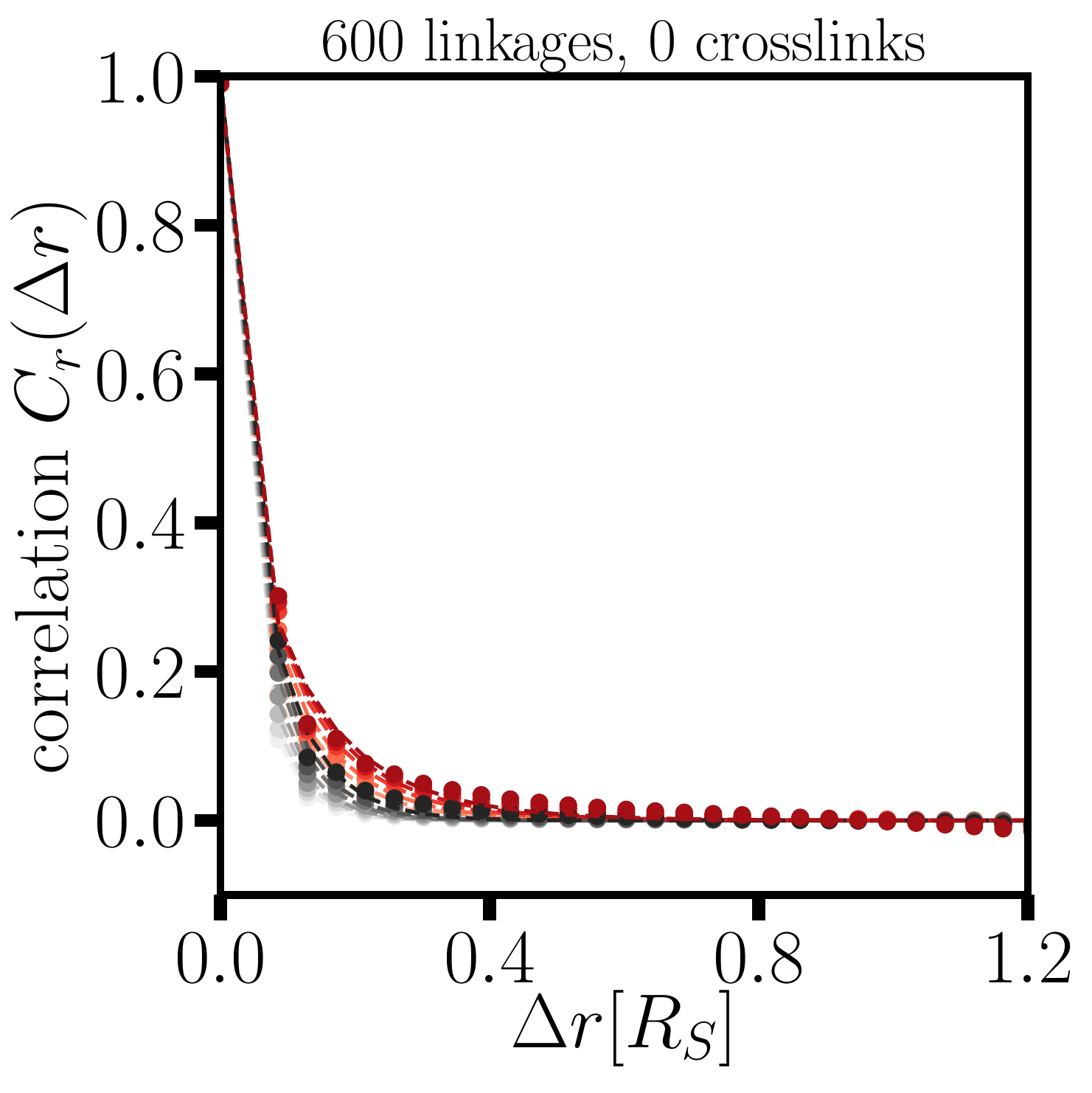}
		\includegraphics[width=0.19\linewidth]{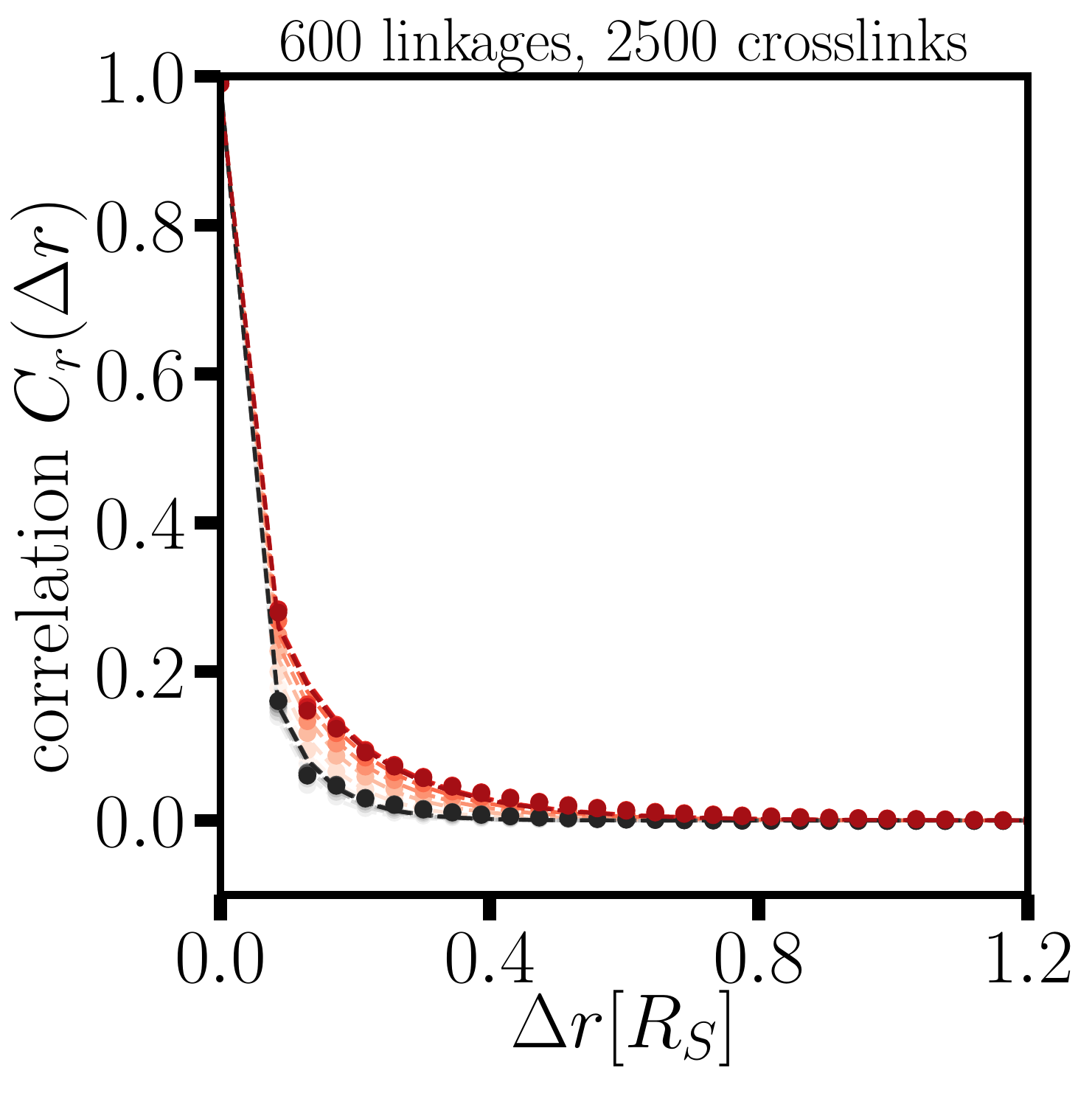}
		
		\includegraphics[width=0.19\linewidth]{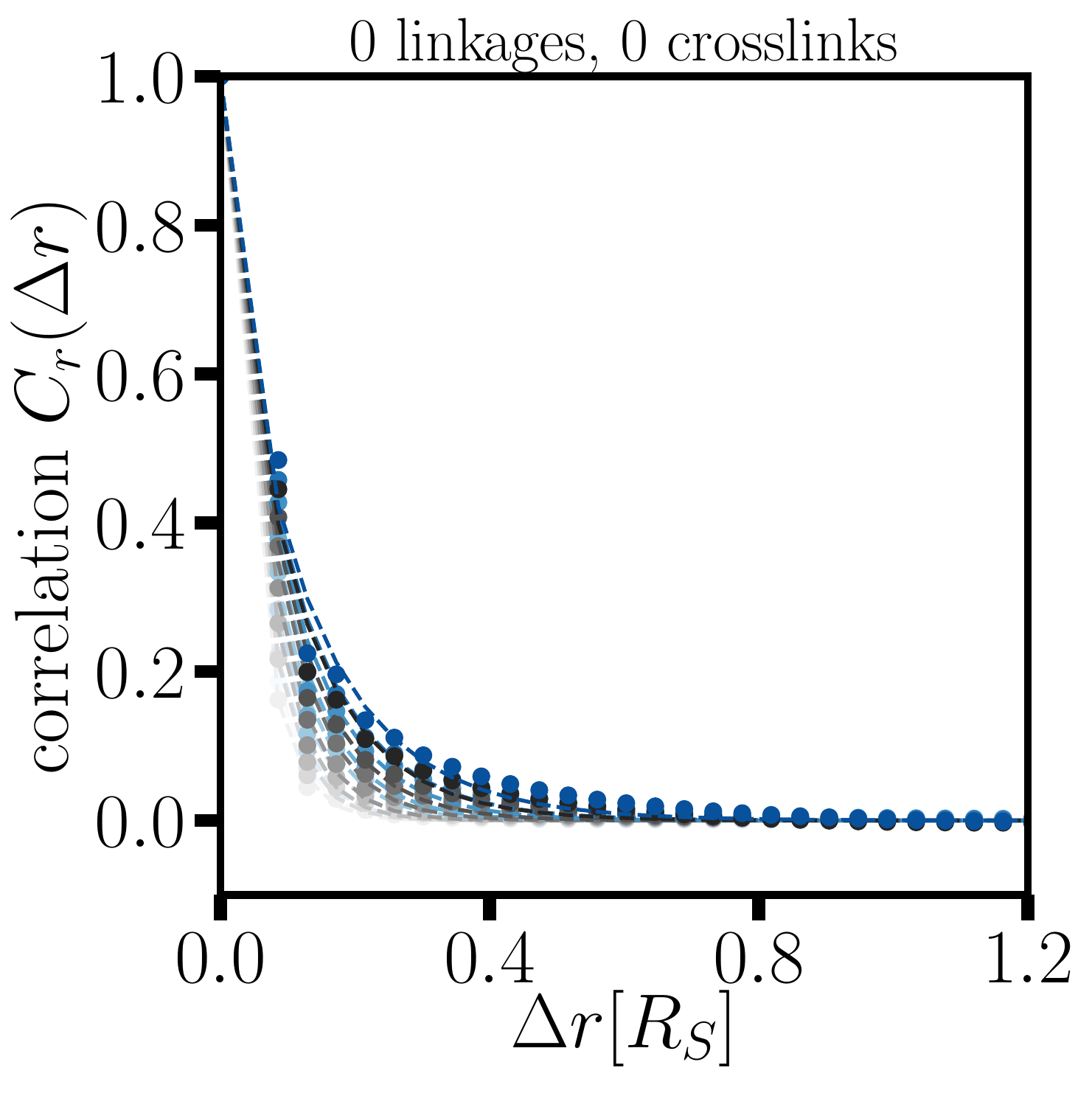}
		\includegraphics[width=0.19\linewidth]{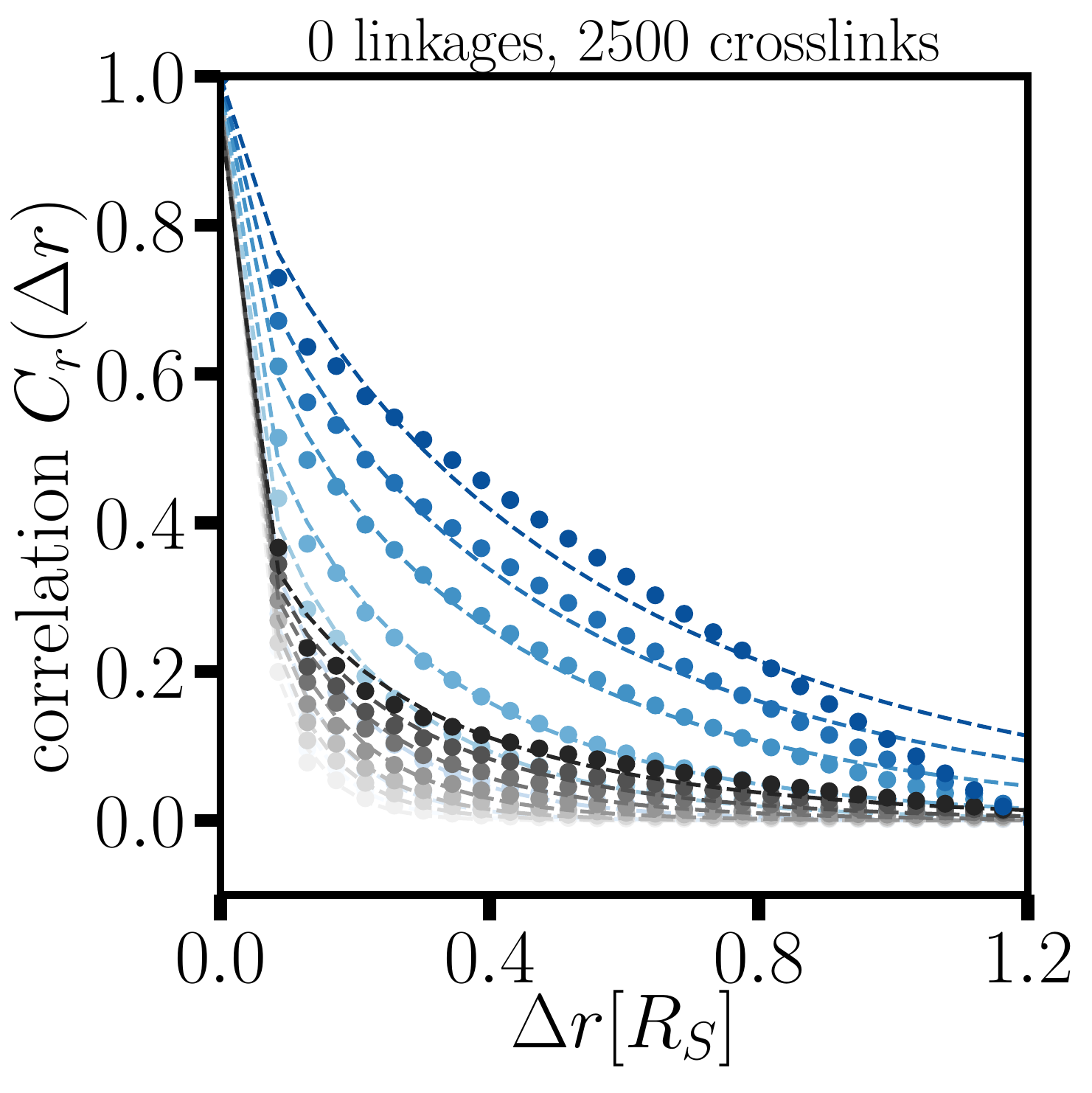}
		\includegraphics[width=0.19\linewidth]{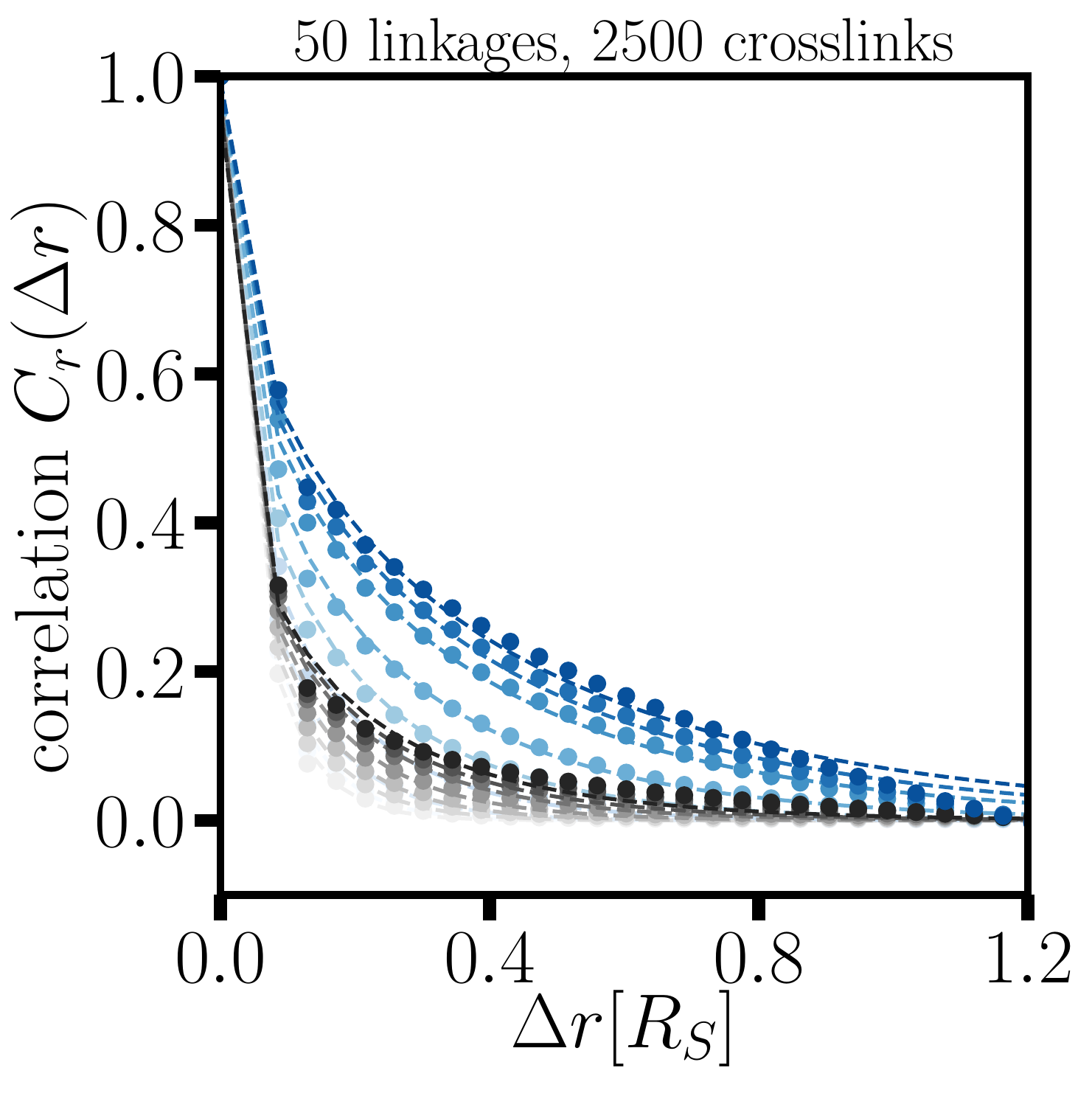}
		\includegraphics[width=0.19\linewidth]{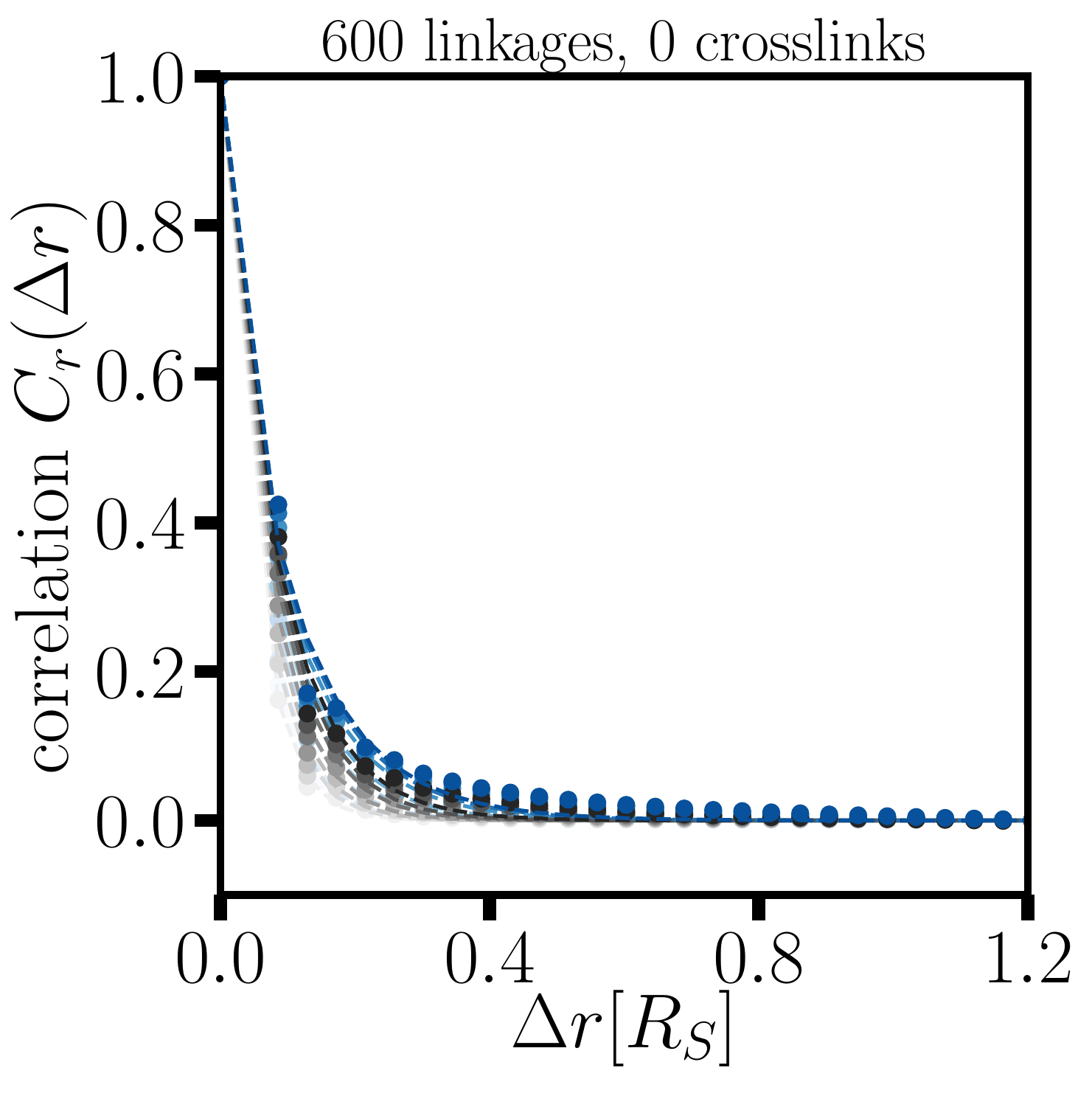}
		\includegraphics[width=0.19\linewidth]{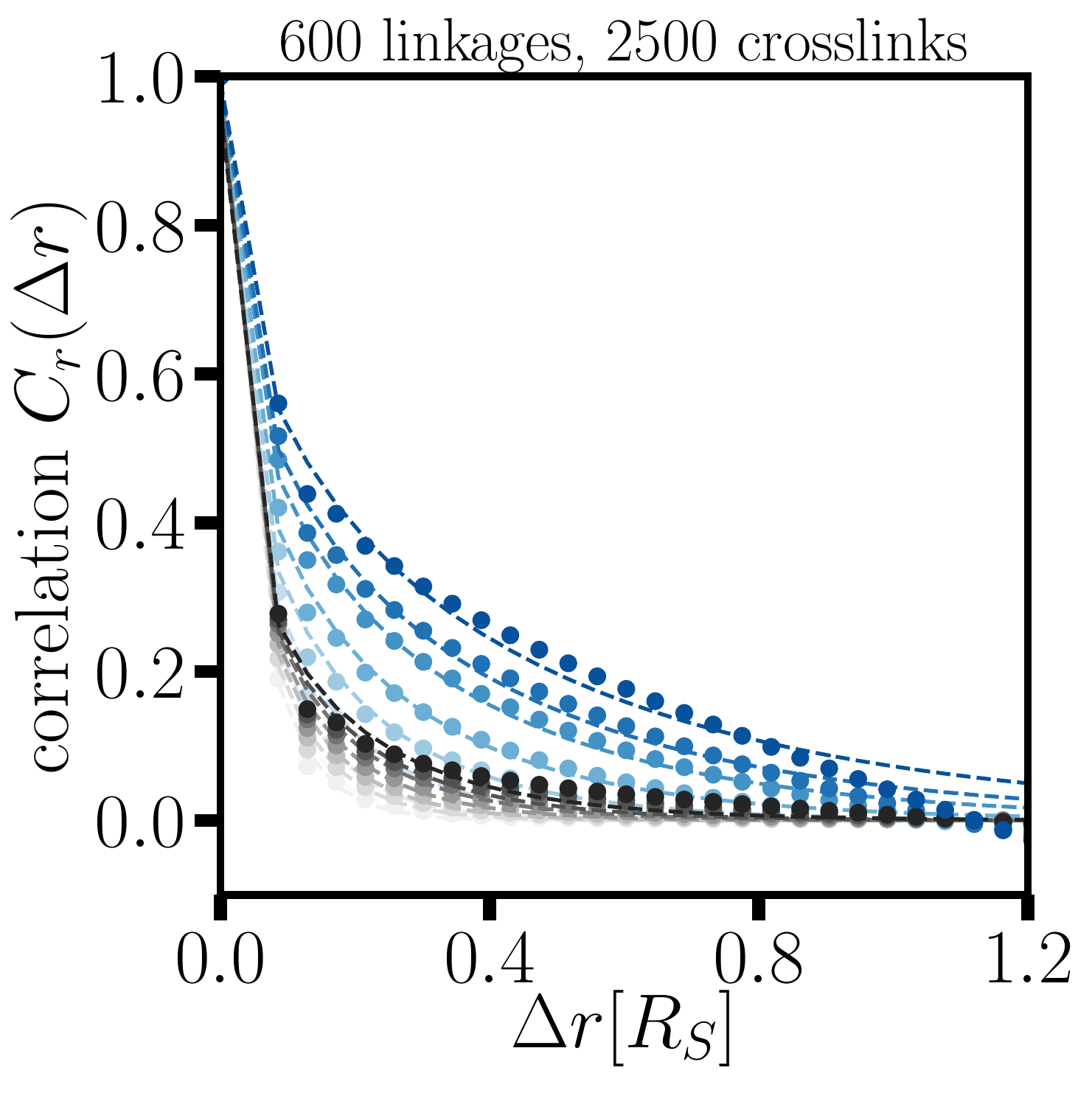}

		\includegraphics[width=0.19\linewidth]{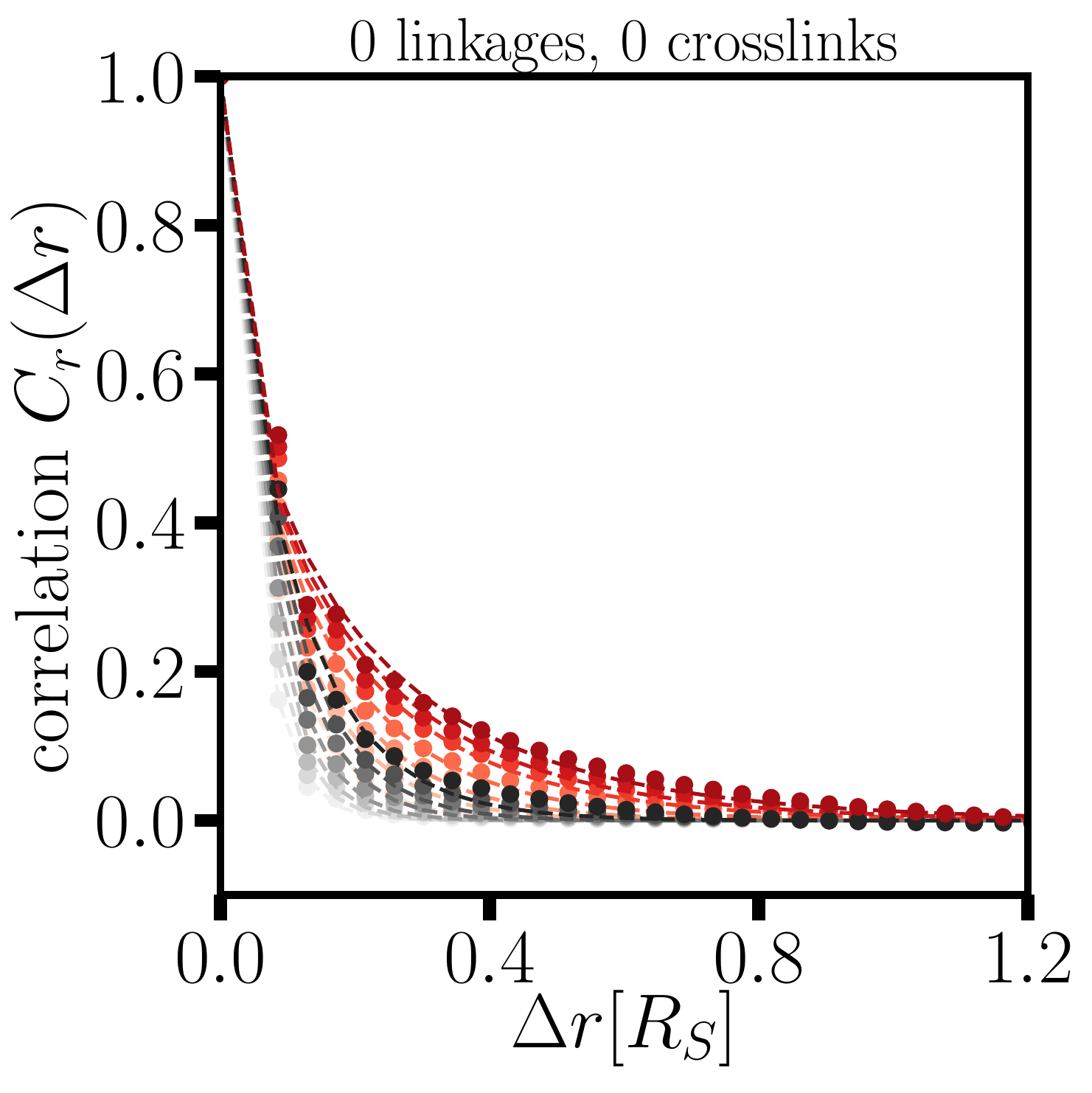}
		\includegraphics[width=0.19\linewidth]{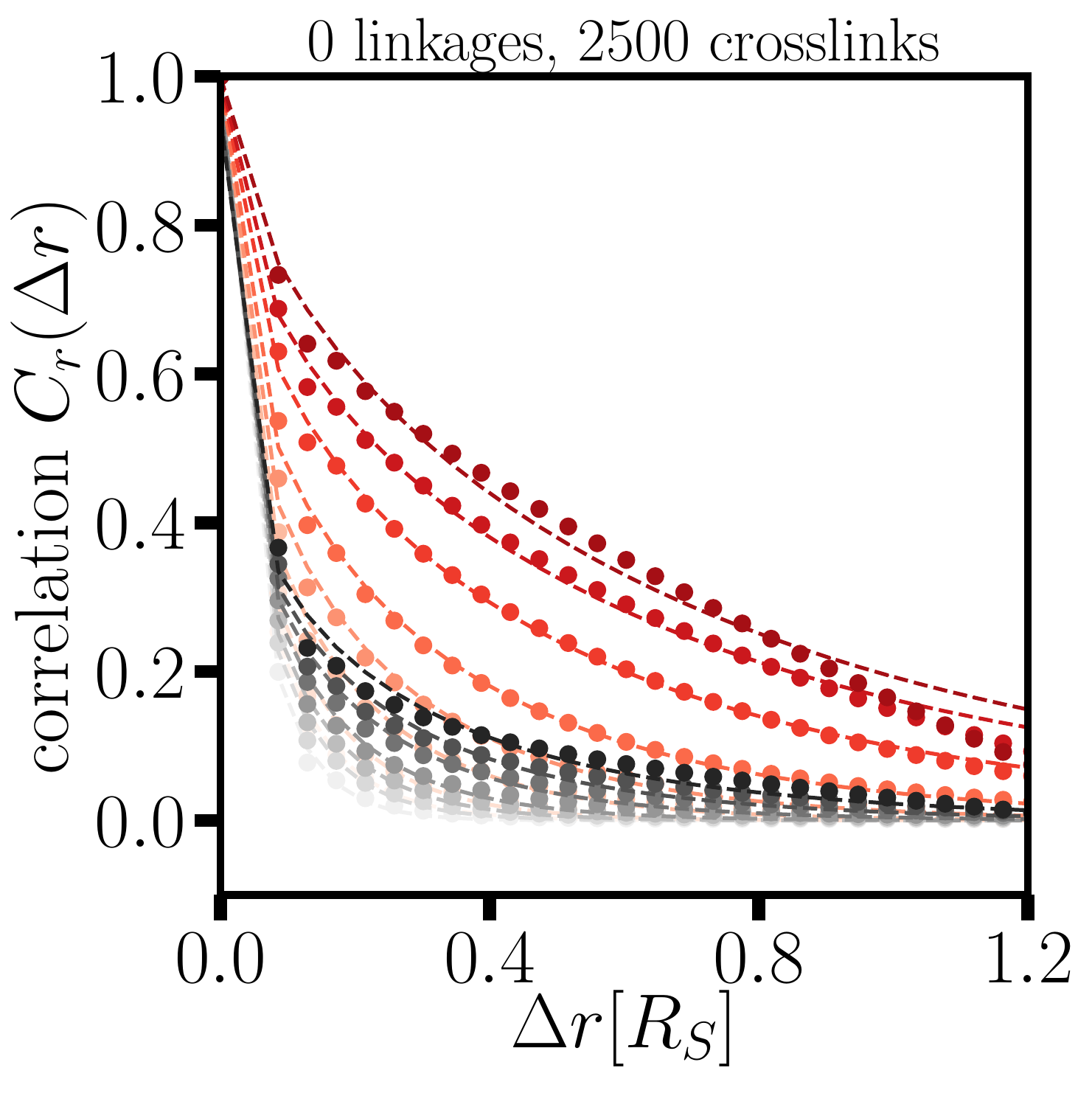}
		\includegraphics[width=0.19\linewidth]{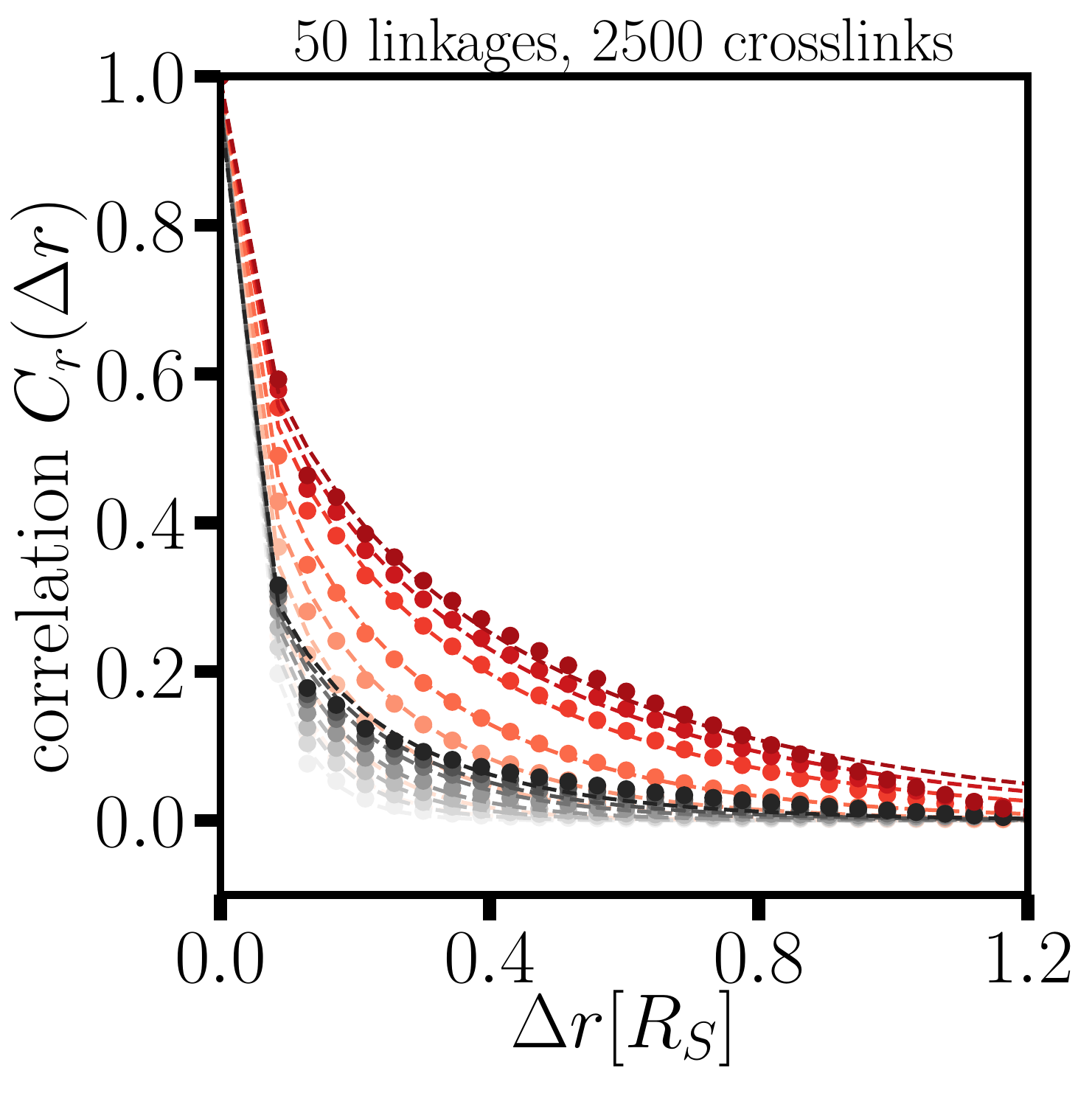}
		\includegraphics[width=0.19\linewidth]{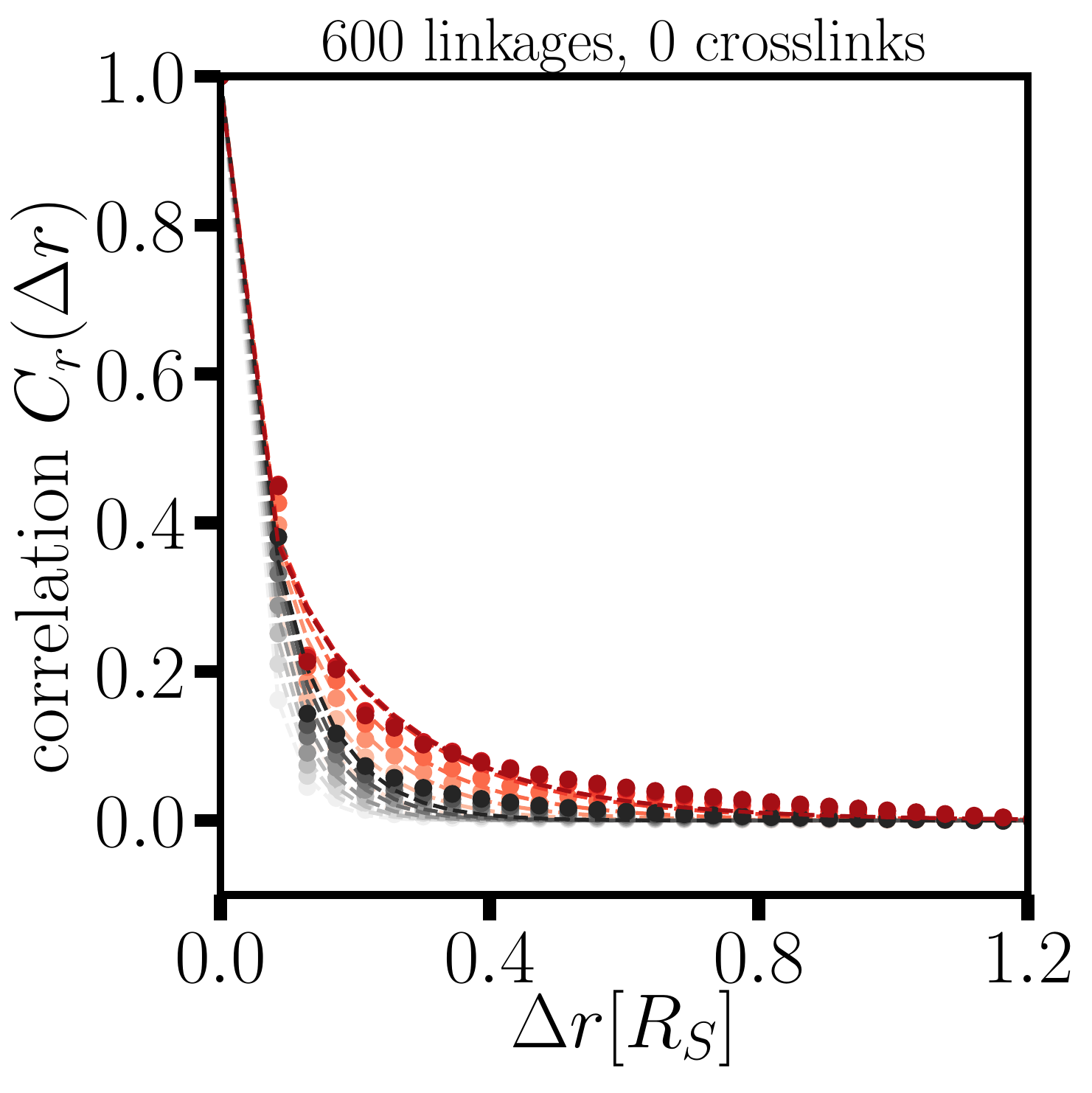}
		\includegraphics[width=0.19\linewidth]{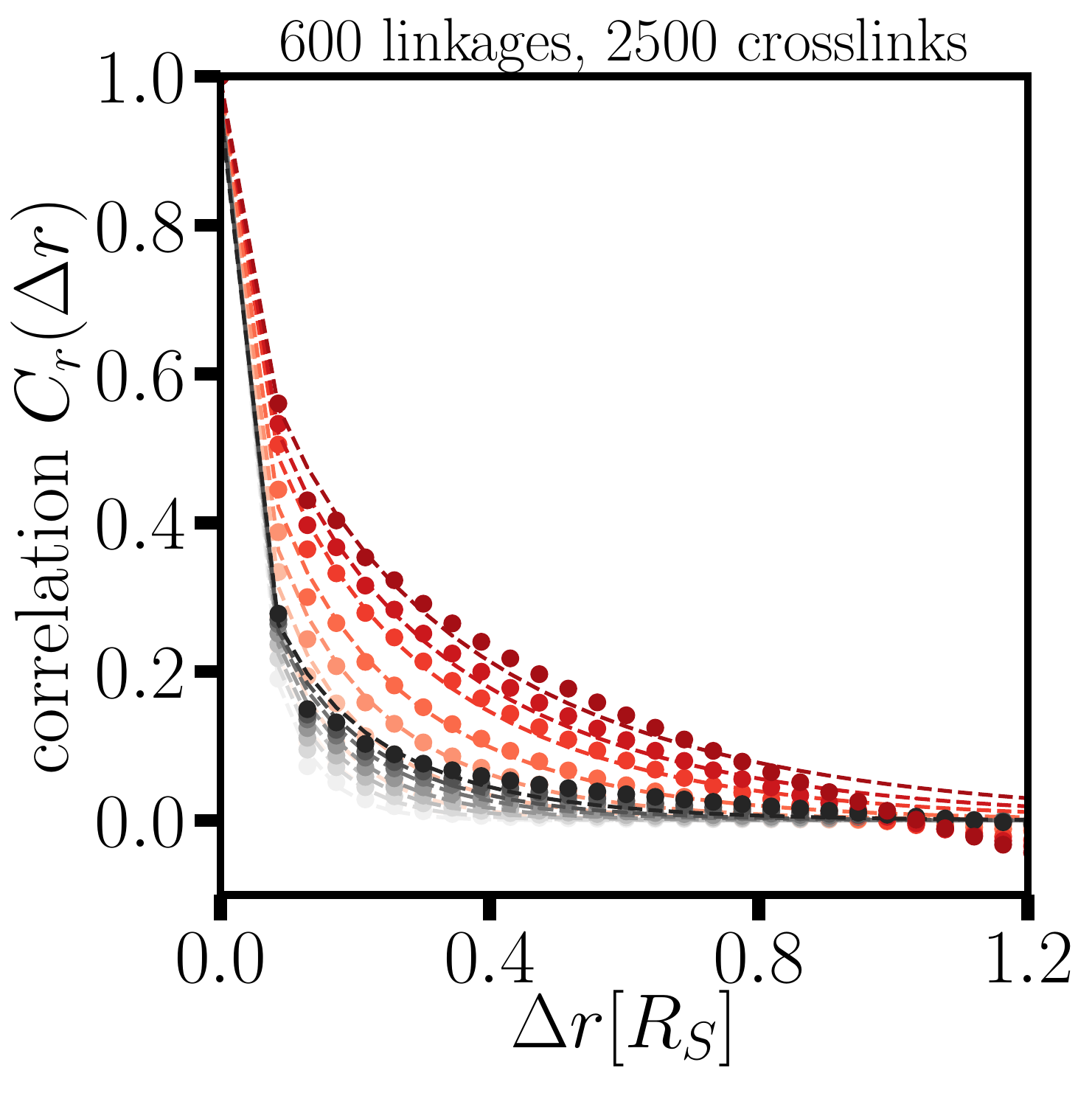}
		
		\includegraphics[width=0.19\linewidth]{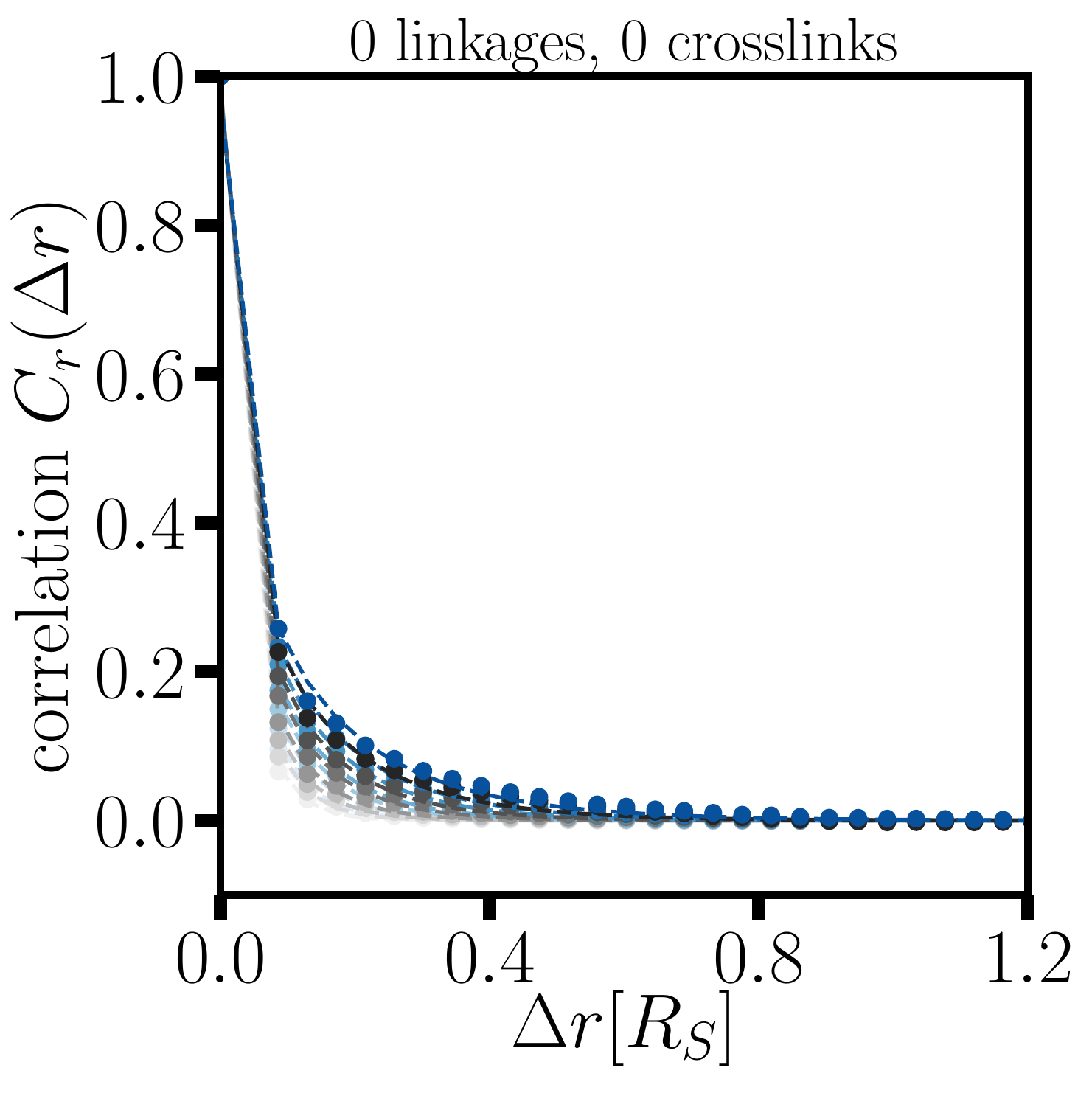}
		\includegraphics[width=0.19\linewidth]{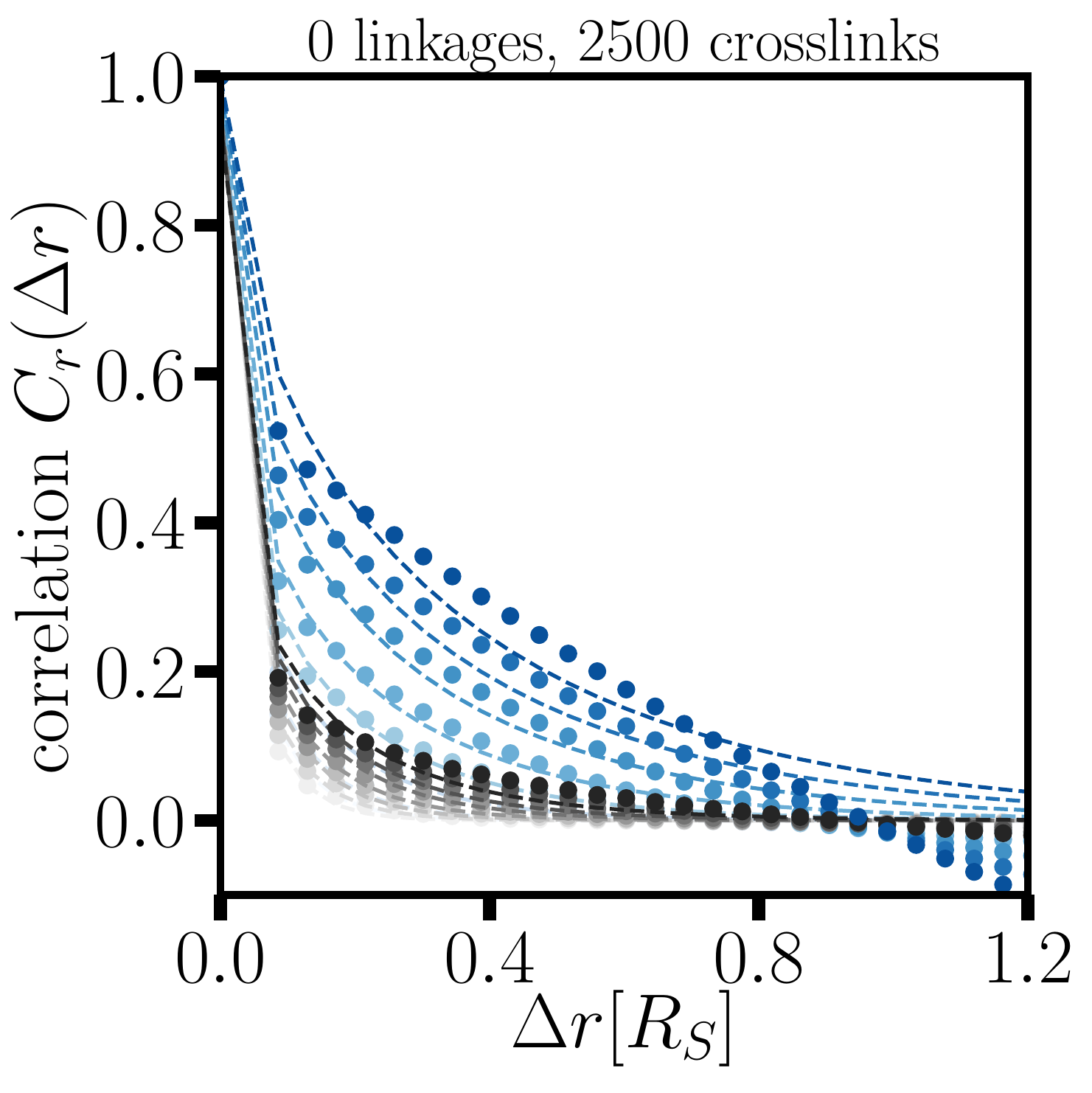}
		\includegraphics[width=0.19\linewidth]{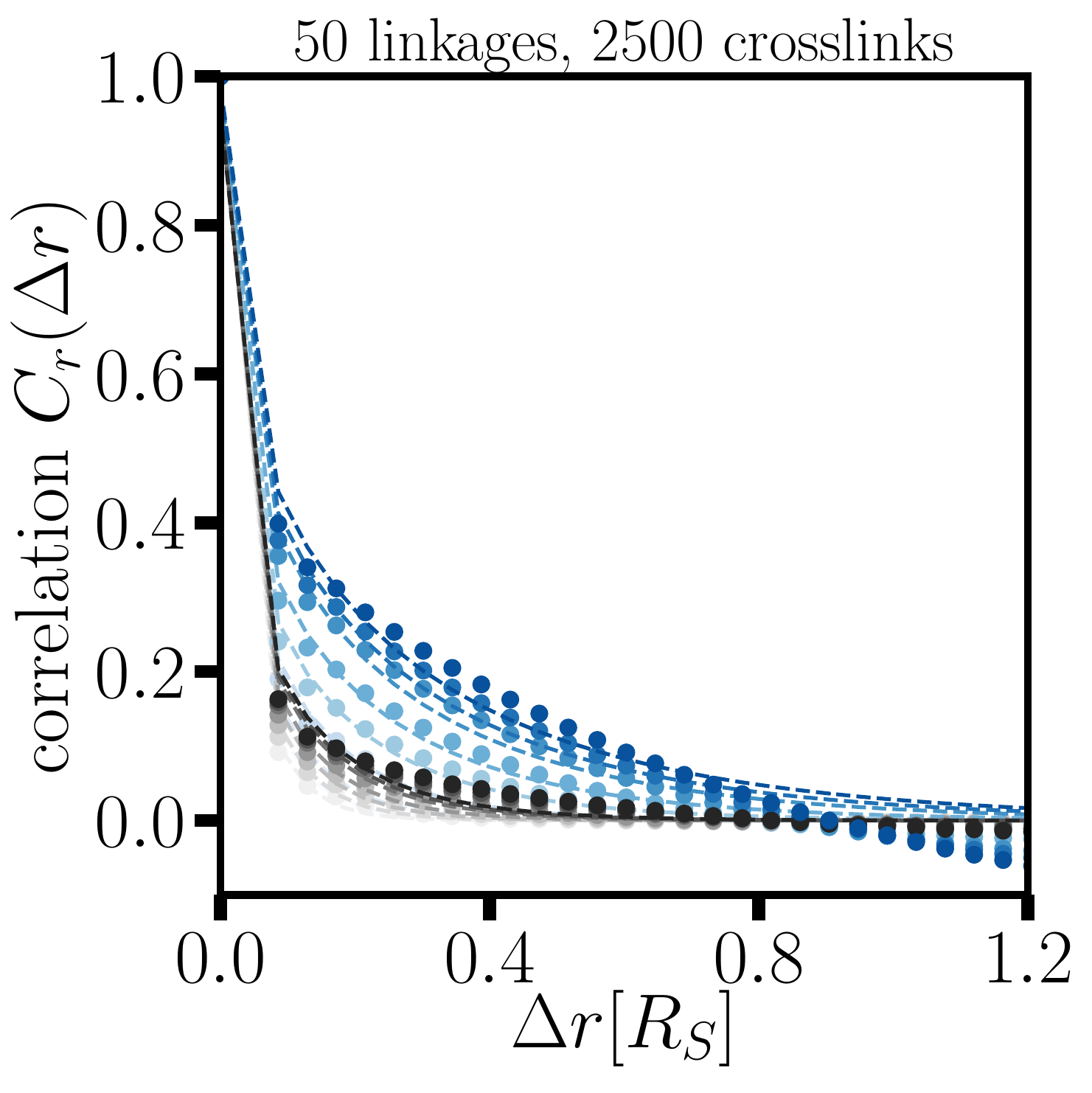}
		\includegraphics[width=0.19\linewidth]{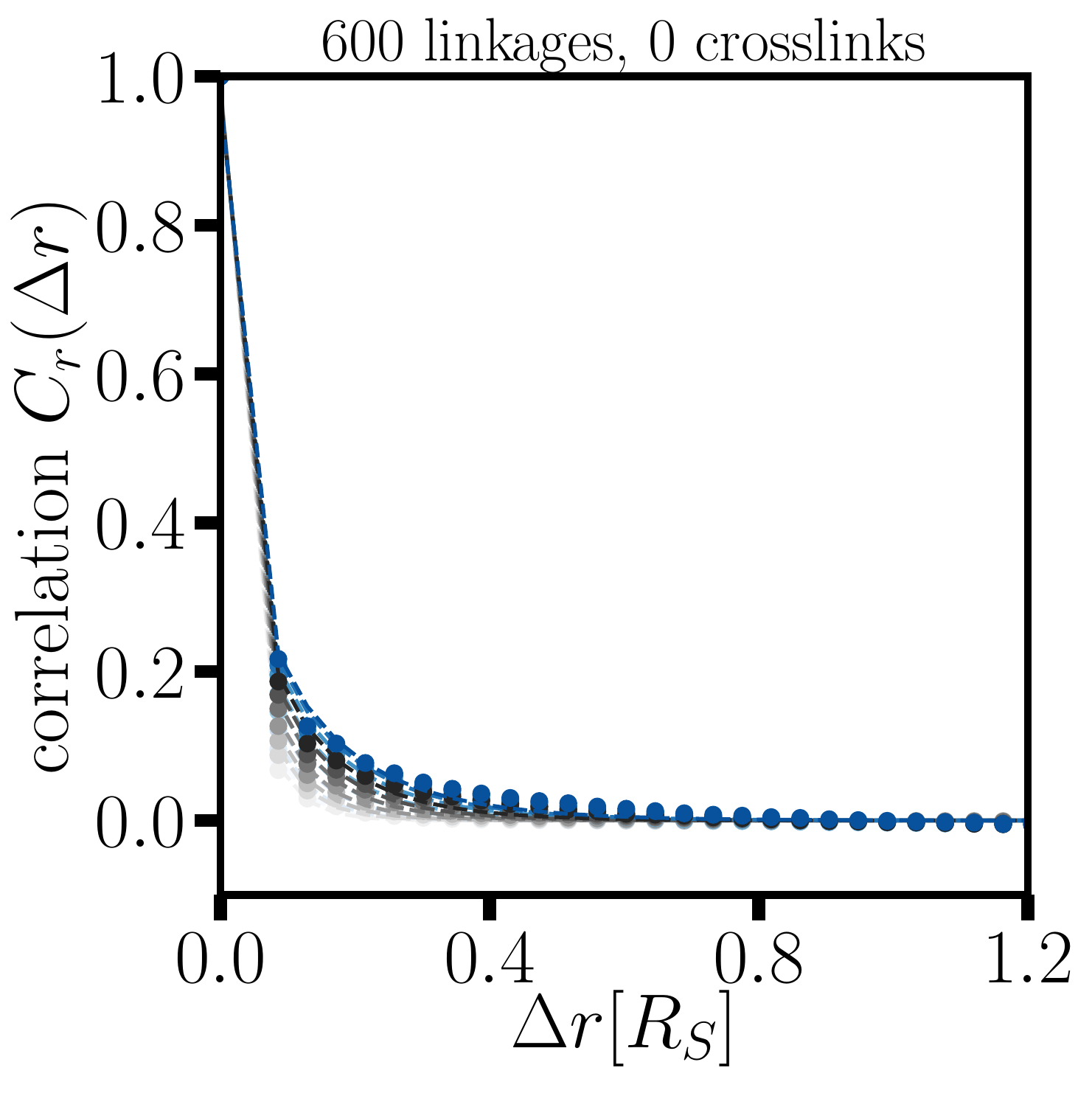}
		\includegraphics[width=0.19\linewidth]{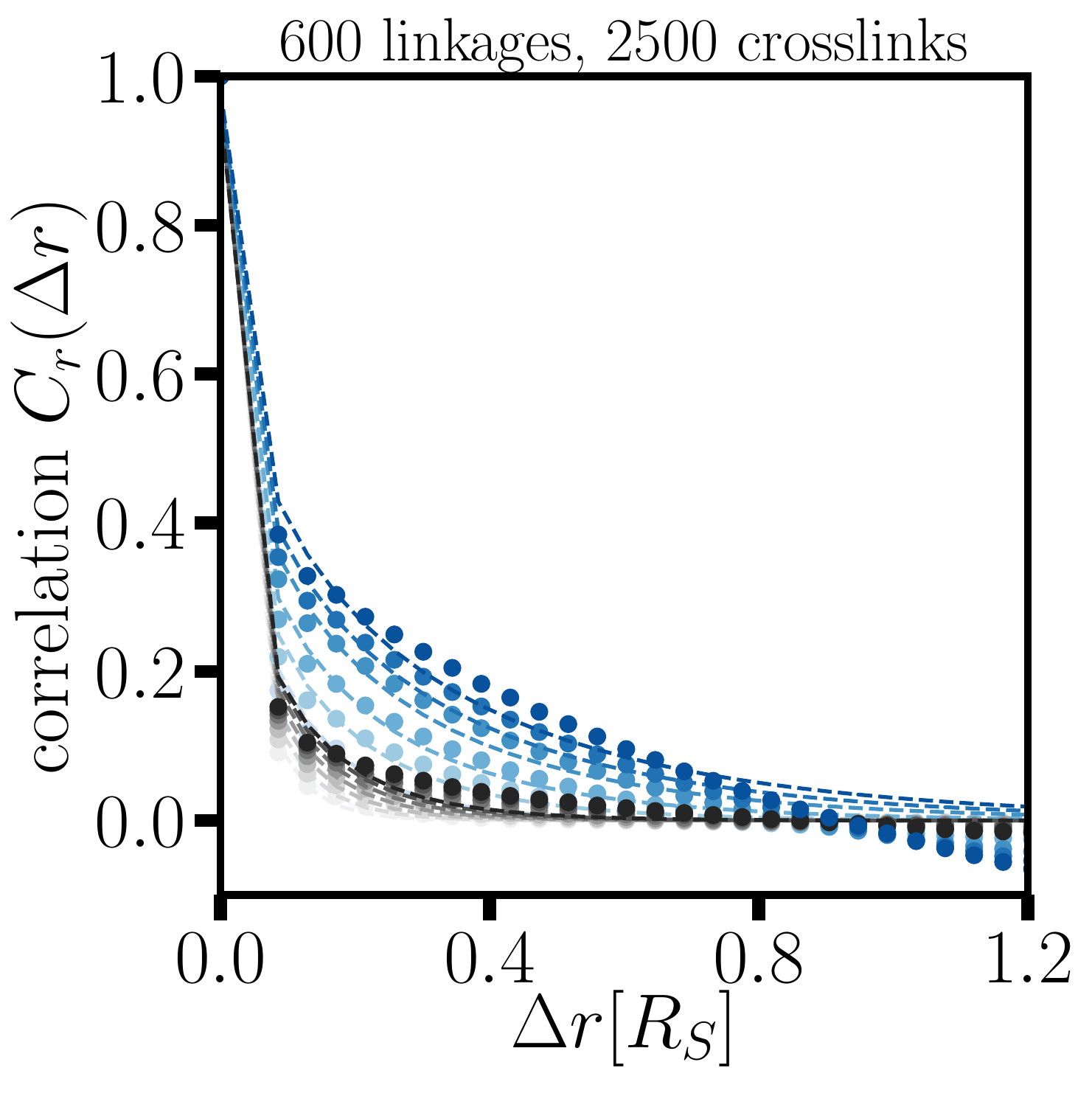}
		
		\includegraphics[width=0.19\linewidth]{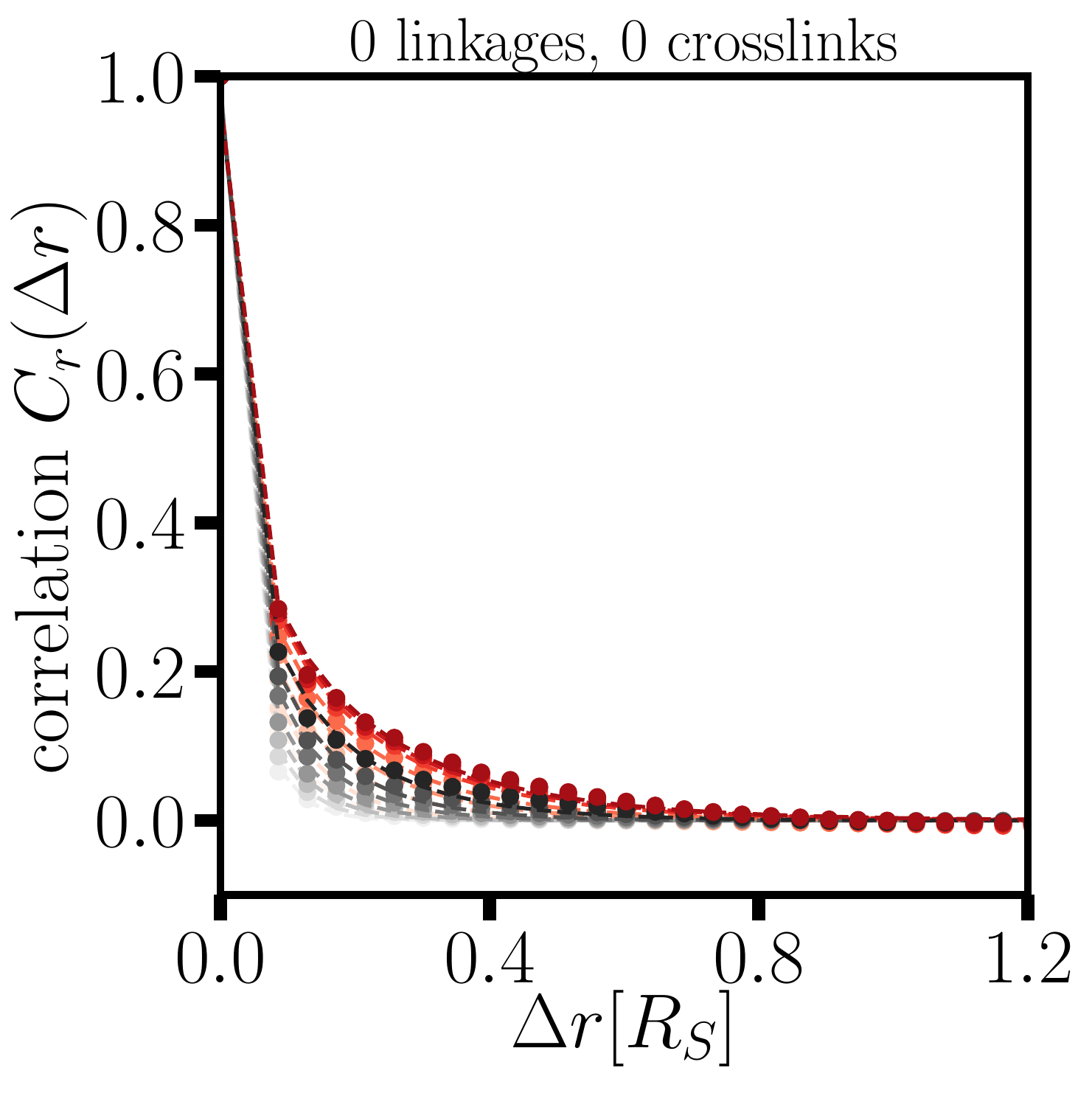}
		\includegraphics[width=0.19\linewidth]{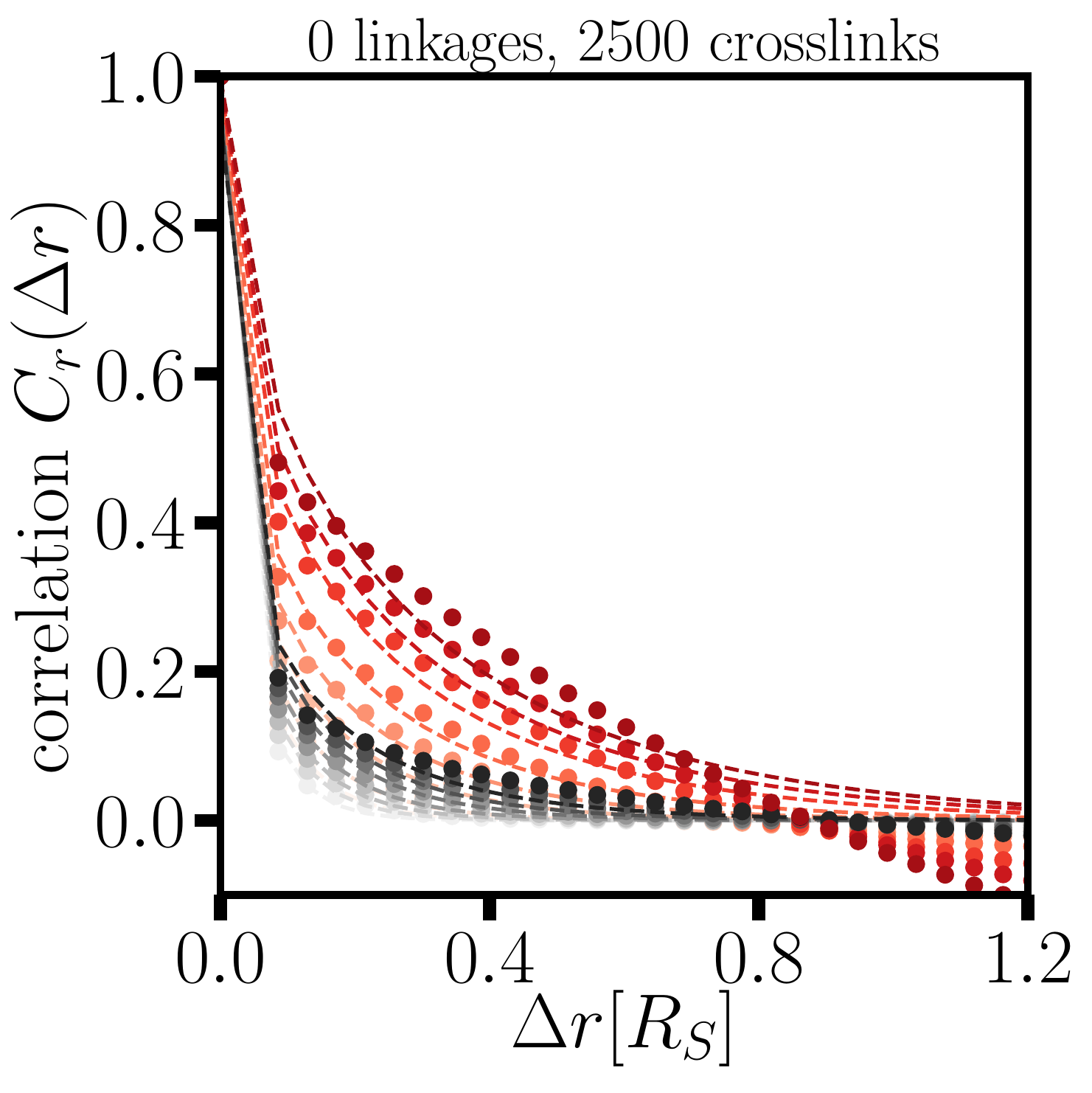}
		\includegraphics[width=0.19\linewidth]{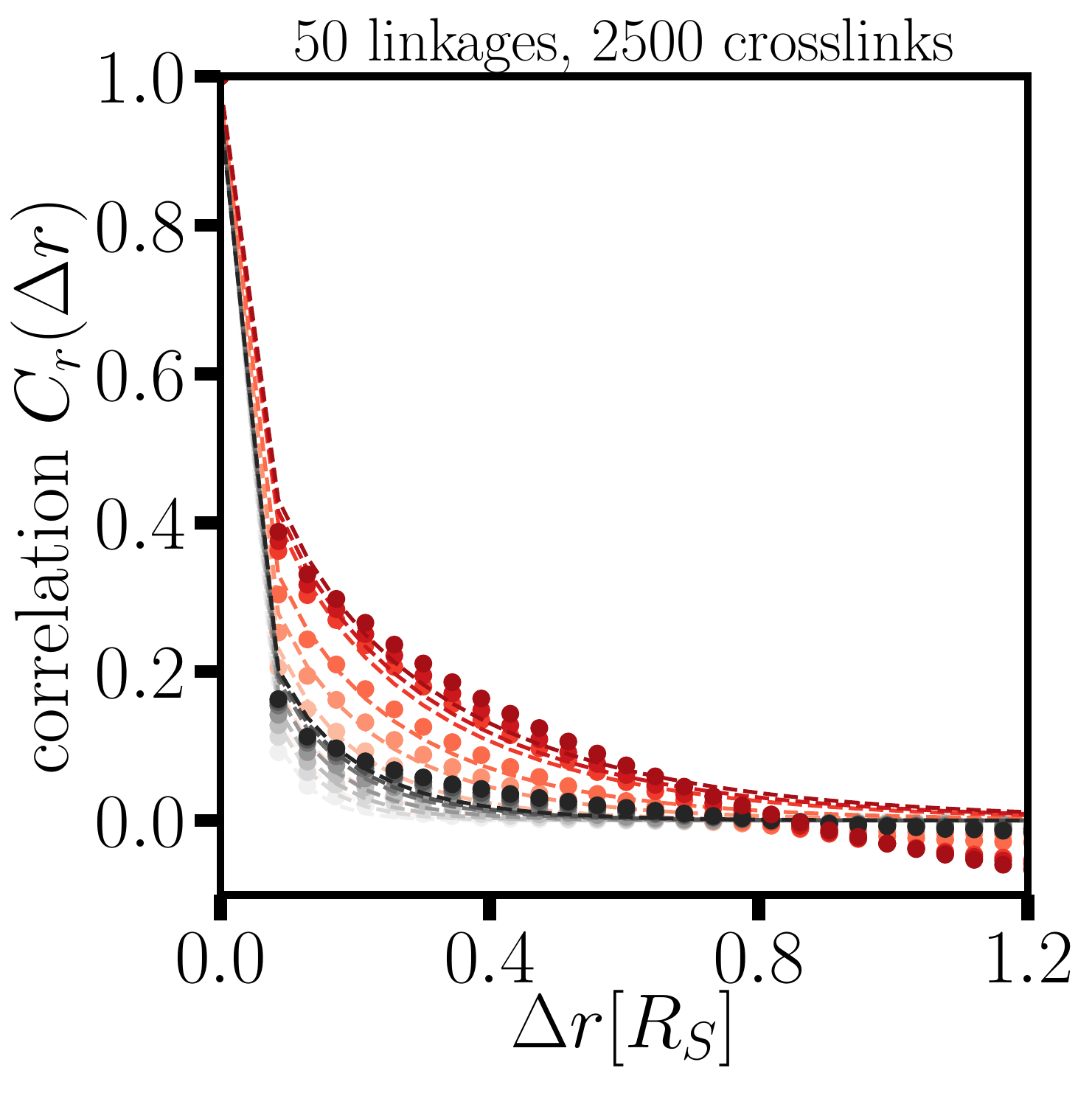}
		\includegraphics[width=0.19\linewidth]{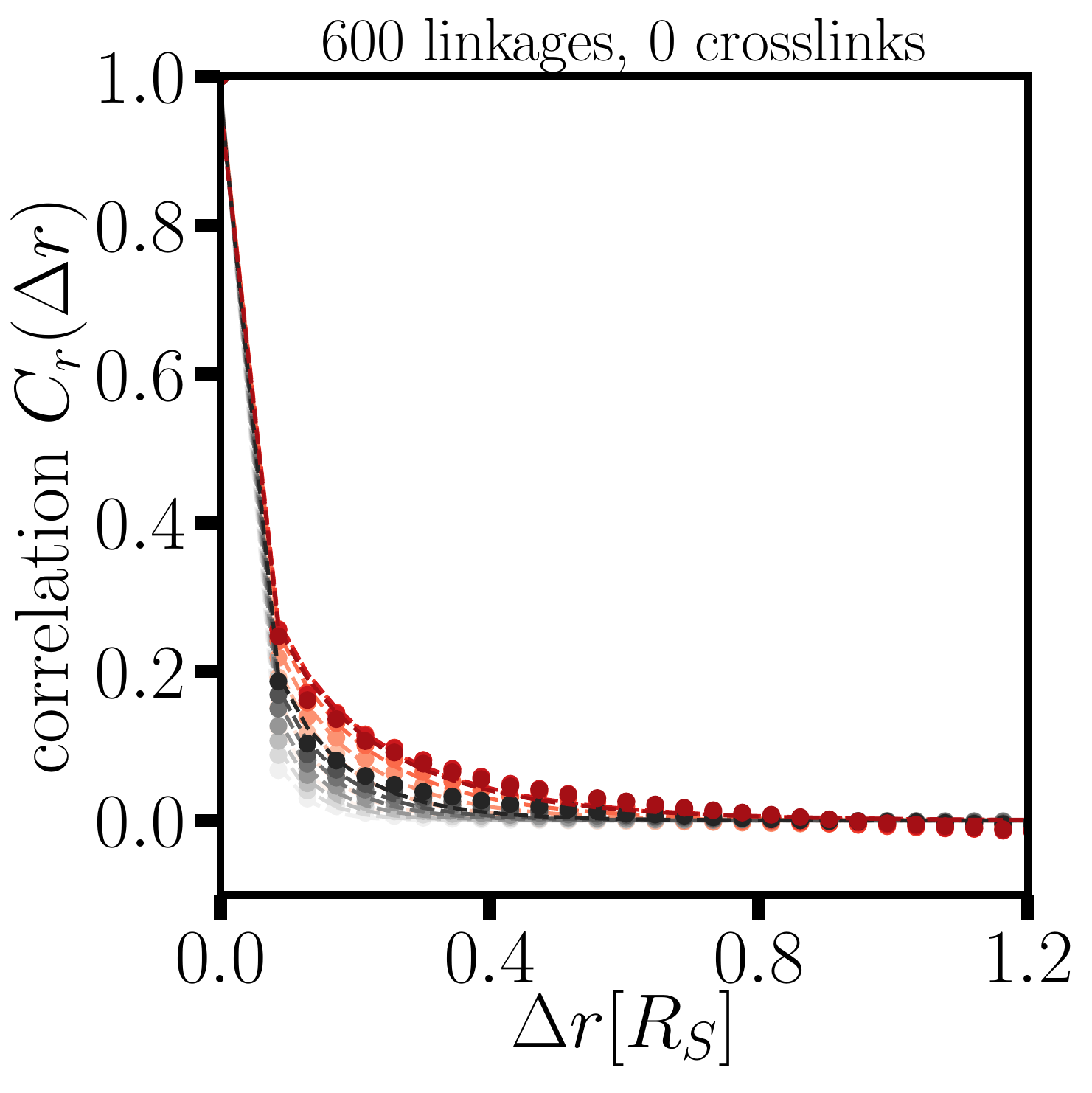}
		\includegraphics[width=0.19\linewidth]{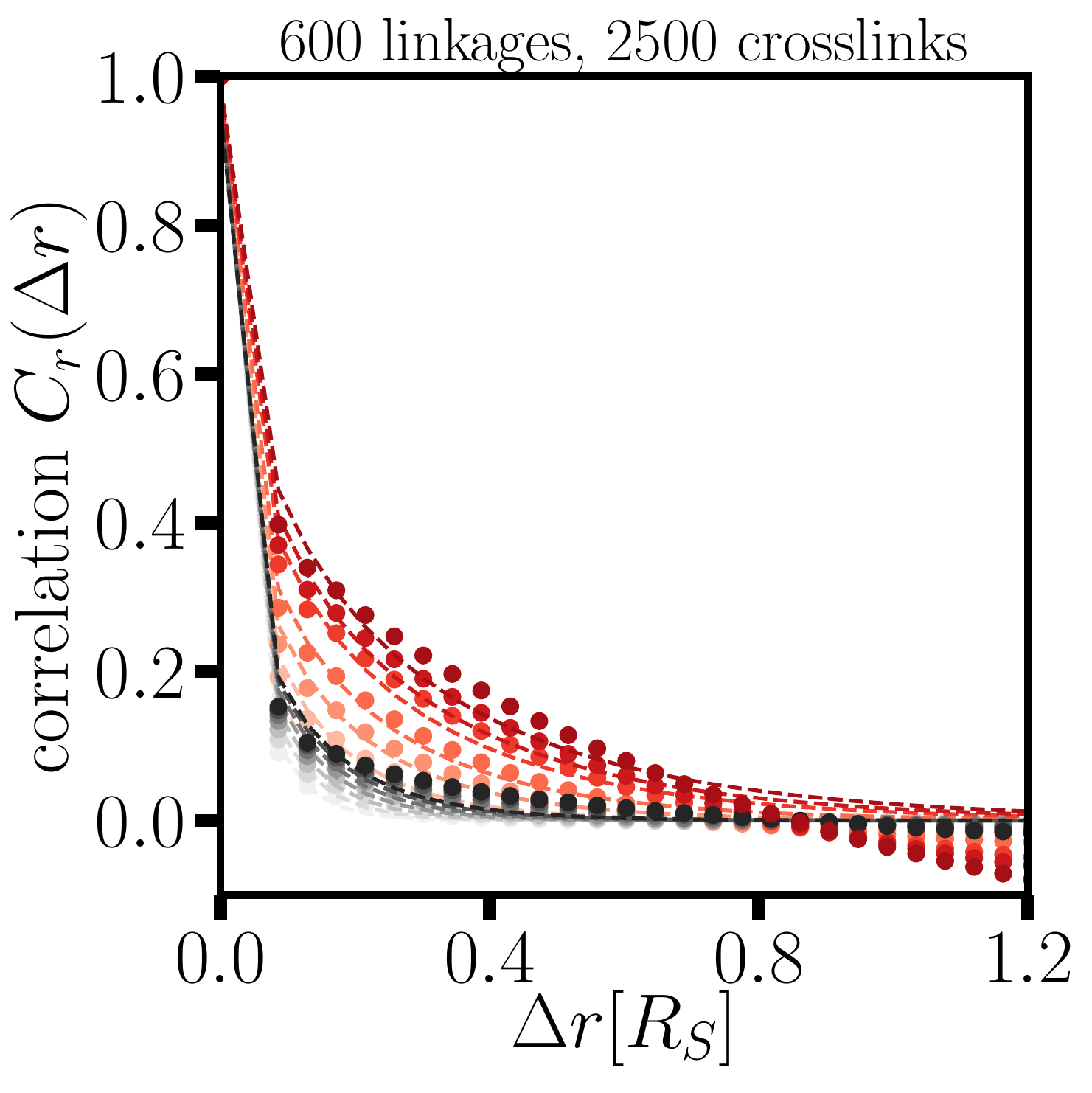}
	\end{center}
	\caption{Correlation functions for $N_C=2500$ and $N_L=50$ (middle column) and four extreme cases (left column: 0 linkages and 0 crosslinks; second from left column: 0 linkages and 2500 crosslinks; second from right column: 600 linkages and 0 crosslinks; right column: 600 linkages and 2500 crosslinks). Top two rows: The three-dimensional  correlation function for the hard shell; Middle two rows: The three-dimensional correlation functions for the soft shell; Bottom two rows: Two-dimensional correlation functions for the soft shell.  Color varies from light to dark as time lag equals $1\,\tau$, $2\,\tau$, $5\,\tau$, $10\,\tau$, $20\,\tau$, $50\,\tau$, $100\,\tau$, $200,\tau$, respectively. Symbols denote the numerical results, while the dashed line represent the fitted correlation functios. Greyscale: passive. Bluescale: active with extensile motors. Redscale: active with contractile motors.}
	\label{fig:correlation_1}
\end{figure*}

In Fig.~\ref{fig:correlation_2}, we plot the correlation length a function of linkage number $N_L$ and crosslink number $N_C$ over the short time window $5\,\tau$ and the long time window $50\,\tau$. We observe that active motors clearly enhance the correlation length. It is also clear that presence of crosslinks also enhance correlation length. The correlation length is larger for the soft shell case. In the soft shell case, without subtracting the diffusion of the center of mass, the correlation length for the long time window spans almost the radius of the system. We note that the correlation length is reduced if we subtract the center-of-mass shell motion; however, it still remains larger than the hard shell case. A quasi-two-dimensional correlation length is computed from a slab-like region and is also shown for potential comparison to experimental results since, in the experiments, the correlated length is extracted using this method. There is not much difference between the three-dimensional correlation length and the two-dimensional correlation length with the center of mass of the shell subtracted. We also show the correlation length as a function of shell stiffness (with the COM of shell subtracted) to demonstrate the direct effect of shell stiffness on the correlated chromatin motion (see Fig. \ref{fig:correlationengthvsstiffness}). 

We consider different motor parameters and types in Figs.~\ref{fig:correlation_3}-\ref{fig:correlationengthvsstiffness}. With faster motor turnover (Fig.~\ref{fig:correlation_3} column 2), correlations are reduced. With faster turnover, the forces from the active monomers change rapidly and become more like uncorrelated active noise. In that case, we expect equilibrium-like behavior, but with a higher effective temperature, as predicted by a previous analysis of active polymers \cite{osmanovic17}.  With slower motor turnover, correlations are increased (Fig.~\ref{fig:correlation_3} column 3). The strength and extent of the correlations also changes with the number of motors, $N_m$, albeit somewhat weakly for appreciable numbers of motors (Figs.~\ref{fig:correlation_4} and \ref{fig:correlationengthvsstiffness}). Decreasing the number of motors from $N_m=400$ (the typical value in the main text) to $N_m=50$ decreases the correlation length by $\sim30\%$ (for time window $\Delta\tau=50$), compared to a $60\%$ decrease from $N_m = 400$ to $N_m =0$ (\textit{i.e.}, the passive system).
We additionally considered motors that exert their forces in a pairwise manner, so that forces exerted by active monomers on nearby monomers are reciprocated by the nearby monomers (Figs.~\ref{fig:correlation_3} columns 4 and 5).  Correlated motions in such systems are largely suppressed, as such systems are either near equilibrium (as in column 4 with pairwise forces and the usual motor turnover time, $\tau_m$) or in equilibrium (as in column 5 with pairwise forces between ``active'' and inactive monomers and no motor turnover, \textit{i.e.}, $\tau_m=\infty$).

\begin{figure*}[h]
	\begin{center}
		\includegraphics[width=0.24\linewidth]{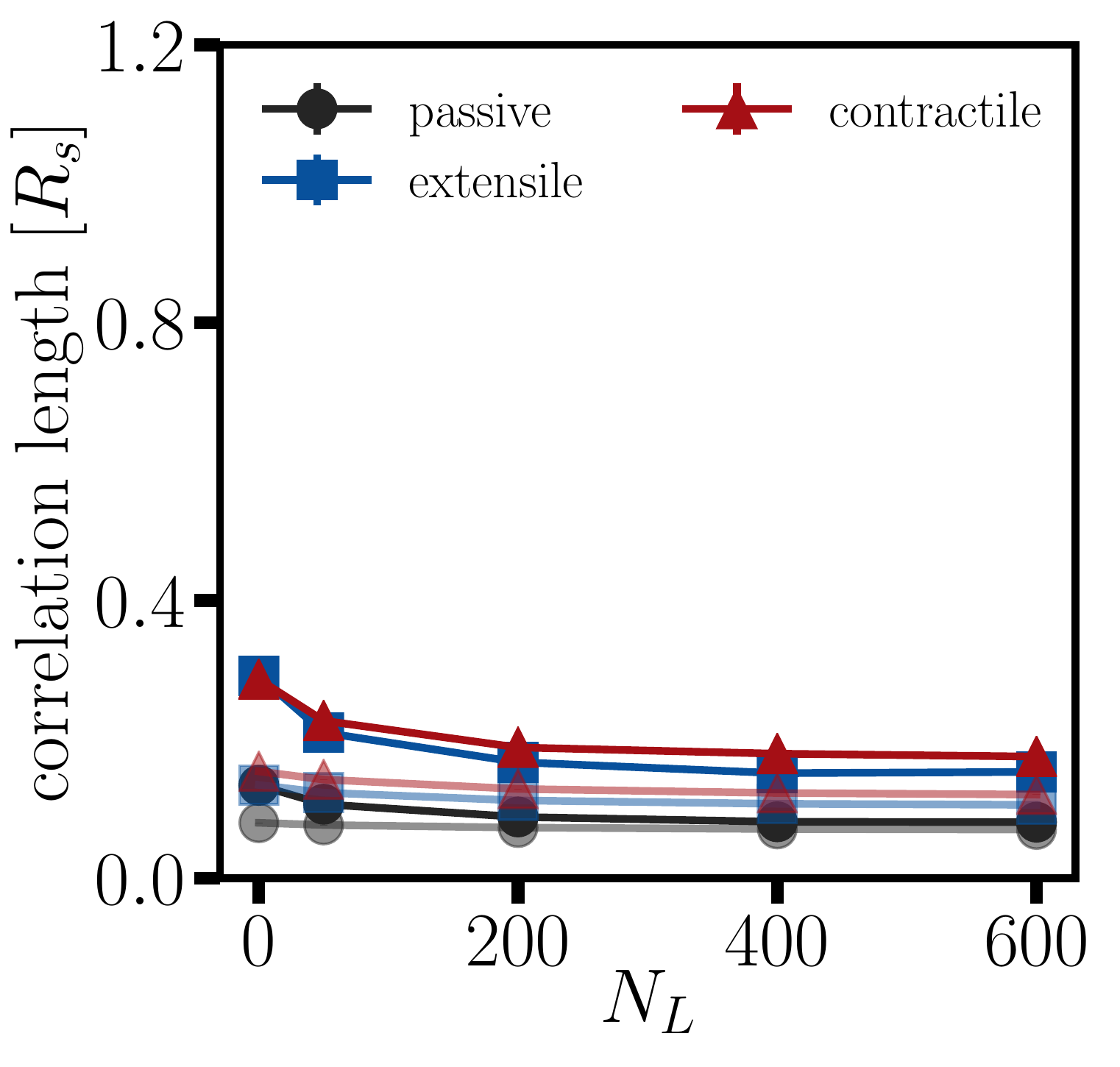}
		\includegraphics[width=0.24\linewidth]{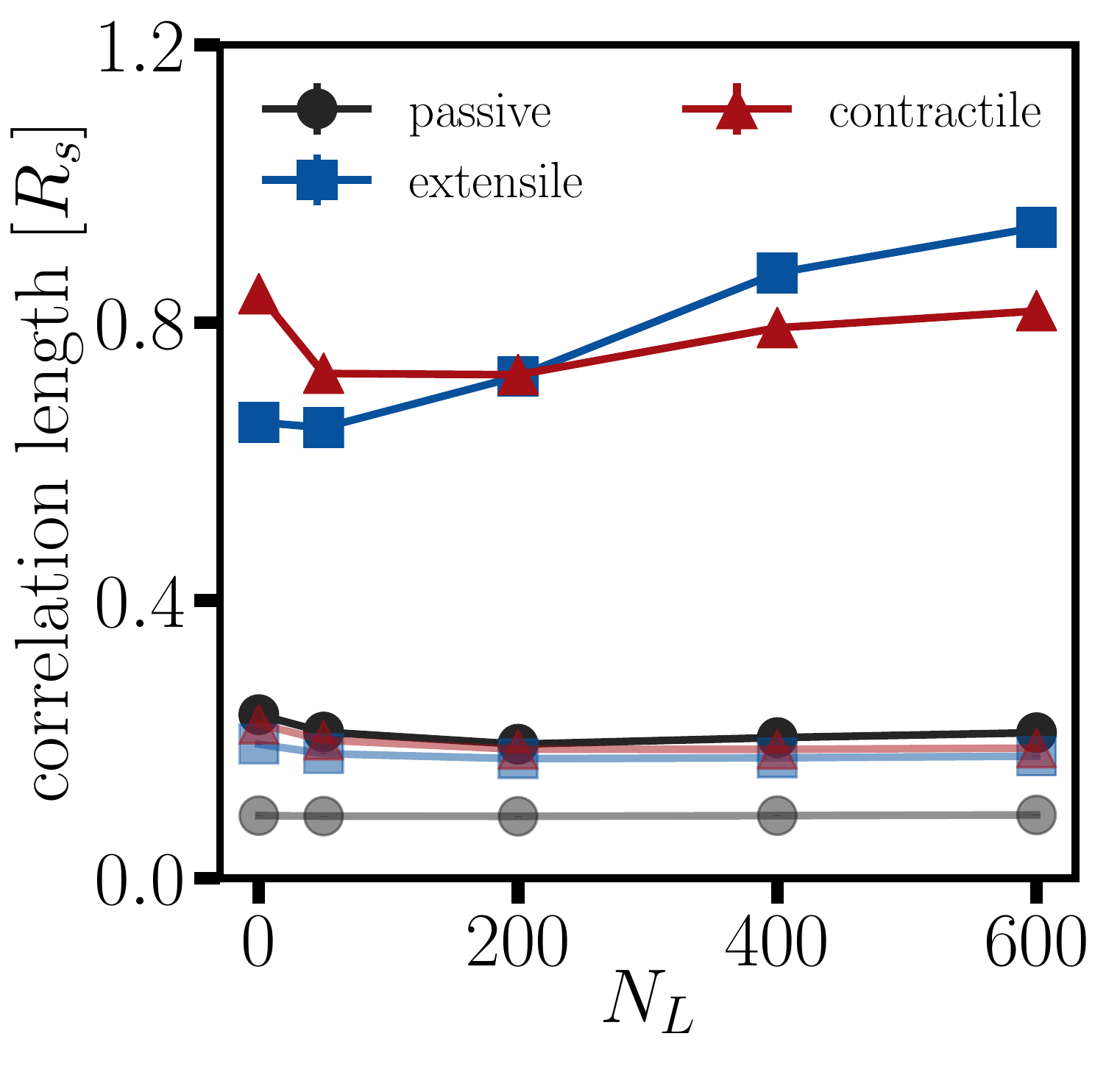}
		\includegraphics[width=0.24\linewidth]{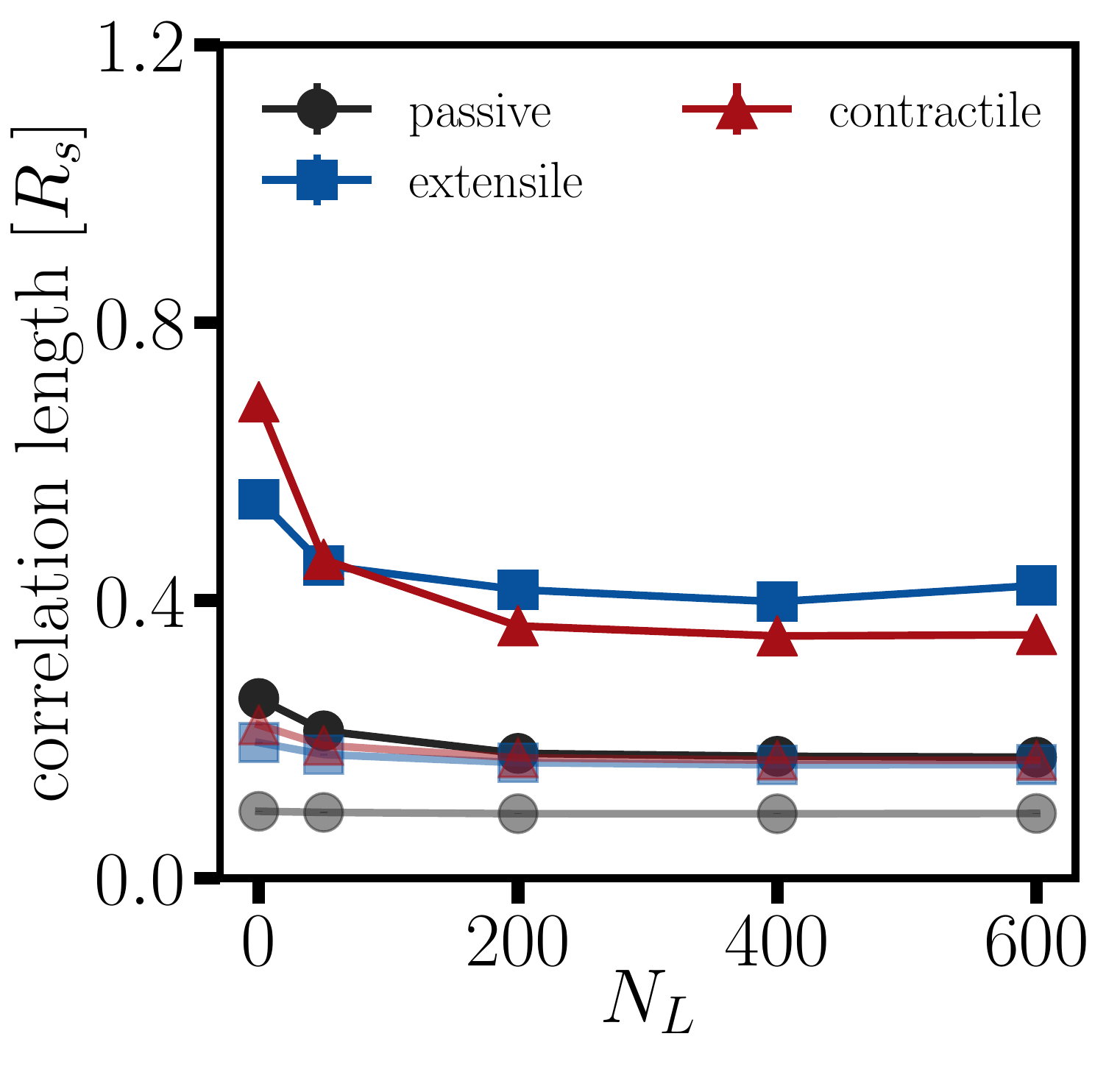}
		\includegraphics[width=0.24\linewidth]{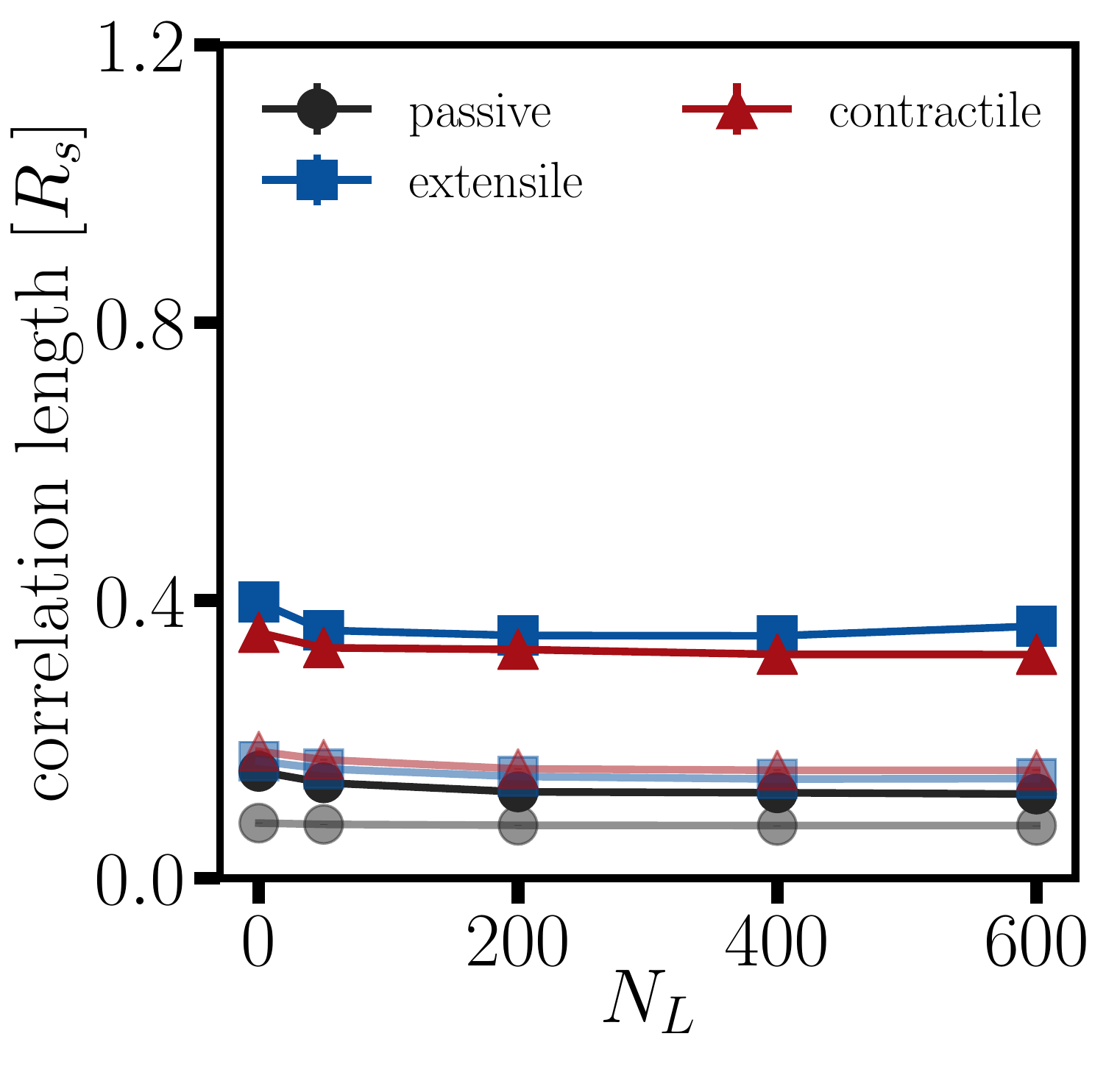}
		\includegraphics[width=0.24\linewidth]{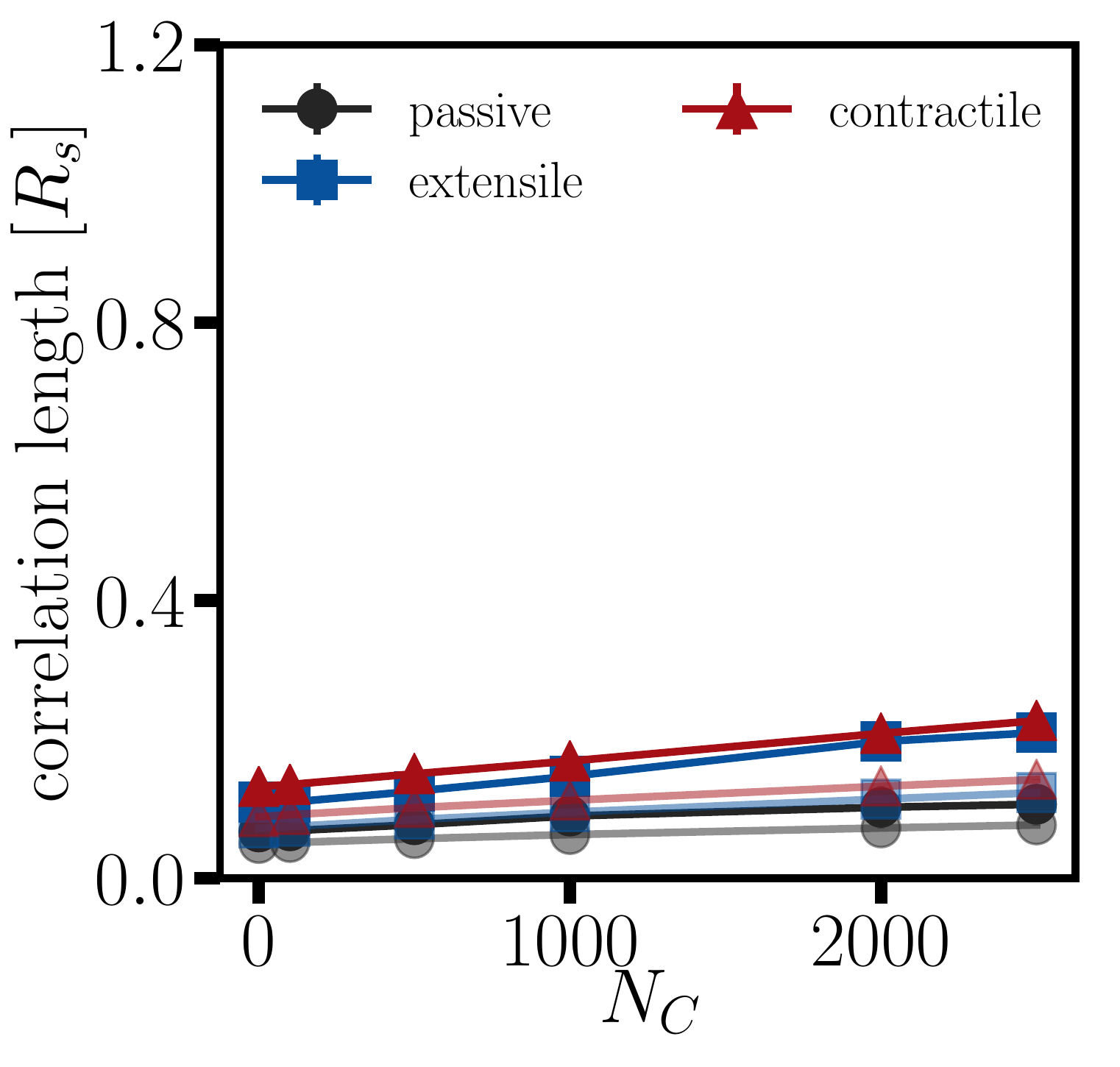}
		\includegraphics[width=0.24\linewidth]{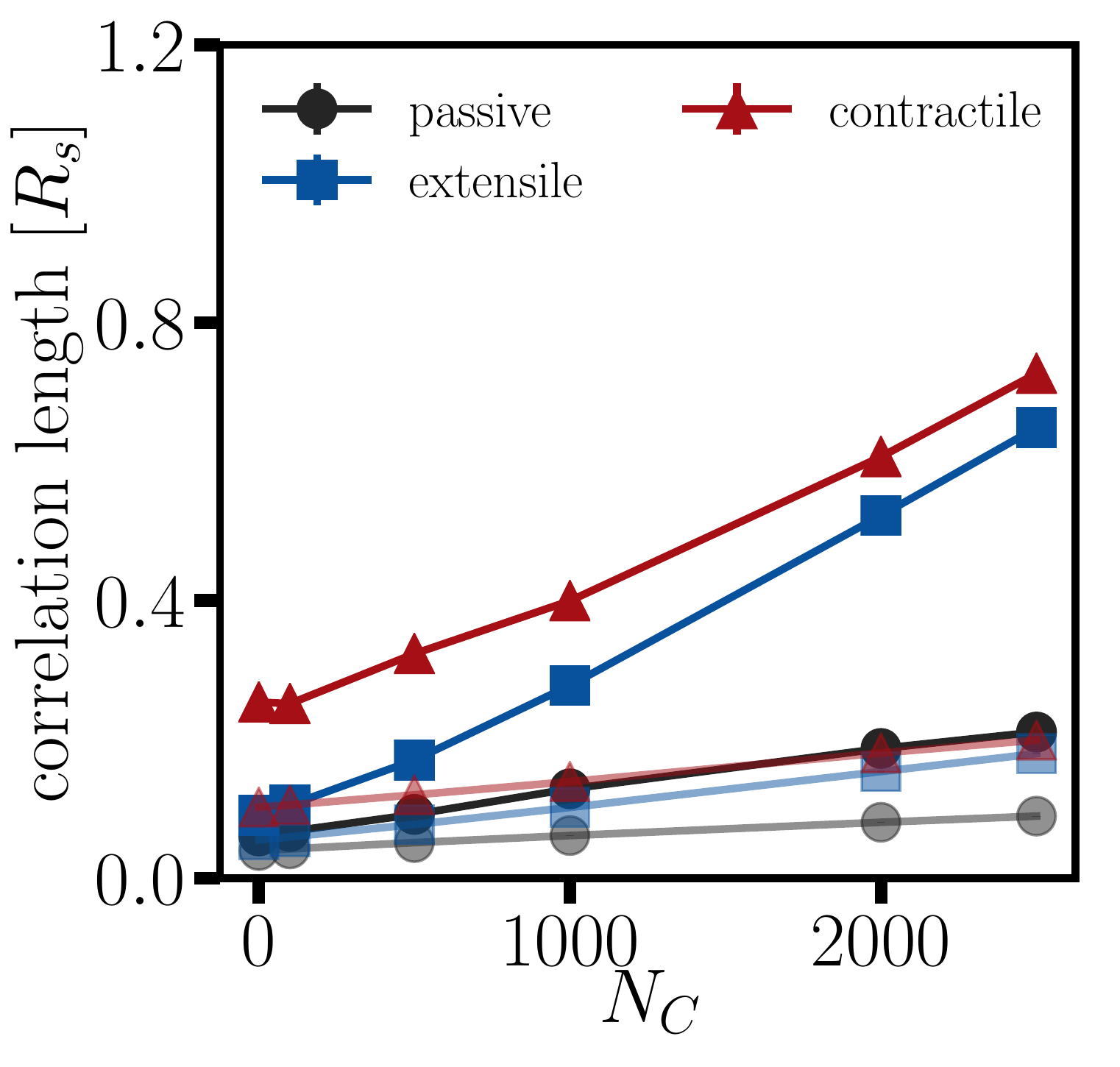}
		\includegraphics[width=0.24\linewidth]{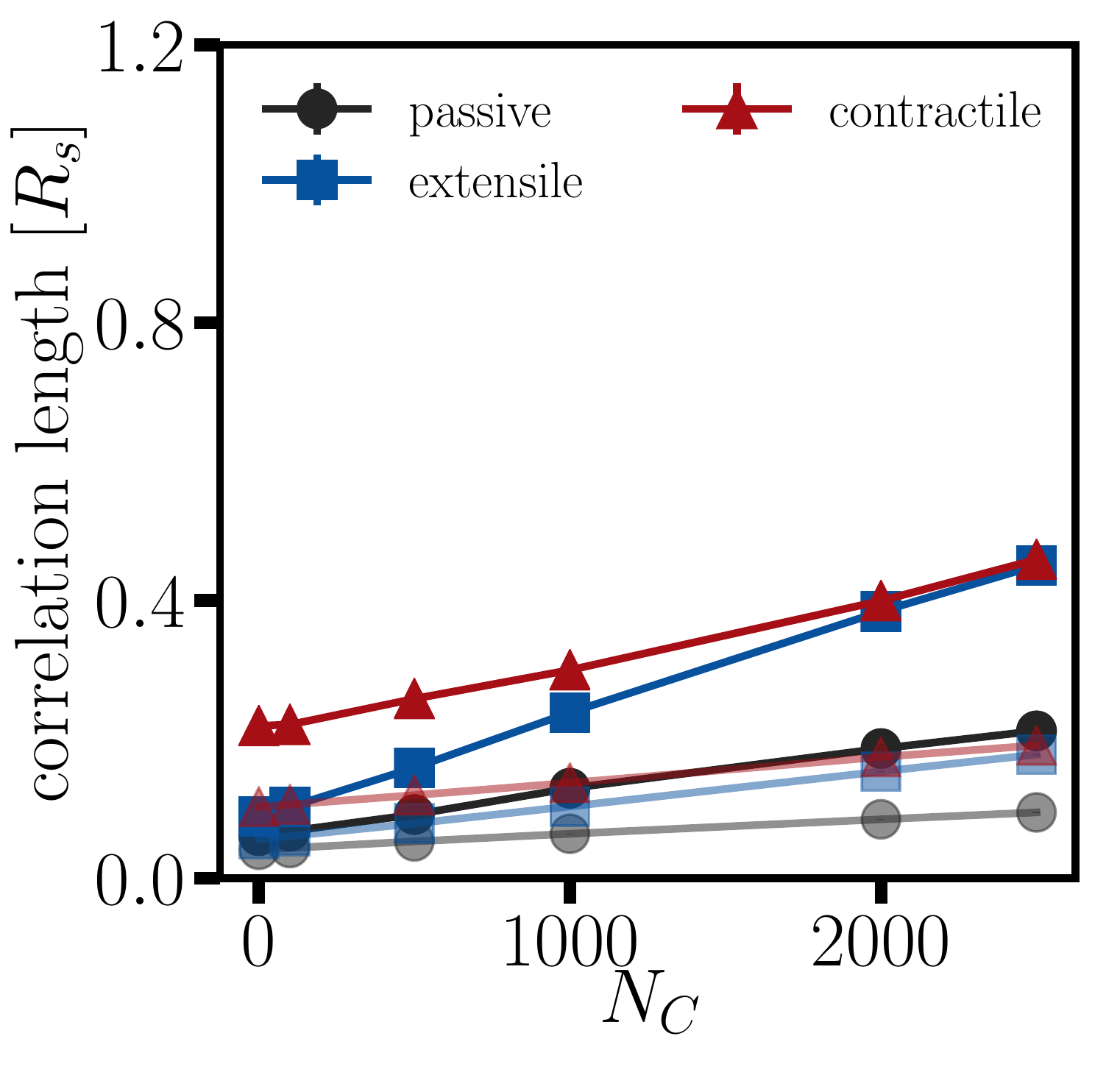}
		\includegraphics[width=0.24\linewidth]{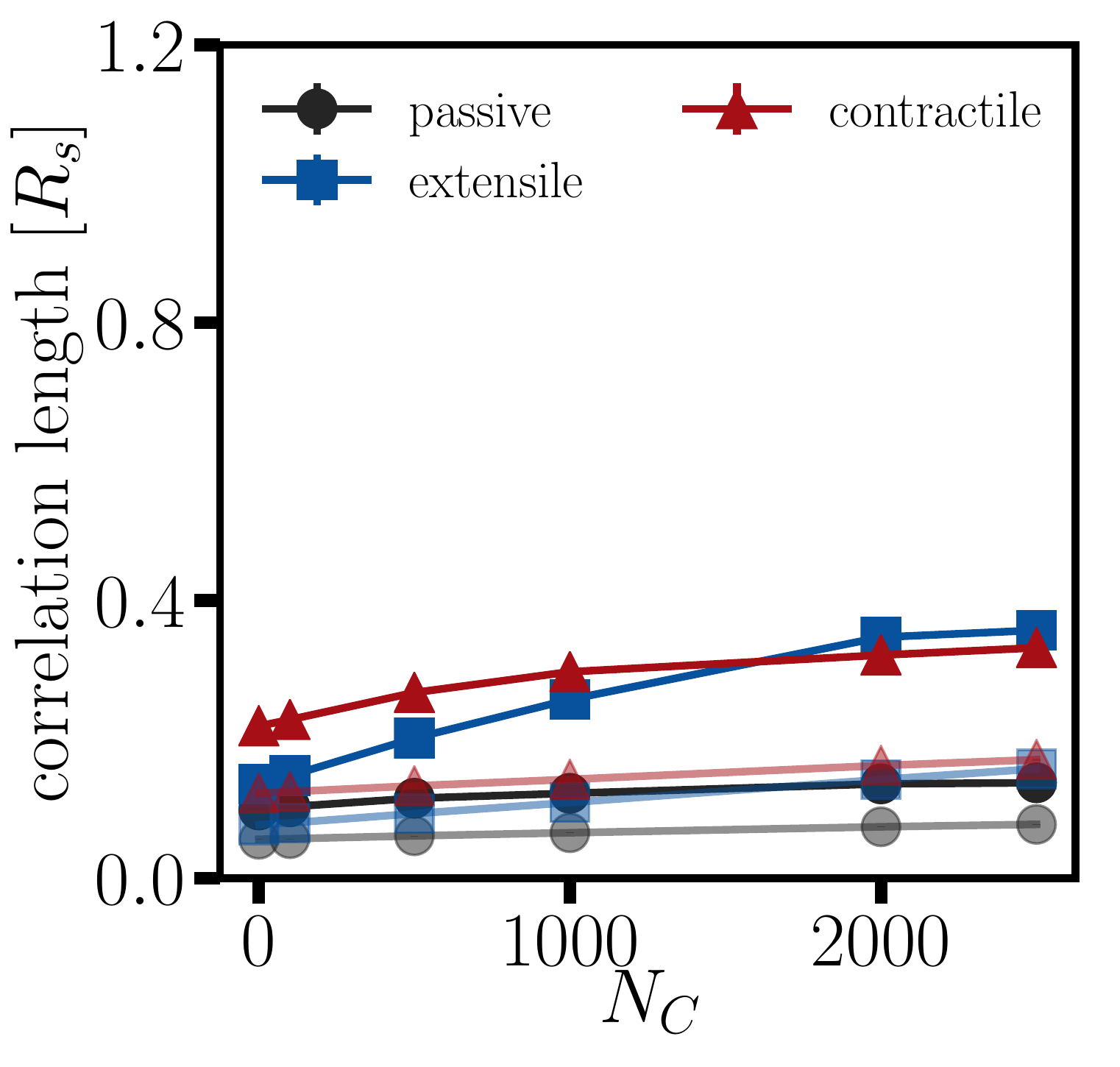}
	\end{center}
	\caption{Plot of correlation length as function of linkage number $N_L$ (top row) or crosslink number $N_C$ (bottom row) for time windows $5\,\tau$ (light) and $50\,\tau$ (dark). From left to right columns: The three-dimensional correlation length for the hard shell; the three-dimensional correlation length for the soft shell; three-dimensional correlation length for the soft shell with the COM motion subtracted; two-dimensional correlation length for  the soft shell with the COM motion subtracted.}
	\label{fig:correlation_2}
\end{figure*}

\begin{figure*}[h]
	\begin{center}
		\includegraphics[width=0.19\linewidth]{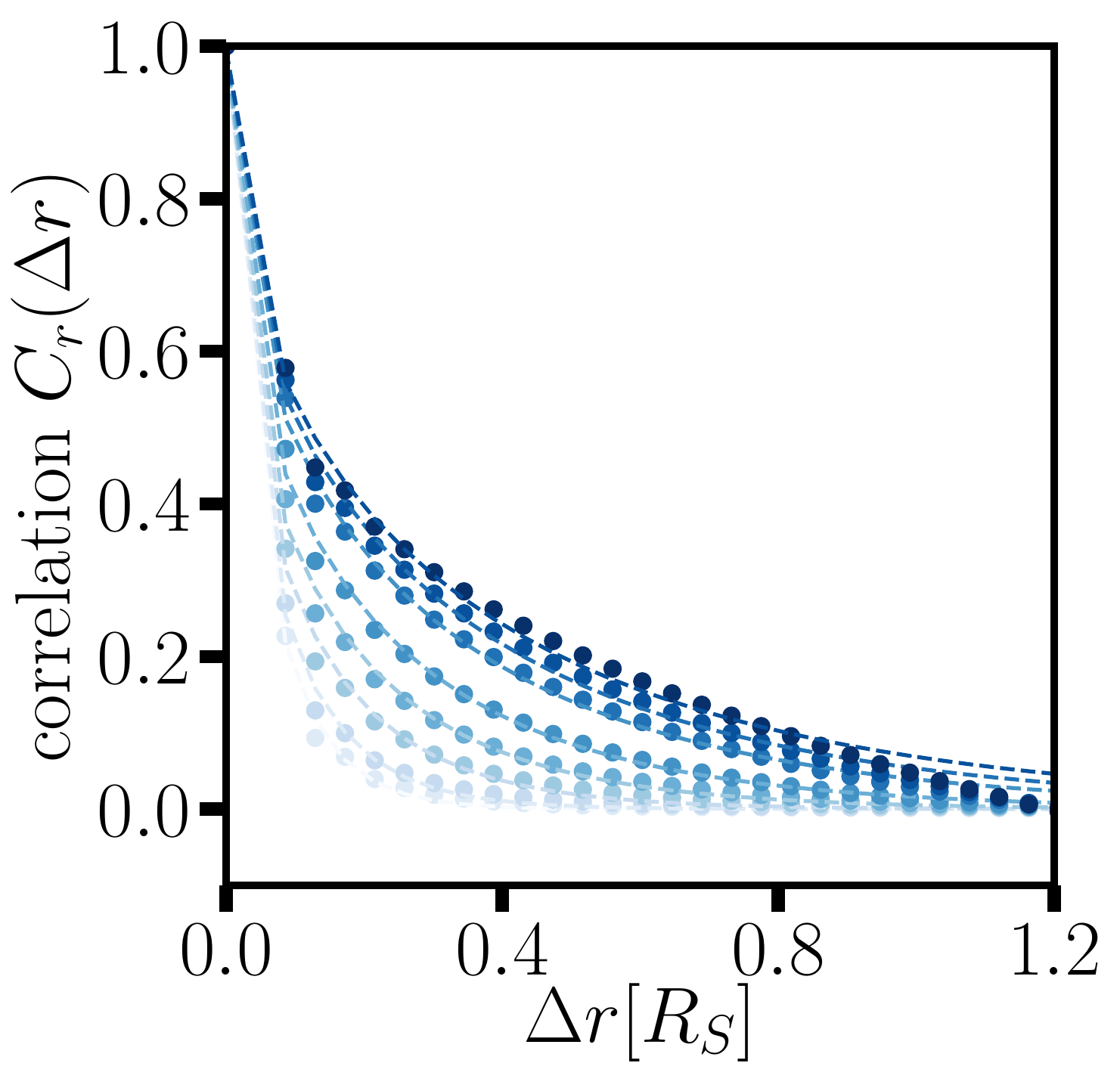}
		\includegraphics[width=0.19\linewidth]{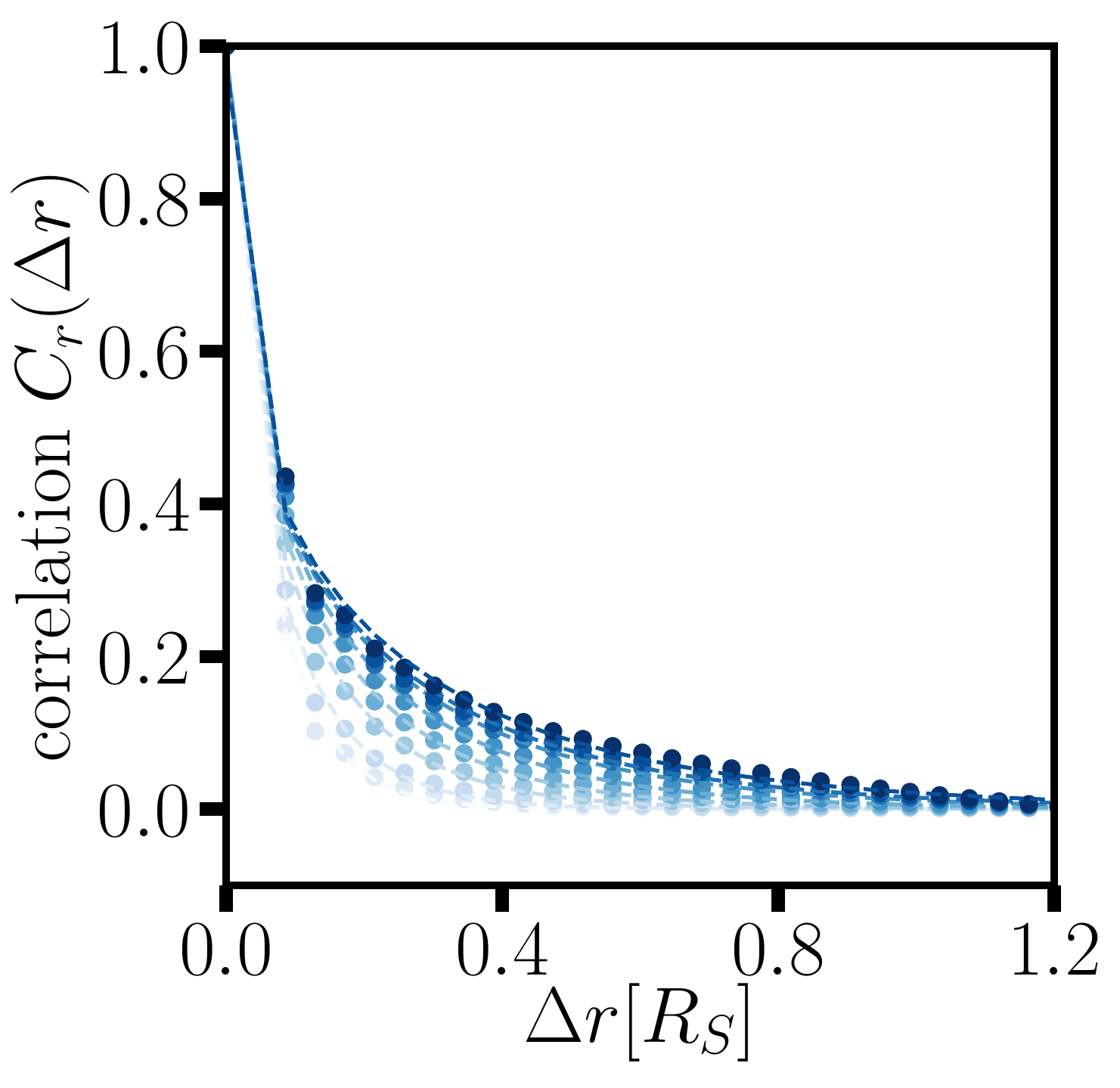}
		\includegraphics[width=0.19\linewidth]{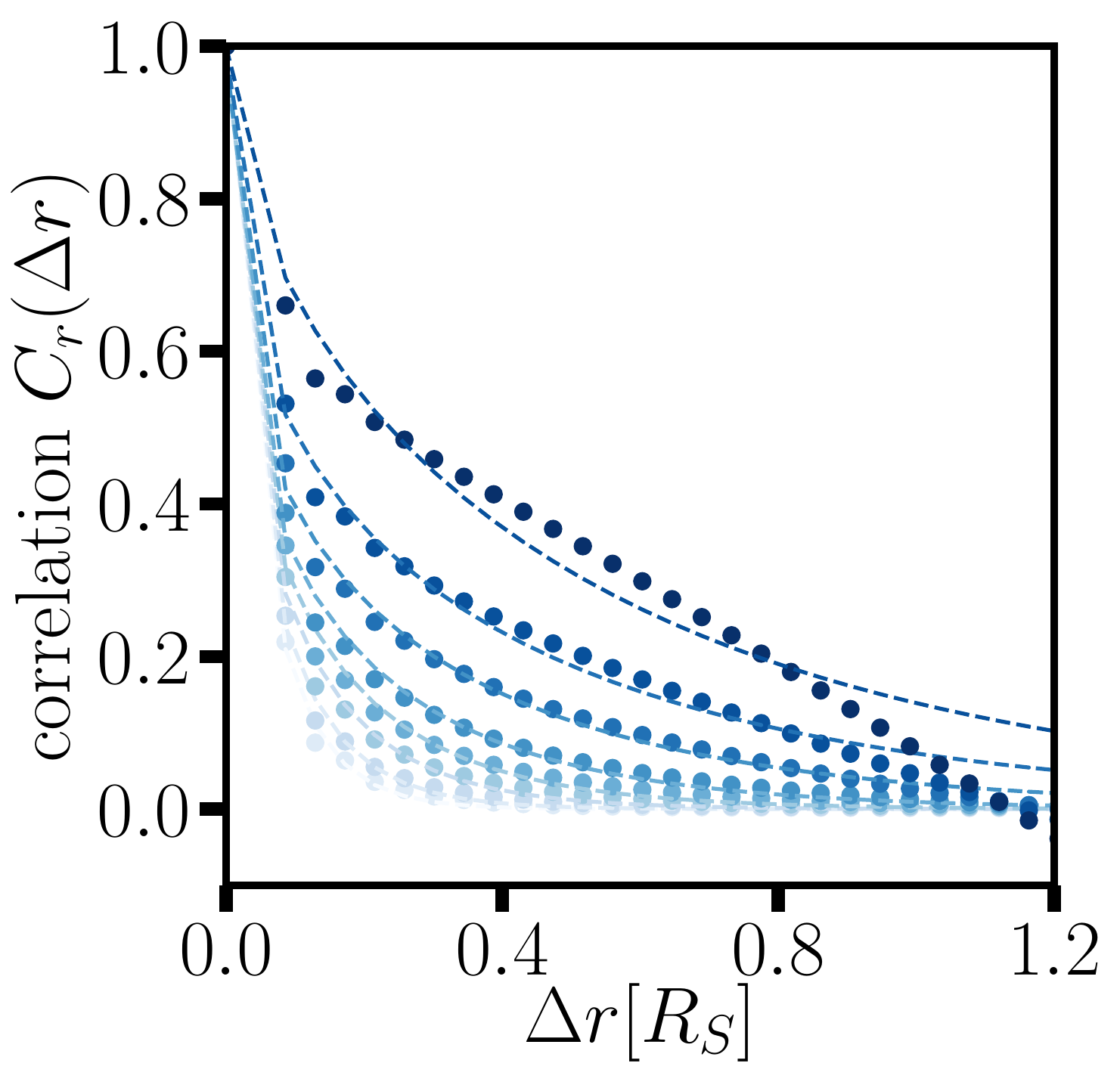}
		\includegraphics[width=0.19\linewidth]{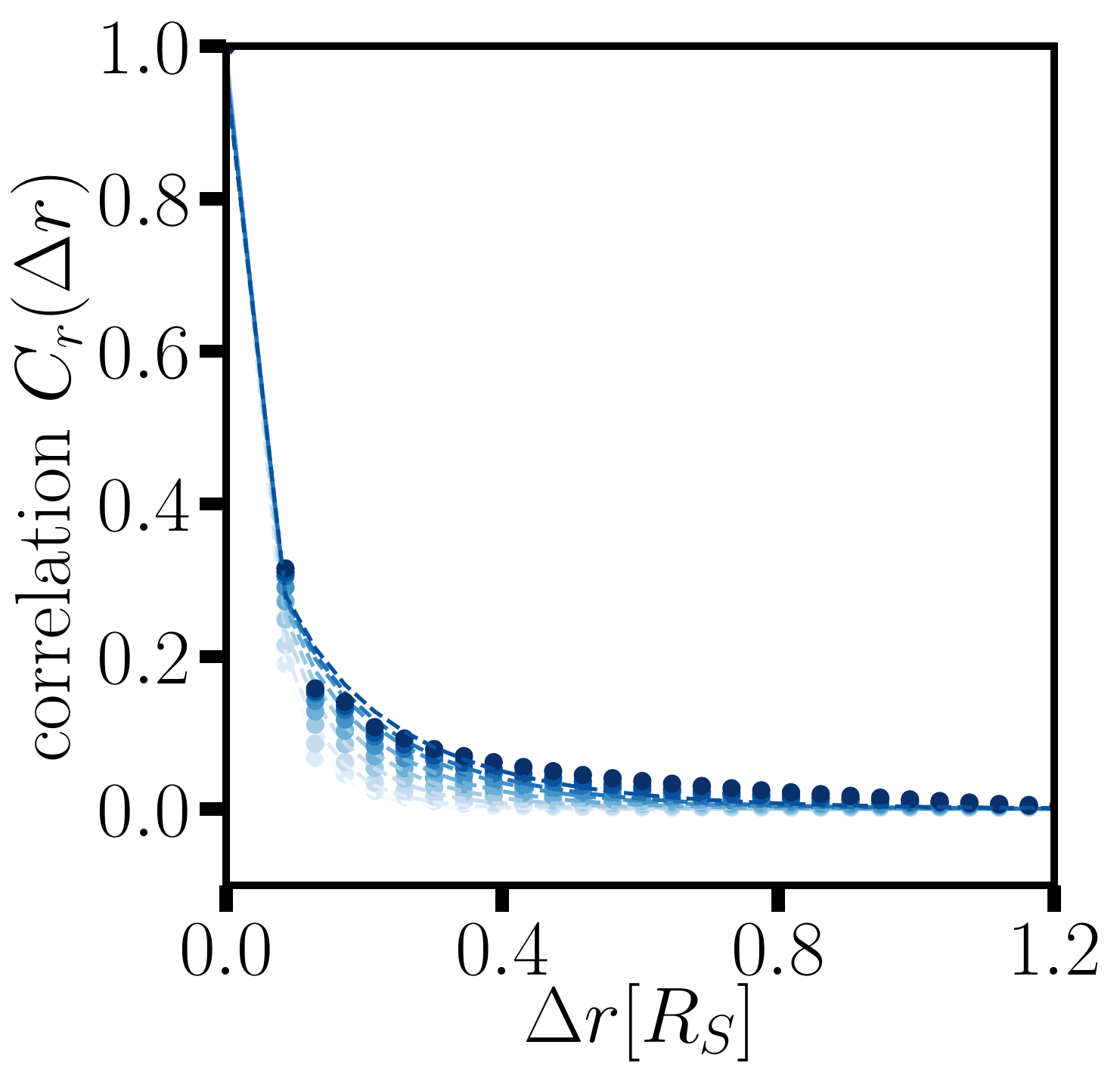}
		\includegraphics[width=0.19\linewidth]{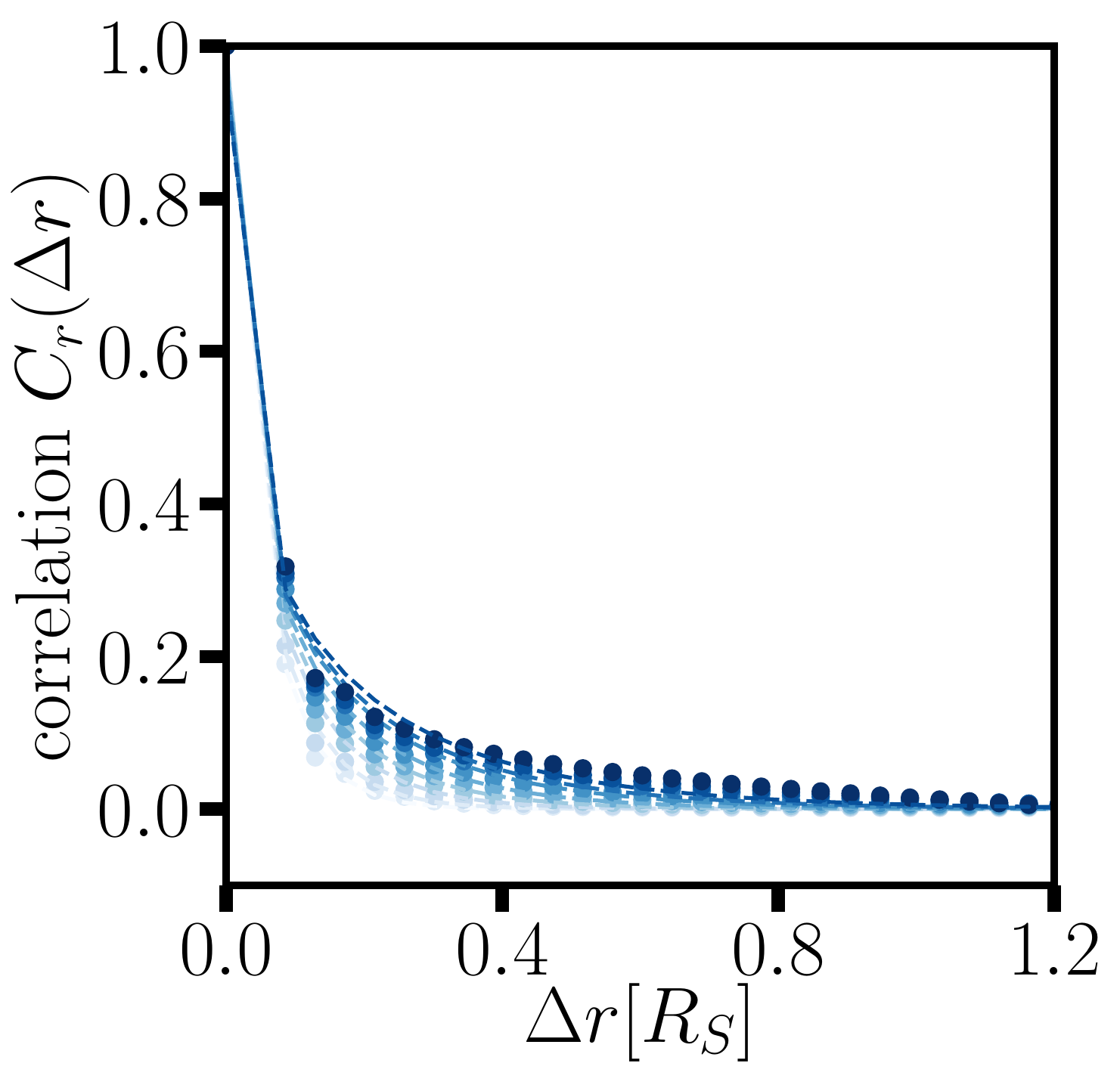}
		\includegraphics[width=0.19\linewidth]{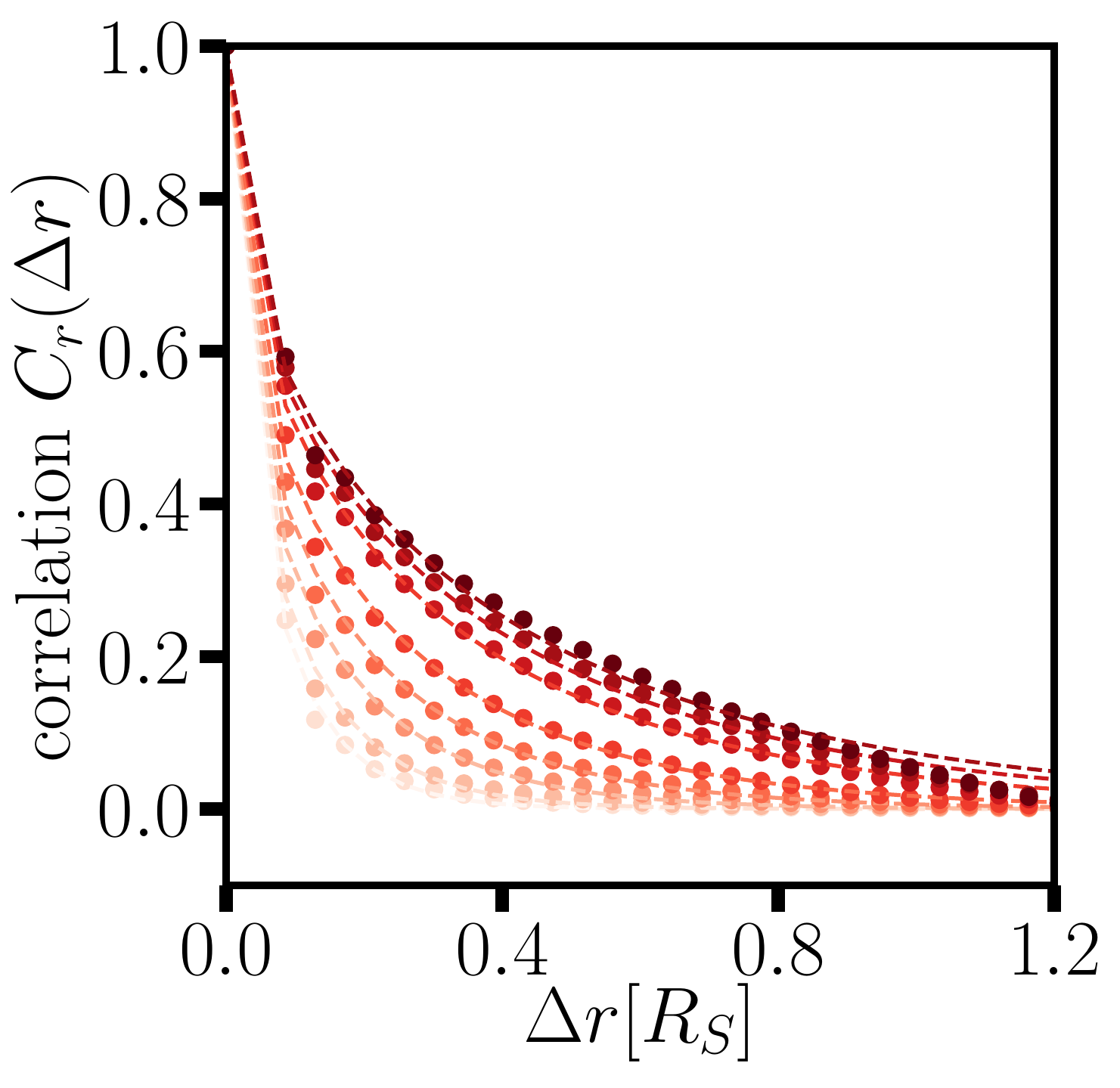}
		\includegraphics[width=0.19\linewidth]{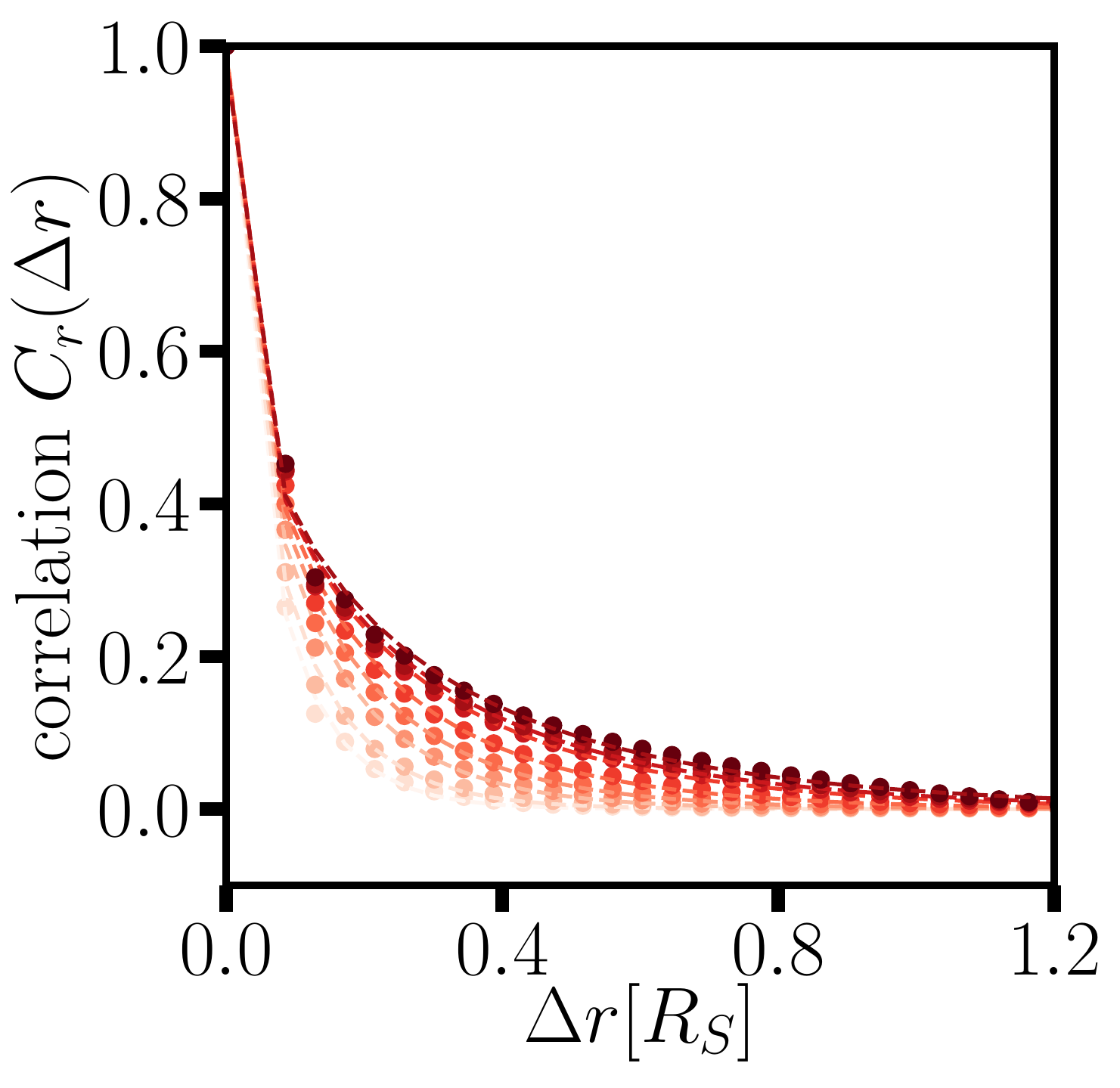}
		\includegraphics[width=0.19\linewidth]{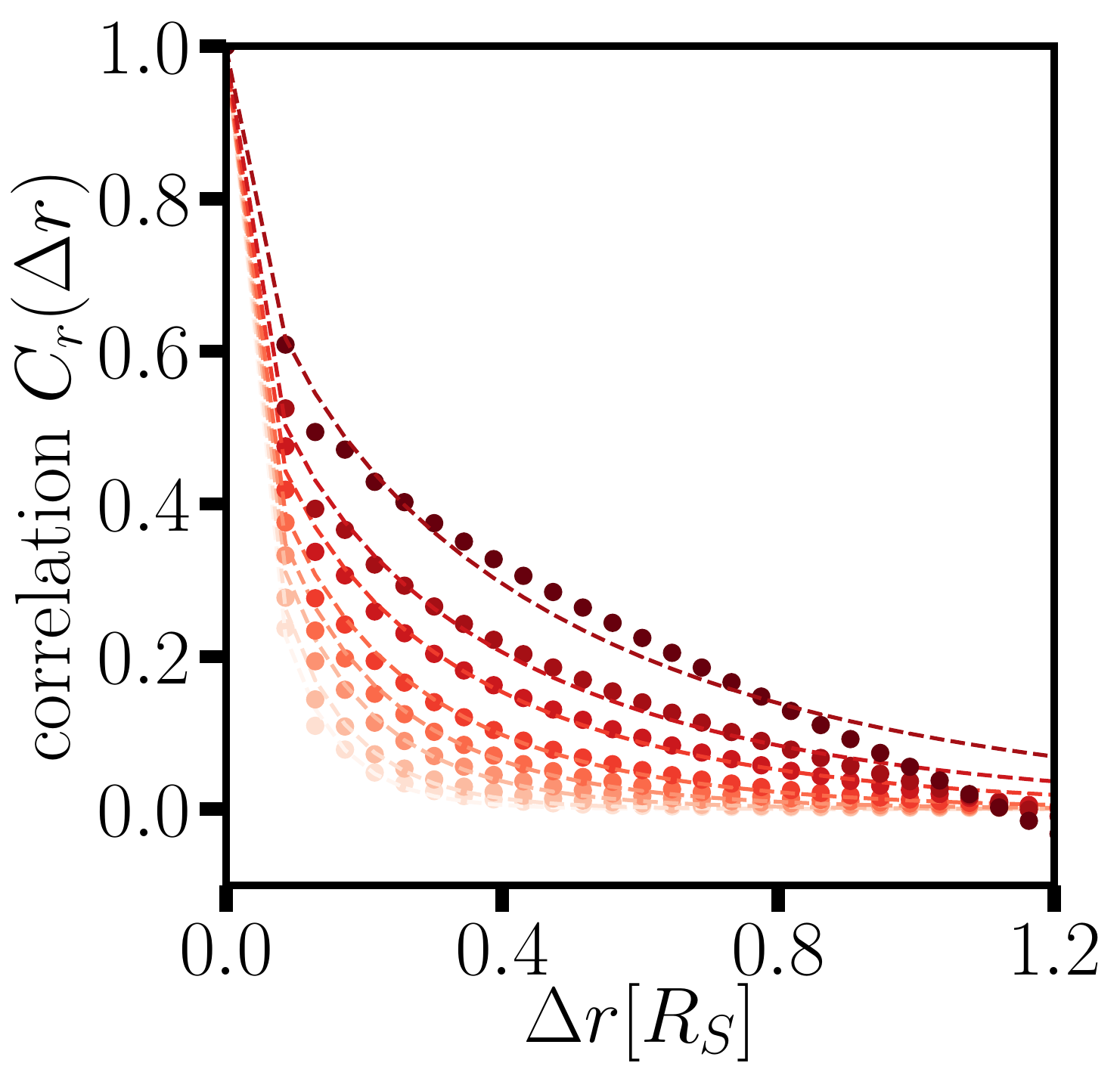}
		\includegraphics[width=0.19\linewidth]{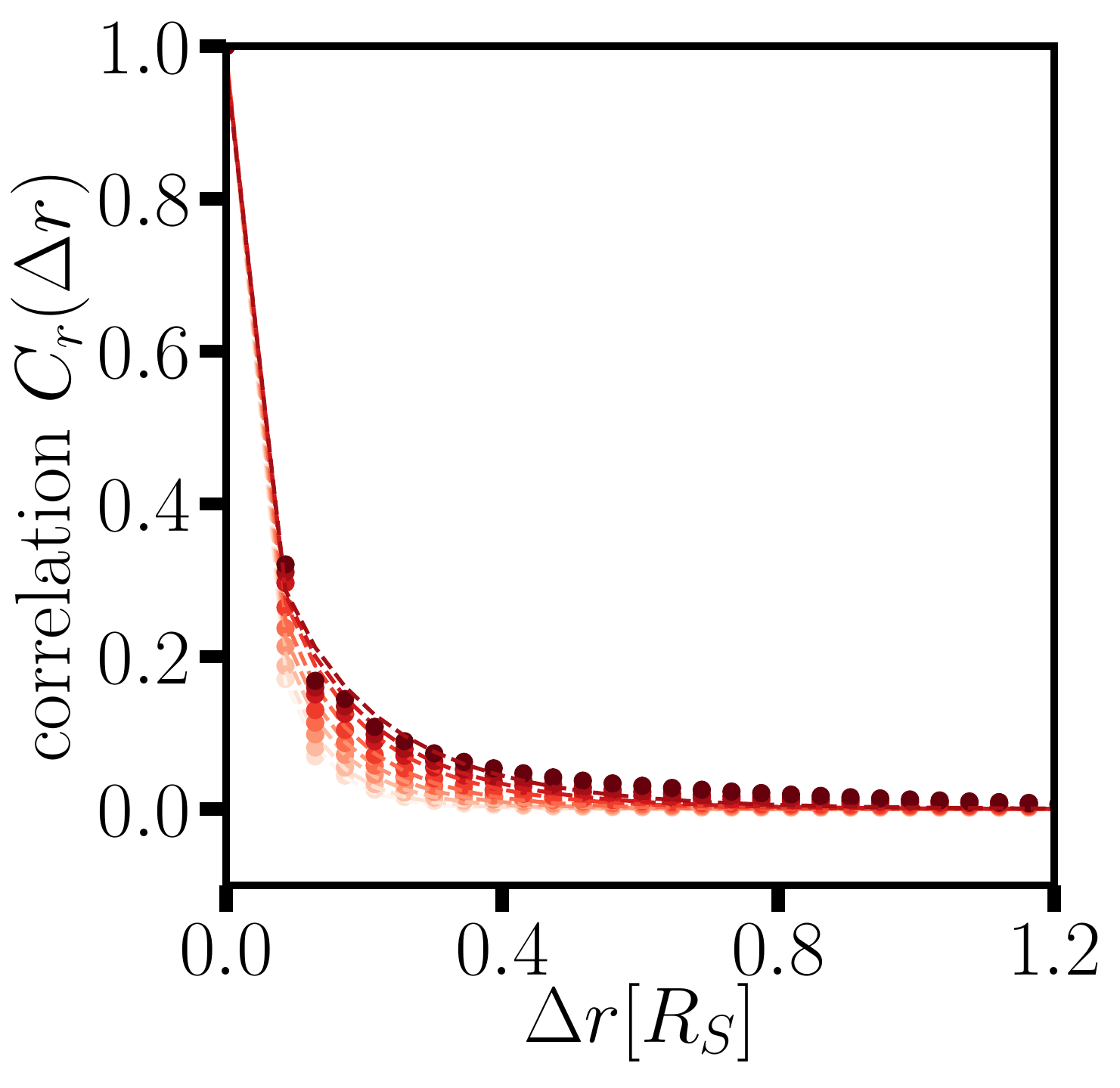}
		\includegraphics[width=0.19\linewidth]{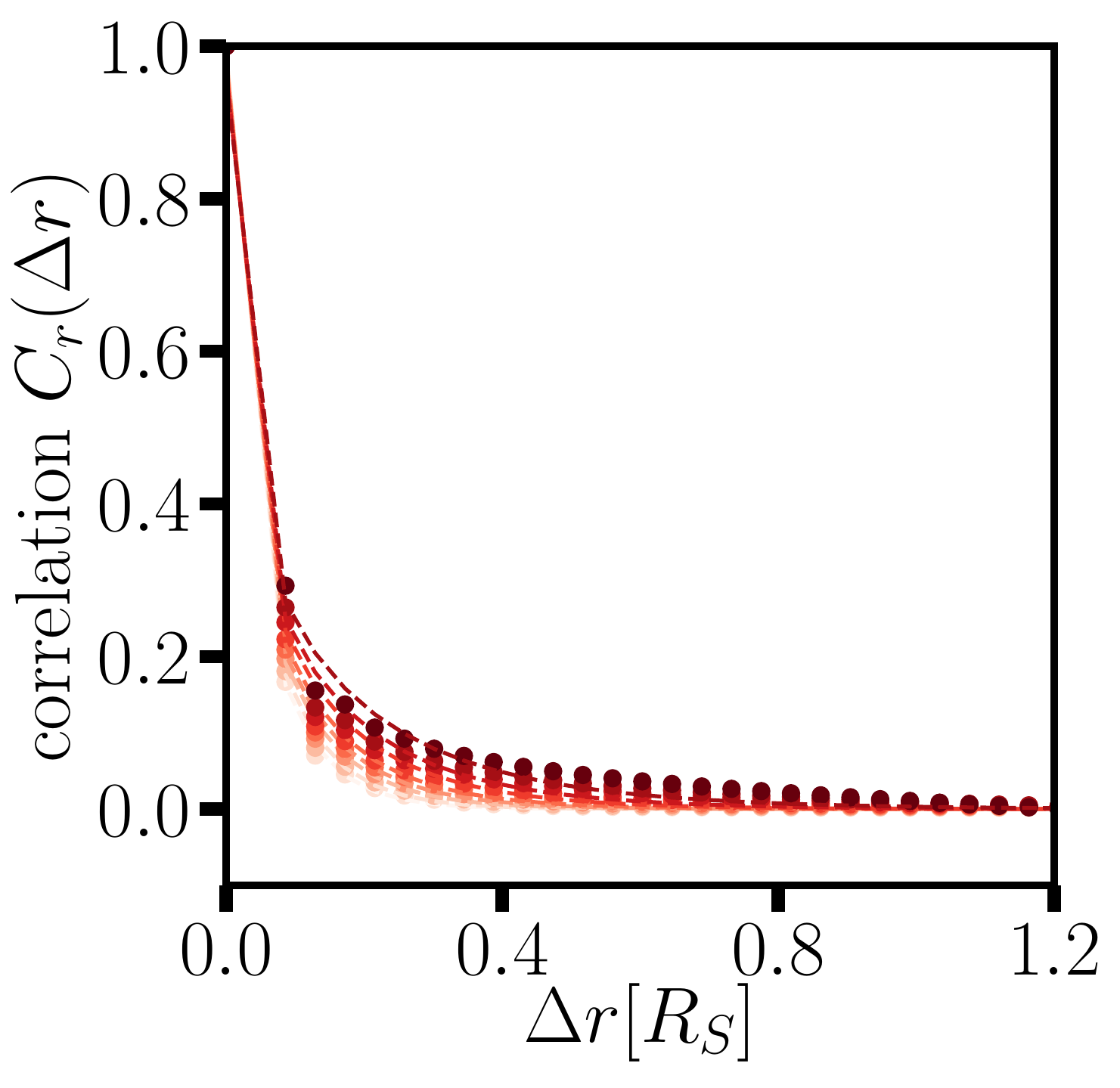}
	\end{center}
	\caption{Correlation functions for $N_C=2500$, $N_L=50$ and $\tau_m=20$ (first column) and four other cases. Column 2: $\tau_m=2$; Column 3: $\tau_m=\infty$; Column 4: $\tau_m=20$ and active forces are pairwise; Column 5: $\tau_m=\infty$ and ``active'' forces are exerted in a pairwise manner (\textit{i.e.}, an equilibrium model).}
	\label{fig:correlation_3}
\end{figure*}

\begin{figure*}[h]
	\begin{center}
		\includegraphics[width=0.22\linewidth]{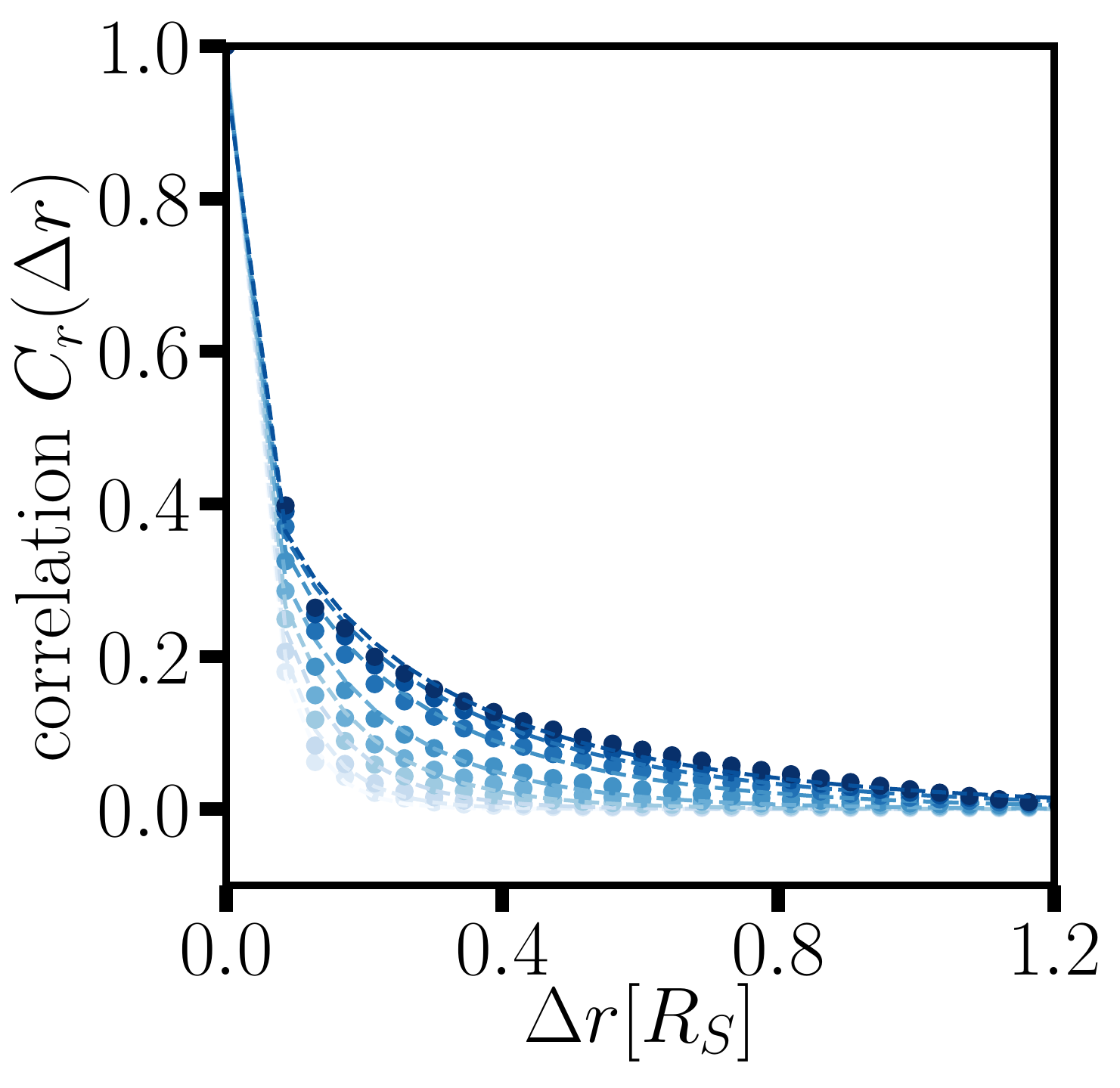}
		\includegraphics[width=0.22\linewidth]{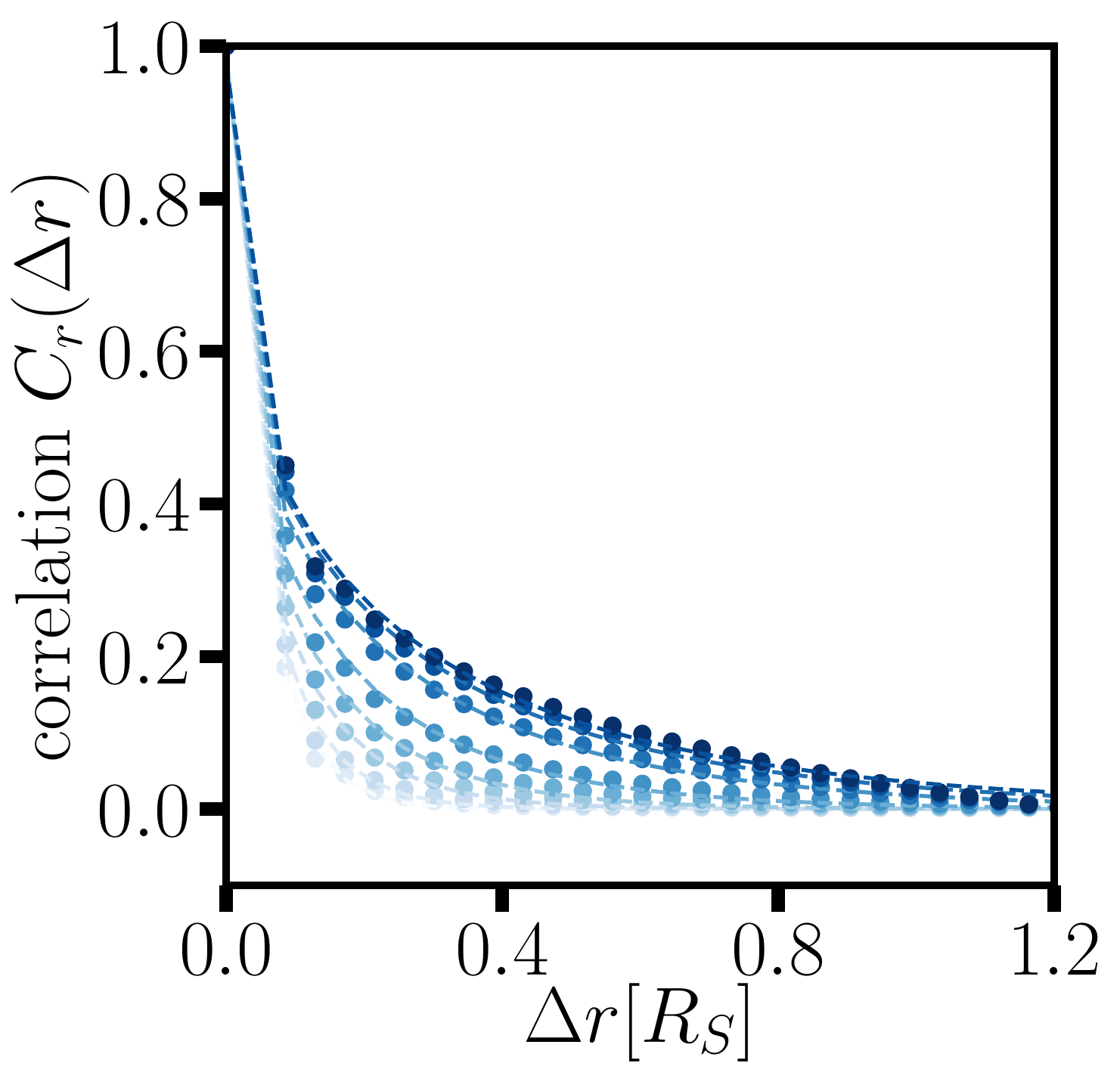}
		\includegraphics[width=0.22\linewidth]{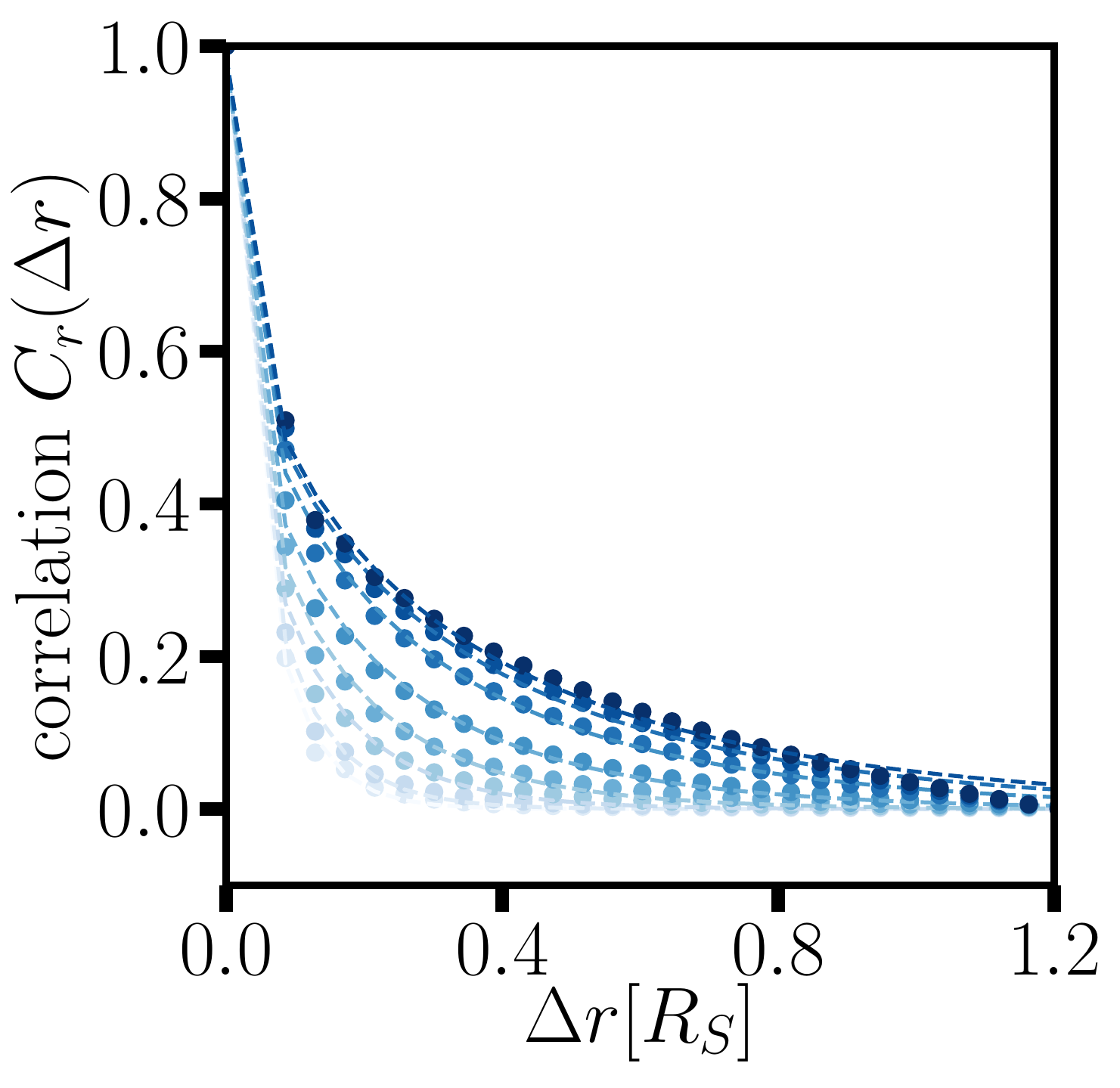}
		\includegraphics[width=0.22\linewidth]{208correlation_notitle.pdf}
		\includegraphics[width=0.22\linewidth]{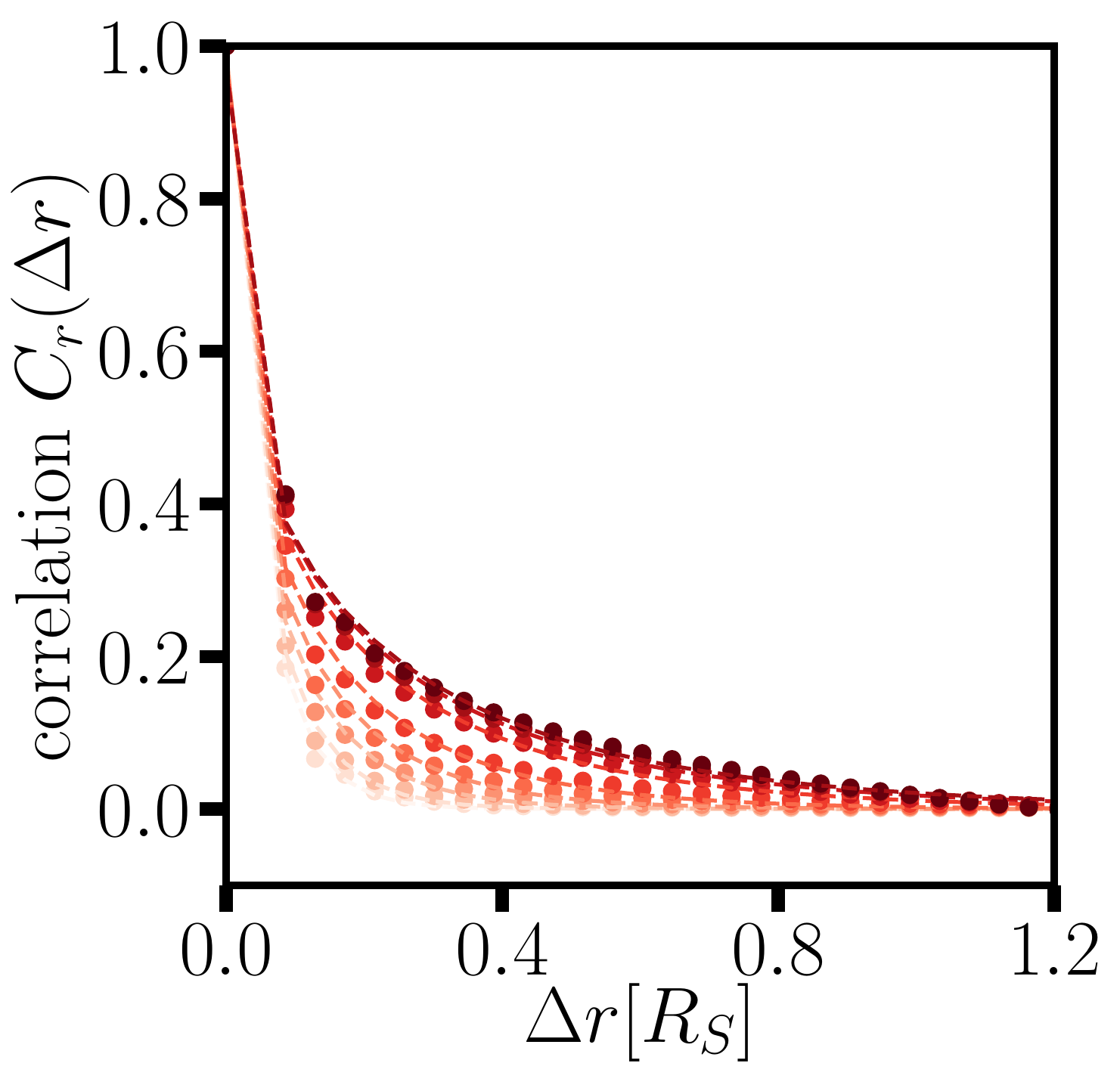}
		\includegraphics[width=0.22\linewidth]{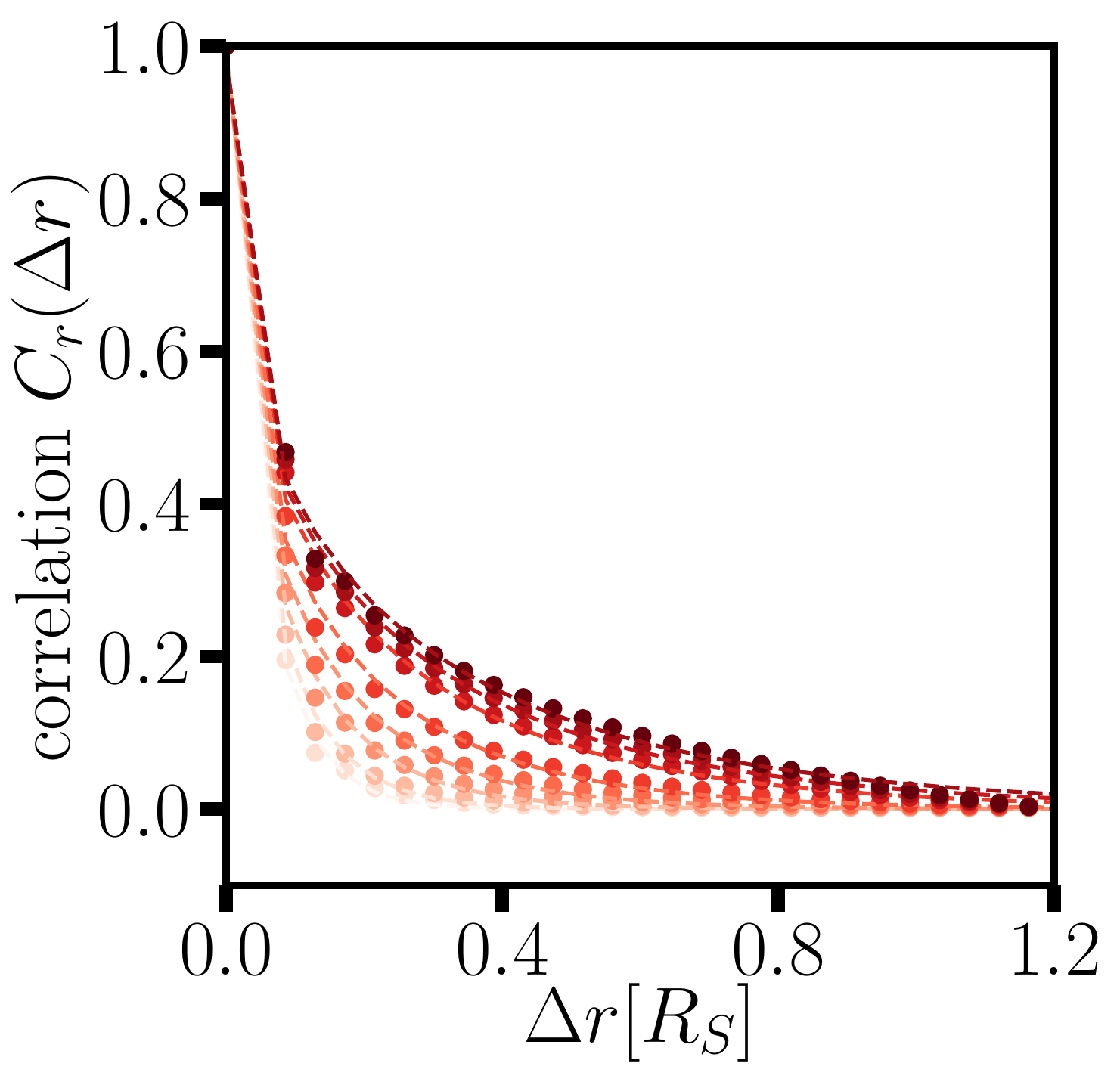}
		\includegraphics[width=0.22\linewidth]{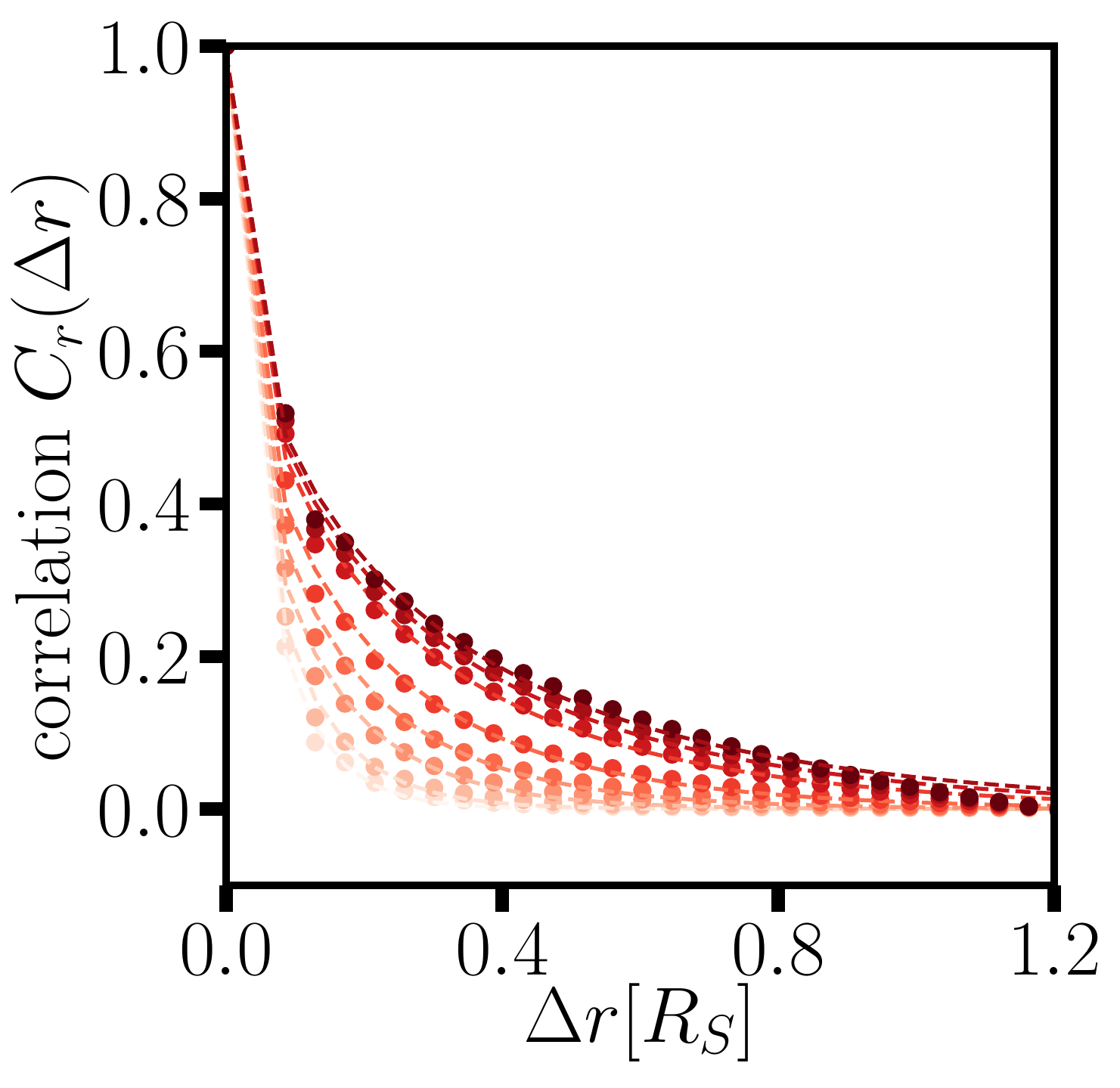}
		\includegraphics[width=0.22\linewidth]{224correlation_notitle.pdf}
	\end{center}
	\caption{Correlation functions for $N_C=2500$, $N_L=50$ and $N_m=400$ (last column) and three other cases. Column 1: $N_m=50$; Column 2: $N_m=100$; Column 3: $N_m=200$.}
	\label{fig:correlation_4}
\end{figure*}

\begin{figure}[h]
	\begin{center}
		\includegraphics[width=0.28\linewidth]{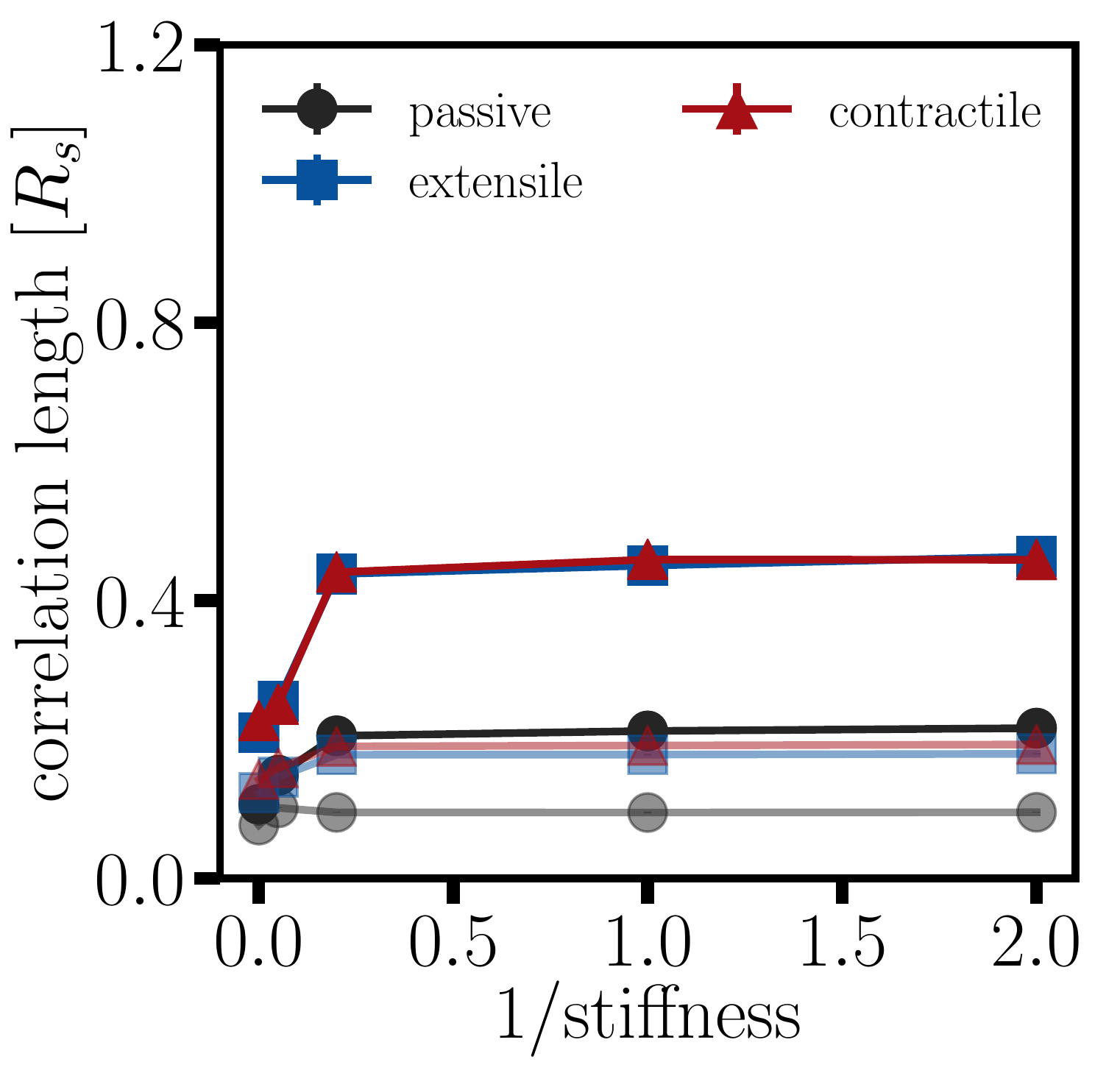}	
		\includegraphics[width=0.28\linewidth]{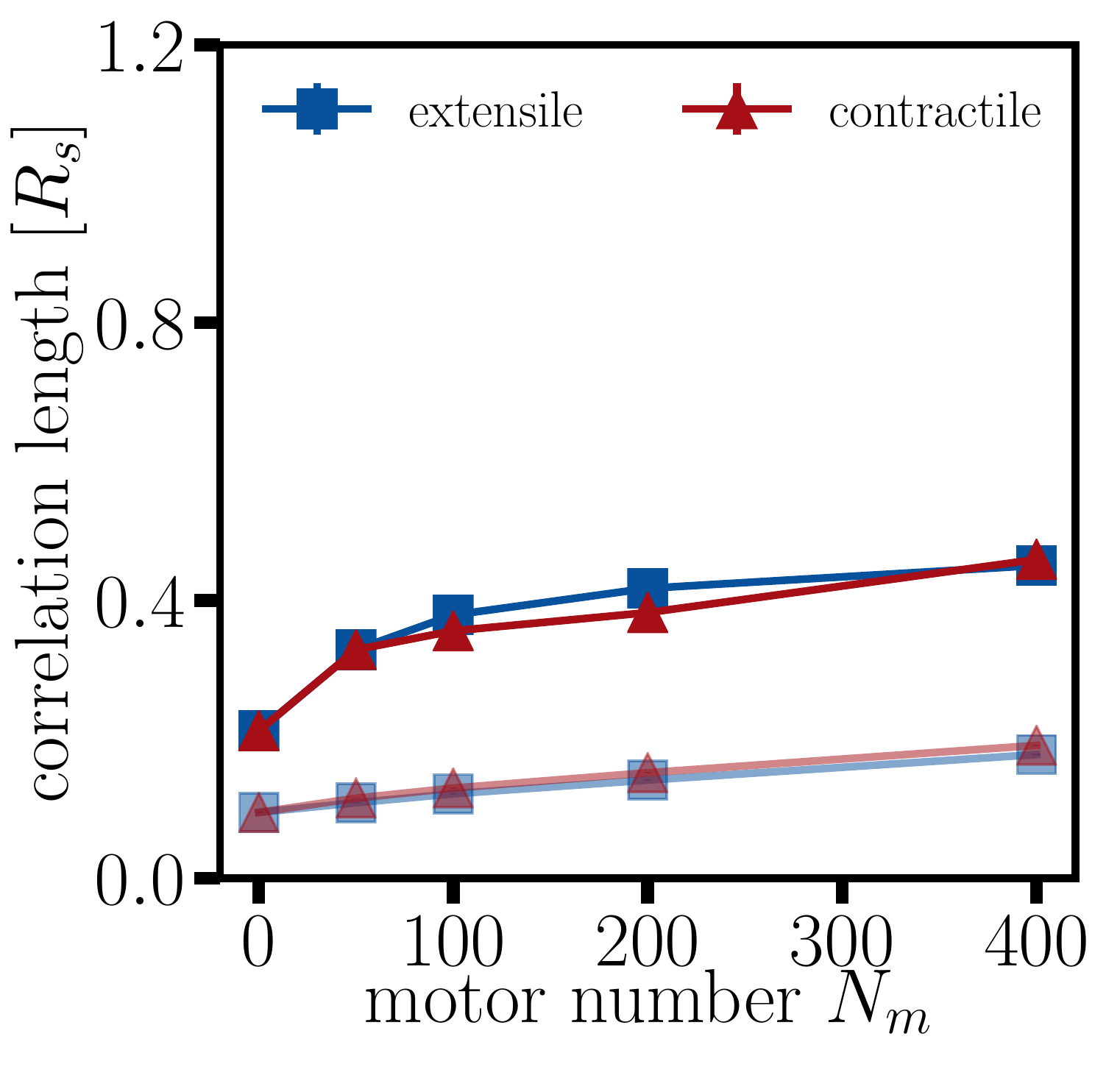}		
	\end{center}
	\caption{Left: plot of the correlation length as a function of shell stiffness for time windows $\Delta\tau=5,50$ (light and dark lines, respectively). Here $N_C=2500$, $N_L=50$, and $M=5$. Right: plot of the correlation length as a function of number, $N_m$, of motors for the same time windows.}
	\label{fig:correlationengthvsstiffness}
\end{figure}

\subsection{Shape fluctuations}
To evaluate shape fluctuations of the shell, we compute fluctuations in two
ways. First, in order to compare with experimental measurements, we
select a random slab through the center and project the coordinates of
the shell monomers in the slab to the plane where slab lies. Then, we
compute the fast-Fourier-transform (FFT) for spatial deviations of these monomers from the average
radius with the deviations with $h_q$ denoting the Fourier transform of the deviation with respect to wavenumber $q$. In Fig.~\ref{fig:shape1}, the power spectrum of the shape fluctuations for the
passive and extensile cases follow a decay exponent of $-2$, as expected for a stretchable membrane or shell~\cite{landau}. The spectrum of the shape
fluctuations increases monotonically with the number of crosslinks. The spectrum varies more dramatically 
with contractile motors as compared to extensile motors. Moreover, the
shape fluctuation spectrum also eventually saturates as a function of chromatin-lamina
linkage number. In Fig.~\ref{fig:shape2} we compute the spectrum of the shape fluctuations as characterized by the spherical harmonic functions (the $Y_{lm}$s with $l$ as the dimensionless spherical wavenumber). We obtain similar trends as in Fig.~\ref{fig:shape1}. Finally, in Fig.~\ref{fig:shape3}, we plot the spectrum for different motor strengths and different shell stiffnesses. 

We also measured shape fluctuations for simulations with different motor turnover times ($\tau_m$; Fig.~\ref{fig:shape4}), numbers of motors ($N_m$; Fig.~\ref{fig:shape5}), and motors that exert forces in a pairwise manner (Fig.~\ref{fig:shape4}). Increasing the turnover time in the active system increases the shape fluctuations.  As with correlations, varying the number of motors only weakly affects the strength of the shape fluctuations.  Simulations with motors that exert pairwise forces exhibit fluctuation spectra similar to that of the passive system.

\begin{figure*}
	\begin{center}
		\includegraphics[width=0.28\linewidth]{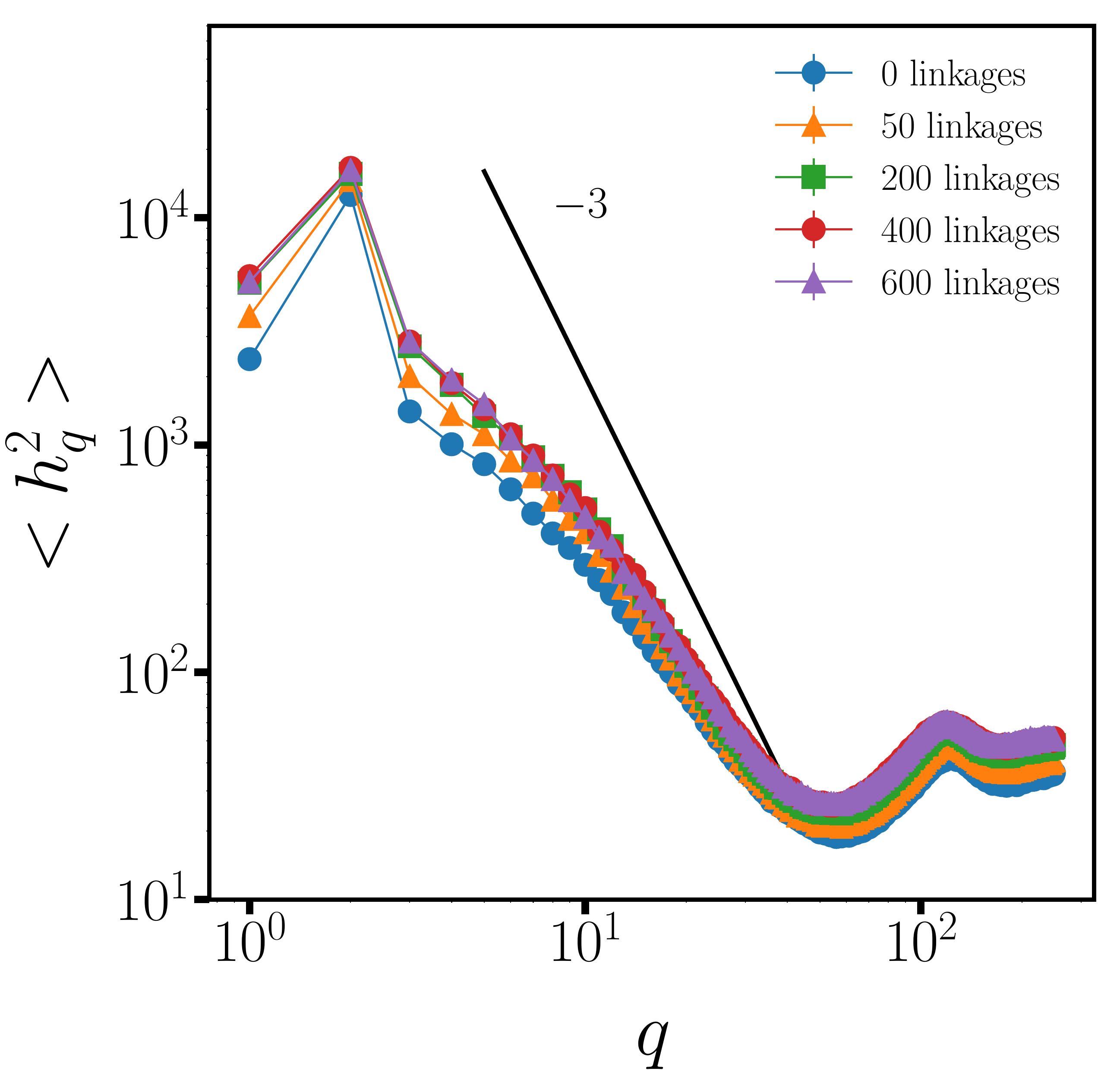}
		\includegraphics[width=0.28\linewidth]{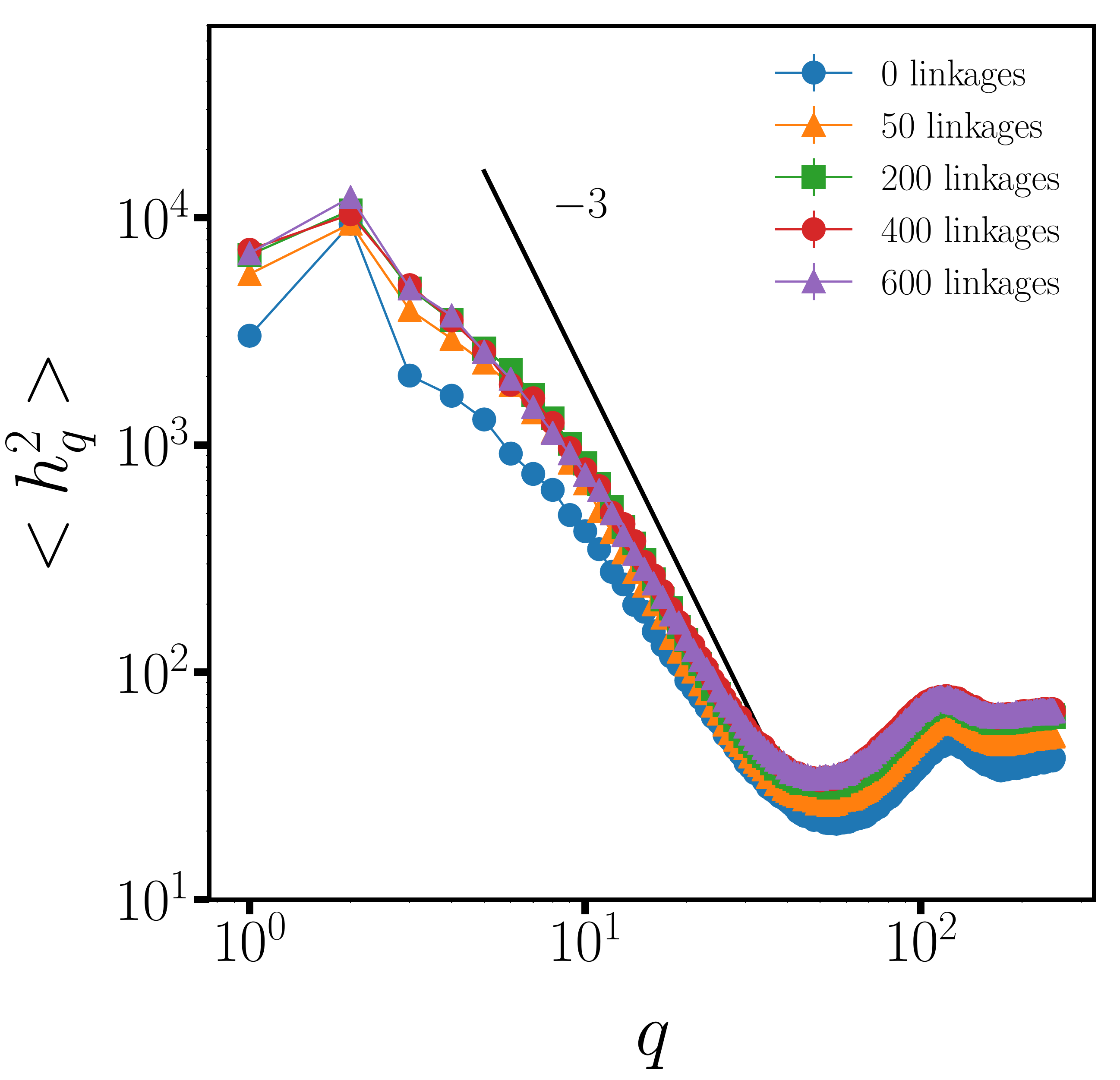}
		\includegraphics[width=0.28\linewidth]{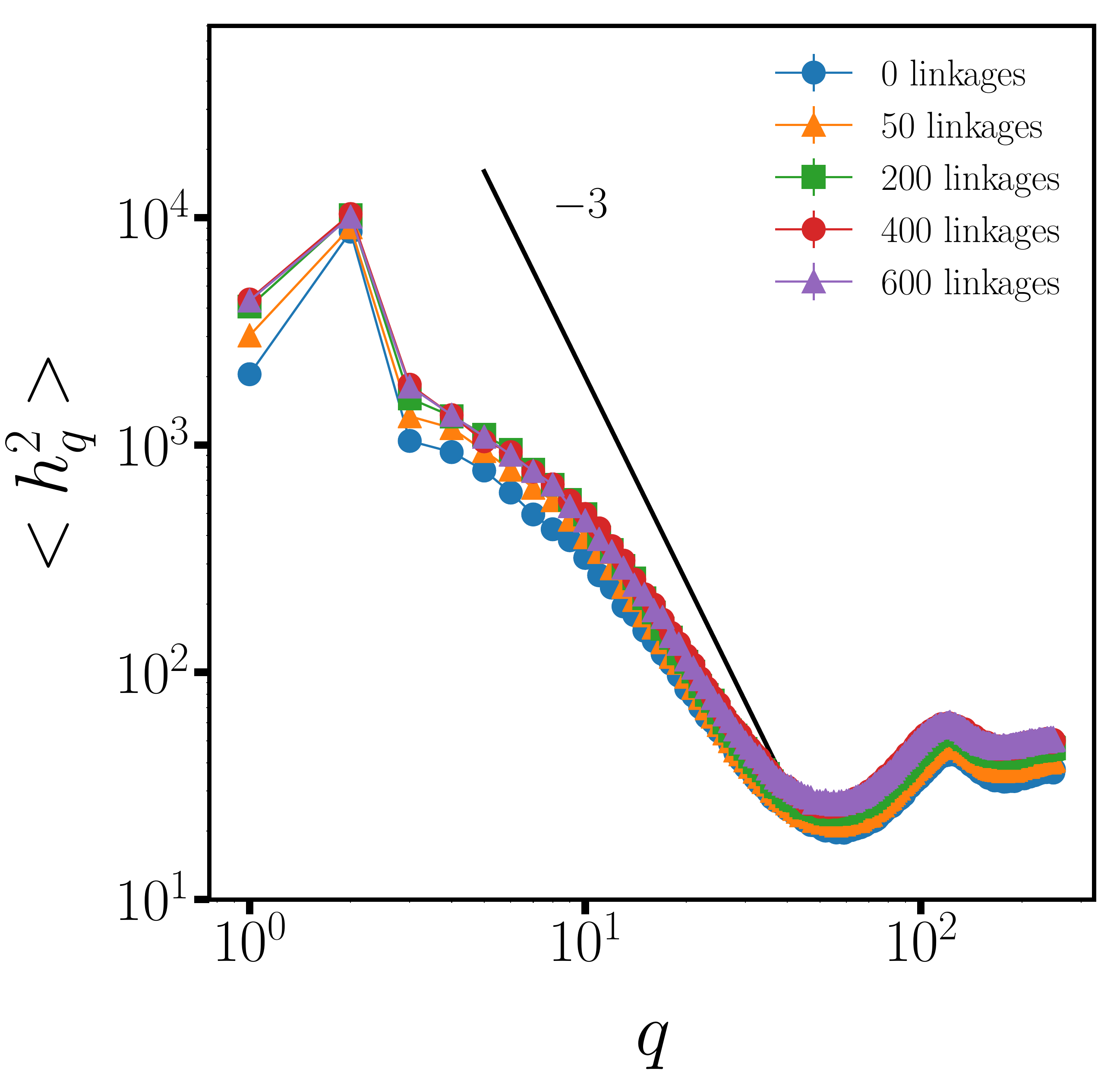}
		\includegraphics[width=0.28\linewidth]{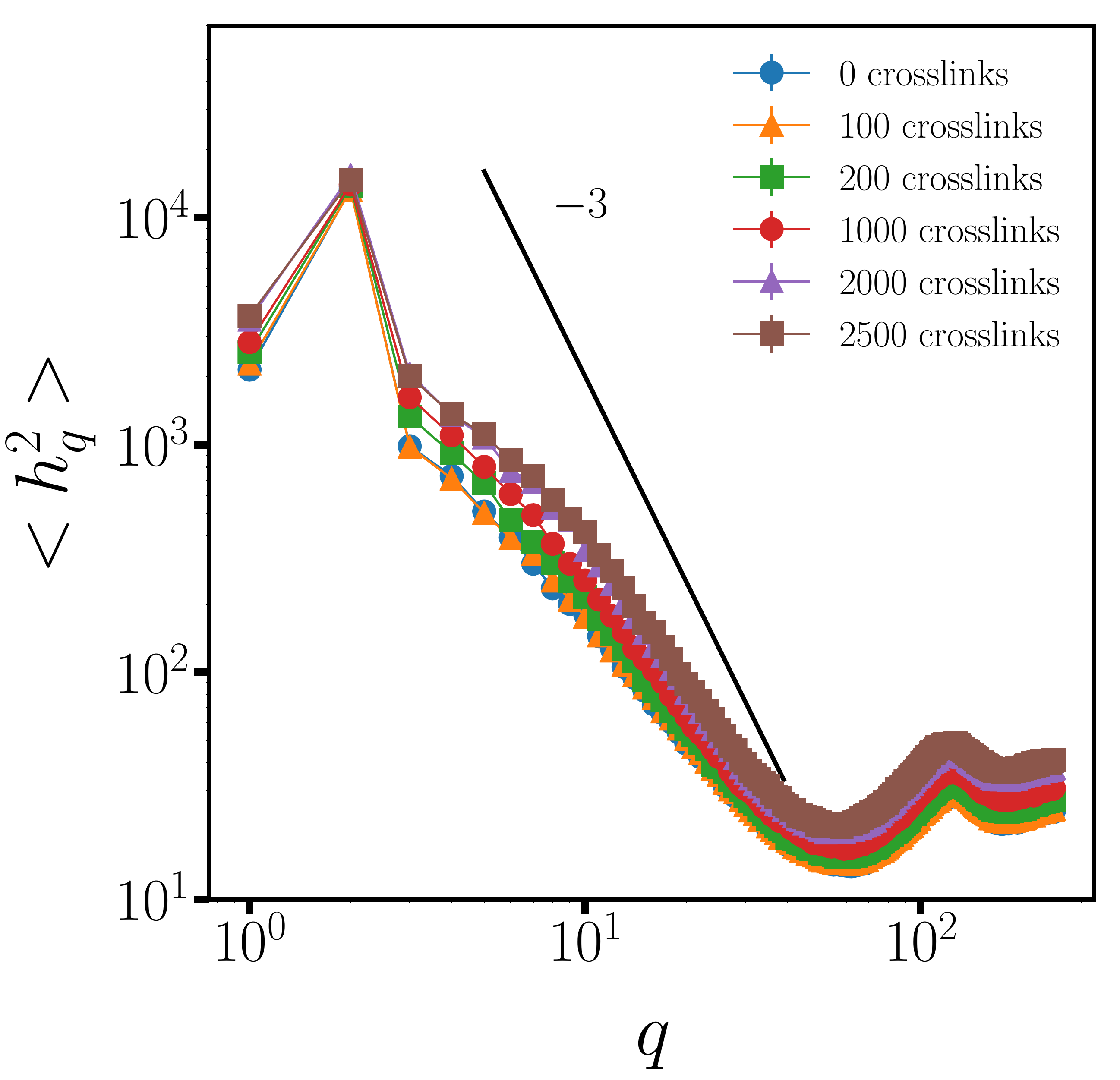}
		\includegraphics[width=0.28\linewidth]{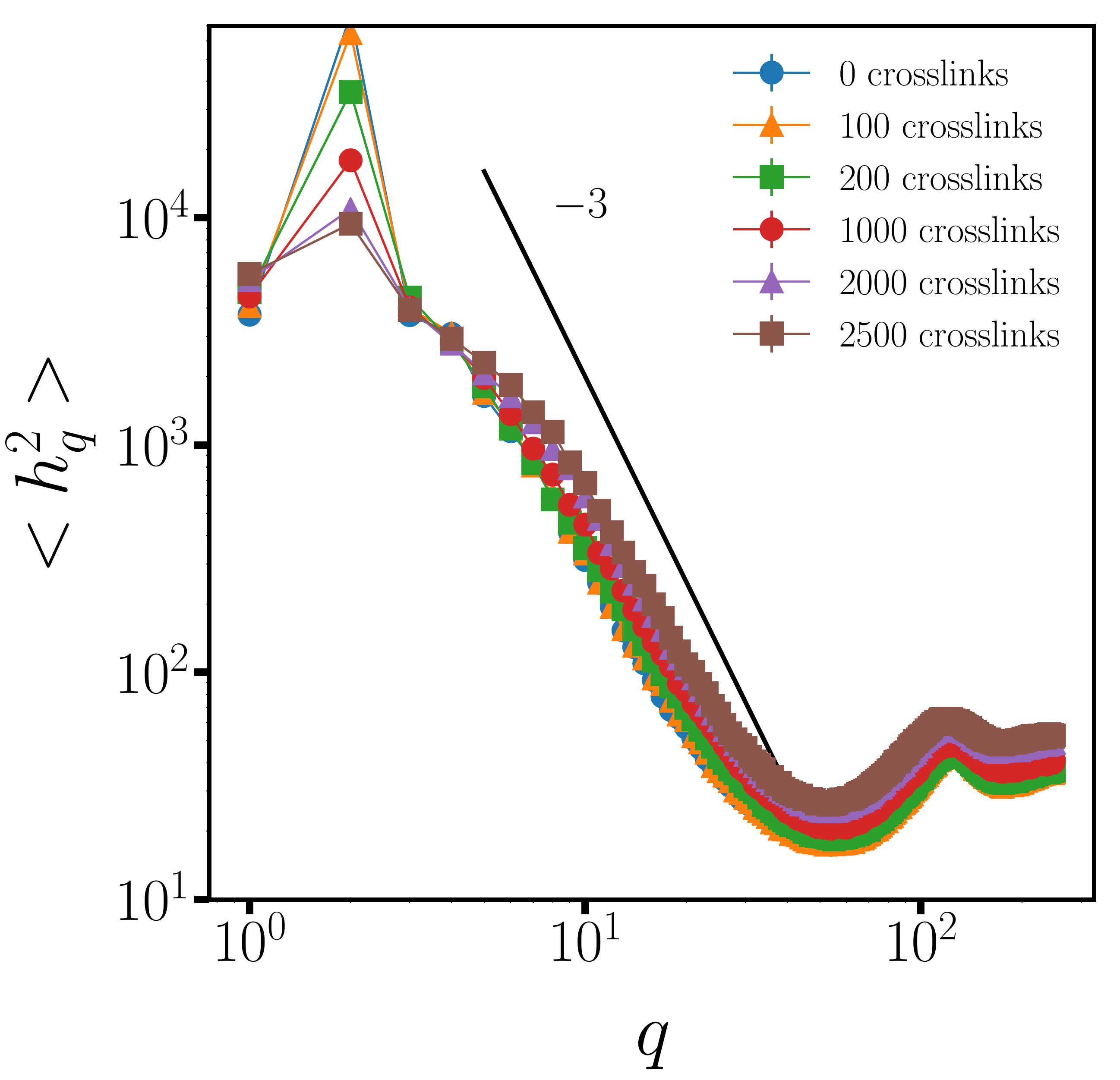}
		\includegraphics[width=0.28\linewidth]{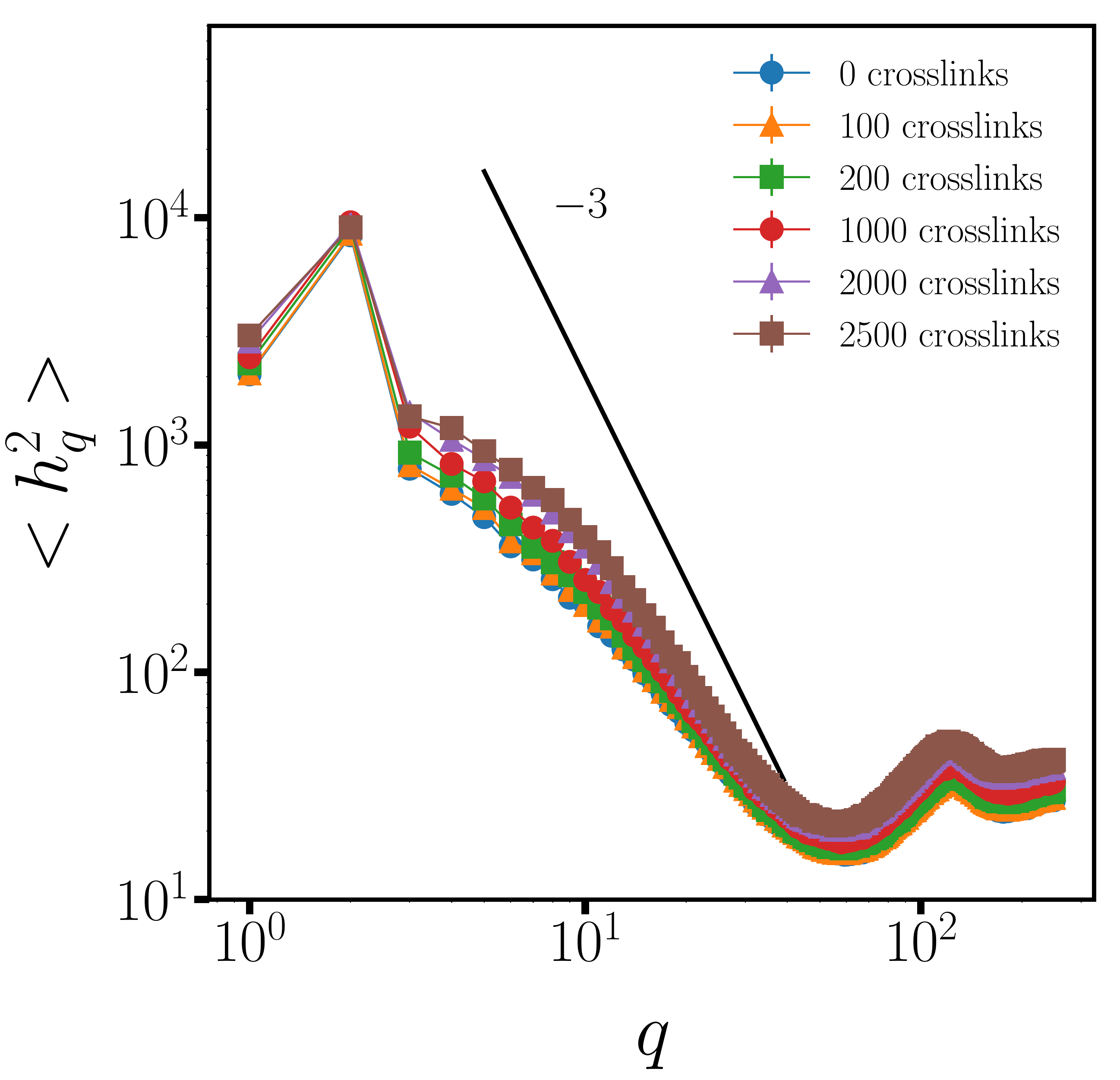}
	\end{center}
	\caption{Power spectrum of the shape fluctuations of a random slab for different boundary linkages (top row) and crosslinks (bottom row). Left column: Extensile motor case. Middle column: Contractile motor case. Right column: Passive case.}
	\label{fig:shape1}
\end{figure*}

\begin{figure*}[h]
	\begin{center}
		\includegraphics[width=0.28\linewidth]{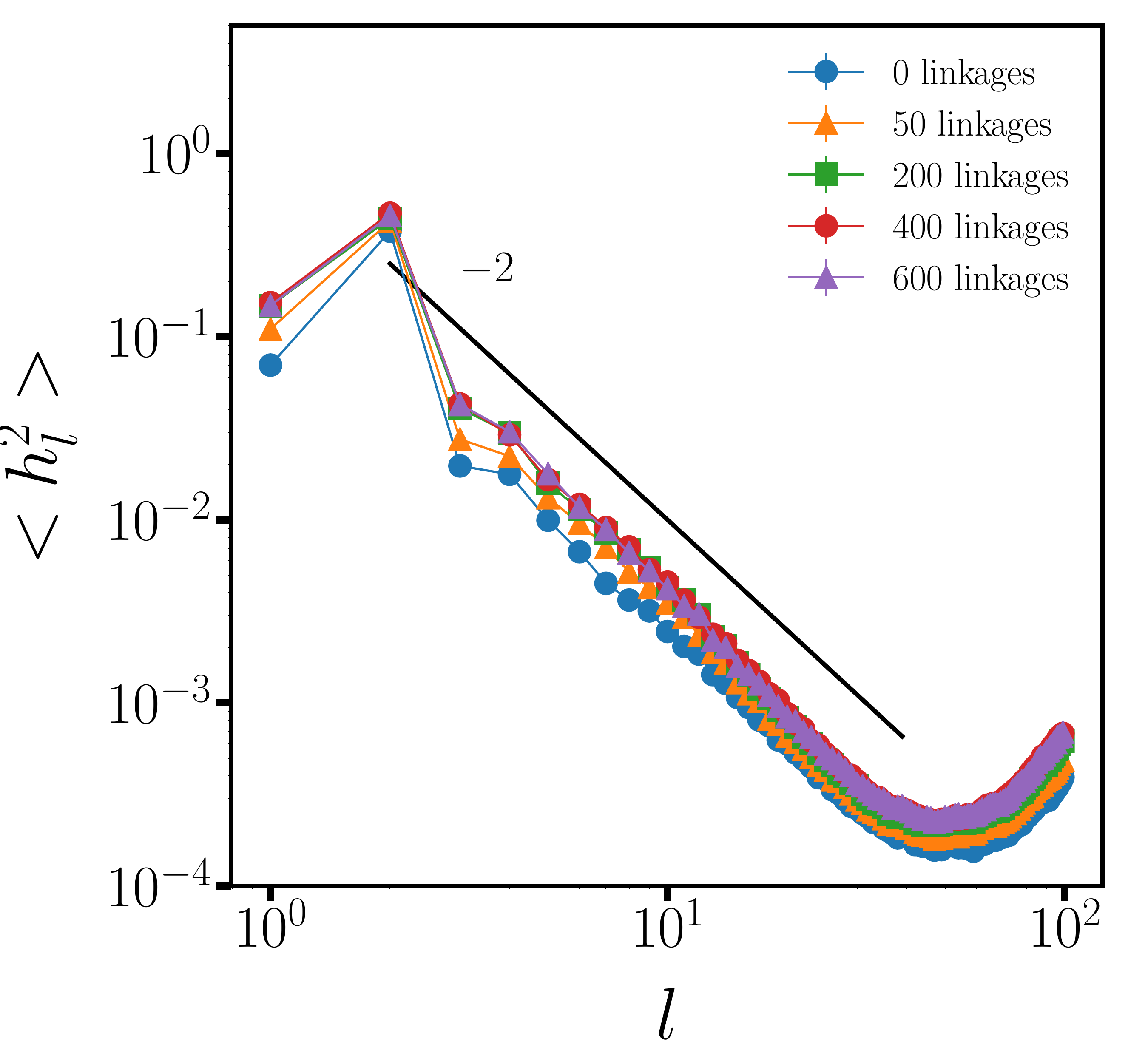}
		\includegraphics[width=0.28\linewidth]{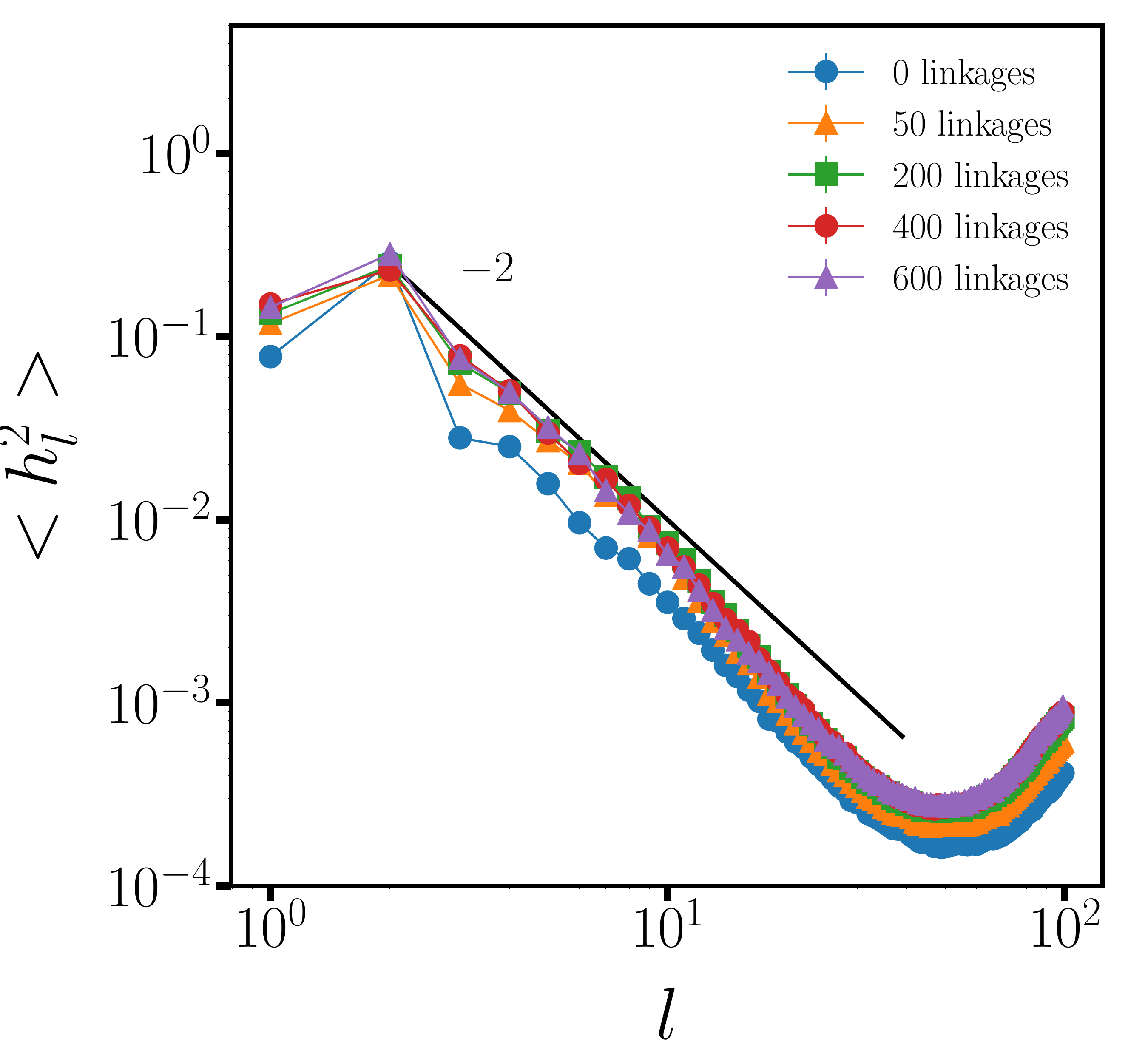}
		\includegraphics[width=0.28\linewidth]{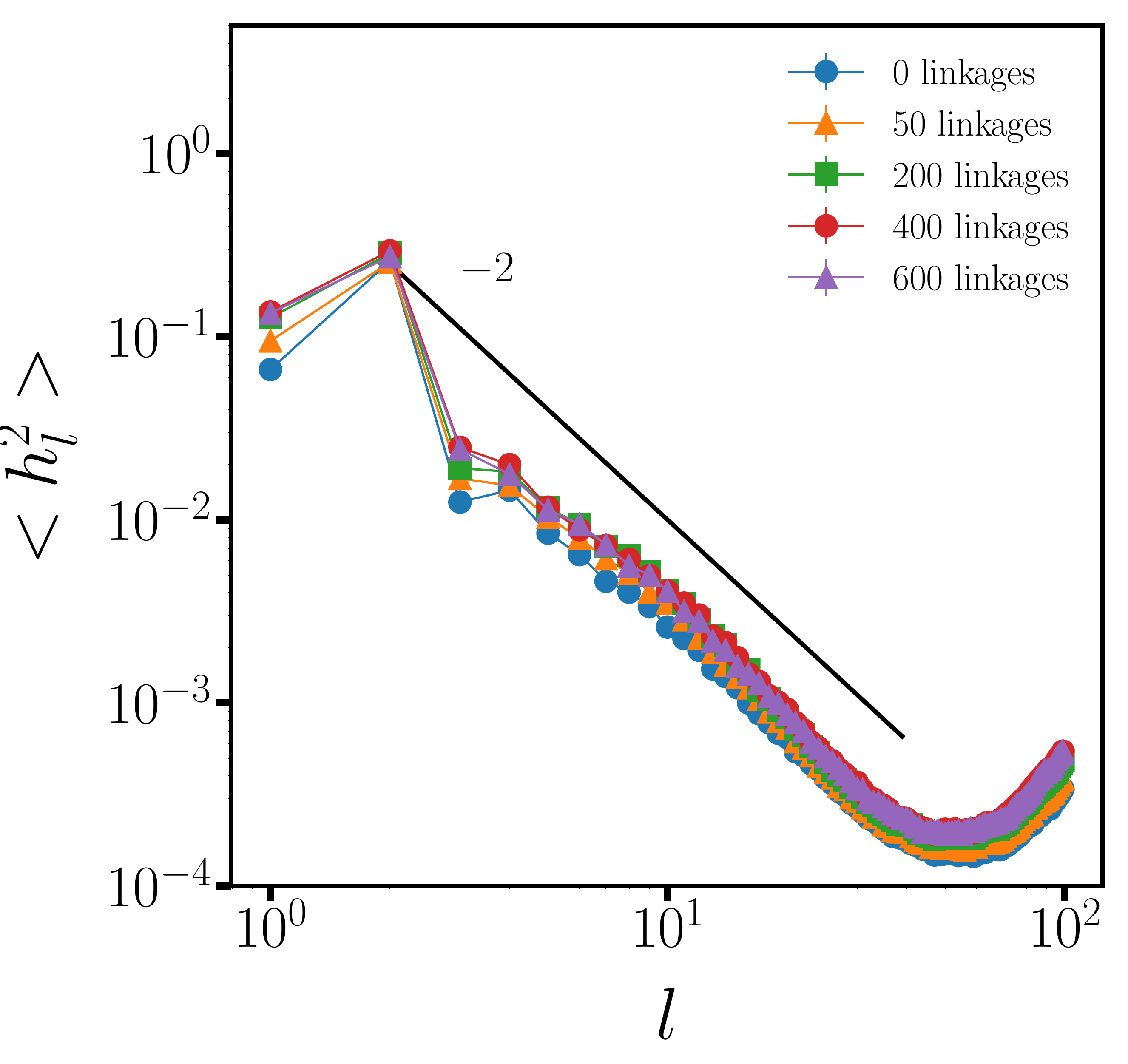}
		\includegraphics[width=0.28\linewidth]{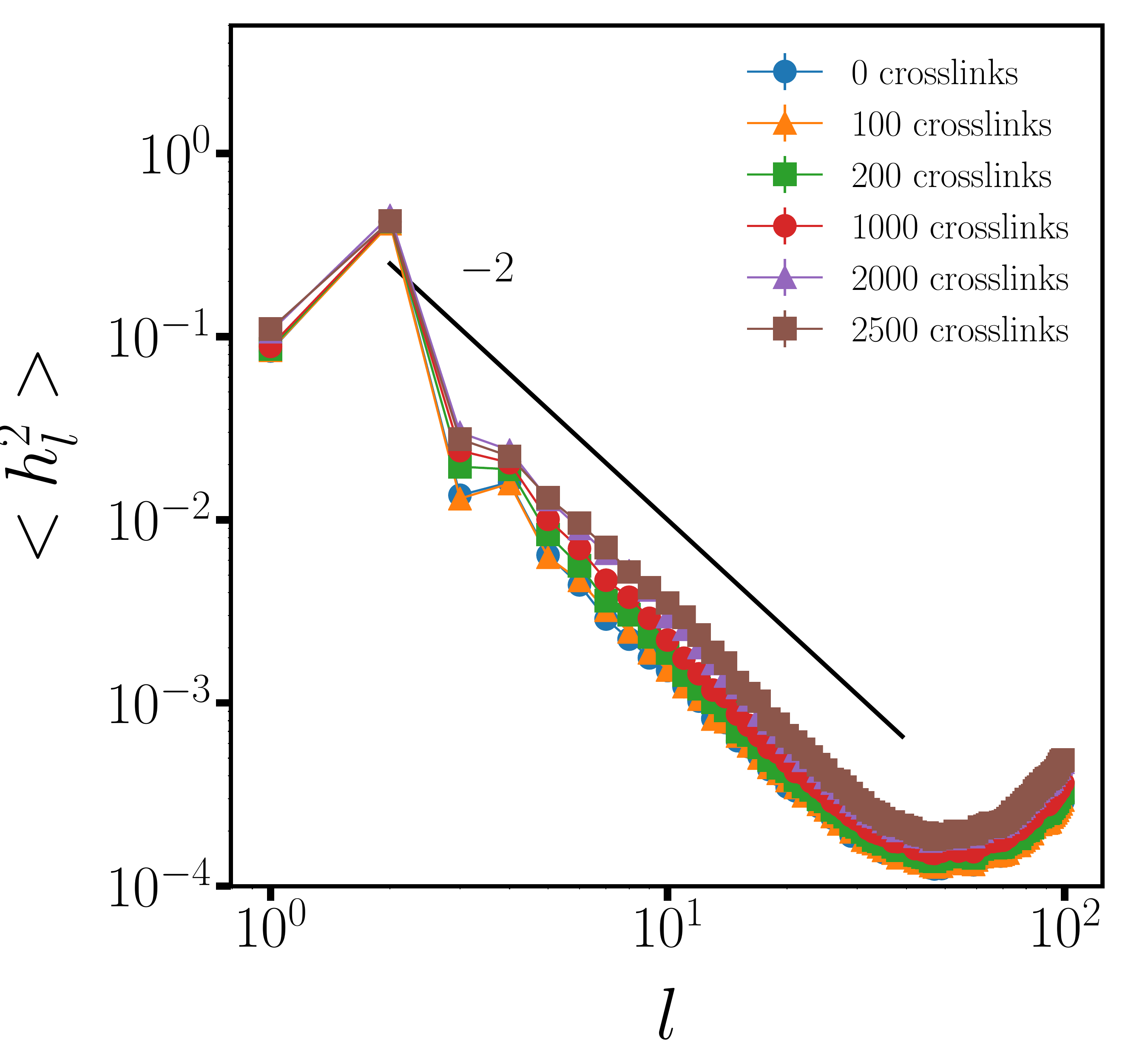}
		\includegraphics[width=0.28\linewidth]{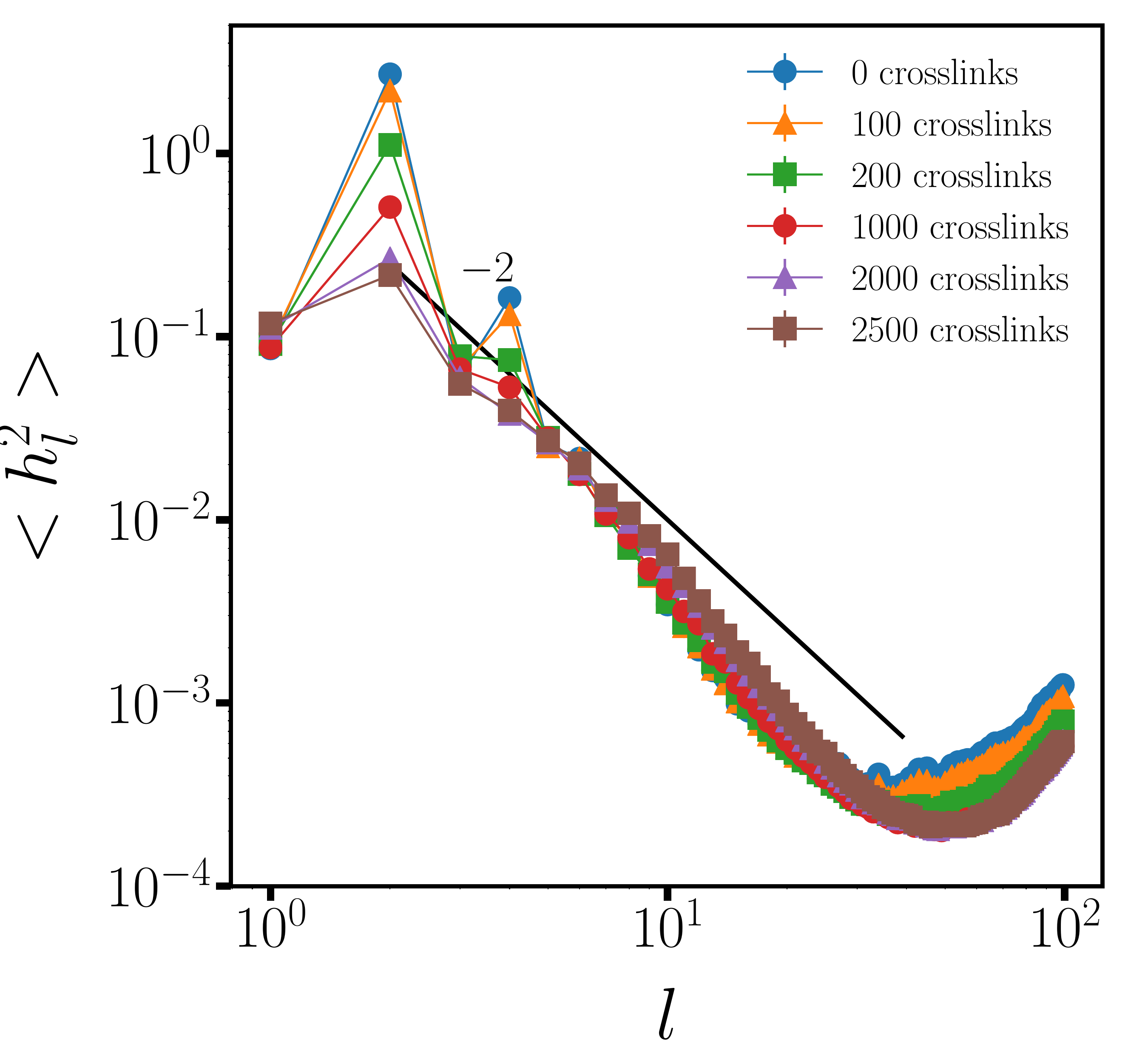}
		\includegraphics[width=0.28\linewidth]{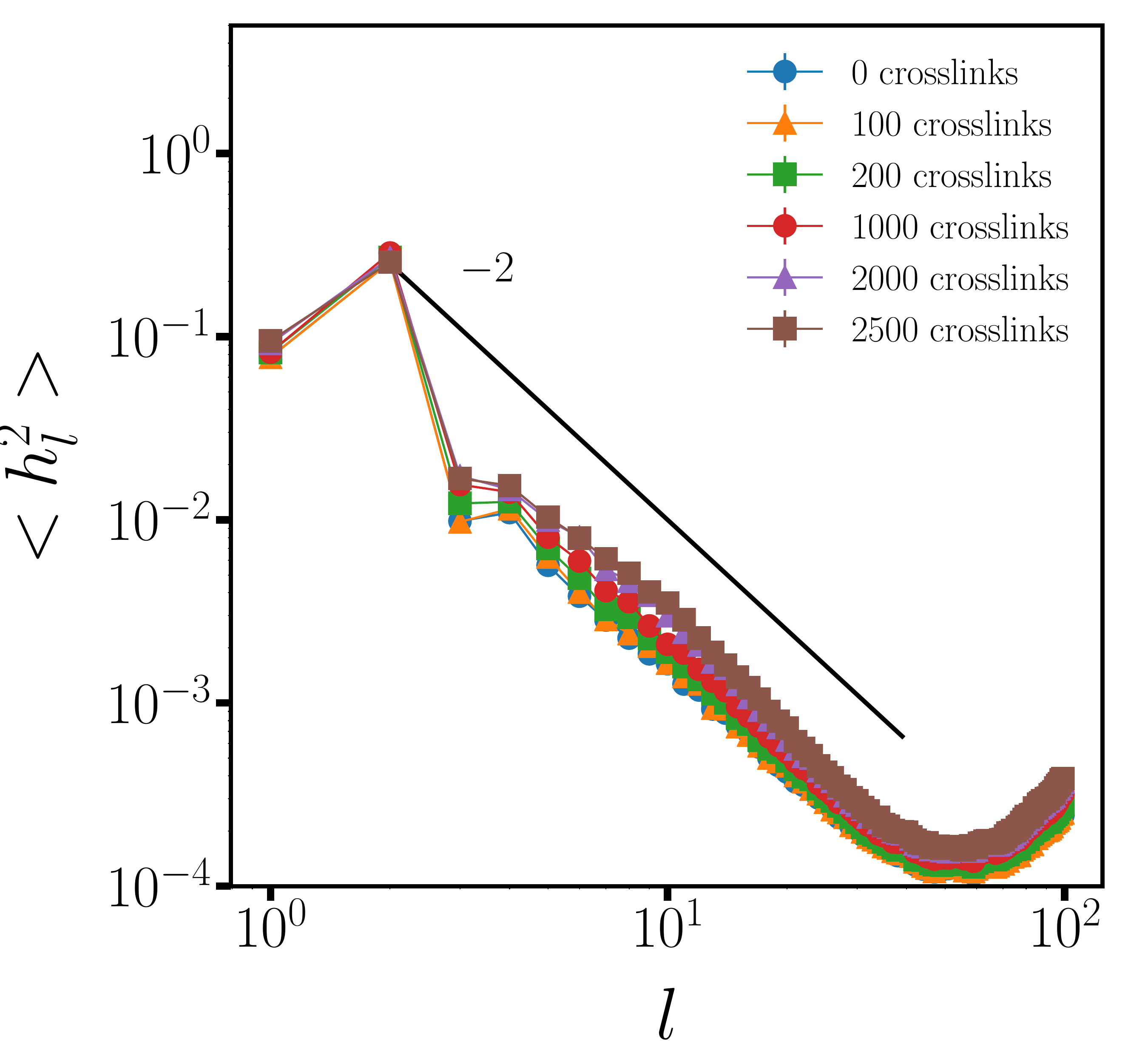}
	\end{center}
	\caption{Power spectrum of the shape fluctuations in spherical harmonics, where $l$ is the dimensionless spherical wavenumber for different chromatin-lamina linkages (top row) or crosslinks (bottom row). Left column: Extensile motor case. Middle column: Contractile motor case. Right column: Passive case.}
	\label{fig:shape2}
\end{figure*}

\begin{figure*}[h]
	\begin{center}
		\includegraphics[width=0.24\linewidth]{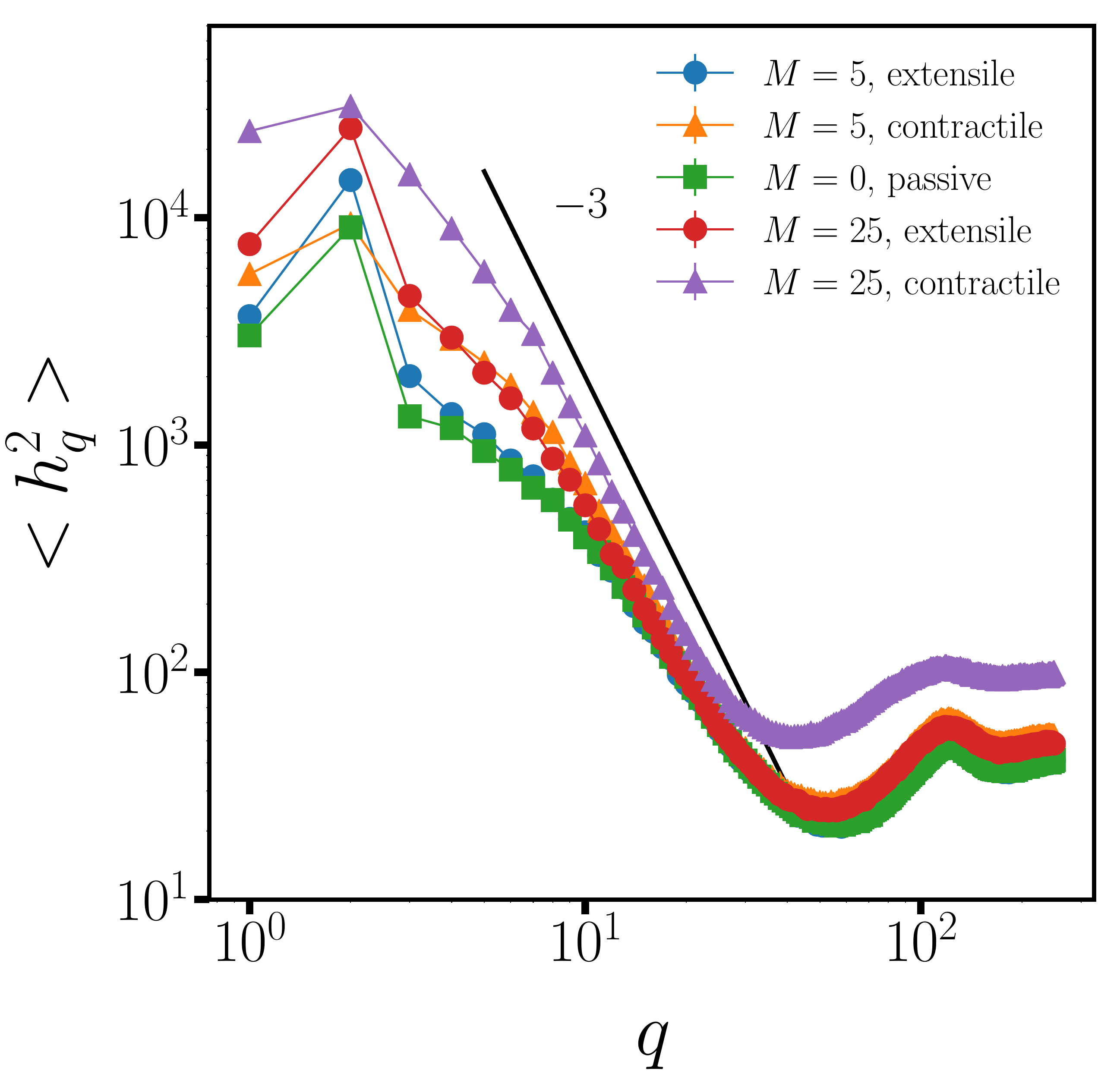}
		\includegraphics[width=0.24\linewidth]{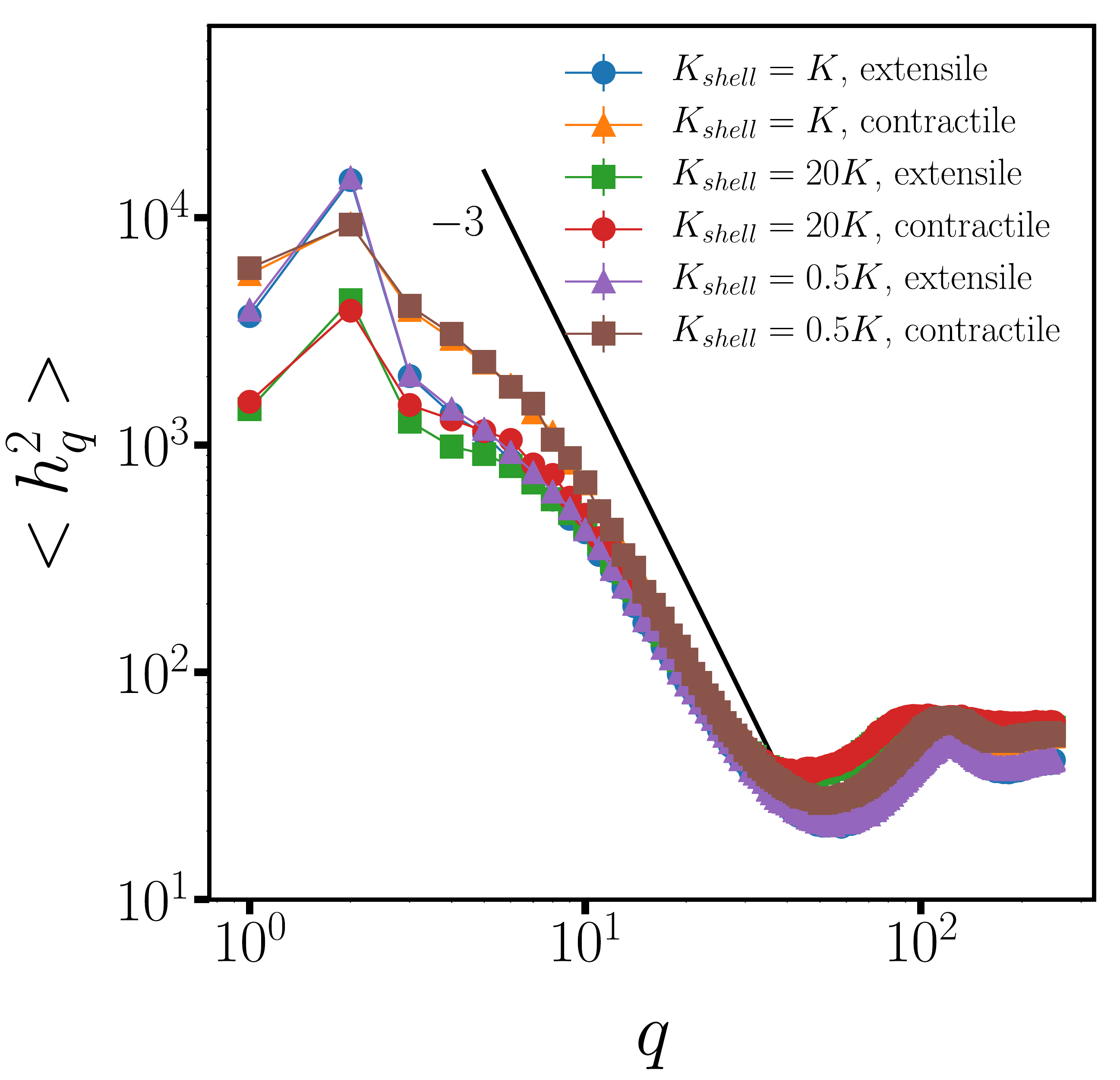}
		\includegraphics[width=0.24\linewidth]{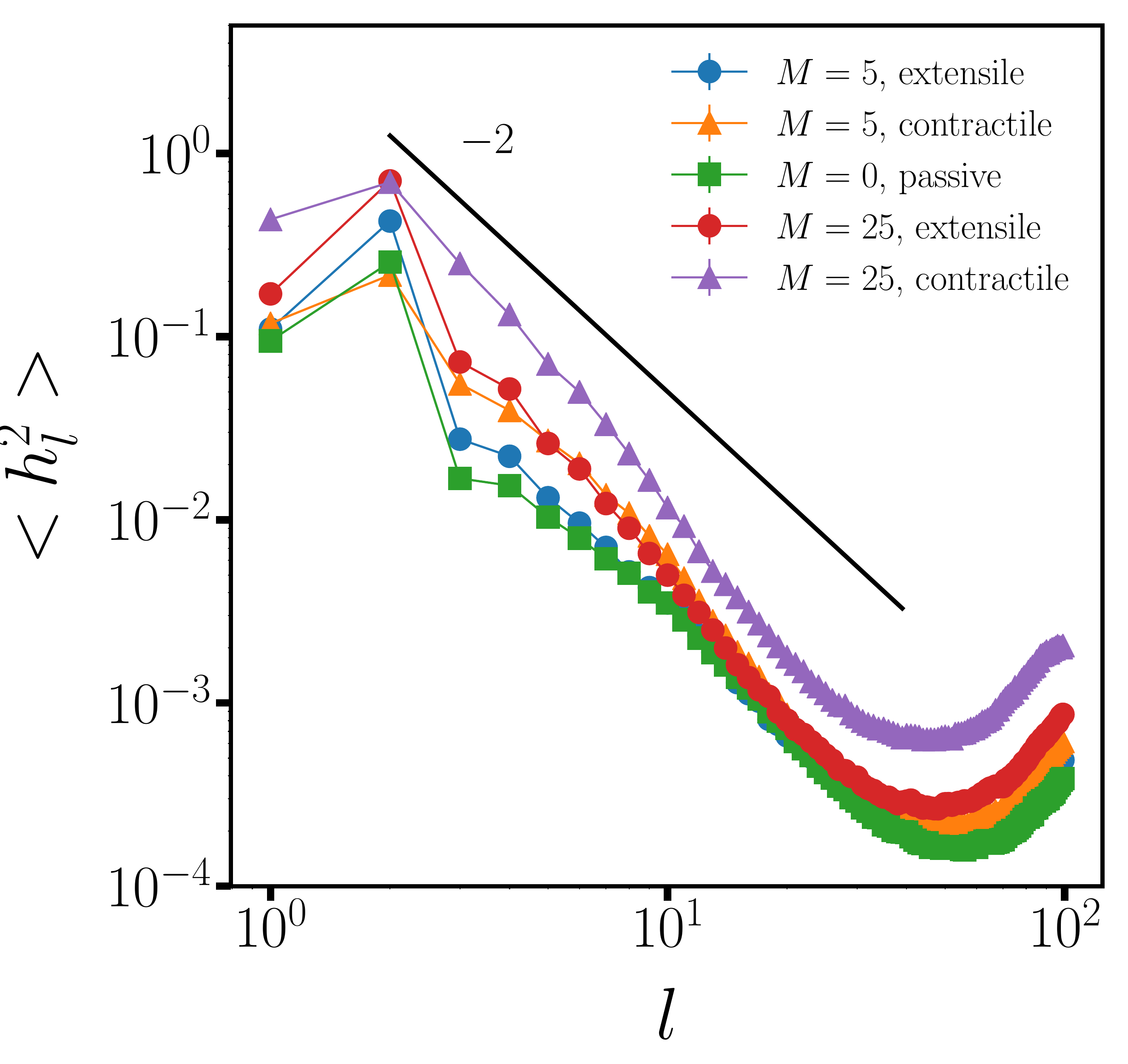}
		\includegraphics[width=0.24\linewidth]{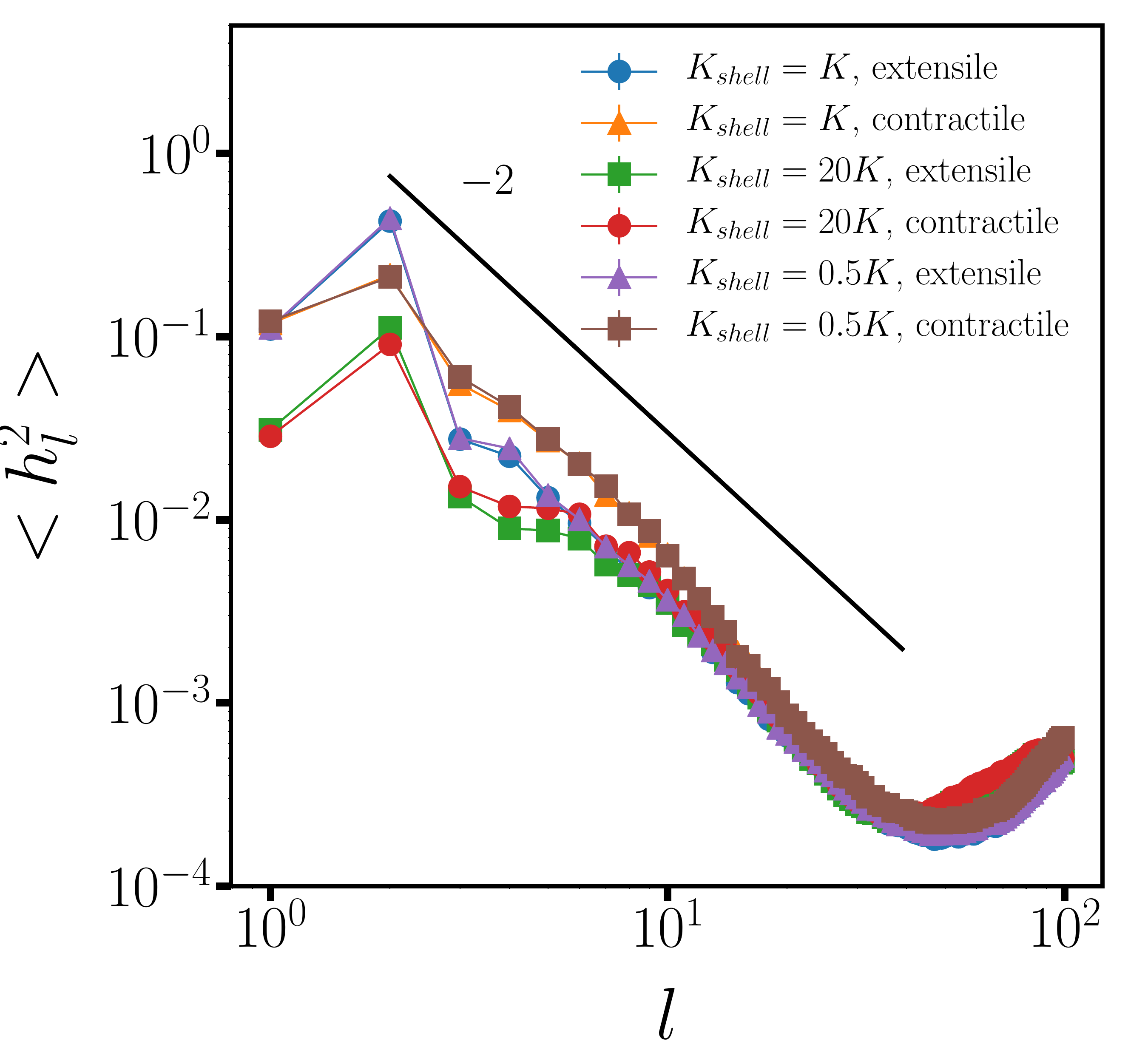}
	\end{center}
	\caption{Power spectrum of the shape fluctuations for different motor strengths or shell stiffnesses. Left two: $q$ plot. Right two: $Y_{lm}$ plot.}
	\label{fig:shape3}
\end{figure*}

\begin{figure*}[h]
	\begin{center}
		\includegraphics[width=0.24\linewidth]{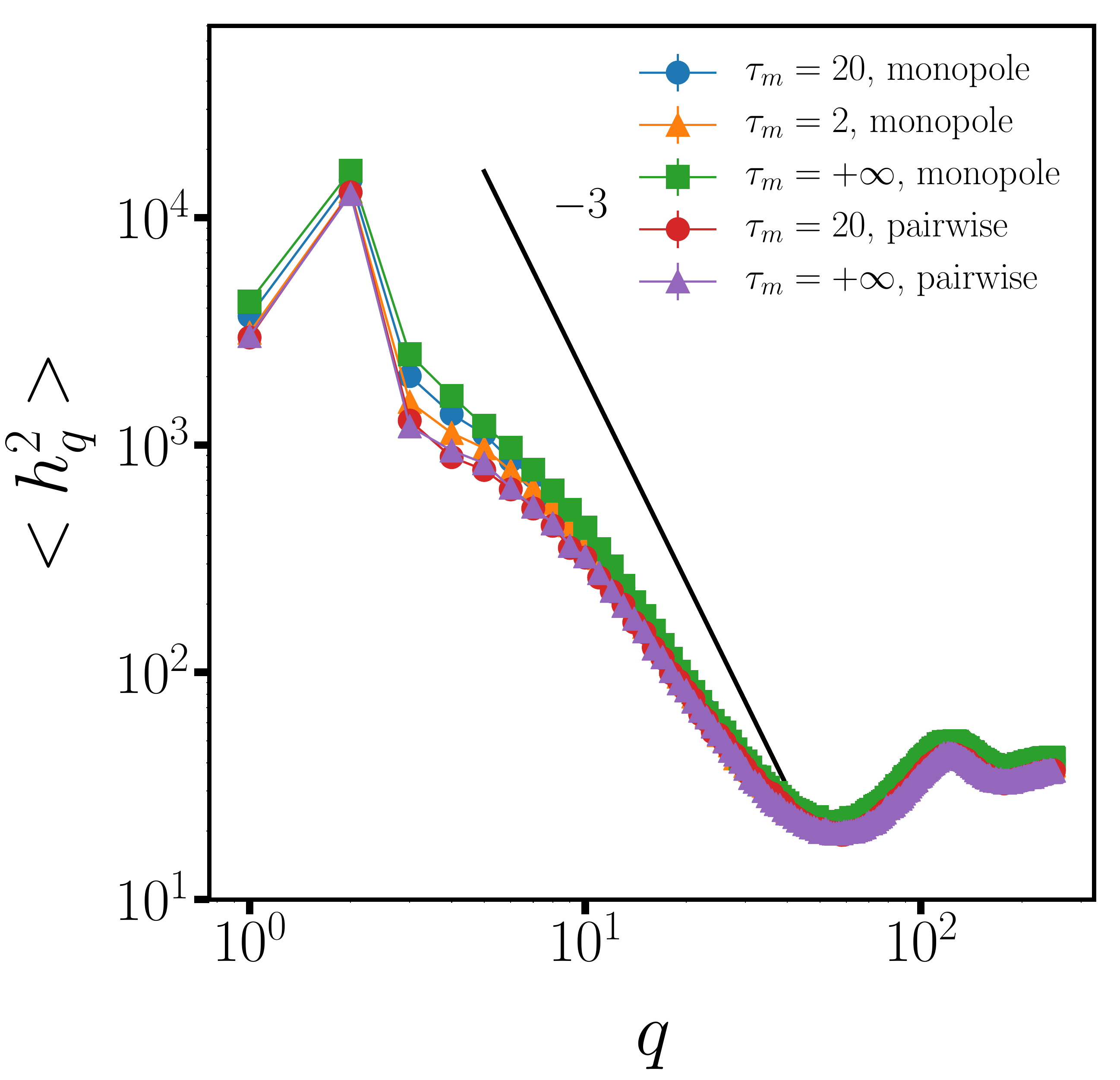}
		\includegraphics[width=0.24\linewidth]{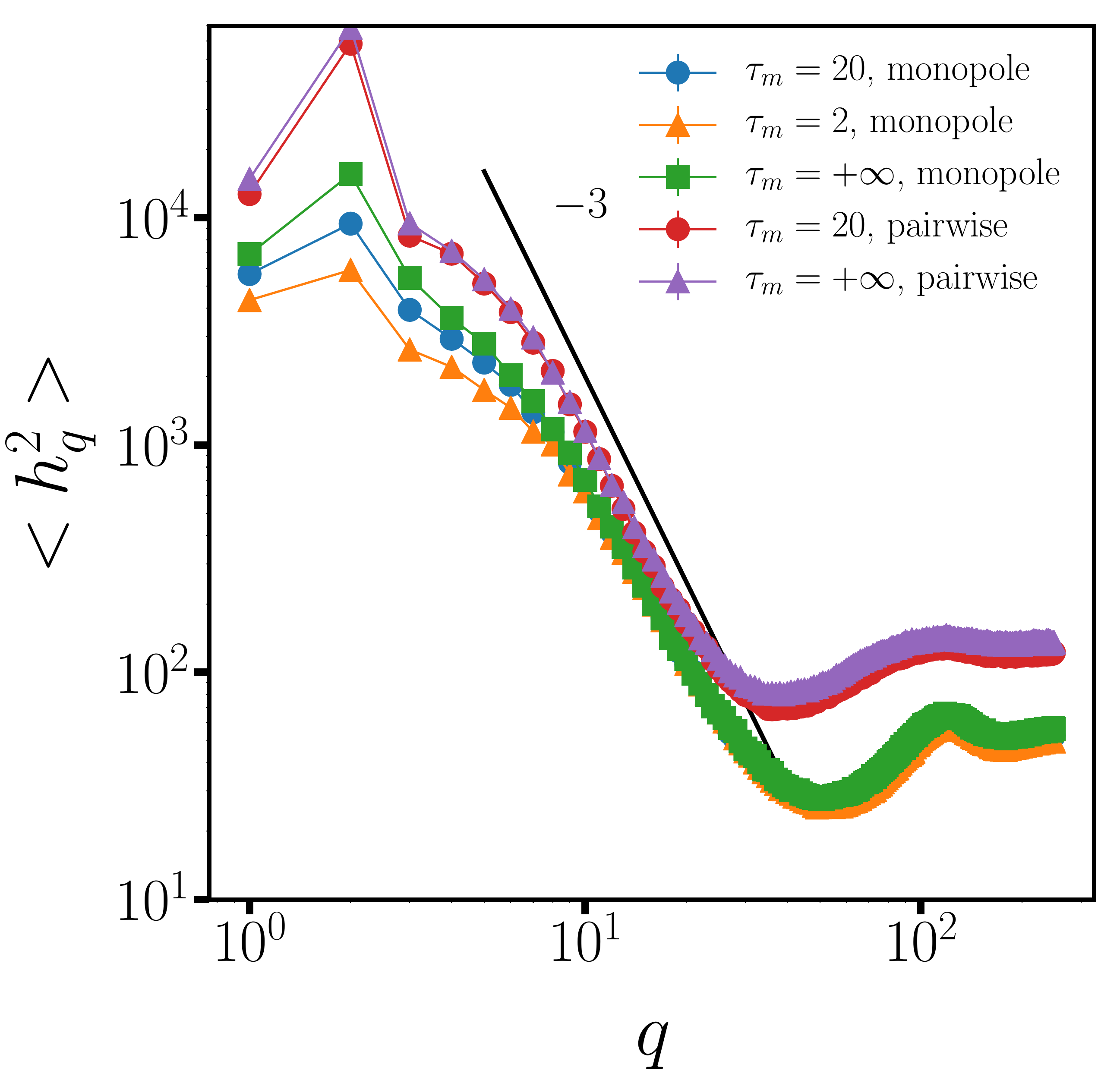}
		\includegraphics[width=0.24\linewidth]{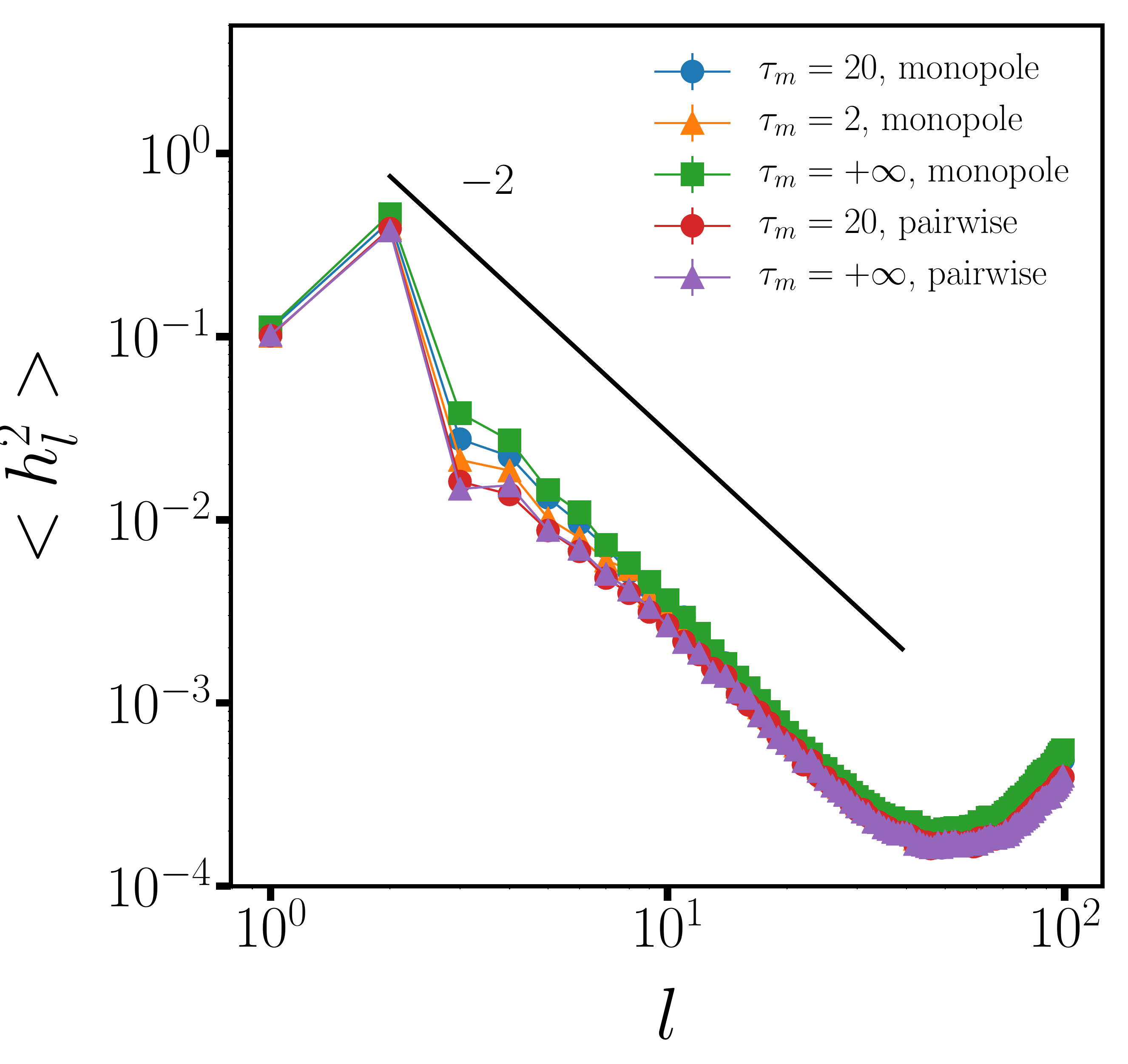}
		\includegraphics[width=0.24\linewidth]{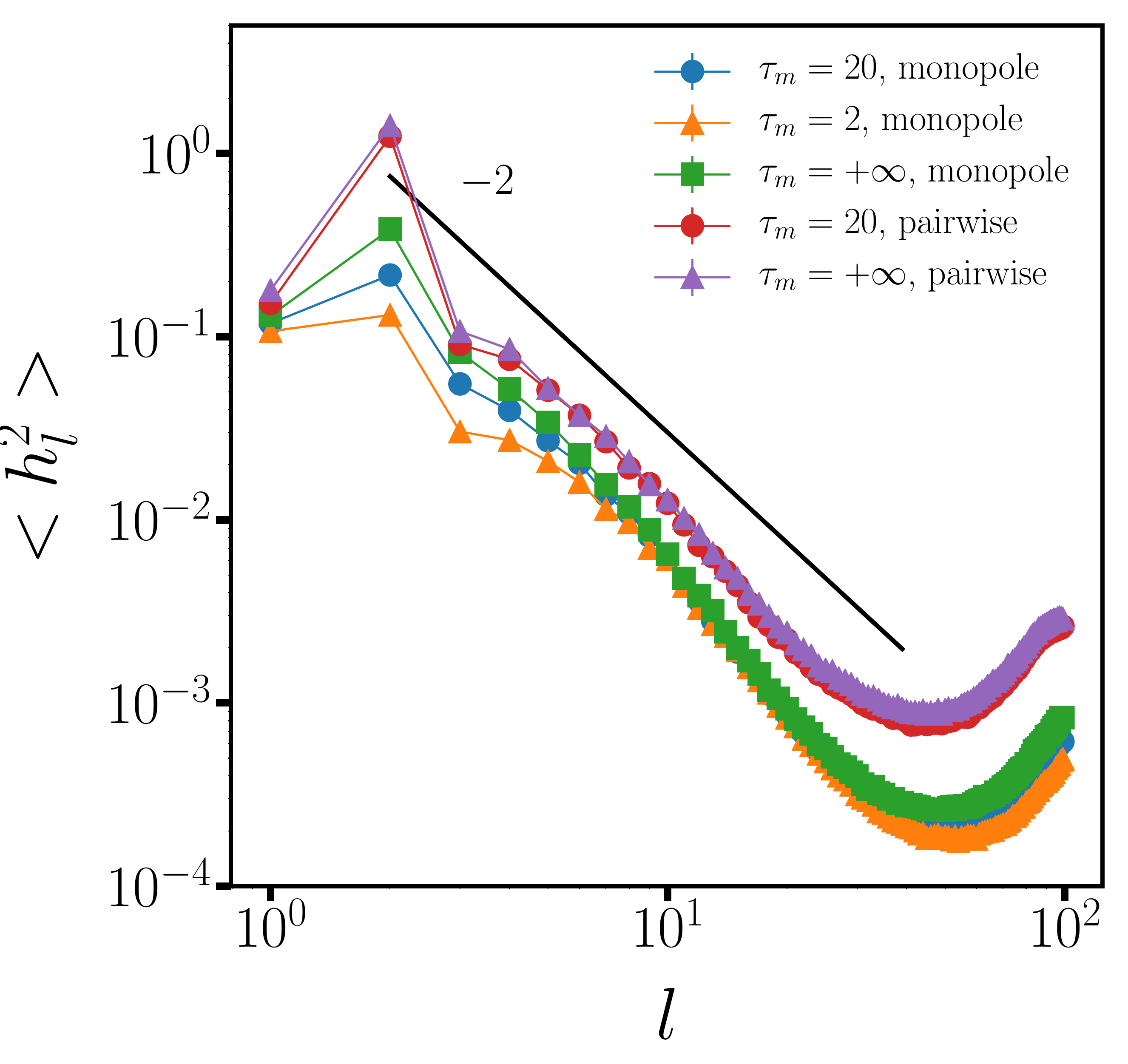}
	\end{center}
	\caption{Power spectrum of the shape fluctuations for different motor turnover times, $\tau_m$, or types of motors. From left to right: $q$ plot for extensile case; $q$ plot for contractile case; $Y_{lm}$ plot for extensile case; $Y_{lm}$ plot for contractile case. In the figure legends, ``monopole'' denotes simulations with active motor monomers that exert unreciprocated forces (as described in the main text), while ``pairwise'' denotes simulations with motor monomers that exert forces on nearby monomers that are reciprocated on the motor.}
	\label{fig:shape4}
\end{figure*}

\begin{figure*}[h]
	\begin{center}
		\includegraphics[width=0.24\linewidth]{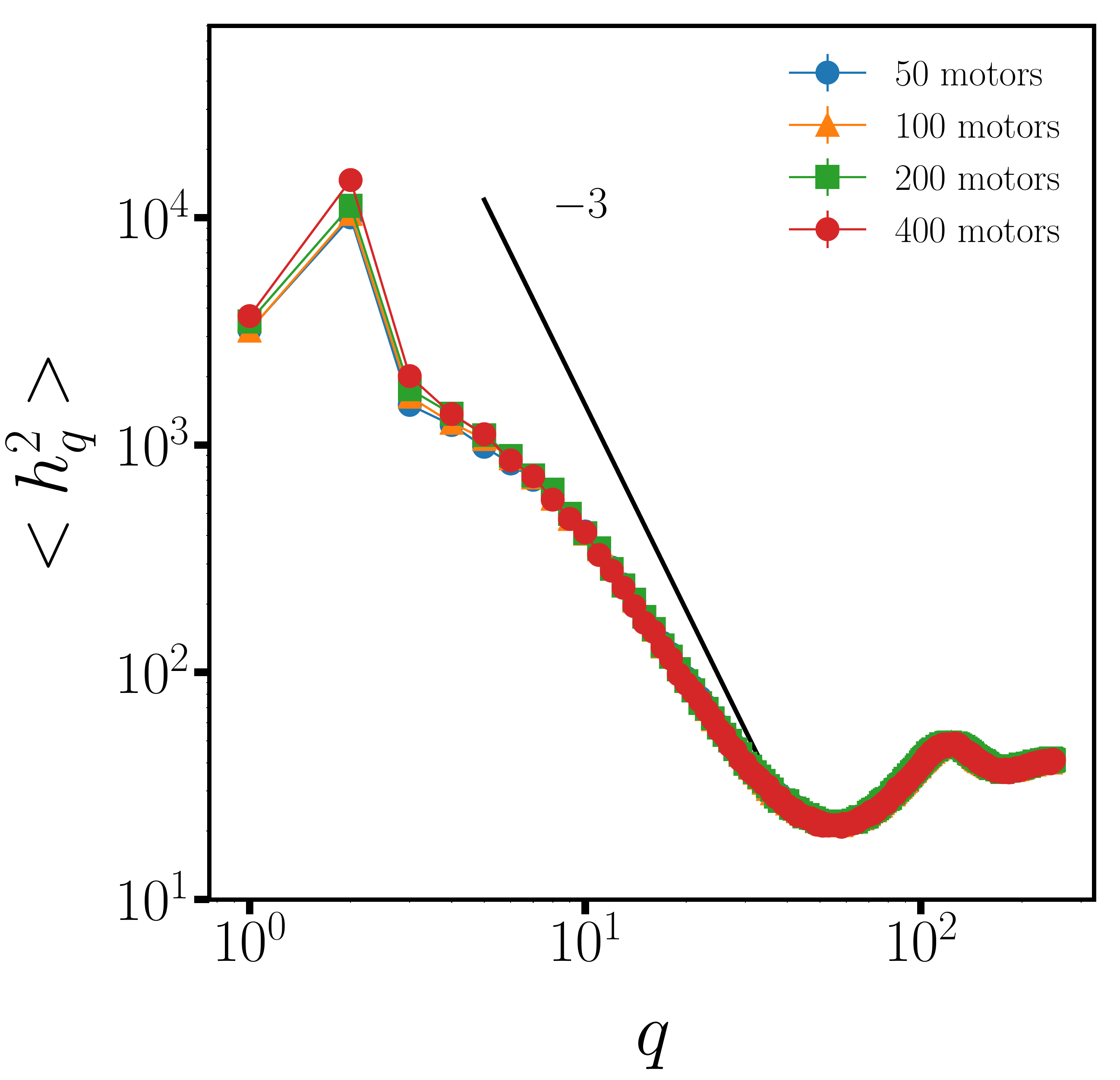}
		\includegraphics[width=0.24\linewidth]{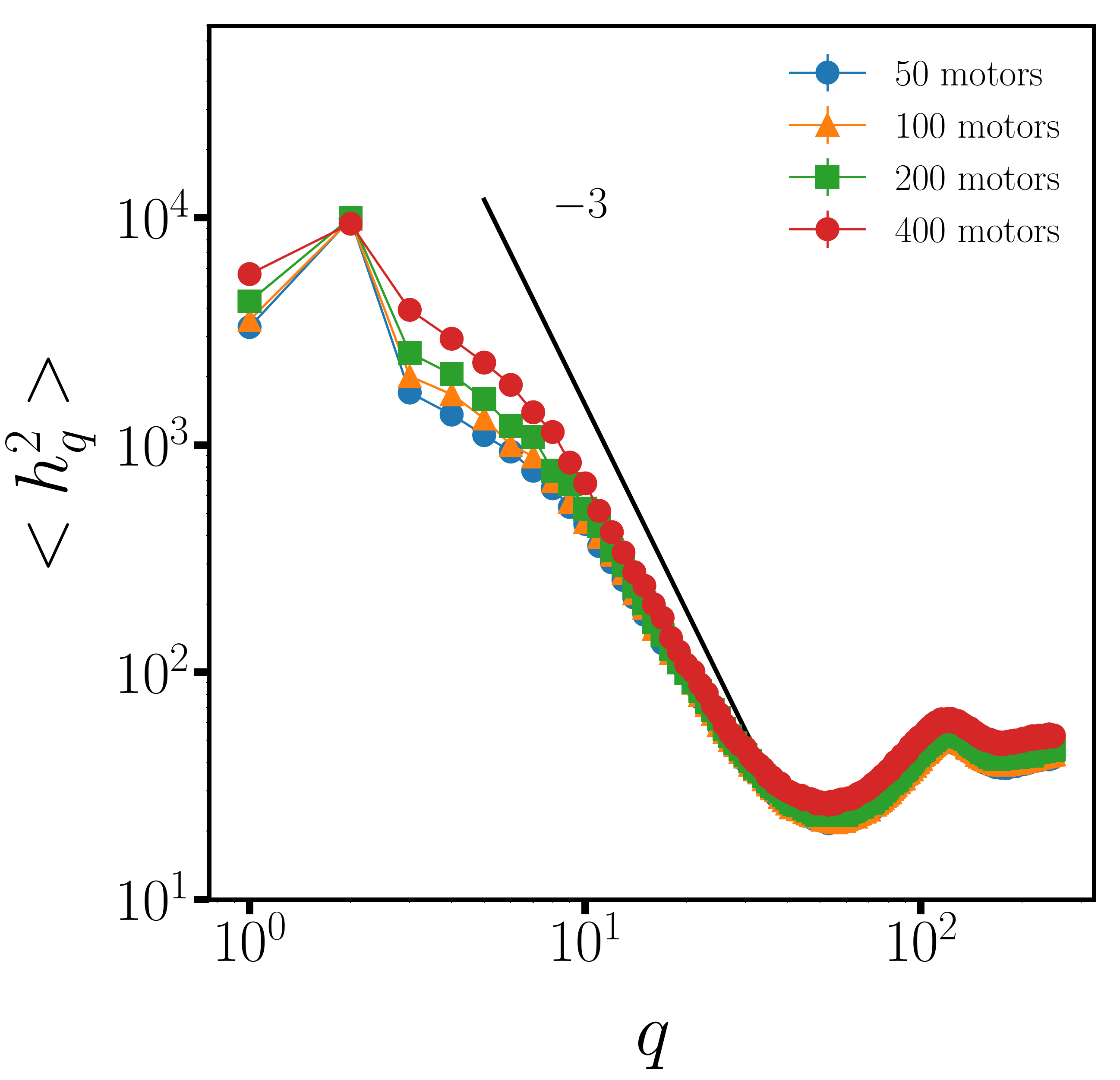}
		\includegraphics[width=0.24\linewidth]{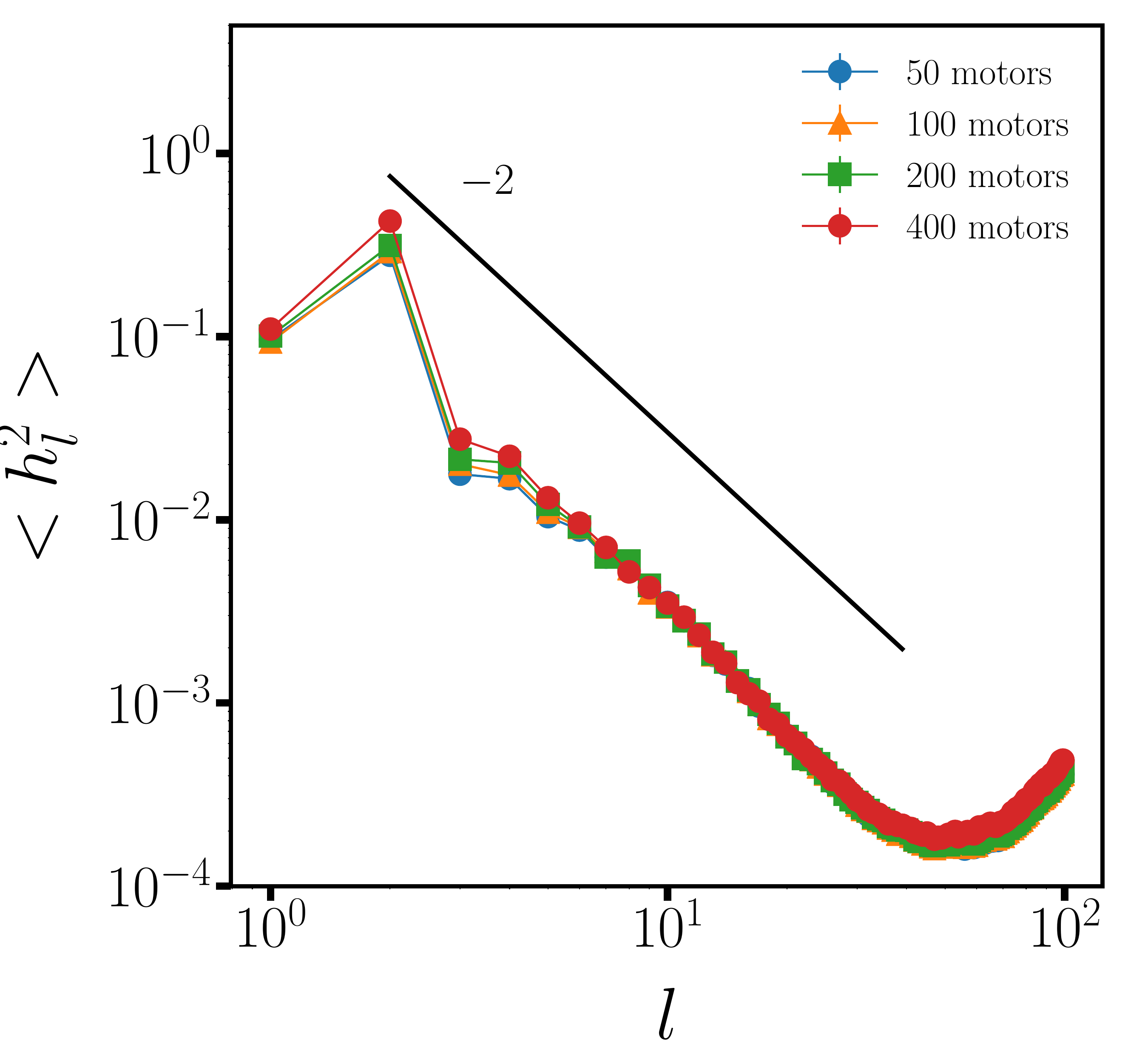}
		\includegraphics[width=0.24\linewidth]{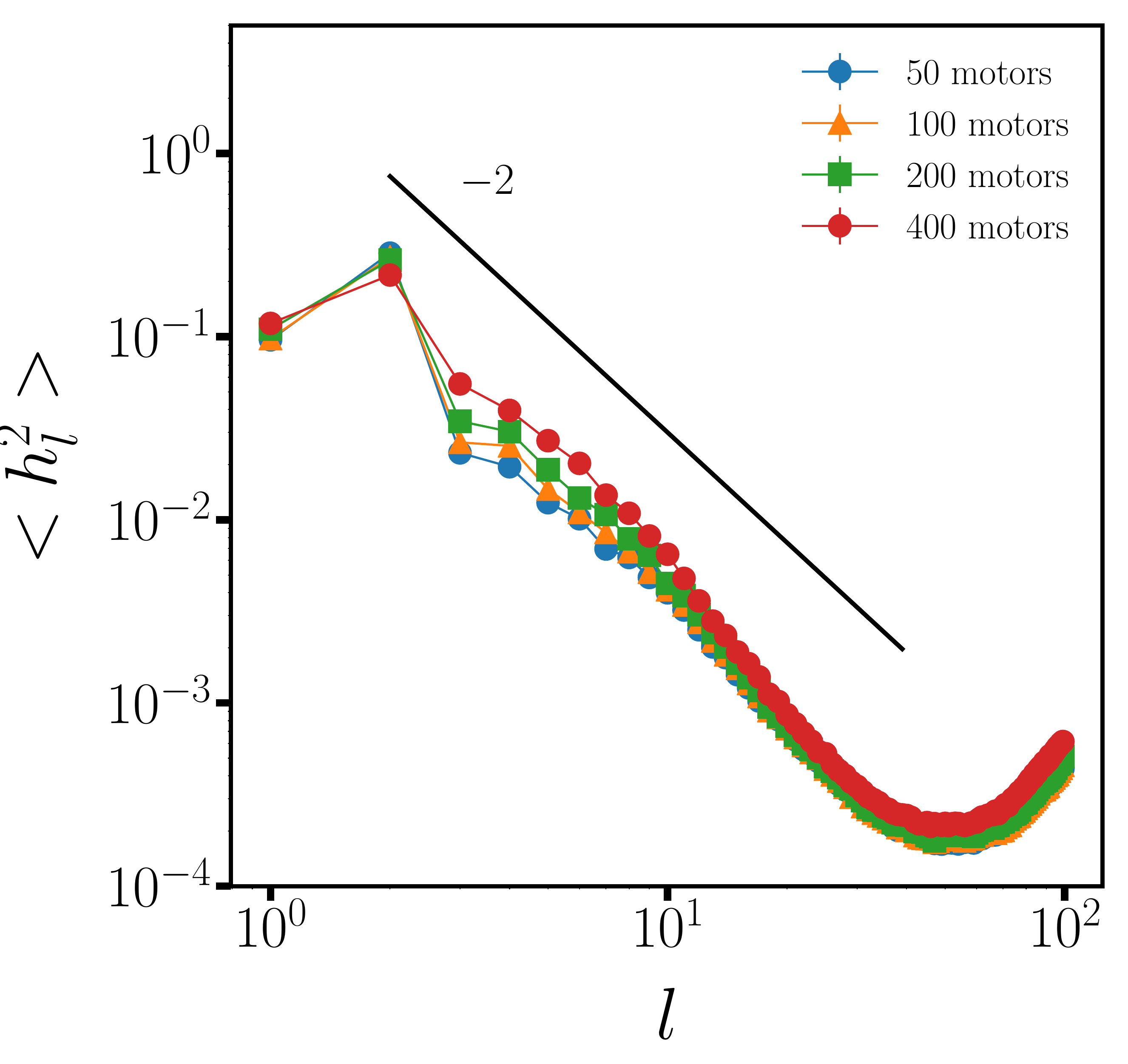}
	\end{center}
	\caption{Power spectrum of the shape fluctuations for different numbers of motors, $N_m$. From left to right: $q$ plot for extensile case; $q$ plot for contractile case; $Y_{lm}$ plot for extensile case; $Y_{lm}$ plot for contractile case.}
	\label{fig:shape5}
\end{figure*}

\section{Experiments}
To measure nuclear shape fluctuations in live cells, 
the wild-type mouse embryonic fibroblasts (MEFs) were kindly provided
by J. Eriksson, Abo Akademi University, Turku, Finland. Cells were
cultured in DMEM with 25 mM Hepes and sodium pyruvate supplemented
with 10\% FBS, 1\% penicillin/streptomycin, and nonessential amino
acids. The cell cultures were maintained at 37 degrees C and 5\% $\mathrm{CO_2}$. 

Cell nuclei were fluorescently labeled by transient transfection with pEGFP-C1-NLS, 48 h before imaging. Cell nuclei were imaged at 2-min increments for 2 h by using wide-field fluorescence with a $40\times$ objective. 
To quantify the structural features of nuclei, we traced the contour, $r(\theta)$, of the NLS-GFP labeled nuclei at each time point. The shape of the nucleus was identified using a custom-written Python script, and its contour was interpolated from 0 to 2$\pi$ by 150 points. Next, the shape fluctuations were calculated as $h(\theta) = r(\theta)-r_0$, where $r_0$ is the average radius for each cell at each time point. The wave number-dependent Fourier modes of the fluctuations, $h_q$, were obtained as Fourier transformation coefficients, as described in Ref~\cite{patteson19}.

The shape fluctuations were quantified for each cell by computing the Fourier mode magnitude square $h^2(q)$ and averaging over each time point. The average shape fluctuations as shown in Fig. 4 in the main text was taken as the average over 15 cells per condition from two independent experiments.

\end{widetext}
\end{document}